\documentclass{aa}  

\usepackage{natbib}
\bibpunct{(}{)}{;}{a}{}{,} 
\usepackage{graphicx}
\usepackage{txfonts}
\begin{document}

   \title{Stellar Population gradients in galaxy discs from the CALIFA survey}

   \subtitle{The influence of bars}

   \author{P. S\'anchez-Bl\' azquez \inst{1}, F. Rosales-Ortega\inst{2,1},
           J. M\'endez-Abreu\inst{3}, I. P\'erez\inst{4}, S.F. S\'anchez\inst{5,6,7},
           S. Zibetti\inst{8},
           A. Aguerri\inst{9}, J. Bland-Hawthorn\inst{10}, C. Catal\'an\inst{11}, R. Cid Fernandes\inst{12},
           de Amorim, A.\inst{6}, A. de Lorenzo-Caceres\inst{3}, 
           J. Falc\'on-Barroso\inst{9}, A. Galazzi\inst{8},  R. Garc\'{\i}a Benito\inst{6}, A. Gil de Paz\inst{10}, 
           R. Gonz\'alez Delgado\inst{7}, B. Husemann\inst{7}, Jorge Iglesias-P\'aramo\inst{5},
           B. Jungwiert\inst{13}, R.A. Marino\inst{11}, I. M\'arquez\inst{6}, D. Mast\inst{14},
           M.A. Mendoza\inst{7}, M. Moll\'a\inst{15}, P. Papaderos\inst{16},
           T. Ruiz-Lara\inst{4}, G. van de Ven\inst{17}, C.J. Walcher\inst{18}, L. Wisotzki\inst{18}}

   \institute{Departamento de F\'{\i}sica Te\'orica, Universidad Aut\'onoma de Madrid, Cantoblanco, E28049, Spain
              \and Instituto Nacional de Astrof\'{\i}sica, \'Optica y Electr\'onica, 72840, Tonantzintla, 
              Puebla, M\'exico
               \and School of Physics \& Astronomy, University of St Andrews, United Kingdom
               \and Departamento de F\'{\i}sica y del Cosmos, Universidad de Granada, Granada, E18071, Spain
               \and Centro Astron\'omico Hispano Alem\'an de Calar Alto (CSIC-MPIA), Almer\'{\i}a, E4004, Spain
               \and Instituto de Astronom\'\i a,Universidad Nacional Auton\'oma de Mexico, A.P. 70-264, 04510, M\'exico DF
               \and Istituto de Astrof\'{\i}sica de Andaluc\'{\i}a (IAA-CSIC), Granada, E18080, Spain
               \and INAF-Osservatorio Astrofisico di Arcetri - Largo Enrico Fermi 5 - I-50125 Firenze, Italy 
               \and Instituto de Astrof\'{\i}sica de Canarias, La Laguna, E38205, Spain
               \and Institute of Astronomy, School of Physics, University of Sydney, NSW 2006, Australia
               \and Departamento de Astrof\'{\i}sica y Ciencias de la Atm\'osfera, Universidad Complutense de Madrid,
               Madrid, E28040, Spain
               \and Departamento de F\'{\i}sica, Universidade Federal de Santa Catarina, PO Box476, 88040-900,
               Florian\'opolis, SC, Brazil
               \and Astronomical Institute, Academy of Science of the Czech Republic, Bocn\'{\i} 1401/1a, 14100 Prague, Czech Republic
               \and Instituto de Cosmologia, Relatividade e Astrof\'{i}sica  ICRA, Centro Brasileiro de Pesquisas F\'{i}sicas, Rua Dr.Xavier
               Sigaud 150, CEP 22290-180, Rio de Janeiro, RJ, Brazil
               \and Departamento de Investigaci\'on B'asica, CIEMAT, Avd Complutense 40, 28040, Madrid, Spain
               \and Centro de Astrof\'{\i}sica da Universidade do Porto, Rua das Estrelas, 4150-762, Porto, Portugal
               \and Max-Planck-Institut f\"ur Astronomie, 69117, Heidelberg, Germany
               \and Leibniz-Institut f\"ur Astrophysik Potsdam (AIP), An der Sternwarte 16, 14482, Postdam, Germany
               }

   \date{Received; accepted }
\titlerunning{Stellar populations in galaxy discs}
\authorrunning{S\'anchez-Bl\'azquez et al.}
 
  \abstract
   {
 While studies of gas-phase metallicity gradients in disc galaxies are common, very little has been done in the
 acquisition of stellar abundance gradients in the same regions. We present here a comparative study of the stellar
 metallicity and age distributions in a sample of 62 nearly face-on, spiral galaxies with and without bars, using
 data from the CALIFA survey.  We measure the slopes of the gradients and study their relation with other properties
 of the galaxies. We find that the mean stellar age and metallicity gradients in the disc are shallow and negative.
 Furthermore, when normalized to the effective radius of the disc, the slope of the stellar population gradients does
 not correlate with the mass or with the morphological type of the galaxies. Contrary to this, the values of both age
 and metallicity at $\sim$2.5 scale-lengths correlate with the central velocity dispersion in a similar manner to the
 central values of the bulges, although bulges show, on average, older ages and higher metallicities than the discs.
 One of the goals of the present paper is to test the theoretical prediction that non-linear coupling between the bar
 and the spiral arms is an efficient mechanism for producing radial migrations across significant distances within
 discs. The process of radial migration should flatten the stellar metallicity gradient with time and, therefore, we
 would expect flatter stellar metallicity gradients in barred galaxies. However, we do not find any difference in the
 metallicity or age gradients in galaxies with without bars.
  We discuss possible scenarios that can lead to this absence of difference.
 }


   \maketitle
%

\section{Introduction}
\label{sec:introduction}

Spatially resolved stellar population studies in the disc region of spiral galaxies are sparse, with exception 
of our own galaxy (see Freeman \& Bland-Hawthorn 2002\nocite{2002ARA&A..40..487F} for a review; Friel et al. 
2002\nocite{2002AJ....124.2693F}; Yong et al. 2006\nocite{2006AJ....131.2256Y}; Carraro et al. 
2007\nocite{2007A&A...476..217C}) and for some nearby galaxies, such as M33 (Monteverde et al. 
1997\nocite{1997ApJ...474L.107M}; Barker et al. 2007\nocite{2007AJ....133.1138B}; Williams et al. 
2009\nocite{2009ApJ...695L..15W}; Cioni 2009\nocite{2009A&A...506.1137C}), M100 (Beauchamp \& Hardy 
1997)\nocite{1997AJ....113.1666B}; M81 (Hughes et al. 1994\nocite{1994ApJ...428..143H}; Davidge 
2006a\nocite{2006PASP..118.1626D}); NGC2403 (Davidge 2007\nocite{2007ApJ...664..820D}); NGC 300 (Vlaji\'c, 
Bland-Hawthorn \& Freeman 2009\nocite{2009ApJ...697..361V}; Kudritzki et al 2008\nocite{2008ApJ...681..269K}; 
Urbaneja et al. 2005\nocite{2005ApJ...622..862U}; Gogarten et al. 2010\nocite{2010ApJ...712..858G}) or M31 
(Worthey et al. 2005\nocite{2005ApJ...631..820W}). For disc galaxies outside the Local Group, stellar 
population gradients have been mainly investigated using colors (e.g., de Jong 
1996\nocite{1996A&A...313..377D}; Peletier \& Balcells 1996\nocite{1996AJ....111.2238P}; Jansen et al. 
2000\nocite{2000ApJS..126..331J}; Bell \& de Jong 2000\nocite{2000MNRAS.312..497B}; MacArthur et al. 
2004\nocite{2004ApJS..152..175M}; Taylor et al. 2005\nocite{2005ApJ...630..784T}; Mu\~noz-Mateos et al. 2007, 
2009\nocite{2009ApJ...703.1569M}\nocite{2007ApJ...658.1006M}; Prochaska et al. 
2010\nocite{2010ApJ...718..392P}). These studies found that disc galaxies tend to be bluer in the external 
parts. This trend has been interpreted as the consequence of a population older and more metal rich in the 
centre compared with that on the external parts. However, there are large discrepancies in the magnitude of 
the stellar population gradients derived by different authors. This is because it is extremely difficult to 
disentangle the effects of age, metallicity and dust extinction (on average $\sim$1 mag in the central 
regions, Ganda et al. 2009\nocite{2009MNRAS.395.1669G}) using only colours. Spectroscopic studies may help to 
alleviate the associated degeneracies, but the low surface brightness of the disc region and the nebular 
emission lines filling some of the most important age-diagnostic absorption lines make the analysis very 
difficult. This is the reason why studies of stellar populations in disc galaxies using spectroscopy are 
scarce, and we are still lacking samples large enough to make statistical statements about the variation of 
age and metallicity as a function of other parameters. Some pioneering works tried to overcome the difficulty 
of measuring low-surface brightness absorption lines in the disc using narrow-band imaging, or performing 
Fabry-P\'erot interferometry with Tunable Filters (Beauchamp \& Hardy 1997\nocite{1997AJ....113.1666B}; Molla, 
Hardy \& Beauchamp 1999\nocite{1999ApJ...513..695M}; Ryder, Fenner \& Gibson 
2005\nocite{2005MNRAS.358.1337R}). Unfortunately, the low spectral resolution and poor S/N ratio of the data 
compromised the results in each case. Furthermore, the works were restricted to the study of just a few 
indices (Mgb, Fe5207, Fe5335) which limited their ability to break the age-metallicity degeneracy. More 
recently, MacArthur et al. 2009\nocite{2009MNRAS.395...28M} and S\'anchez-Bl\'azquez et al. 
2011\nocite{2011MNRAS.415..709S}) have measured star formation histories as a function of radius using 
long-slit spectroscopy.  However, long-slit data limit the number of spectra to add in the external parts and, 
therefore, the signal-to-noise. As a consequence, these analysis concentrated in the study of the bulges and 
inner discs. Both studies found very mild gradients in both, age and metallicity in the disc region.

 In the last year, a series of papers have been published using 2D data from the VENGA (Blanc et al. 2010), 
the PINGS (Rosales-Ortega et al. 2010\nocite{2010MNRAS.405..735R}) and the CALIFA (S\'anchez et al.\ 2012) 
surveys where the star formation history of a sample of nearby galaxies of all morphological types has been 
studied in detail (Yoachim et al. 2010\nocite{2010ApJ...716L...4Y};
Yoachim et al.\ (2012)\nocite{2012ApJ...752...97Y};
S\'anchez-Bl\'azquez et al. (2014)\nocite{2014MNRAS.437.1534S};
 P\'erez et al. 2013\nocite{2013ApJ...764L...1P}; Gonz\'alez Delgado et al.\ 2013\nocite{2013arXiv1310.5517G}; Cid Fernandes 
et al\ 2013, 2014\nocite{2014A&A...561A.130C}\nocite{2013arXiv1307.0562C}). These papers show the potential 
of these data to overcome previous difficulties in the derivation of stellar population properties. In the 
present paper, we want to concentrate in the spatially resolved properties of the disc in spiral galaxies. In 
particular, we want to test the predictions that bars are very efficient agents in producing radial movements 
of stars across the disc without heating the disc (Minchev \& Famaey 2010).

The idea that stars in galactic discs may migrate radially across significant distances have received a lot of 
attention in the literature over the last few years (Sellwood \& Binney 2002\nocite{2002MNRAS.336..785S}; 
Ro{\v s}kar et al. 2008ab\nocite{2008ApJ...684L..79R}\nocite{2008ApJ...675L..65R}; Berentzen et al.\ 
2007\nocite{2007ApJ...666..189B}; Minchev \& Famaey 2010\nocite{2010ApJ...722..112M}; Minchev et al.\ 
2011\nocite{2011A&A...527A.147M}, 2012ab\nocite{2012A&A...548A.126M}; Grand et al. 
2012ab\nocite{2012MNRAS.421.1529G}\nocite{2012MNRAS.426..167G};
 Comparetta \& Quillen 2012\nocite{2012arXiv1207.5753C}; Di Matteo et
 al. 2013\nocite{2013A&A...553A.102D}; Brunetti et al. 2011\nocite{2011A&A...534A..75B}; Shevchenko 
2011\nocite{2011ApJ...733...39S}; Kubryk, Prantzos \& Athanassoula 2013; \nocite{2013MNRAS.436.1479K}; 
Minchev, Chiappini \& Martig 2014ab\nocite{2014arXiv1401.5442M}\nocite{2014arXiv1401.5796M}). The reason is 
the possible important consequences that this process may have in a number of studies as the solar 
neighbourhood age-metallicity relation, the metallicity distribution function, the evolution of the metallicity 
gradient, the composition of the thick disc, and the reconstruction of star formation histories from present 
day observations of stellar populations (Ro{\v s}kar et al. 2008b\nocite{2008ApJ...684L..79R}; Sch\"onrich \& 
Binney 2009\nocite{2009MNRAS.399.1145S}). These observations are used to constrain, among others, the amount 
of inflow and the role of feedback in chemical evolution models. Therefore, if we aim to understand the physics 
of disc formation, it is essential to understand the importance of stellar radial migrations. The idea that 
stars may not remain at the radii of birth was already suggested by Wielen (1977)\nocite{1977A&A....60..263W},
to explain the metallicity of the Sun with respect to other stars of the same age in our solar neighbourhood. 
Because of a general increase in the non-circular motion with age, older stars oscillate more than young ones. 
However, for observationally constrained velocity dispersions, variations in galactocentric radius of at most 
few kpc are expected. A new mechanism was proposed by Sellwood \& Binney (2002) \nocite{2002MNRAS.336..785S} 
that can induce strong radial movements without significantly heating the disc due to the exchange of angular 
momentum at the corotation resonance (CR) of spiral arms. This mechanism can produce radial movements on the 
stars of several kpc in a few hundred Myr while keeping them in circular orbits. Because this mechanism 
involves CR, the spiral arms have to be transient or, otherwise, stars would get trapped in horseshoe orbits 
(Sellwood \& Binney 2002)\nocite{2002MNRAS.336..785S}.

On the other hand Minchev \& Famaey (2010)\nocite{2010ApJ...722..112M} (see also Minchev et al. 
2011\nocite{2011A&A...527A.147M}, 2012ab\nocite{2012A&A...548A.126M}\nocite{2012A&A...548A.127M}) proposed 
that spiral structure interacting with a central bar could be an extremely efficient mechanism for radial 
migration in galactic discs. Although angular momentum changes are more important at the vicinity of the 
corotation radius of each individual perturber, the non-linear coupling between the bar and spiral waves
 (e.g., Tagger et al.\ 1987\nocite{1987ApJ...318L..43T}; Sygnet et
 al. 1988\nocite{1988MNRAS.232..733S}) 
can make this mechanism effective over
the entire galactic disc. Furthermore, being non-linear, this way of
mixing can be significantly more efficient at increasing the angular
momentum
than transient spirals alone and works with both, short- and
long-lived
spirals. In fact, Comparetta \& Quillen (2012)\nocite{2012arXiv1207.5753C} showed that, even if
patterns are long-lived, radial migration can result from short-lived
density
peaks arising from interference among density waves overlapping in
radius.

Bars are very common structures; the most recent optical studies indicate that approximately half of all 
massive and nearby discs galaxies contain bars (Barazza et al.\ 2008\nocite{2008ApJ...675.1194B}; Aguerri et 
al. 2009\nocite{2009A&A...495..491A}). This fraction increases in dust-penetrating near-infrared wavelengths 
(Eskridge et al. 2000\nocite{2000AJ....119..536E}). The mass of the galaxy seem to be the main physical 
parameter regulating the bar fraction (e.g., M\'endez-Abreu et al. 2010, 
2012\nocite{2012ApJ...761L...6M}\nocite{2010ApJ...711L..61M};
 Nair \& Abraham 2010\nocite{2010ApJ...714L.260N}). 
Given the fact that bars are very common and that our own Galaxy hosts a bar, it is extremely important
to understand the influence of this structure in shaping the disc.

One of the predicted observational consequences of stellar radial migrations in the disc is a flattening of 
the chemical abundance profiles of the stellar component (e.g., Friedli et al. 
1994\nocite{1994ApJ...430L.105F}; Friedli 1998\nocite{1998ASPC..147..287F}; Ro{\v s}kar et al. 2008b; Loebman 
et al.\ 2011\nocite{2011ApJ...737....8L}; Brunetti, Chiappini \& Pfenniger 2011\nocite{2011A&A...534A..75B}; 
Di Matteo et al.\ 2013\nocite{2013A&A...553A.102D}), especially for those of the old stars (although this 
depend of the velocity dispersion of the stars when they were formed as stellar migration is less efficient 
for dynamically heated populations -- see Minchev, Martig \& Chiappini 2014 for a discussion--). Therefore, a 
way to test the importance of the mechanism proposed by Minchev \& Famaey (2010) is to compare the metallicity 
profiles of galaxies with and without bars.

To do this, in this paper, we make a comparison of the age and stellar metallicity gradients in a sample of 
low-inclination spiral galaxies with and without bars from the CALIFA survey (S\'anchez et al. 
2012\nocite{2012A&A...538A...8S}). Sect.~\ref{sec:sample} presents the sample, Sect.~\ref{sec:analysis} the 
derivation of the star formation histories, mean ages and metallicities and Sect.~\ref{sec:results} our 
results. In Sect.~\ref{sec:evolution} we analyse the metallicity gradients for stellar populations of different 
ages and Sect.~\ref{sec:bulge_disk} shows briefly the relation between the stellar populations in the disc and 
the bulge. In Sect.~\ref{sec:discussion} we briefly discuss the possible scenarios that can explain our results 
while Sect.~\ref{sec:conclusions} presents our conclusions.

\section{Sample and data reduction}
\label{sec:sample}
The data for the present study  are taken from  the Calar Alto Legacy Integral Field Area survey (CALIFA)
(S\'anchez et al. 2012\nocite{2012A&A...538A...8S}). The survey is observing a statistically
well defined sample of $\sim$600 galaxies in the local Universe using 250 observing nights
with PMAS/PPAK integral field
spectrophotometer mounted at the Calar Alto 3.5m telescope. The targets are randomly selected
from the mother sample that comprises 939 galaxies from the SDSS DR7 (Abazajian et al. 2009\nocite{2009ApJS..182..543A}). The main selection
criteria are the angular isophotal size (45$'' < $D$_{25}$  $<80''$, where D$_{25}$ is the isophotal diameter
in the SDSS r-band) and the proximity of the galaxies $(0.005 < z < 0.03$).
PPAK offers a combination of
extremely wide field of view ($> 1$ arcmin$^2$) with a high filling factor in one single
pointing (65\%).  A dithering scheme with three pointing  has been adopted in CALIFA
in order to cover
the complete field-of-view  of the central bundle and to increase the spatial resolution of the data.
The spectra cover the range 3400-7300~\AA~in two overlapping setups, one in the
red (3745-7300~\AA) at a spectral resolution of R=850 (V500 setup) and one in the blue (3400-4750~\AA) at
R$\sim$1650 (V1200 setup). In the present paper we have combined the two setups by degrading the spectral
resolution of  the blue part of the spectra to match the resolution of the red using a wavelength-dependent
Gaussian smoothing kernel.
The exposure time is fixed for all the observed objects. For the V500 setup a single exposure
of 900s per pointing of the dithering scheme is taken while for the V1200 setup 3 or 2 exposures of
600s or 900s, respectively, are obtained per pointing.
The data reduction is explained in detail in S\'anchez et al.\ (2012a)\nocite{2012A&A...538A...8S} and
Husemann et al.\ (2013)\nocite{2013A&A...549A..87H}. The basic reduction tasks include cosmic ray rejection, optimal
extraction, flexure correction, wavelength and flux calibration and sky subtraction. Finally, all three pointings are combined using a
flux conserving inverse distance weighting scheme (see the original paper for details), to reconstruct the final data cube, with a one
arcsec spatial sampling.

The selection criteria to build the sample (diameter and redshift) are such that the selected objects represent
a wide range of galactic properties such as morphological types, luminosities, stellar masses and colors. Further
details on the selection criteria effects and a detailed characterization on the CALIFA mother sample are explained
in S\'anchez et al.\ (2012).

For this study, we selected galaxies that are morphologically classified as discs, without signs
of recent interactions and with inclination lower than 60 degrees.
The morphological classification will be
presented in Walcher et al. (2014, in preparation), and is based on an eyeball classification
by 5 members of the collaboration using r and i band SDSS images.
The constrain in inclination is given by the difficulty in detecting the bar presence
in highly inclined galaxies.
This restrictions lead us with a sample of 62 galaxies, 28 of which are unbarred, 25 strongly barred (B) and 9 weakly barred (AB).
Although the  presence of the bar is determined visually we also performed an ellipse analysis
to calculate its size and strength (M\'endez-Abreu et al.\ 2014, in
preparation). The strength of the bar, which measure the contribution of the bar
to the total galaxy potential, is derived using:
\begin{equation}
f_{\rm bar}=\frac{2}{\pi}(\arctan(1-\epsilon_{\rm bar})^{-1/2}-\arctan(1-\epsilon_{\rm bar})^{+1/2})
\end{equation}
where $\epsilon_{\rm bar}$ is the intrinsic mean ellipticity in the bar region.
This parameter was defined by Abraham \& Merrifield (2000)\nocite{2000AJ....120.2835A} and it is correlated 
with the most widely used parameter $Q_g$, defined by Buta \& Block
(2001)\nocite{2001ApJ...550..243B}
 (Laurikainen et al. 2007\nocite{2007MNRAS.381..401L}). The size of the bar has been calculated as the mean value of those determined
with three different methods: (i) the maximum of the ellipticity, (ii) the minimum of the
ellipticity and the (iii) change in the position angle of the ellipses in more than 5 degrees
(see M\'endez-Abreu et al.\ 2014 for details).

In the present study we  also  use the stellar mass determined by fitting SED models to galaxy photometry
with {\tt Kcorrect} (Bekeiraite et al. 2014, in preparation). These masses agree extremely well
with the stellar masses derived with full spectral fitting using {\tt STARLIGHT} 
(Gonz\'alez Delgado et al. 2013).
We will also use the effective radius of the disc (radius enclosing
half of the light of the disc component, i.e., excluding the bulge, r$_{\rm eff}$ hereafter).
This radius is calculated using an analysis of the azimuthal surface brightness profile
derived through an  isophotal analysis of the SDSS imaging survey g-band images
(York et al. 2000\nocite{2000AJ....120.1579Y}).
In the region dominated by the disc, the profile
is fitted with a pure exponential profile, using the classical formulae:
\begin{equation}
I=I_0 \exp[-(r/r_d)].
\end{equation}
The effective radius of the disc is related to the scale-length through $r_{\rm eff} =$ 1.67835r$_d$
 and is very well correlated with the total effective radius obtained with the curve of growth
(see appendix A of S\'anchez et al. 2013b for more details
about how the disc  radius is determined).
Table \ref{tab:sample} show the main properties of our final sample.
\begin{table*}
\centering
\small
\begin{tabular}{lllrrrrrrrr}
\hline
         &\multicolumn{1}{c}{CALIFA} &
         \multicolumn{1}{c}{Morph}&
         \multicolumn{1}{c}{r$_{\rm bar}$ (``)}&
         \multicolumn{1}{c}{f$_{\rm bar}$} &
         \multicolumn{1}{c}{log M$_{*}$} &
         \multicolumn{1}{c}{r$_{\rm eff}$($''$) }  &
         \multicolumn{1}{c}{t-type} &
         \multicolumn{1}{c}{Inc} \\
\hline\hline
IC1256   & 856   &  Sb(AB)  &  10.6   &  0.20   &   9.873 &  17.3      & 3.3  & 56.2\\
IC1683   & 043   &  Sb(AB)  &  11.2   &  0.03    & 10.517 &  12.0      &2.7   & 56.5\\
NGC0001  & 008   &  Sbc(A)  &  --     &    --    & 10.656 &  17.7      &3.1   & 37.9\\
NGC0036  & 010   &  Sb(B)   &  24.2    &  0.30  & 10.914     &  31.7      &3.0   & 51.5 \\
NGC0160  & 020   &  Sa(A)   &  --            &    --        & 10.919     &  37.6      &$-0.4$& 57.9\\
NGC0214  & 028   &  Sbc(AB) &  15.2       &  0.08       & 10.872     &  17.3      & 5.0  & 49.9\\
NGC0234  & 031   &  Sc(AB)  &  --            &     --       & 10.597     &  18.4      &5.3   & 32.3\\
NGC0257  & 033   &  Sc(A)   &  --            &      --      & 10.795     &  19.4      &5.8   & 56.2\\       
NGC0776  & 073   &  Sb(B)   &  18.2         &   0.26      & 10.720     &  18.0      &2.5   & 48.8\\
NGC0496  & 045   &  Scd(A)  &                &              &            &            &      &     \\
NGC1167  & 119   &  S0(A)   &  --            &     --       & 11.129     &  29.8      &$-2.4$& 42.3 \\
NGC1645  & 134   &  S0a(B)  &  16.6       &   0.29     & 10.780     &  24.9      &$-0.9$& 57.5\\
NGC2253  & 147   &  Sbc(B)  &  14.5            &     --   & 10.547     &   8.4      &5.8   & 30.4\\
NGC2347  & 149   &  Sbc(AB) &  14.6            &     --   & 10.759     &  18.5      &3.1   & 52.4\\
NGC2906  & 275   &  Sbc(A)  &  --            &     --       & 10.288     &  16.1      &5.9   & 60.9\\
NGC2916  & 277   &  Sbc(A)  &  --            &     --       & 10.421     &  25.3      &3.1   & 56.7\\
NGC3106  & 311   &  Sab(A)  &  --            &     --       & 11.001     &  22.7      &$-1.9$&24.4 \\
NGC3300  & 339   &  S0a(B)  &  17.0        &   0.26    & 10.526     &  11.5      &$-1.8$&60.0 \\
NGC3614  & 388   &  Sbc(AB) &  32.0            &     --   &  9.937     &  41.1      &5.2  & 45.1\\
NGC3687  & 414   &  Sb(B)   &  15.2        &  0.19     & 10.040     &  20.1      &3.8  & 24.3\\
NGC4047  & 489   &  Sbc(A)  &  --            &     --       & 10.539     &  16.5       &3.2 & 39.5\\
NGC4185  & 515   &  Sbc(AB) &  21.0        &   0.03    & 10.577     &  27.9      &3.7  & 52.1\\
NGC4210  & 518   &  Sb(B)   &  20.8        &   0.33    & 10.193     &  17.0      &3.0  & 45.2\\
NGC4470  & 548   &  Sc(A)   &  --           &     --        &  9.853     &  12.6       &1.4 & 52.6\\
NGC5000  & 608   &  Sbc(B)  &  27.7       &   0.37     & 10.545     &  18.2      & 3.8 & 55.0\\
NGC5016  & 611   &  Sbc(A)  &  --            &       --     & 10.093     &  19.0     & 4.4  & 44.4\\
NGC5205  & 630   &  Sbc(B)  & 21.4         &   0.28    &  9.728     &  21.1      & 3.5 & 50.3\\
NGC5218  & 634   &  Sab(B)  & 18.5         &   0.31    & 10.469     &  17.6      &3.1  & 58.9\\
NGC5378  & 676   &  Sb(B)   & 34.2         &   0.25    & 10.335     &  24.6      &1.0  & 55.5\\
NGC5394  & --    &  Sbc(B)  & 25.3             &     --   &  9.873     &  19.0      & 3.1 & 43.6\\
NGC5406  & 684   &  Sb(B)   & 22.9         &   0.29    & 11.005     &  20.4       &3.9 & 29.0\\
NGC5614  & 740   &  Sa(A)   & --             &      --      & 10.976     &  27.5      & 1.7 & 19.2\\
NGC5633  & 748   &  Sbc(A)  & --             &      --      & 10.097     &  12.6      &3.2  & 52.2\\
NGC5720  & 764   &  Sbc(B)  & 11.2         &   0.23    & 10.847     &  19.9      &3.0  & 51.7\\
NGC5732  & 768   &  Sbc(A)  &  --            &     --       &  9.792     &  16.2      &4.0  & 56.5\\
NGC5784  & 778   &  S0(A)   &  --            &     --       & 10.985     &  22.3      &$-2.0$& 42.1\\
NGC6004  & 813   &  Sbc(B)  &  20.4        &   0.32    & 10.467     &  21.7      &4.9  & 19.8\\
NGC6063  & 823   &  Sbc(A)  &  --           &    --        &   9.908     &  20.7       &5.9 & 54.1\\
NGC6154  & 833   &  Sab(B)  &  31.8        &   0.28    & 10.734     &  20.0      &1.0  & 54.0\\
NGC6155  & 836   &  Sc(A)   &  --            &     --       &  9.958     &  13.8      &5.2  & 48.2\\
NGC6301  & 849   &  Sbc(A)  &  --            &     --       & 10.929     &  26.4      &5.9  & 54.3   \\
NGC6497  & 863   &  Sab(B)  &  15.6        &    0.25   & 10.316      &  17.7     &3.1  & 51.9\\
NGC6941  & 869   &  Sb(B)   &  18.6        &    0.23   & 10.862      &  23.0     &3.2  & 45.2\\
NGC7025  & 874   &  S0a(A)  &  --            &     --       & 11.063      &  23.0     &1.0  & 47.5\\
NGC7321  & 887   &   Sbc(B) &  14.1        &    0.22   & 10.984      &  23.0     &3.1  & 48.7\\
NGC7489  & 898   &  Sbc(A)  &  --            &     --       & 10.483     &   14.6     & 6.4 & 58.2\\
NGC7549  & 901   &  Sbc(B)  &  31.2            &     --   & 10.539     &   11.9     & 5.9 & 42.1\\ 
NGC7563  & 902   &  Sa(B)   &  25.9        &    0.28  &  10.753     &   10.0     & 1.0 & 50.5\\
NGC7591  & 904   &  Sbc(B)  &  13.6       &    0.18  &   10.768      &   19.9     & 3.6 & 56.5\\
NGC7653  & 915   &  Sb(A)   &  --           &     --      &   10.486      &   17.6     & 3.1 & 29.2\\
NGC7671  & 916   &  S0(A)   &  --           &     --      &   10.786       &   14.0     &$-2.0$& 60.0\\
NGC7782  & 931   &  Sb(A)   &  --            &     --      &  11.096      &    26.0     & 3.0 & 59.7\\
UGC00005  & 002   & Sbc(A)   & --            &     --      &  10.883     &    16.1     & 3.9 & 59.6\\
UGC00036  & 007   & Sab(AB)  & --           &    --      &    10.781      &    16.2      & 1.0& 57.9\\ 
UGC03253  & 146   & Sb(B)    & 17.4        & 0.25     &  10.397    &     18.2     & 3.0& 53.9\\
UGC07012  & 486   & Scd(AB)  & 4.6             &    --   &   9.010 &     18.4     & 5.7&59.2\\ 
UGC08234  & 607   & S0(A)    & --            &    --      &   11.061     &     9.0     &$-0.1$&56.6\\
UGC10205  & 822   & S0a(A)   & --            & --         &   10.947      &     17.9    & 1.0 & 59.8\\
UGC11649  & 872   &  Sab(B)  & 23.3         & 0.25   &   10.467     &      18.2    & 1.0& 30.3\\
UGC11680  & 873   &  Sb(B)   & 12.9             & --    &   10.892    &     33.6    & 3.0 &41.5\\ 
UGC12224  & 891   &  Sc(A)   & --             &  --       &    9.751     &     31.5     & 5.0 & 34.3\\
UGC12816  & 930   &  Sc(A)   & --             &   --      &    9.649     &     20.2     & 5.8& 53.0\\
\hline
\end{tabular}
\caption{(1) Galaxy name; (2) CALIFA identifier; (3) Visual morphological classification;
(4) Bar radius obtained as the mean value between three methods: (i) the maximum and the (ii) minimum
of the ellipticity and the (iii) radius at which the position angle change by more than 5 degrees.
(5) Strength of the bar derived as explained in the text; (6) Stellar masses (in logarithmic scale) derived
by a SED fitting using {\tt Kcorrect}; (7) Effective radius of the disc in arcsec. (8)
t-type classification using the modified RC3 classifiers; (9) Inclination (in degrees), purely based on the 
ellipticity.
\label{tab:sample}}
\end{table*}

\section{Analysis}
\label{sec:analysis}
We spatially bin by means
of the centroidal Voronoi tessellation algorithm of Cappellari \& Copin (2003)\nocite{2003MNRAS.342..345C}
to ensure a minimum signal-to-noise of 40 ({\it per} \AA) at 5800\AA, necessary for a reliable determination
of the stellar population properties\footnote{The same analysis has
  been repeated for azimuthally binned spectra and the results remain unaltered}.
Pre-processing
steps include spatial masking of foreground/background sources, very low signal-to-noise  spaxels and bad pixels.

\subsection{Emission line cleaning, and determination of radial
  velocity and line broadening}
\label{sec:gandalf}
On the binned spectra, we run the code
{\tt GANDALF} (Sarzi et al. 2006\nocite{2006MNRAS.366.1151S}).
{\tt GANDALF} fits, simultaneously, the absorption and emission lines, treating the latter as
additional Gaussians. In a first step, emission lines are masked and the absorption line spectrum
is fitted with the penalized pixel-fitting {\tt pPXF} (Cappellari \& Emsellem
2004\nocite{2004PASP..116..138C}), using as templates
the stellar population models of Vazdekis et al. (2010) (V10 hereafter) \nocite{2010MNRAS.404.1639V}
based on the MILES stellar library
(S\'anchez-Bl\'azquez et al. 2006\nocite{2006MNRAS.371..703S}; Cenarro et al.
2007\nocite{2007MNRAS.374..664C}; Falc\'on-Barroso et al. 2011\nocite{2011A&A...532A..95F})
\footnote{The stellar population models and the stellar library are publicly available at
http://miles.iac.es.}. In this step, radial velocities and absorption line broadening\footnote{The velocity 
dispersions for these galaxies are
calculated using the data from the V1200 dataset (Falc\'on-Barroso et al. 2014, in preparation). 
What we derive here is the total broadening of the lines, including the
instrumental and the doppler broadening.} ($\sigma$, hereafter)
for the stellar component are derived. The best values of velocity and $\sigma$ and the best template
mix are then used as initial values for the calculation of emission lines using {\tt GANDALF}. Emission
lines equivalent widths, radial velocities and $\sigma$ for the gaseous component are derived in this
second step. The fit allows for low-order Legendre polynomial in order to account for small differences
in the continuum shape between the pixel spectra and the templates. The best fitting template mix is
determined by a $\chi^2$ minimization in pixel space. Emission lines  were subtracted from the
observed spectra. Figure~\ref{fig:fit} shows, as an example, the spectrum
of the central spaxel for the galaxy  IC1256, before and after subtracting the emission lines.

\subsection{Star formation histories}
\label{sec:setckmap}

Disc galaxies certainly show  complex star formation histories. 
In the last decade, numerical  techniques
have been developed to derive the whole evolution of the star formation history with
time using as much information as possible from the spectra
(Heavens et al. 2000\nocite{2000MNRAS.317..965H}; Panter et al. 2003\nocite{2003MNRAS.343.1145P};
Cid Fernandes et al.\ 2005\nocite{2005MNRAS.358..363C};
Ocvirk et al. 2006ab\nocite{2006MNRAS.365...46O}\nocite{2006MNRAS.365...74O};
Tojeiro et al. 2007\nocite{2007MNRAS.381.1252T}; Koleva et al. 2009\nocite{2009MNRAS.396.2133K}
--see, .e.g, Walcher et al.\ 2011\nocite{2011Ap&SS.331....1W}; S\'anchez et al.\ 2012\nocite{2012A&A...546A...2S};
Rosales-Ortega et al.\ 2012\nocite{2012ApJ...756L..31R}).

To derive star formation histories we used the code
{\tt STECKMAP} (STEllar Content and Kinematics via Maximum A Posteriori likelihood, Ocvirk et al.
2006ab\nocite{2006MNRAS.365...46O}\nocite{2006MNRAS.365...74O}) on the emission
line-cleaned spectra (see previous section).
{\tt STECKMAP} projects the observed spectrum onto a temporal sequence of models of single
stellar populations to determine the linear combination that fits the observed
spectrum best.
The weights of the various components of this linear
combination indicate the stellar content of the population. As  templates, we use the stellar
population models by V10\nocite{2010MNRAS.404.1639V} based on the MILES library (S\'anchez-Bl\'azquez et
2006). We chose those models with a Kroupa Universal IMF (Kroupa 2001)\nocite{2001MNRAS.322..231K}\footnote{This IMF is a multi-part power-law IMF, which is
similar to the Salpeter (1955)\nocite{1955ApJ...121..161S} IMF for stars of masses above 0.5~M$_{\sun}$, but with a decreasing contribution of
lower masses by means of two
flatter segments.} a range of ages and metallicities from 63~Myr to
17.8~Gyr and  $-2.32<[Z/H]<+0.2$ respectively.

\begin{figure*}
\centering
\resizebox{0.5\textwidth}{!}{\includegraphics[angle=0]{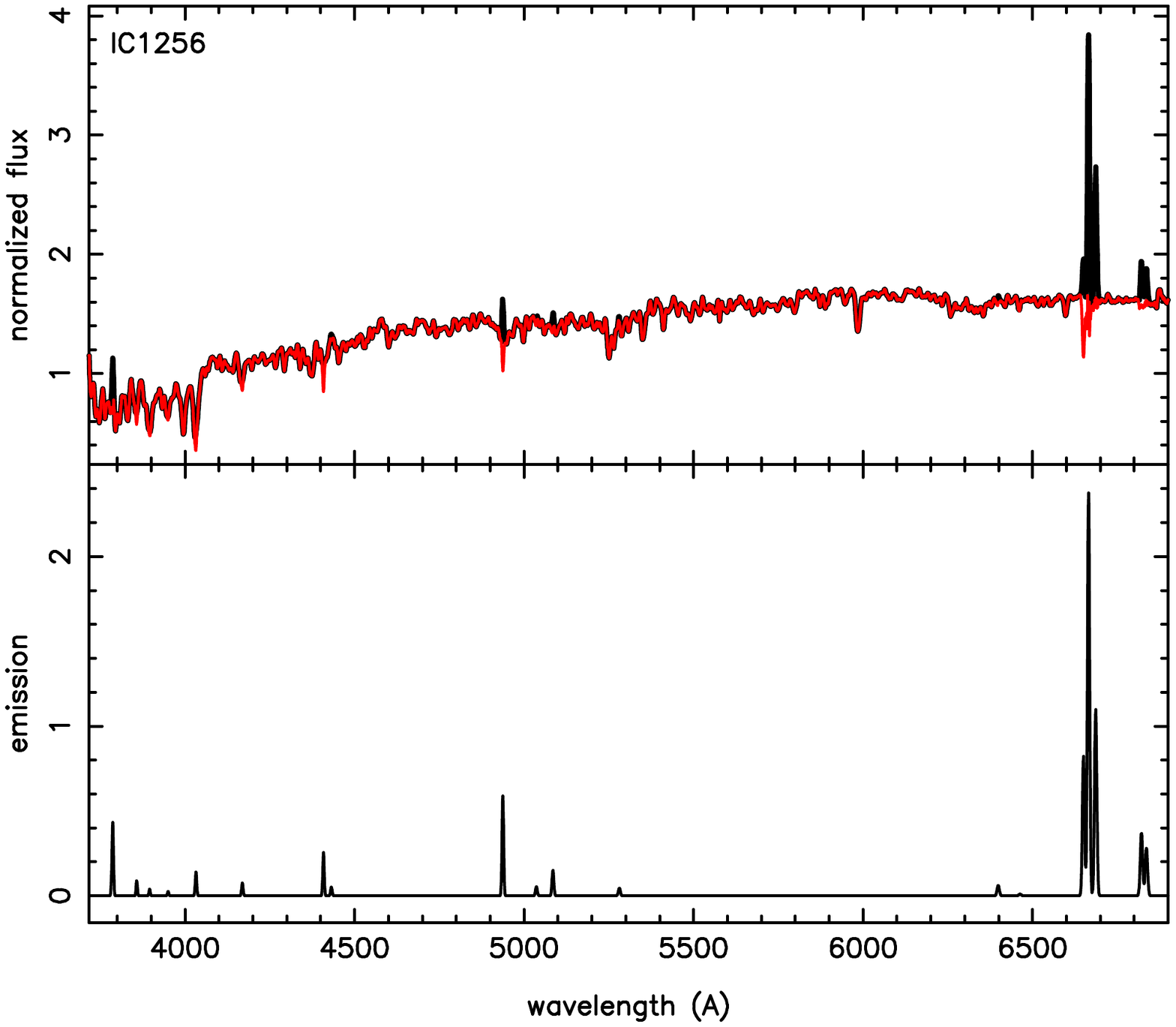}}
\resizebox{0.5\textwidth}{!}{\includegraphics[angle=0]{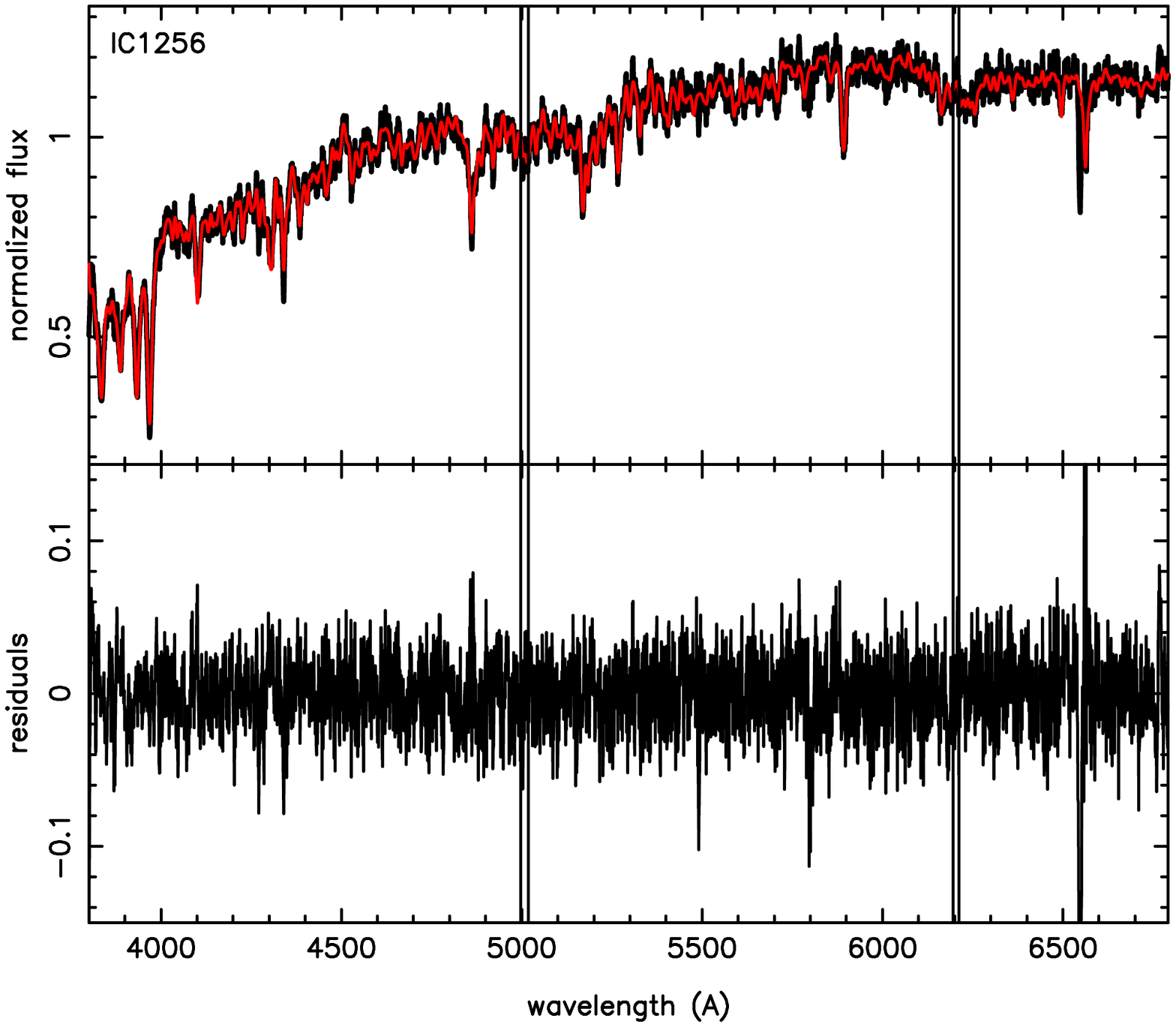}}
\caption{Left panel-- original observed central spectrum (central spaxel) of IC1256 (black line) and the one corrected
from emission using {\tt GANDALF} (red line). Bottom panel--  emission line spectrum.
Righ panel-- top panel-- Fit to the central spectra of the galaxies from the sample (red line).
Bottom panel: residuals from the fit. Masked areas are indicated with vertical lines.\label{fig:fit}}
\end{figure*}

{\tt STECKMAP} is a Bayesian method that simultaneously recovers the kinematic and stellar
population properties via a maximum a {\it posteriori} algorithm. It has been extensively tested
and used in a variety of applications. It is a public tool and can be obtained at
{\tt http://astro.u-strasbg.fr/$\sim$ocvirk/}.  The method is not parametric and does not make any
{\it a priori} assumption regarding the shape of the star formation history. The only
condition that {\tt STECKMAP} imposes is that the different unknowns, namely the stellar age
distribution, the age-metallicity relation and the line-of-sight velocity distributions  or the
broadening function have to be smooth in order to avoid extreme oscillating solutions that are not
robust and most likely unphysical. The function to minimize is defined as:
\begin{equation}
Q_{\mu}=\chi^2 (s(x,Z,g)) + P_{\mu}(x,Z,g),
\end{equation}
which is a penalized $\chi^2$, where $s$ is the modeled spectrum resulting from the
age distribution $x$, the age-metallicity relation ($Z$) and the broadening function ($g$)
\footnote{Note that {\tt STECKMAP} fits different broadening function for different ages as different
populations are expected to have different velocity dispersions (e.g., House et al. 2011).}
The penalization $P_{\mu}$ can be written as: $P_{\mu}(x,Z,g) = \mu_x P(x) + \mu_Z P(Z) + 
\mu_v P(g)$, where the function $P$ gives high values for solutions with strong oscillations
(i.e., a rapid variation of the metallicity with age or a noisy broadening function) and
small values for smoothly varying solutions. Adding the penalization $P$ to the
function $Q$ is exactly equivalent to inject {\it a priori} information into the problem.
In practice, this is like imposing an {\it a priori} probability density to the solution
as $f_{\rm prior}(x)=\exp(-\mu_x P(x))$. For this work we define $P$ as a quadratic function
of the unknown $x$, involving a kernel $L$. We use a Laplacian smoothing kernel of the
age distribution and a gradient kernel for the age-metallicity relation, as in Ocvirk
(2010\nocite{2010ApJ...709...88O}, see Ocvirk et al. 2006a\nocite{2006MNRAS.365...46O} for details). Choosing the right values of the smoothing parameters
$\mu_{x,Z,v}$ is not a trivial problem. In principle, one could choose the values giving the
smaller $\chi^2$ in the fit, but this usually yields a wide range of smoothing parameters,
spanning typically 3-4 decades, in which the fit is acceptable. In any case, although the
detailed shape of the derived star formation history can be affected by this choice, the
range of smoothing parameters we are using neither change the overall interpretation of the
star formation history nor the mean age and metallicity values, which is what are are going
to use across the paper. We are using in this work, $\mu_{x}=1$ and $\mu_z=1$.
In the present analysis we have not fitted simultaneously the star
formation histories and the kinematics. Instead, for the kinematic we adopt 
the solution of  {\tt pPXF} that we obtained in the correction from emission line (see Sec.~\ref{sec:gandalf}). The reasons are
explained in detail in Appendix B of S\'anchez-Bl\'azquez et al. (2011)\nocite{2011MNRAS.415..709S}. Basically,
the existing degeneracy between the metallicity and the
velocity dispersion (Koleva et al. 2008\nocite{2008MNRAS.385.1998K}) biases the mean-weighted metallicites if
both parameters are fitted at the same time.
One important detail of this work is that we do not use the continuum in our derivation
of the star formation history. This is to avoid spurious results due to possible flux calibration
errors or extinction. To do this, we multiply the model by a smooth non-parametric transmission
curve, representing the instrumental response multiplied by the interstellar extinction. This curve
has 30 nodes, spread uniformly along the wavelength range, and the transmission curve is obtained by
spline interpolating between the nodes. The latter are treated as additional parameters and adjusted
during the minimization procedure. By using this curve to remove the continuum, we do not have to
correct from extinction, as dust extinction does not change the equivalent width of the absorption lines
(MacArthur 2005\nocite{2005ApJ...623..795M}).
We fit the whole wavelength range of the data, masking the regions affected by
sky residuals or others defects of the detectors.
Figure~\ref{fig:fit} shows the integrated spectrum of IC~1256 together with the best fit obtained with {\tt STECKMAP}.

\begin{figure}
\centering
\resizebox{0.6\textwidth}{!}{\includegraphics[angle=0]{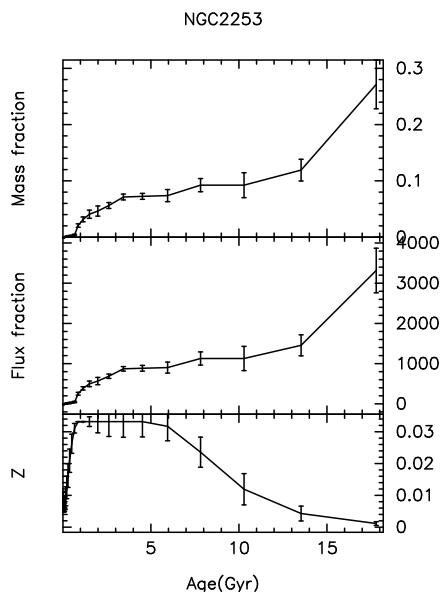}}
\caption{Typical output of {\tt STECKMAP} for the central spectrum of NGC~2253.
The top panel shows the mass fraction of stars of different ages, the middle panel,
the flux fraction and the bottom panel the evolution of metallicity Z. The error bars
represent the root-mean-square dispersion  from the mean, of a series of 250 Monte Carlo
simulations where each pixel of the spectrum is perturbed with noise following a Gaussian distribution of width
given by the error spectrum. \label{steckmap_output}}
\end{figure}

The typical {\tt STECKMAP} outputs give the proportion of stars at each age that are contributing to the
observed flux and to the stellar mass and the evolution of the metallicity
with time. Figure~\ref{steckmap_output} shows an example of these outputs for the central spectrum of NGC~6155.
It is easy to see from these outputs that our results imply that
  20-30\% of the stars have ages older 
than the most recent estimates of the  age of the Universe. 
This is a well known problem in the stellar population models (e.g., Vazdekis et al. 2001\nocite{2001ApJ...549..274V}; 
Schiavon et al. 2002\nocite{2002ApJ...580..873S}; Maraston et al.
2011\nocite{2011MNRAS.418.2785M}, V10). Several possible solutions
have been proposed  to alleviate this so-called zero point problem, 
and, in fact, several groups have successfully reconcile the age of
globular clusters with $\Lambda$CDM cosmology 
(see Krauss \& Chaboyer 2003\nocite{2003Sci...299...65K});
Percival \& Salaris (2009)\nocite{2009ApJ...703.1123P} showed that systematic uncertainties associated with the three fundamental stellar atmospheric 
parameters might have a non negligible impact on the resulting SSP line-strengths. In particular, a relatively
small offset in the effective temperature of 50-100~K, which is of the order of the systematic errors in the conversion 
from temperature to colours used in the V10 models used here (Alonso, Arribas \& Mart\'{\i}inez-Roger 1996\nocite{1996A&A...313..873A}) may change the age of a 14 Gyr stellar 
population by 2-3 Gyr and alleviate the zero point problem (see V10 for more details).
Isochrones that have into account effects such as $\alpha$-enhancement and diffusion of heavy elements also help 
to alleviate the problem (Vazdekis et al. 2001). Different ages are
also obtained when using  isochrones with  different  opacities (e.g.,
Bruzual \& Charlot 2003\nocite{2003MNRAS.344.1000B}). For example, the
use of Padova (2000, Girardi et al. 2000\nocite{2000A&AS..141..371G}) (which are those used by V10) gives older ages 
than Padova (1994, Bertelli et al. 1994\nocite{1994A&AS..106..275B}).
Schiavon et al. (2002)\nocite{2002ApJ...580..873S} also stressed the importance of 
correctly modeling the luminosity function at the level of the giant branch  for the particular problem of globular clusters. He showed, for 47 Tuc, that if the 
observed luminosity 
function is used rather than the theoretical ones, the spectroscopic and the colour-magnitude diagram 
ages coincide and are lower than the age 
of the Universe.

However, despite these possibilities, there is no physical reason to prefer Padova (1994) 
to Padova (2000), which include improved physics. There is also no reason  to chose a
different temperature scale than the one we chose in our models  to transform from
the theoretical to the observational plane. On the other hand, we
cannot determine, observationally, the   luminosity  function of 
red giant branch stars in our galaxies. Our models make use of the latest and, in our opinion, best possible ingredients, despite this choice leads to very old ages. 
We prefer to use the whole range of model (i.e., we use SSPs with
ages older than the age of the Universe)  as, otherwise  we could be biasing the results artificially. In any case, the present study is comparative and so the age scale 
should not affect our main  conclusions.
We can obtain a mean log(age) and metallicity weighting with
the light or with the mass of each population as:
\begin{eqnarray}
<\log q>_{MW} = \frac{\sum_i mass(i)\log q_i}{\sum_i mass(i)},\\
<\log q>_{LW} = \frac{\sum_i flux(i)\log q_i}{\sum_i flux(i)},
\end{eqnarray}
where $q$ is the physical parameter we want to estimate, i.e., age or
metallicity, and $mass(i)$ and $flux(i)$ are, respectively, the reconstructed mass and flux
contributions of the stars in the $i$-th age bin, as returned by {\tt STECKMAP}.
When present, young stars are very luminous in
the optical range, therefore, they will contribute more to the
light-weighted values. This means that the light-weighted values of age will be
strongly biased towards the youngest stellar component ages. The mass weighted values will be less biased towards the age and
metallicity of the youngest components however,
they are also more uncertain. This is  especially true when the contribution by mass of the old stars is large, 
as these stars are not very luminous and, therefore, their contribution to the observed spectrum is small.
We have to note here that the average values of age and metallicities can change considerably if we
add them logarithmically or linearly (S\'anchez-Bl\'azquez et al.\ 2011; Gonz\'alez Delgado et al.\ 2013).

\begin{figure*}
\resizebox{0.30\textwidth}{!}{\includegraphics[angle=0]{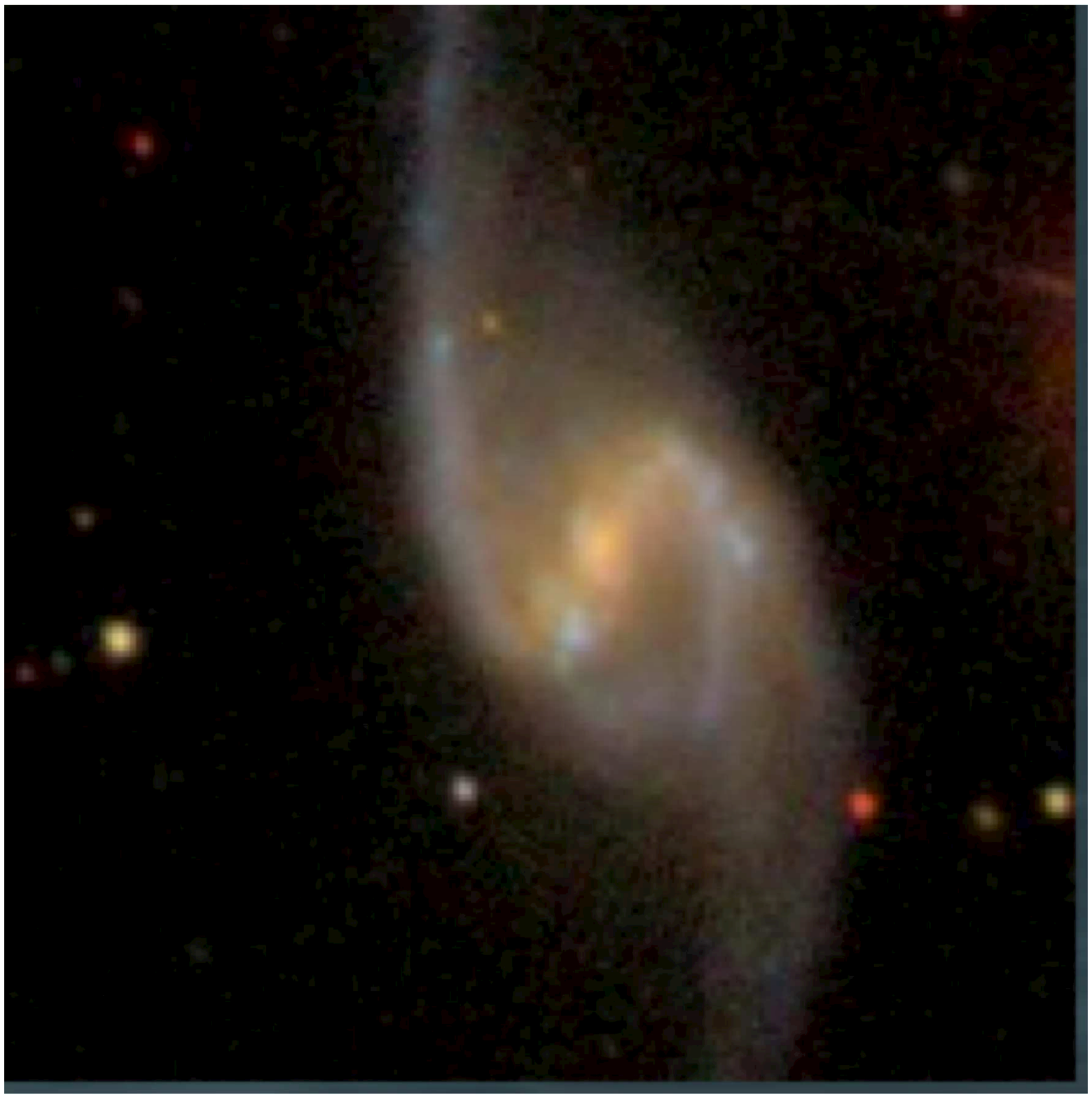}}

\resizebox{0.42\textwidth}{!}{\includegraphics[angle=-90]{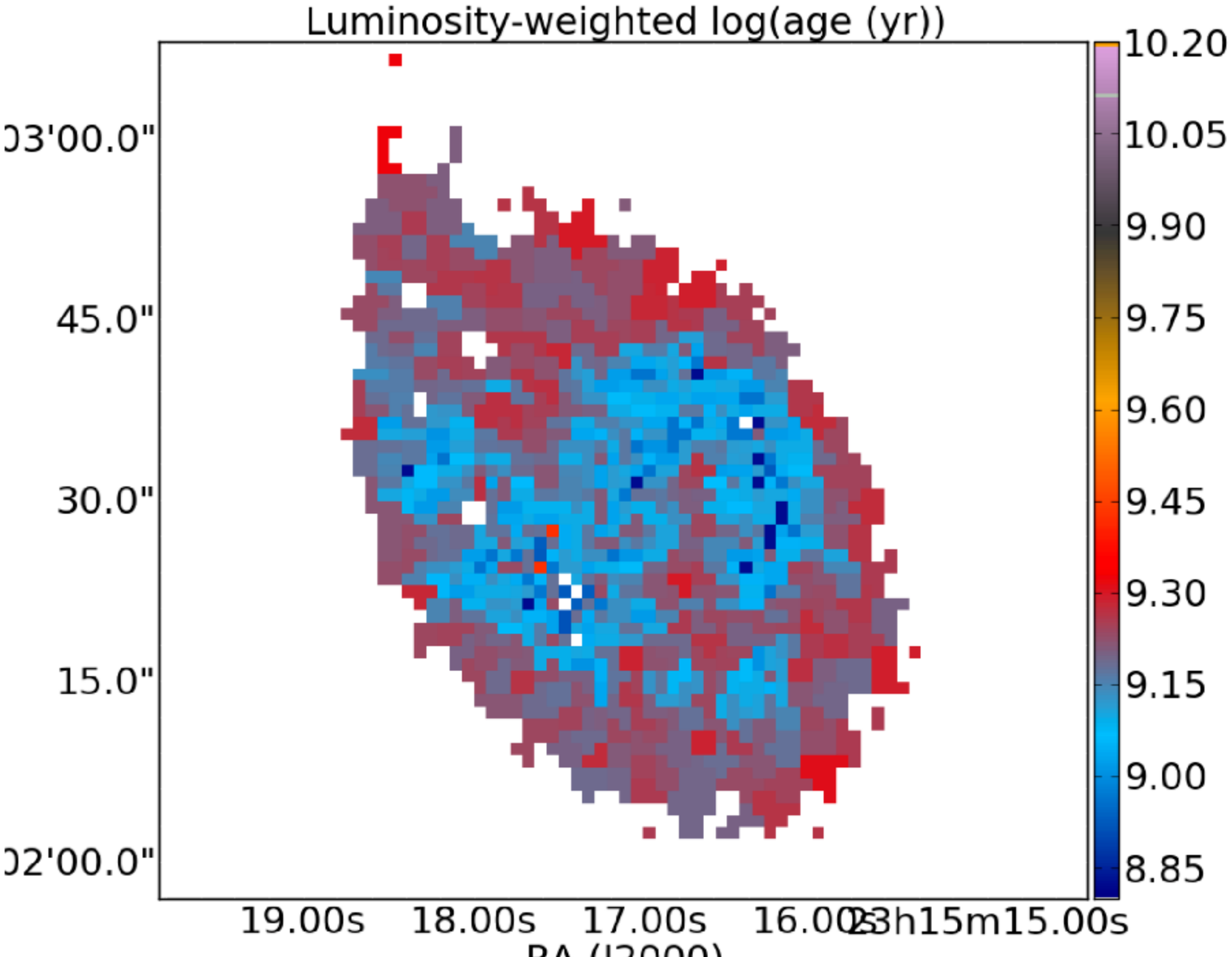}}
\resizebox{0.42\textwidth}{!}{\includegraphics[angle=-90]{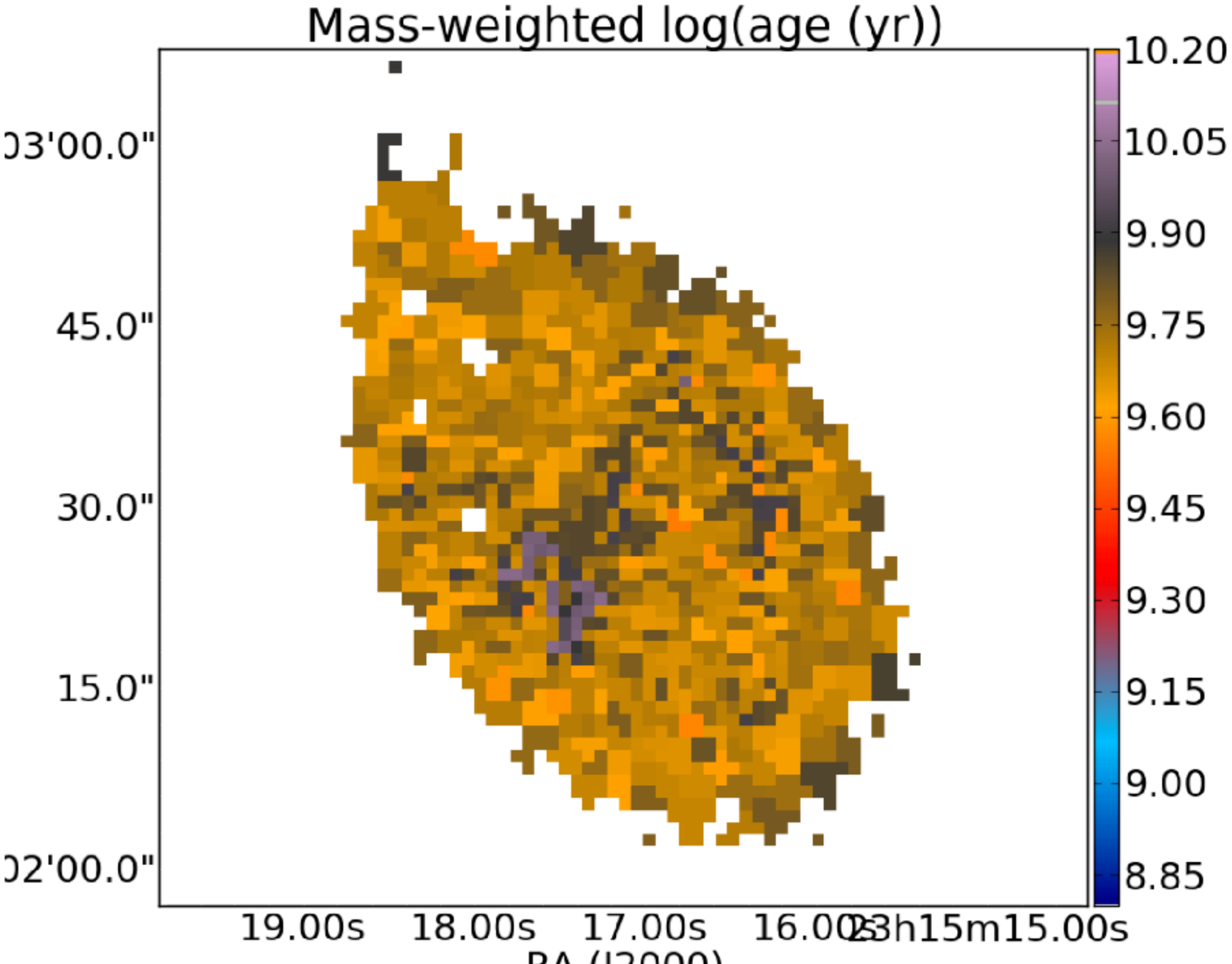}}
\resizebox{0.42\textwidth}{!}{\includegraphics[angle=-90]{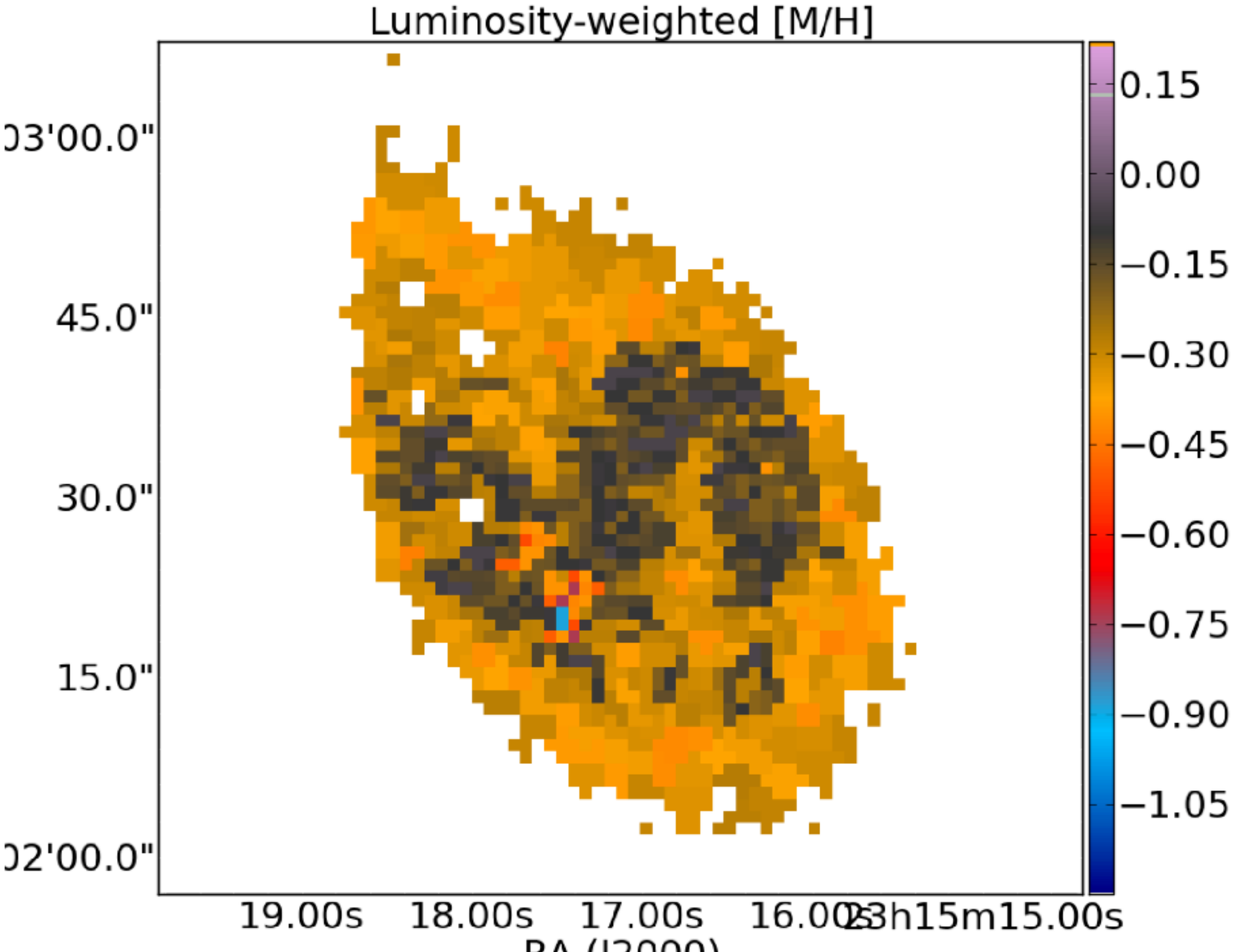}}\hspace{2.5cm}
\resizebox{0.42\textwidth}{!}{\includegraphics[angle=-90]{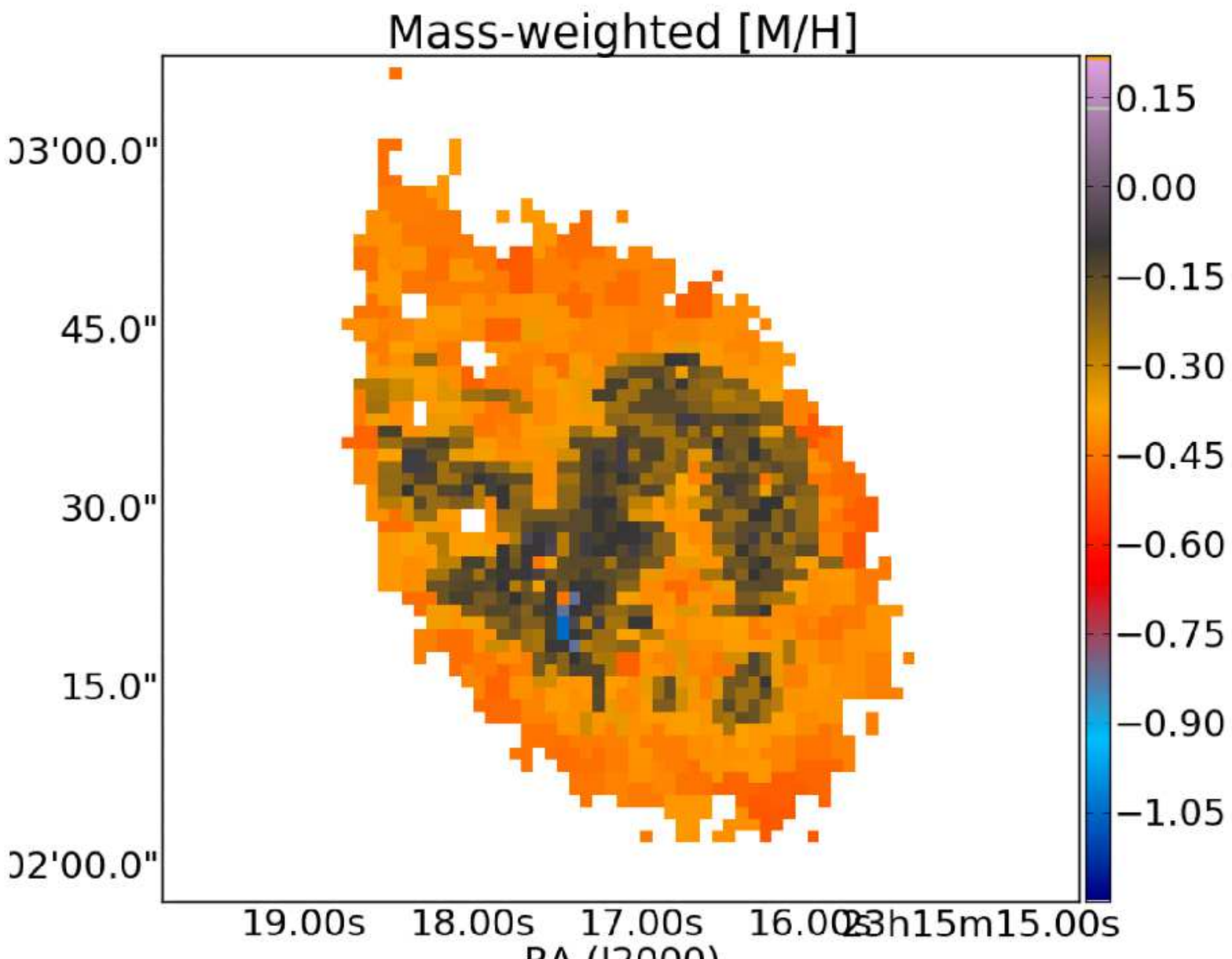}}
\caption{Mean log(age) and metallicity maps for one of the galaxies of our sample,
NGC~7549. The maps for the rest of the sample are available in the electronic
version of the manuscript.\label{fig:maps}}
\end{figure*}

We have obtained maps of the mean log(age) and metallicities weighting with both, light
and mass, for all the galaxies of our sample.
Figure~\ref{fig:maps} shows an example for one galaxy of our sample, NGC~7549.
The rest of the maps  are available in the electronic
version of this paper.

Note that the detailed spatially resolved values for individual galaxies are method and 
model dependent. To check the extent of this dependence,  we have compared our results with those obtained using {\tt
STARLIGHT}  (Cid Fernandes et al. 2005) together with a  set of models combination of V10 and  
Gonz\'alez Delgado et al. (2005)
for young stars (see P\'erez et al. 2013 \& Gonz\'alez Delgado et al. 2014).  
While none of the results presented here depend on the 
method used, details on individual galaxies may change. A detailed
description of the differences in the derived stellar population
parameters using several methods and 
stellar population models will be presented in a future paper. 

\section{Results}
\label{sec:results}
\subsection{Stellar population gradients}

To visualize the variation of
the stellar population trends with the radius of the galaxies, we average, from the
Voronoi binned image, the values of age and
metallicities along isophotes with the  ellipticity and position angle
of each galaxy (all the ellipses for a given galaxy with the same ellipticity and orientation).

{\bf Age gradients:} Figures~\ref{fig:grad_age1}, \ref{fig:grad_age2} and \ref{fig:grad_age3}
in appendix~\ref{appendix2}
 show the variation of the mean age, weighted with both, the mass and the
light, for the sample of barred, unbarred and weakly barred galaxies respectively.
The radius has been normalized to the effective radius of the disc.

In this paper we are only considering the gradients in the disc region. 
To obtain the radius where the disc light starts to dominate over the bulge,
we analyzed the one-dimensional surface brightness profiles, fitted the disk component with an exponential law and
estimated the contribution of the bulge as the flux above the exponential fit. 
The $r_{\rm disc 0}$ is defined as the one at which the surface brightness level coincides with
that of the exponential component of the fit.
We used the SDSS r-band imaging data,
available for the whole sample. 

The first thing that can be seen in the figures is that the mass-weighted age gradients
are very flat and the mass-weighted mean ages are, in most cases, around $\sim$ 10~Gyr,
even in the most external parts sampled by our data.
This is in agreement with results
from previous studies using long-slit spectroscopy (e.g., MacArthur et al. 2009; S\'anchez-Bl\'azquez et al. 2011), although
these were based on a reduced number of galaxies (8 and 4 respectively) and did not reach radii as large as 
those in the present study.
This result is also in agreement with studies of resolved stellar populations in nearby disk galaxies,
which find that {\it all} galaxies host a large percentage of old stellar populations at {\it all radii}
(Williams et al. 2009ab\nocite{2009ApJ...695L..15W}\nocite{2009AJ....137..419W}; Gogarten et al. 2010\nocite{2010ApJ...712..858G};
Barker et al. 2007\nocite{2007AJ....133.1138B}), although we do not reach radii as large as those sampled in studies of resolved
stellar populations.

\begin{table*}
\centering
\begin{tabular}{lrr|rr}
       & \multicolumn{2}{c}{$[Z/H]$} &  \multicolumn{2}{c}{loge Age}\\
       &  Mass-weighted   & Luminosity weighted &  Mass-weighted  & Luminosity weighted    \\
       &    slope         &     slope           &   slope         &  slope\\
IC1256 &   $ 0.25\pm0.05$   & $ 0.17\pm 0.03$     & $-0.01\pm 0.01$ & $-0.04\pm 0.07$\\
IC1683 &   $-0.25\pm0.02$   & $-0.17\pm 0.02$     & $-0.31\pm 0.02$ & $-0.02\pm 0.03$\\
NGC0001& $ 0.23\pm0.18$   & $ 0.01\pm 0.05$     & $ 0.20\pm 0.04$ & $-0.10\pm 0.12$\\
NGC0036& $ 0.08\pm0.05$   & $ 0.02\pm 0.07$     & $ 0.10\pm 0.05$ & $-0.26\pm 0.07$\\
NGC0160& $ 0.06\pm0.02$   & $ 0.02\pm 0.01$     & $-0.17\pm 0.02$ & $-0.19\pm 0.03$\\
NGC0214& $ 0.16\pm0.10$   & $ 0.23\pm 0.06$     & $-0.04\pm 0.07$ & $ 0.07\pm 0.16$\\
NGC0234& $-0.08\pm0.01$   & $ 0.81\pm 0.01$     & $-0.08\pm 0.01$ & $-0.08\pm 0.01$\\
NGC0257& $-0.004\pm0.027$ & $ 0.02\pm 0.01$     & $-0.14\pm 0.02$ & $-0.23\pm 0.05$\\
NGC0776& $ 0.00\pm0.02$   & $ 0.01\pm 0.01$     & $ 0.15\pm 0.03$ & $-0.05\pm 0.03$\\
NGC1167& $ 0.02\pm0.10$   & $-0.03\pm 0.13$     & $-0.17\pm 0.04$ & $-0.34\pm 0.07$\\
NGC1645& $ 0.02\pm0.08$   & $ 0.10\pm 0.09$     & $-0.18\pm 0.02$ & $-0.26\pm 0.10$\\
NGC2253& $-0.03\pm0.01$   & $ 0.01\pm 0.01$     & $-0.03\pm 0.01$ & $-0.01\pm 0.01$\\
NGC2347& $-0.10\pm0.03$   & $ 0.15\pm 0.05$     & $-0.20\pm 0.03$ & $-0.14\pm 0.04$\\
NGC2906& $-0.14\pm0.01$   & $-0.14\pm 0.01$     & $-0.011\pm 0.005$&$-0.14\pm 0.01$\\
NGC2916& $-0.11\pm0.07$   & $ 0.06\pm 0.09$     & $-0.05\pm 0.08$ & $-0.38\pm 0.05$\\
NGC3106& $ 0.11\pm0.02$   & $ 0.06\pm 0.09$     & $-0.12\pm 0.04$ & $-0.58\pm 0.09$\\
NGC3300& $-0.06\pm0.01$   & $-0.04\pm 0.01$     & $ 0.038\pm 0.003$&$ 0.029\pm 0.004$\\
NGC3614& $-0.11\pm0.04$   & $-0.05\pm 0.08$     & $-0.02\pm 0.04$ & $-0.52\pm 0.07$\\
NGC3687& $-0.21\pm0.02$   & $ 0.12\pm 0.02$     & $-0.01\pm 0.01$ & $ 0.15\pm 0.02$\\
NGC4047& $-0.49\pm 0.08$  & $ -0.25\pm 0.06$    & $ 0.37\pm 0.04$ & $ 0.43\pm 0.03$\\
NGC4185& $-0.16\pm 0.06$  & $-0.24\pm 0.05$     & $ 0.22\pm 0.04$ & $-0.00\pm 0.08$\\
NGC4210& $-0.16\pm0.02$   & $-0.15\pm 0.01$     & $-0.09\pm 0.02$ & $-0.04\pm 0.03$\\
NGC4470& $ 0.19\pm 0.02$  & $ 0.13\pm 0.02$     & $-0.10\pm 0.03$ & $ 0.01\pm 0.02$\\
NGC5000& $ 0.08\pm0.45$   & $ 0.08\pm 0.36$     & $-0.01\pm 0.19$ & $ 0.04\pm 0.14$\\
NGC5016& $-0.02\pm 0.02$  & $ 0.07\pm 0.01$     & $-0.13\pm 0.06$ & $-0.17\pm 0.04$\\
NGC5205& $-0.18\pm0.01$   & $-0.25\pm 0.01$     & $ 0.09\pm 0.01$ & $-0.06\pm 0.02$\\
NGC5218& $-0.04\pm0.01$   & $-0.02\pm 0.01$     & $ 0.17\pm 0.02$ & $-0.03\pm 0.01$\\
NGC5378& $-0.30\pm0.06$   & $-0.24\pm 0.03$     & $-0.02\pm 0.04$ & $-0.21\pm 0.06$\\
NGC5394& $-0.02\pm0.01$   & $-0.02\pm 0.01$     & $-0.03\pm 0.02$ & $ 0.02\pm 0.01$\\
NGC5406& $-0.02\pm0.01$   & $-0.02\pm 0.01$     & $-0.01\pm 0.02$ & $-0.09\pm 0.02$\\
NGC5614& $ 0.14\pm 0.04$  & $ 0.05\pm 0.03$     & $ 0.07\pm 0.04$ & $-0.02\pm 0.06$\\
NGC5633& $-0.06\pm 0.02$  & $-0.01\pm 0.02$     & $ 0.08\pm 0.03$ & $ 0.17\pm 0.02$\\
NGC5720& $-0.14\pm0.09$   & $-0.20\pm 0.09$     & $ 0.05\pm 0.04$ & $-0.36\pm 0.07$\\
NGC5732& $-0.21\pm0.08$   & $ 0.10\pm 0.06$     & $ 0.21\pm 0.11$ & $ 0.29\pm 0.19$\\
NGC5784& $-0.117\pm0.005$ & $-0.09\pm 0.03$     & $-0.08\pm 0.02$ & $-0.08\pm 0.02$\\
NGC6004& $-0.08\pm0.01$   & $-0.08\pm 0.01$     & $ 0.0045\pm 0.002$&$ 0.033\pm 0.005$\\
NGC6063& $-0.13\pm 0.05$  & $-0.07\pm 0.04$     & $ 0.05\pm 0.02$ & $-0.16\pm 0.03$\\
NGC6154& $-0.01\pm 0.04$  & $-0.02\pm 0.02$     & $-0.11\pm 0.02$ & $-0.51\pm 0.04$\\
NGC6155& $ 0.01\pm 0.04$  & $-0.01\pm 0.03$     & $ 0.06\pm 0.01$ & $ 0.10\pm 0.02$\\
NGC6301& $ 0.00\pm 0.01$  & $ 0.02\pm 0.01$     & $-0.05\pm 0.01$ & $-0.16\pm 0.01$\\
NGC6497& $ 0.01\pm 0.02$  & $-0.16\pm 0.02$     & $ 0.00\pm 0.02$ & $-0.21\pm 0.03$\\
NGC6941& $ 0.02\pm 0.04$  & $ 0.01\pm 0.02$     & $ 0.02\pm 0.05$ & $-0.20\pm 0.04$\\
NGC7025& $-0.20\pm 0.07$  & $-0.23\pm 0.08$     & $ 0.00\pm 0.02$ & $-0.01\pm 0.05$\\
NGC7321& $ 0.13\pm 0.048$ & $ 0.22\pm 0.03$     & $ 0.01\pm 0.01$ & $-0.05\pm 0.02$\\
NGC7489& $-0.50\pm 0.11$  & $-0.21\pm 0.09$     & $-0.29\pm 0.04$ & $-0.58\pm 0.07$\\
NGC7549& $-0.06\pm 0.02$  & $-0.05\pm 0.01$     & $-0.008\pm 0.005$ & $0.03\pm 0.01$\\
NGC7563& $-0.05\pm 0.02$  & $-0.04\pm 0.01$     & $ 0.026\pm 0.006$ & $-0.01\pm 0.01$\\
NGC7591& $-0.01\pm 0.01$  & $ 0.04\pm 0.01$     & $-0.02\pm 0.02$  & $-0.08\pm 0.01$\\
NGC7653& $-0.11\pm 0.06$  & $ 0.04\pm 0.01$     & $-0.10\pm 0.01$  & $-0.27\pm 0.01$\\
NGC7671& $-0.19\pm 0.01$  & $-0.16\pm 0.01$     & $ 0.07\pm 0.07$  & $ 0.08\pm 0.01$\\
NGC7782& $-0.03\pm 0.03$  & $-0.04\pm 0.02$     & $-0.06\pm 0.02$  & $-0.15\pm 0.06$\\
UGC00005&$-0.22\pm 0.03$  & $-0.045\pm 0.003$   & $ 0.03\pm 0.01$  & $0.03\pm 0.01$\\
UGC00036&$-0.27\pm 0.08$  & $-0.26\pm 0.03$     & $ 0.11\pm 0.03$  & $-0.27\pm 0.05$\\
UGC03253&$-0.10\pm 0.05$  & $-0.21\pm 0.06$     & $-0.09\pm 0.08$  & $-0.46\pm 0.06$\\
UGC07012&$ 0.06\pm 0.18$  & $-0.11\pm 0.19$     & $ 0.25\pm 0.08$  & $-0.02\pm 0.23$\\
UGC08234&$ 0.01\pm 0.01$  & $0.004\pm 0.004$    & $-0.06\pm 0.02$  & $-0.03\pm 0.01$\\
UGC10205&$-0.02\pm 0.16$  & $-0.01\pm 0.13$     & $ 0.13\pm 0.12$  & $ 0.18\pm 0.13$\\
UGC11649&$-0.43\pm 0.19$  & $-0.47\pm 0.18$     & $ 0.03\pm 0.06$  & $-0.09\pm 0.12$\\
UGC11680&$-0.07\pm 0.24$  & $ 0.03\pm 0.13$     & $-0.07\pm 0.19$  & $ 0.12\pm 0.32$\\
UGC12224&$ 0.00\pm 0.05$  & $0.14\pm 0.12$      & $-0.38\pm 0.09$  & $-0.20\pm 0.10$\\
UGC12816& $-0.05\pm 0.01$ & $ 0.000\pm 0.002$   &$-0.009\pm 0.002$  & $-0.018\pm 0.004$\\
\hline
\end{tabular}
\caption{Luminosity- and Mass-weighted log(age) and metallicity gradient
  calculated as the slope of a linear fit of the form $a+b log
  (r/r_{\rm eff}$ for our sample of galaxies.\label{tab:gradients}} 
\end{table*}

\begin{table*}
\centering
\begin{tabular}{lrr|rr}
       & \multicolumn{2}{c}{$[Z/H]$} &  \multicolumn{2}{c}{loge Age}\\
       &  Mass-weighted   & Luminosity weighted &  Mass-weighted  & Luminosity weighted    \\
       &    grad         &     grad           &   grad         &  grad\\
IC1256 &   $  0.08\pm 0.05$ & $  0.09\pm 0.03$ & $-0.01\pm 0.01$ & $-0.12  \pm 0.08$\\
IC1683 &   $-0.28\pm 0.02$ & $-0.01\pm 0.02$ & $-0.34\pm 0.02$ & $-0.07  \pm 0.03$\\
NGC0001& $ 0.14\pm 0.07$ & $-0.01\pm 0.05$ & $  0.05\pm 0.04$ & $-0.09  \pm 0.12$\\
NGC0036&  -- & -- & --& --\\
NGC0160&  -- & -- & --& --\\
NGC0214& $  0.08 \pm 0.10$ & $0.05 \pm 0.06$ &  $-0.01\pm 0.07$ & $-0.06 \pm 0.16$\\
NGC0234& $-0.22\pm  0.01$ & $0.10 \pm 0.01$ &  $-0.15\pm 0.01$ & $-0.08 \pm 0.01$\\
NGC0257& $  0.11\pm  0.03$ & $0.07\pm  0.014$ &$-0.09 \pm 0.02$ & $-0.20\pm 0.05$\\
NGC0776& $0.01 \pm 0.02$ & $0.05 \pm 0.01$ & $-0.10 \pm 0.03$ & $-0.07\pm 0.03$\\
NGC1167& -- & -- & --& --\\
NGC1645&  -- & -- & --& -- \\
NGC2253& $-0.01 \pm 0.03$ & $-0.07 \pm 0.03$ & $-0.01 \pm 0.01$ & $-0.14 \pm 0.01$\\
NGC2347& $-0.10 \pm 0.03$ & $  0.06 \pm 0.04$ & $-0.07\pm  0.03$ & $-0.06\pm 0.04$\\
NGC2906& $-0.05\pm 0.01$ & $-0.03\pm 0.01$ & $0.005 \pm 0.005$ & $-0.057 \pm 0.007$\\
NGC2916& $-0.16 \pm 0.07$ & $-0.13\pm 0.11$ & $-0.15\pm 0.08$ & $-0.54  \pm 0.05$\\
NGC3106&  -- & -- & --& -- \\
NGC3300& $-0.050\pm   0.006$ & $-0.036\pm 0.005$ & $0.018 \pm 0.003$ & $0.007\pm  0.004$\\
NGC3614&  -- & -- & --& -- \\
NGC3687& $-0.11 \pm 0.03$ & $0.01 \pm 0.02$ & $-0.10\pm 0.01$ & $-0.20\pm 0.02$\\
NGC4047& $-0.28\pm 0.08$ & $-0.16\pm 0.06$ & $0.28 \pm 0.04$& $0.23\pm  0.03$\\
NGC4185& $-0.10 \pm 0.06$ & $-0.17\pm 0.05$ & $0.03\pm 0.04$ & $-0.15\pm 0.08$\\
NGC4210& $-0.15\pm 0.02$ & $-0.15\pm 0.01$ & $-0.07\pm 0.02$ & $-0.32\pm 0.03$\\
NGC4470& $0.14\pm 0.02$ & $0.12\pm 0.02$ & $-0.08\pm 0.03$ & $0.03\pm 0.02$\\
NGC5000& $-0.02\pm 0.14$ & $0.10\pm 0.07$ & $-0.20\pm 0.11$ & $-0.15\pm 0.19$\\
NGC5016& $0.01\pm 0.02$ & $0.009\pm 0.009$ & $-0.09\pm 0.06$ & $-0.09\pm 0.04$\\
NGC5205& $-0.164\pm 0.005$ & $-0.04\pm 0.01$ & $-0.01\pm 0.01$ & $-0.23\pm 0.02$\\
NGC5218& $-0.09\pm 0.01$& $-0.027\pm 0.004$& $-0.25\pm 0.02$& $-0.11\pm 0.01$\\
NGC5378& $  0.24\pm 0.12$& $0.01\pm 0.06$& $-0.77\pm 0.04$ & $-0.42\pm 0.08$\\  
NGC5394& $-0.04\pm 0.01$ & $-0.02\pm 0.01$ & $-0.08\pm 0.02$ & $-0.02\pm 0.01$\\
NGC5406& $-0.07\pm 0.01$ & $-0.09\pm 0.01$ & $0.003 \pm 0.020$ & $-0.17 \pm 0.02$\\
NGC5614& $ 0.18\pm 0.04$ & $0.24 \pm 0.03$ & $0.15\pm 0.04$ & $-0.11 \pm 0.06$\\
NGC5633& $-0.07 \pm 0.02$ & $-0.02\pm 0.02$ & $0.03\pm 0.02$ & $0.09\pm 0.02$\\
NGC5720& $-0.14\pm 0.12$ & $-0.25\pm 0.22$ & $0.02\pm 0.05$ & $-0.37\pm 0.07$\\
NGC5732& $-0.09\pm 0.20$ & $0.04\pm 0.13$ & $0.04\pm 0.11$ & $0.18\pm 0.19$\\
NGC5784& $-0.08\pm 0.01$ & $-0.06\pm 0.03$ & $0.10\pm 0.02$ & $0.13\pm 0.02$\\
NGC6004& $-0.09\pm 0.01$ & $-0.091\pm 0.004$ & $0.000 \pm 0.002$ & $-0.0652 \pm 0.0054$\\
NGC6063& $-0.18 \pm 0.05$ & $-0.16\pm 0.04$ & $-0.05\pm 0.02$ & $-0.31\pm 0.03$\\
NGC6154& $-0.02\pm 0.04$ & $-0.03\pm 0.04$ & $-0.08\pm 0.03$ & $-0.29\pm 0.05$\\
NGC6155& $-0.05\pm 0.04$& $-0.001\pm 0.026$ & $-0.04\pm 0.01$ & $-0.01\pm 0.02$\\ 
NGC6301& $  0.028\pm 0.009$ & $0.08\pm 0.02$ & $-0.05\pm 0.01$ & $-0.13 \pm 0.01$\\
NGC6497& $-0.00\pm 0.02$ & $-0.07\pm  0.02$ & $-0.04\pm 0.02$ & $-0.15\pm 0.03$\\
NGC6941& $-0.02\pm 0.03$ & $-0.10\pm 0.04$ & $ 0.03\pm 0.05$ & $-0.35\pm 0.05$\\ 
NGC7025& $-0.13\pm 0.07$ & $-0.14\pm 0.08$ & $-0.04\pm 0.02$ & $-0.11\pm 0.05$\\ 
NGC7321& $-0.21\pm 0.05$ & $-0.21\pm 0.03$ & $0.06\pm 0.01$ & $-0.16\pm 0.02$\\
NGC7489& $-0.60\pm 0.07$ & $-0.24\pm 0.09$ & $-0.37\pm 0.04$ & $-0.63\pm 0.07$\\
NGC7549& $0.01\pm 0.02$ & $0.006\pm 0.009$ & $0.004\pm 0.005$ & $-0.01\pm 0.01$\\
NGC7563& $-0.003\pm 0.014$ & $ 0.00\pm 0.01$ & $0.001 \pm 0.009$ & $-0.002 \pm 0.01$\\
NGC7591& $-0.039 \pm 0.013$ & $-0.015\pm 0.013$ & $-0.015\pm 0.021$ & $-0.045\pm 0.013$\\
NGC7653& -- & -- & --& -- \\
NGC7671& $-0.18\pm 0.01$ & $-0.15\pm 0.01$ & $0.07\pm 0.01$ & $0.08\pm 0.01$\\
NGC7782& $-0.14 \pm 0.01$ & $-0.11\pm 0.01$ & $-0.11\pm 0.01$ & $-0.28\pm 0.03$\\
UGC00005&$-0.17\pm 0.03$ & $-0.033\pm 0.003$ & $0.02 \pm 0.01$ & $0.01\pm 0.01$\\
UGC00036&-- & -- & --& -- \\
UGC03253&-- & -- & --& -- \\
UGC07012&-- & -- & --& -- \\
UGC08234& $-0.02 \pm 0.01$ & $-0.008 \pm 0.004$ & $0.02 \pm 0.02$ & $0.01 \pm 0.01$\\
UGC10205&$-0.09 \pm 0.16$ & $ 0.00\pm 0.13$ & $0.00\pm 0.12$ & $ 0.00 \pm 0.13$\\
UGC11649&$-0.51\pm 0.17$ & $-0.52\pm 0.16$ & $0.03\pm 0.05$ & $-0.26\pm 0.04$\\
UGC11680&-- & -- & --& -- \\
UGC12224&-- & -- & --& -- \\
UGC12816&-- & -- & --& -- \\
\hline
\end{tabular}
\caption{Luminosity- and Mass-weighted log(age) and metallicity gradient
  calculated as the difference of the metallicity and age between
  1.5~$r_{\rm eff}$ and $r_{\rm disc 0}$.\label{tab:gradients2}} 
\end{table*}

{\bf Metallicity gradients:} it has been well established that in a disk growing inside-out, metallicity decreases
from the center of the galaxy outwards (Goetz \& Koeppen 1992\nocite{1992A&A...262..455G};
Matteucci \& Francois
1989\nocite{1989MNRAS.239..885M}). Figure~\ref{fig:metgrads1}, \ref{fig:metgrads2}
and \ref{fig:metgrads3} shows
the variation of the light- and mass-weighted metallicity gradients for our sample of barred, unbarred
and weakly barred galaxies respectively. It can be seen that, in general, we find this decrease of metallicity
with radius, although with some exceptions.
The light-weighted metallicity (biased towards the values of the youngest stars) is always larger than the 
mass weighted value (which largely reflects the metallicity of the old stars), showing the metallicity 
evolution of the galaxy, where new stars formed from gas enriched by the  previous stellar generations.
The slopes of the gradients are, on the other hand, very similar for the mass- and light-weighted components.
We will analyse in more detail the  metallicity gradient in populations of different
ages in Sec~\ref{sec:evolution}.

\subsection{Stellar population gradients}
We  quantify the variation of  age and metallicity with radius in two
different ways:
\begin{enumerate}
\item method 1: we perform a linear fit in the radial region where
the surface brightness profile is dominated by the disc.
\item method 2: we measure the stellar parameters difference between 
1.5~r$_{\rm eff}$ (2.5 scale lengths of the disc) and  $r_{\rm disc 0}$, which correspond the radius at which the light starts being
dominated by the disc.
\end{enumerate}

For barred galaxies, numerical simulations predict that  the flattening
of the metallicity gradients due to the non linear coupling of a bar and spiral arms is 
more prominent beyond the bar corotation, which causes variations in the slope of the gradient
at this radius  (e.g., Friedli 1998;
Minchev et al. 2013; Di Matteo et al. 2013).  Therefore, for the method 1 we use 
only those values at radii larger than the bar corotation.  
However, in the majority of the galaxies we do not
see any variation of the slope at this radius and, therefore, we perform the linear fit in the
whole disc region.
Note that we have not calculated the corotation radius of the bars. 
However,  recent results by Aguerri et al. (2013) -- using
the Tremaine \& Weinberg (1984)  method to calculate bar pattern speed -- show that the  mean value of 
the ratio  R$_{\rm CR}$/R$_{\rm bar}$ is around 1 (where
$R_{\rm CR}$ is the corotation radius and R$_{\rm bar}$ the length of the bar) and
that there is not much dependence with the morphological type. Therefore, we will use the bar length as an orientative position of the bar corotation.
Throughout the paper we designate  the slope of the linear fits:
$[$Z/H$]$=a+b (r/r$_{\rm eff}$) and log Age=a+b (r/r$_{\rm eff}$) with  grad[Z/H] and grad Age respectively.

For  method~2  we always measure the
difference between  the metallicity (and age) at 
$r_{\rm disc 0}$ and at 1.5~$r_{\rm eff}$, independently of the presence
or not of the bar. 

\begin{figure}
\centering
\resizebox{0.5\textwidth}{!}{\includegraphics[angle=0]{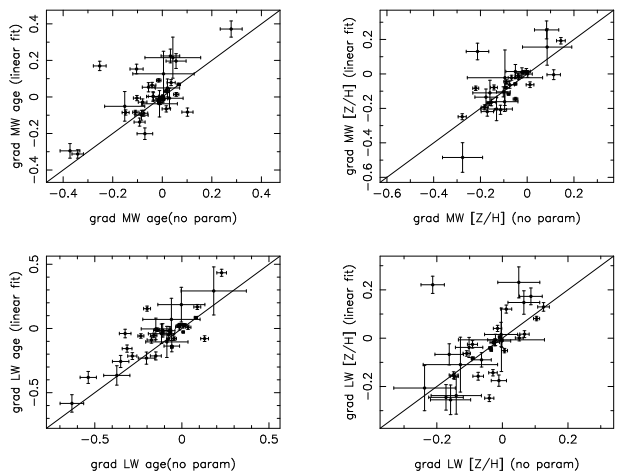}}
\caption{Comparison of the stellar population gradients calculated
  with a linear fit (y-axis) and calculated as the difference between
  the  parameters (age or metallicity)  at 1.5~$r_{\rm eff}$ and at $r_{0\rm disc}$ 
(see text for details).\label{comparison_methods}}
\end{figure}

Figure~\ref{comparison_methods} shows the comparison of the gradients
obtained  with both methods. As can be
seen, the values  are compatible within the errors.
Tables~\ref{tab:gradients} and \ref{tab:gradients2} 
list all the values calculated with method~1 and 2 respectively.
In the second case
we only consider those galaxies for which we reach 1.5~$r_{\rm eff}$.
For the rest of the paper we will use the 
gradients calculated with the method~2, as they are more readily reproducible and do not assume any
shape on the variation of the stellar parameters with radius. 

\begin{figure}
\centering
\resizebox{0.5\textwidth}{!}{\includegraphics[angle=0]{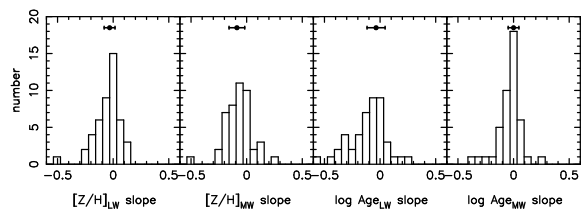}}
\caption{Histograms showing the distribution of stellar population
  gradients. From left to right: Luminosity-weighted metallicy,
  mass-weighted metallicity, luminosity weighted age and mass
  weighted age. The dot with the error bars show the value of the
  mean and its error. \label{histogram_mean}}
\end{figure}
Figure~\ref{histogram_mean} shows the distribution of the stellar
population  gradients for the complete sample of galaxies.
The mean age gradient for the whole sample is $-0.036\pm 0.010$ and
$0.000\pm 0.006$ dex/r$_{\rm eff}$ when averaging with the light and the
mass respectively, while the mean light- and mass-weighted
metallicity gradients are $-0.032\pm 0.006$ dex/r$_{\rm eff}$
and $-0.087\pm  0.008$ dex/r$_{\rm eff}$ respectively.
 We stress here that these values represent the gradients in the disc
regions, excluding the bulge.
As can be seen, the gradients in both quantities are very shallow.

\subsection{Comparison between barred and unbarred galaxies}
\label{sec:comparison}

\begin{table}
\begin{tabular}{lrlll}
gradient          & Mean & RMS   & RMS$_{exp}$ & N$_{\rm gal}$  \\
\hline
MW age (barred)   & 0.011  & 0.040 & 0.001     & 28   \\
MW age (unbarred) &$-0.008$& 0.044 & 0.002     & 27  \\
\hline
LW age (barred)   &$-0.017$&  0.059& 0.002    & 28 \\
LW age (unbarred) &$-0.051$&  0.103& 0.002    & 27\\
\hline
MW [Z/H] (barred)  &$-0.092$& 0.067 & 0.002   & 25\\
MW [Z/H] (unbarred)&$-0.063$& 0.083 & 0.003   & 23\\
\hline
MW [Z/H] (barred) &$-0.040$ & 0.068 & 0.002  & 26\\
MW [Z/H] (unbarred) &$-0.013$ & 0.043 & 0.002 & 25\\
\hline
\end{tabular}
\caption{Mean values for the gradient slope in the different
stellar population parameters for barred and unbarred galaxies.
Third column indicate the root-mean-square dispersion among the mean
values and forth column the dispersion expected by the measured errors
in the gradients. Last column indicate the number of galaxies used to
calculate these values.\label{tab:means}}
\end{table}

\begin{figure*}
\centering
\resizebox{0.47\textwidth}{!}{\includegraphics[angle=0]{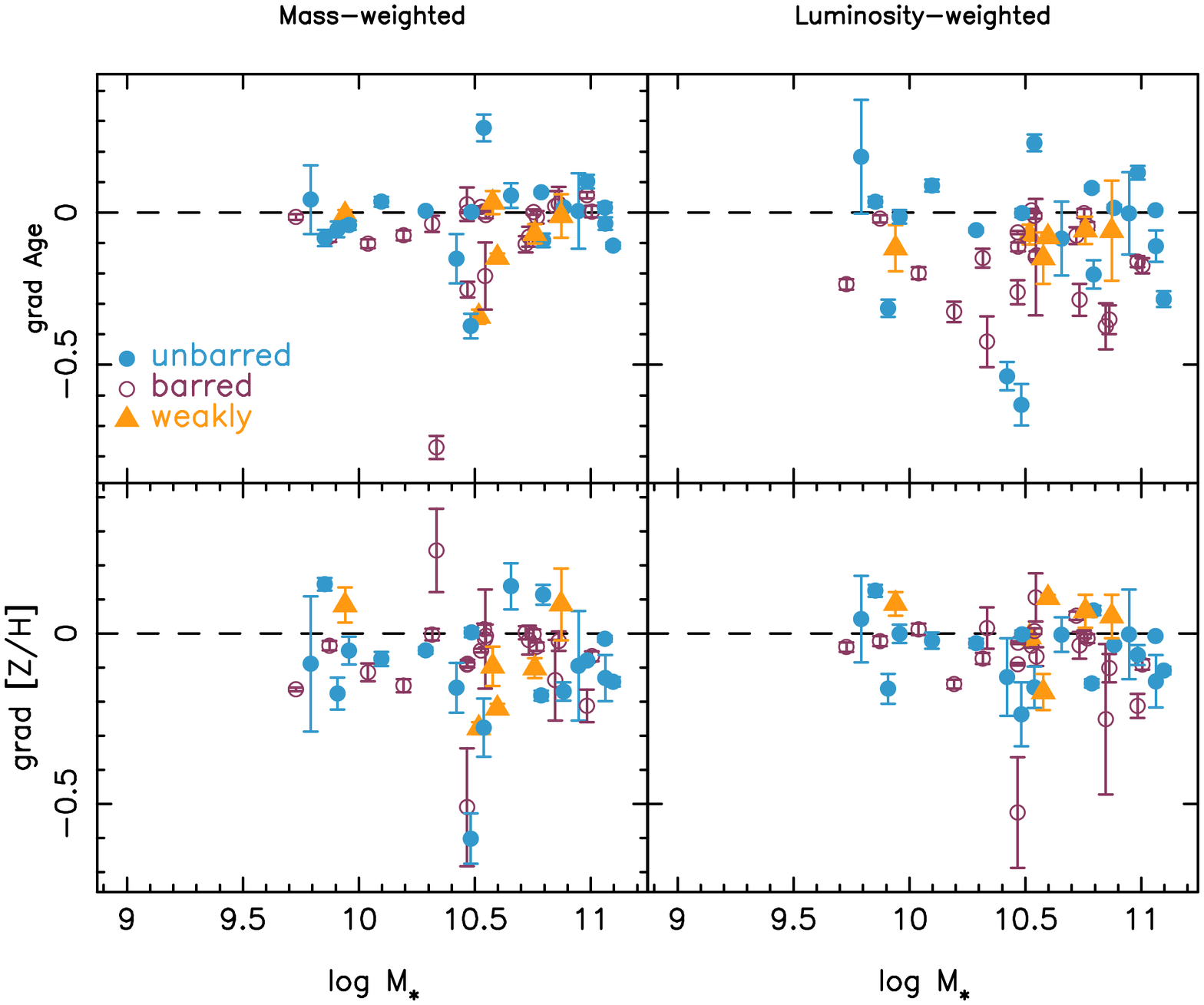}}
\resizebox{0.47\textwidth}{!}{\includegraphics[angle=0]{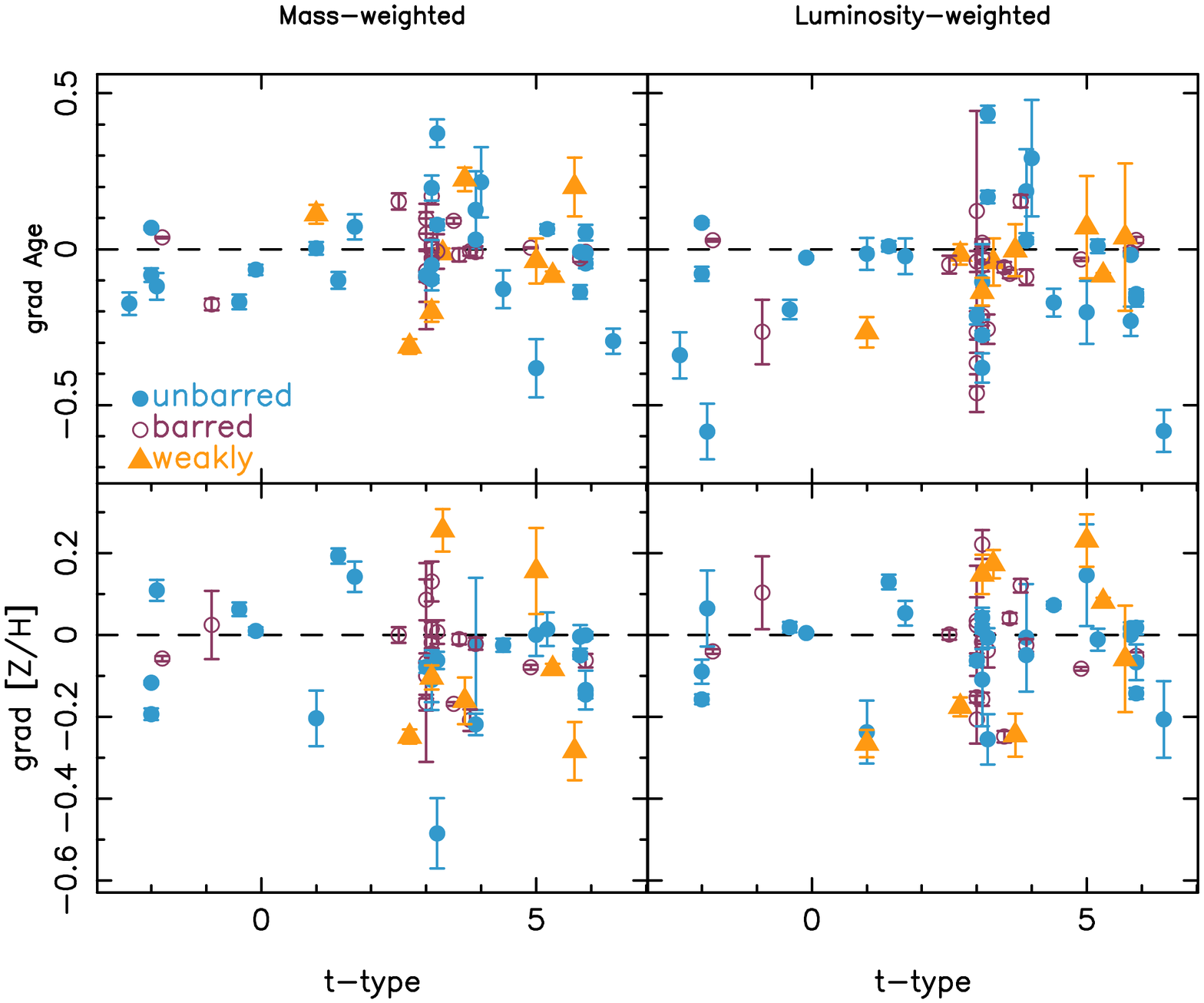}}
\caption{Age (top panels) and metallicity (bottom panels) gradients as a function of the
stellar mass and the t-type for our sample of barred (red) and unbarred (blue) galaxies. We
have also represented weakly barred galaxies in orange. Both, mass- (left panels)
and light-weighted (right panel) values
are represented.\label{fig:comp}}
\end{figure*}

The main goal of the present work is to quantify the influence of bars in shaping 
the metallicity gradient in disc galaxies.
Figure~\ref{fig:comp} shows the metallicity gradients (in the disc region) as a function of the stellar 
mass and the morphological t-type.

\begin{figure*}
\centering
\resizebox{0.57\textwidth}{!}{\includegraphics[angle=0]{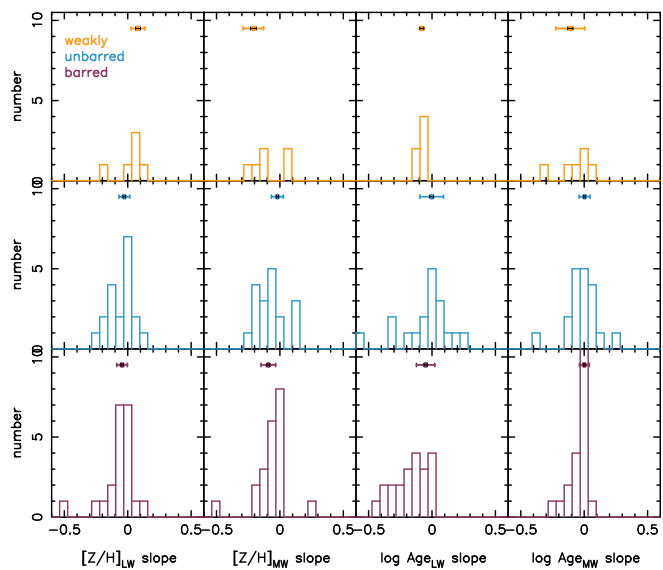}}
\caption{Histograms of the slope of the gradients of : luminosity-weighted [Z/H] (first
column); mass-weighted metallicity (second column); luminosity-weighted age (3rd column)
and mass-weighted age (4th column), for the different subsamples: weakly barred (top raw); unbarred
(middle raw) and barred galaxies (bottom raw). the point on the top of the diagram indicate the mean
value (weighted with the errors) and the error bar the RMS dispersion. The white error bar show the dispersion expected by the
errors.\label{fig_hist}}
\end{figure*}

As can be seen in the figure, contrary to the prediction of numerical simulations
we do not find any difference in the values of the slope
between the gradients of barred and unbarred galaxies.
Table~\ref{tab:means} lists the mean values of the gradients for the different subsamples.
To visualize this in an easier way we plot, in Fig~\ref{fig_hist},
the histograms showing the slope of the gradients in the different parameters
for the three subgroups (barred, unbarred and weakly barred galaxies).
A t-test shows that the difference between the mean values of barred and unbarred galaxies are not
significant at a 95\% probability level. This is true for the age and the metallicity (both for the 
mass-weighted and the luminosity-weighted values).

It is interesting to note, also, the lack of correlation between the slope of the
gradients and the stellar mass or the t-type of the galaxies. There may be a weak correlation between the luminosity
weighted metallicity gradient and the stellar mass, in the sense that more  massive galaxies have
a steeper gradient, but this correlation is not statistically significant.
This lack of trends is similar to that found in
the {\it gas-phase} metallicity  (Diaz 1989\nocite{1989epg..conf..377D};
S\'anchez et al. 2012b\nocite{2012A&A...546A...2S}; S\'anchez et al. 2014) when the metallicity gradient is
normalized to a physical scale of the disc (e.g., the R$_{25}$ or the disc scale-length).

\begin{figure}
\centering
\resizebox{0.47\textwidth}{!}{\includegraphics[angle=0]{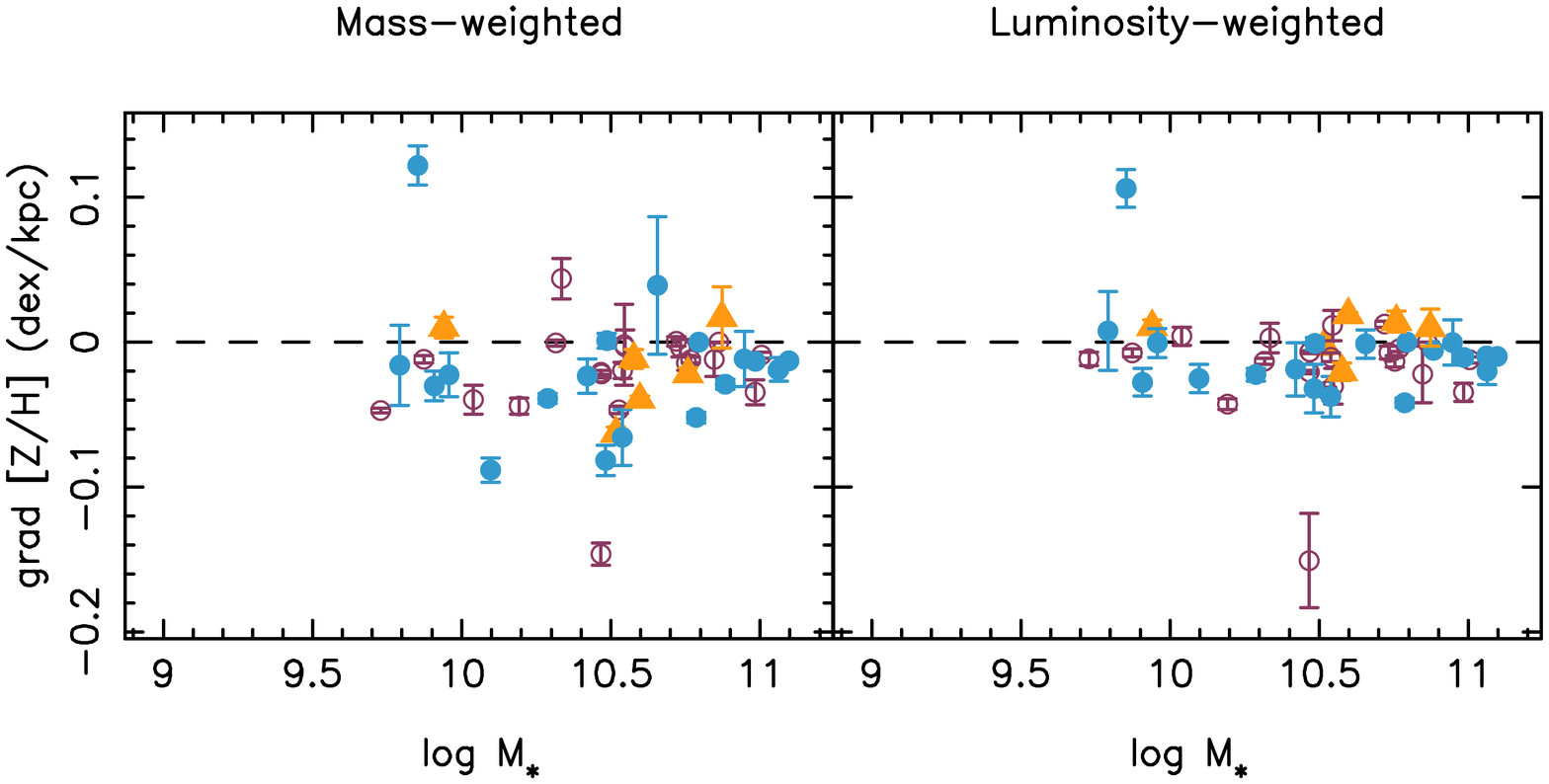}}
\caption{Mass-(left) and luminosity-weighted (right) gradients vs. stellar mass. The slope of the gradient
has been calculated in dex/kpc.\label{fig:kpc}}
\end{figure}

On the other hand, some studies have found a correlation between the gas-phase metallicity
gradient (measured in physical scales, dex/kpc) and the morphological type or the 
mass, with early-type, and more  massive spirals showing
shallower slopes (e.g., McCall et al. 1985\nocite{1985ApJS...57....1M}; Vila-Costas \& Edmunds
1992\nocite{1992MNRAS.259..121V}, Vila-Costas \& Edmunds\nocite{1992MNRAS.259..121V};
Oey \& Kennicutt 1993\nocite{1993ApJ...411..137O}).
To see if a trend  between the metallicity gradient and the mass is visible in our data 
when the gradient is measured in dex/kpc, we repeated the fits without normalizing to the effective radius
of the disc and transforming the radius to kpc. The results are shown in Fig.~\ref{fig:kpc}.
As can be seen, even in this case, we do not see clear trends between the stellar metallicity gradient and the mass.

\begin{figure}
\centering
\resizebox{0.47\textwidth}{!}{\includegraphics[angle=0]{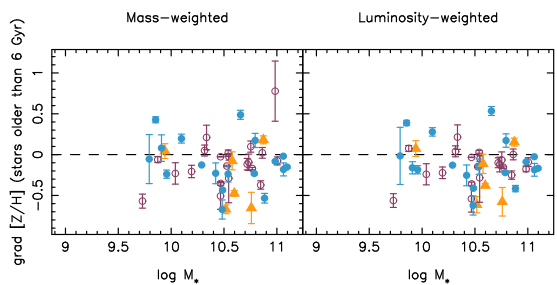}}
\caption{Mass- (left panel) and luminosity-weighted (right panel) stellar metallicity gradient as a
function of total stellar mass for stars older than 6~Gyr. The colors of the symbols indicate if the galaxies are barred, unbarred
or host a weakly bar, as indicated in the label. \label{fig:grad_old}}
\end{figure}

Another prediction of numerical  simulations is that the flattening of the metallicity gradient
due to radial mixing is stronger
for the old populations, as they have had more time to migrate along the disc. Because one of the
{\tt STECKMAP} outputs is the age-metallicity relation, we are, in principle, able to obtain a metallicity
gradient for the old stars.
We have obtained the mean metallicity for stars with ages$>$6 Gyr.
This is what is shown in Fig.~\ref{fig:grad_old}.
The mean luminosity-weighted values of the metallicity gradients are $-0.04\pm 0.06$ and
$-0.06\pm 0.07$ for unbarred and barred galaxies respectively (the
error represent the RMS dispersion). A t-test comparing the mean values show, however,  that the
differences are not statistically significant.

However, the  expectation for a flatter metallicity gradient
in old stars is based on predictions of idealised numerical simulations
where the disc is evolved in isolation  (e.g., Minchev et al. 2012). Recently, it has been shown by
Minchev et al. (2014) -- using zoom cosmological simulations of disk
galaxies presented in Martig et al. (2012)-- that the large vertical velocity dispersion of 
old stars (which were born hot at high redshift,
 or get heated early-on due to strong merger activity) prevent
 strong migration efficiency in this population (see also Brunetti et al. 2001).
 It worth noting, nevertheless, that 
 the vertical velocity dispersion of the stars in
cosmological and non-cosmological simulations is still considerably larger than what is observed
in our own Galaxy, even for the oldest stars 
(House et al. 2011\nocite{2011MNRAS.415.2652H})

We have test if these results are reliable by testing if 
{\tt STECKMAP} is able to recover the age metallicity relation. We performed a series
of experiments where a synthetic spectra of simulated SFHs with different metallicity evolution
were introduced as input in {\tt STECKMAP}, previously degraded to the resolution of our data
and adding noise to reach our Voronoi limit of 40. These tests are shown in appendix~\ref{appendix1}.
As can be seen, although the scatter is larger at old ages,  we can
reproduce the input trend and recover  the age-metallicity relation for
synthetic  data with similar characteristics than the data used in this study.

\subsection{Comparison of the gradients with the bar properties}
\begin{figure}
\centering
\resizebox{0.57\textwidth}{!}{\includegraphics[angle=0]{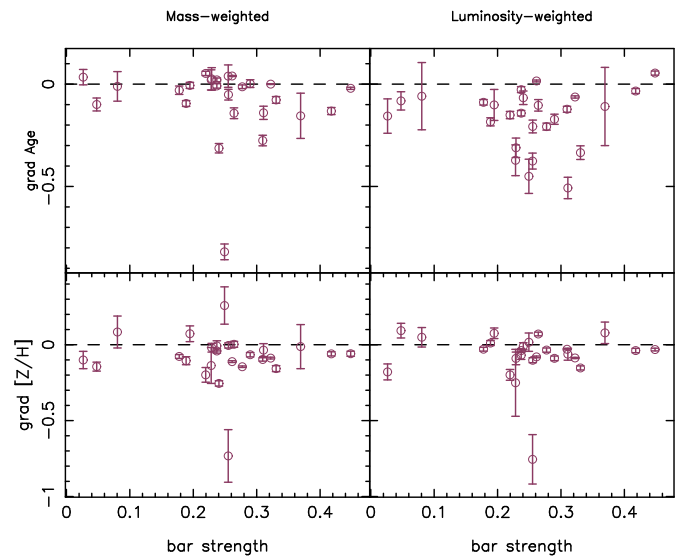}}
\caption{Comparison of the metallicity gradients (measured in dex/r$_{\rm eff}$) of barred
galaxies {\it vs} the  strength of the bar.\label{fig:barproperties}}
\end{figure}

The changes in angular momentum predicted in numerical simulations are 
larger for stronger bars (e.g., Minchev \& Famaey 2010).
 Figure~\ref{fig:barproperties}
shows the metallicity and age gradient {\it vs.} the strength of the bar.
As can be seen, we do not find any correlation between the slope of the stellar population properties
and the properties of the bar.
\section{Metallicity gradients for stellar populations of different ages}
\label{sec:evolution}
Inside-out formation may lead to some
specific evolution of the abundance gradients. How do the abundance gradients evolve within the
disk? 
While most galactic chemical evolution
models are able to reproduce the present-day radial distribution of several chemical elements
derived from a sample objects representative of the present-day composition of the interstellar
medium, they generally disagree on the predicted behavior of its time evolution (compare, e.g.,
the models of  Tosi 1988\nocite{1988A&A...197...33T};
Chiappini et al. 1997\nocite{1997ApJ...477..765C}; Chiappini et al.
2001\nocite{2001ApJ...554.1044C} -- which predict a
steeping of the gradients with time, with those of Moll\'a et al.\
1997\nocite{1997ApJ...475..519M}; Portinari \& Chiosi
1999\nocite{1999A&A...350..827P} and Hou et al. 2000\nocite{2000A&A...362..921H} -- which predict a flattening with time). The main differences between
the two $''$types$''$ of models are the efficiency of the enrichment processes in the inner and
outer regions of the disc and the degree of enrichment of the infall material. The same happens
with fully cosmological hydrodynamical simulations. Pilkington et
al. (2012)\nocite{2012A&A...540A..56P}\nocite{2013A&A...554A..47G} analysed the metallicity
gradient and its evolution with time in a suite of
25 cosmological simulations of disc galaxies with similar properties to the Milky Way finding that, while the
majority of the models predict radial gradients today which are consistent with those observed in late-type discs, they evolve
in different fashion. They also found that the main differences came from the radial efficiency of the star
formation rate that are controlled, mainly, by the different prescriptions of feedback (Gibson et al. 2013).

\begin{figure}
\centering
\resizebox{0.47\textwidth}{!}{\includegraphics[angle=0]{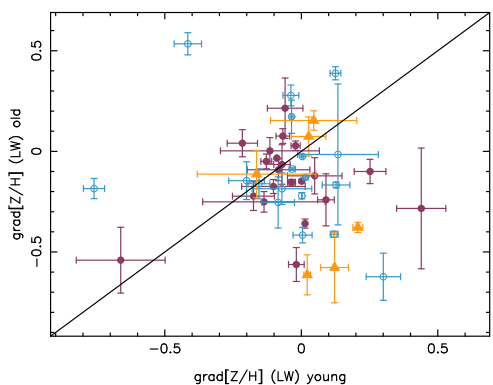}}
\caption{Comparison of the metallicity gradient for the stellar population with an age
older and younger than 6~Gyr.\label{fig:compa_zl}}
\end{figure}

\begin{figure*}
\centering
\resizebox{0.33\textwidth}{!}{\includegraphics[angle=0]{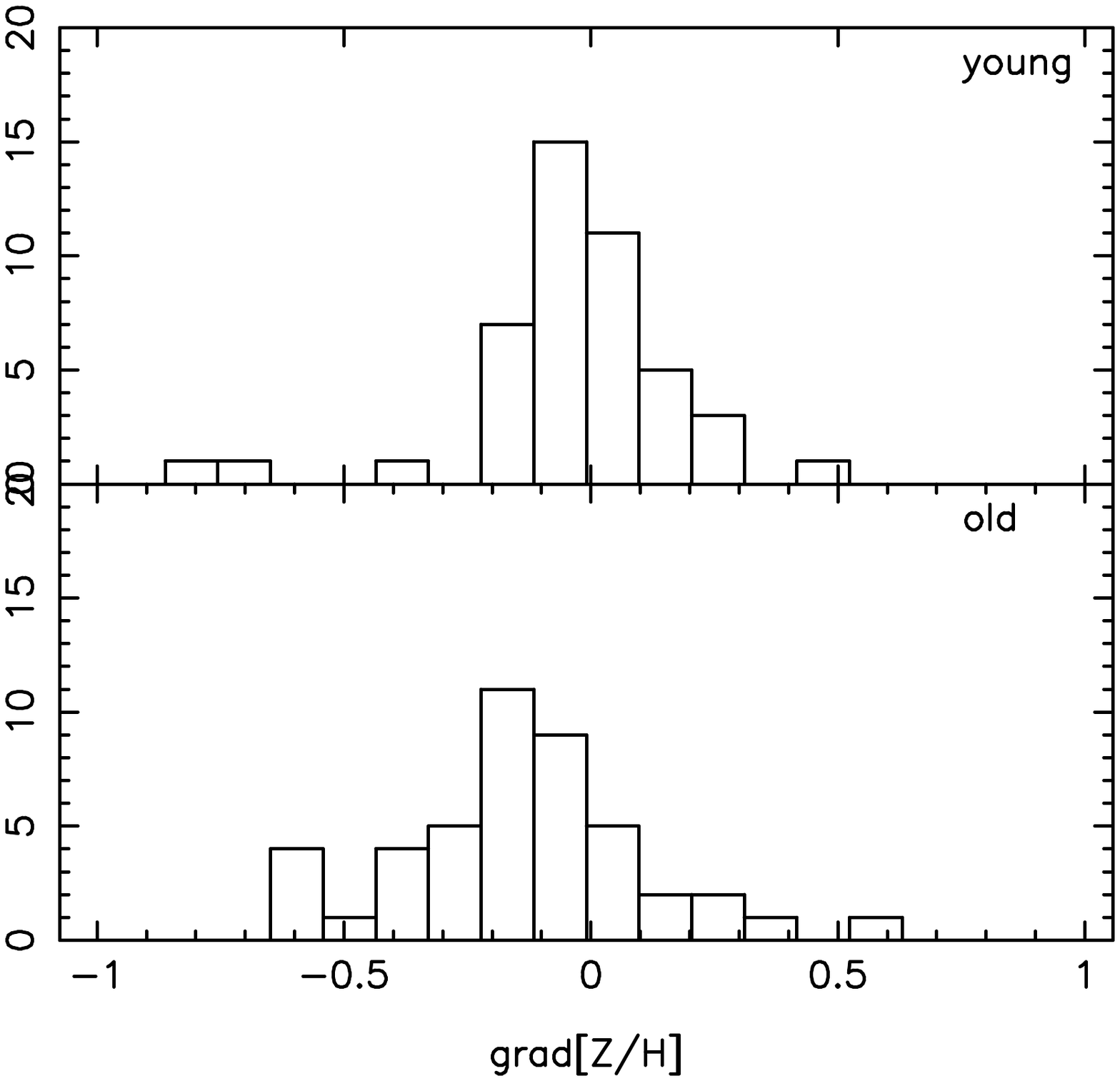}}
\resizebox{0.33\textwidth}{!}{\includegraphics[angle=0]{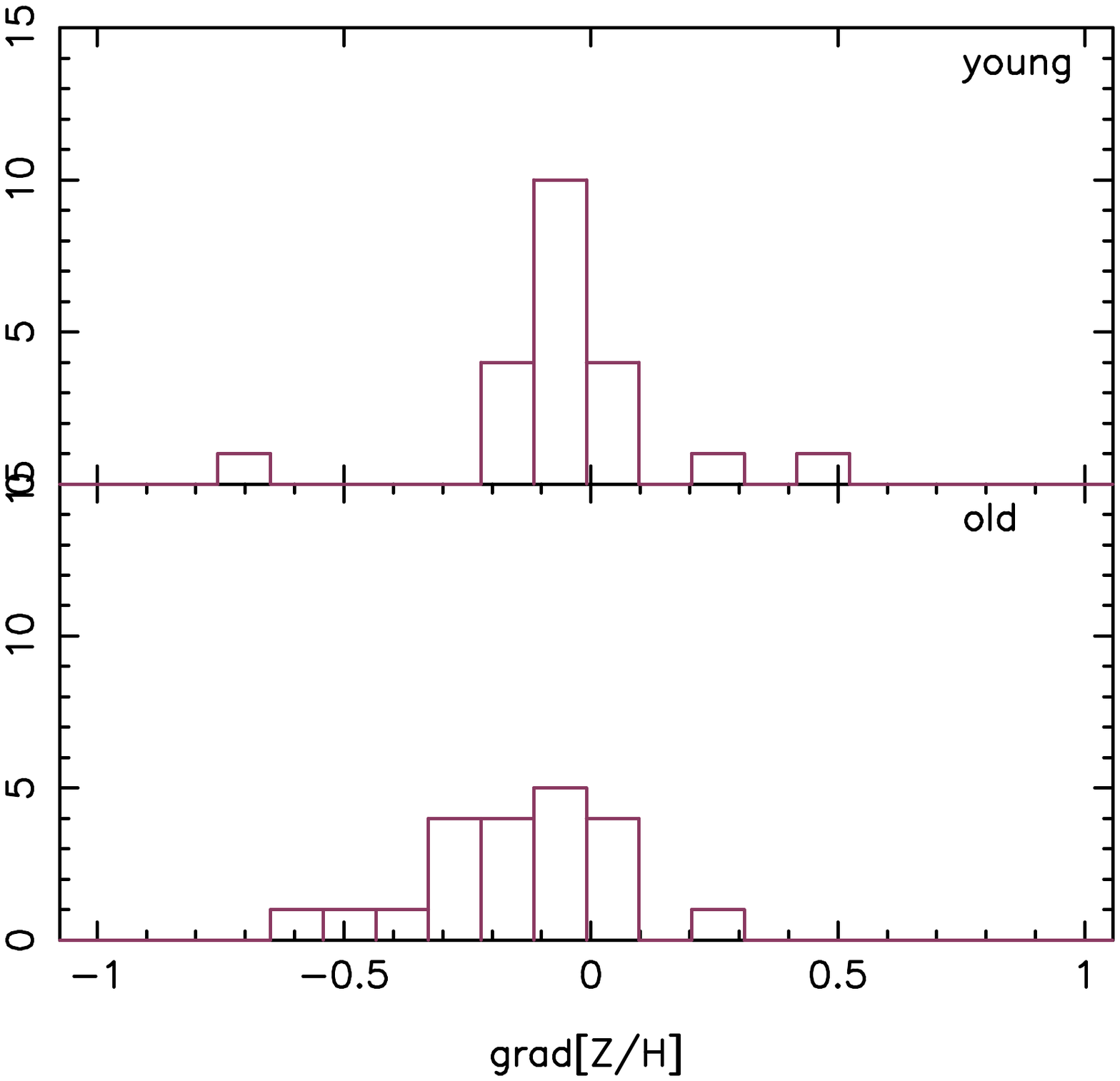}}
\resizebox{0.33\textwidth}{!}{\includegraphics[angle=0]{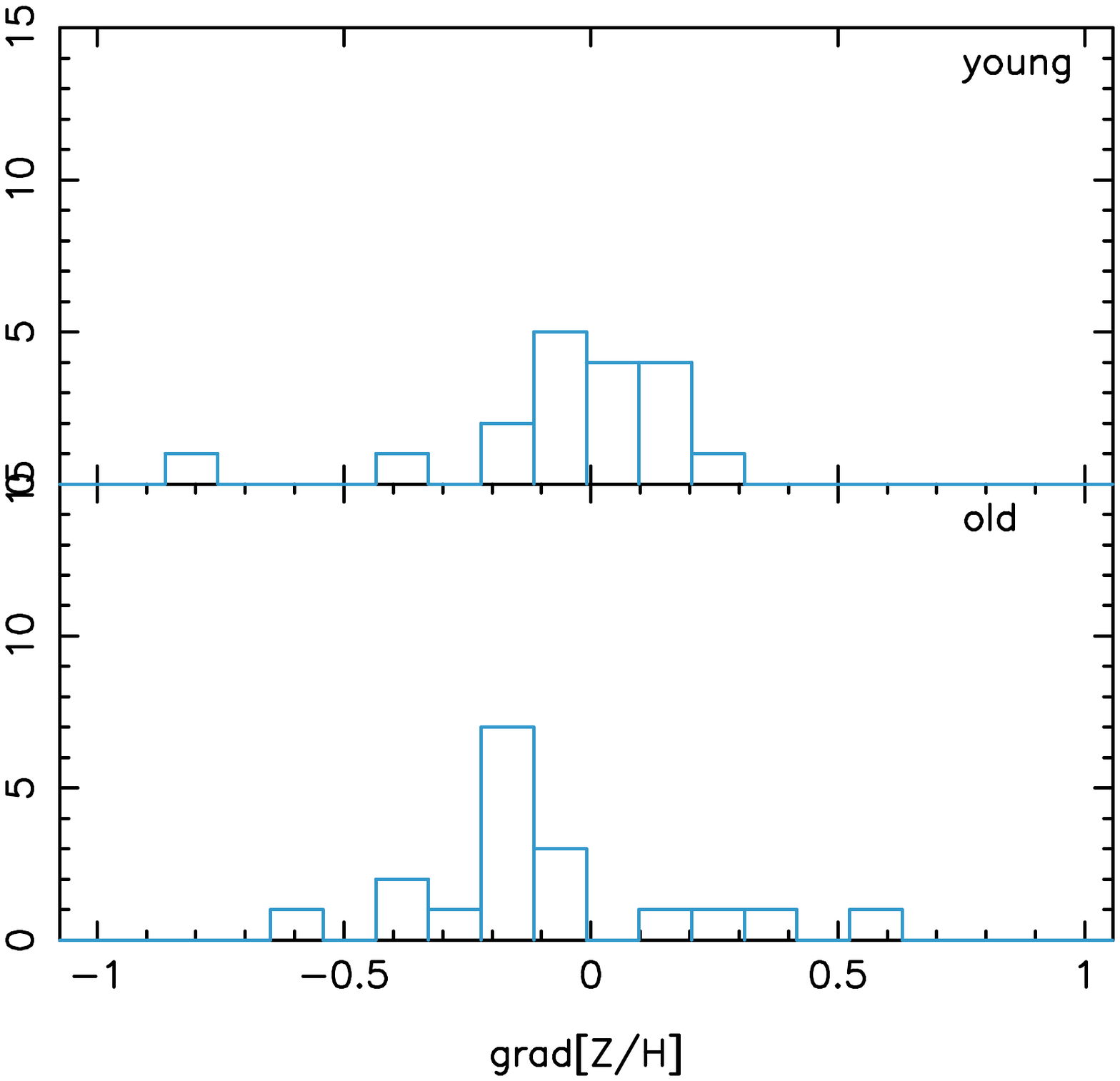}}
\caption{Mean luminosity-weighted metallicity gradients for stellar populations 
with ages older and younger than 6~Gyr. Black histogram represent the whole sample
while the blue and the red histograms show the results for unbarred and barred galaxies respectively. \label{hist_grad_age}}
\end{figure*}

Up to now, these predictions could not  be constrained due to the lack of observational results.
Because the output of {\tt STECKMAP} gives us the whole evolution of the metallicity with time,
we are  able to obtain the metallicity gradient for population of different
ages (see appendix~\ref{appendix1} for tests supporting the reability
of the results).  With the same methodolgy that we
used to obtain gradients  for the whole population, we measure the metallicity gradient
for stars with ages greater than 6~Gyr and  lower than 2~Gyr.
Figure~\ref{fig:compa_zl} compares the slopes of these two populations
for the same galaxies while, in Fig.~\ref{hist_grad_age}
we show the distribution of the slopes.
As can be seen, the metallicity gradient for
the young component is flatter (and, in many cases, positive) compared with that of the old
component.  However. the differences are not statistically significant.
The mean values are
($0.002\pm 0.208$)  and ($-0.005\pm 0.262$) dex/r$_{\rm eff}$ for the young and the old component
respectively, where the quoted error represent the RMS dispersion.
Flatter abundance gradients in the younger population are also found
in the Milky Way
(Friel et al.\ 2002\nocite{2002AJ....124.2693F};
Chen, Hou \& Wang 2003\nocite{2003AJ....125.1397C}; Maciel, Costa \& Uchida 2003\nocite{2003A&A...397..667M};
Daflon \& Cunha 2004\nocite{2004ApJ...617.1115D}),
although recent results by Maciel \& Costa (2013)\nocite{2013arXiv1308.1884M} do not detect variations in the slope of the
metallicity gradient measured in planetary nebulae of different ages.
We do not find, either, any difference in the mean of the gradients
for the old and young populations between   galaxies with and without bars.

\section{Relation between the stellar populations of  discs and bulges}
\label{sec:bulge_disk}
\begin{figure}
\centering
\resizebox{0.53\textwidth}{!}{\includegraphics[angle=0]{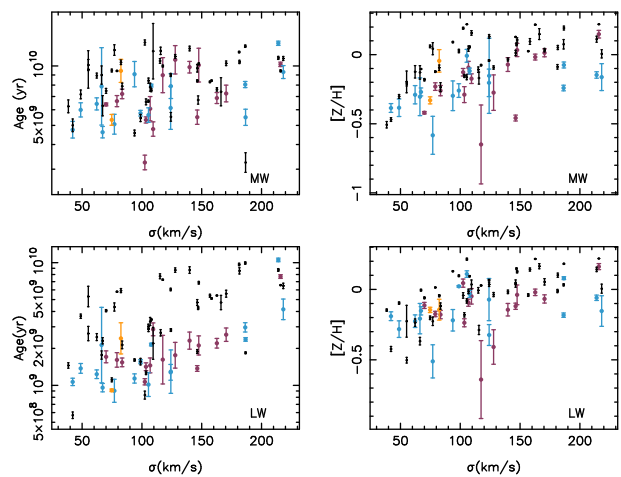}}
\caption{Relation between the luminosity- and mass-weighted age and
  metallicity values calculated at 1.5 r$_{\rm eff}$ and the centr1al velocity
  dispersion. The meaning of the colors is the same as in previous
  figures. We have overplotted, with black symbols, the values
  calculated in the central spaxel. \label{disk_sigma}}
\end{figure}

We have seen, in previous sections, that the stellar population
gradients in the disc region of our sample do not
correlate with other properties of the galaxies. This does not mean
that the characteristic values of the stellar age and metallicities in
the disc regions do not correlate with other
properties. Figure~\ref{disk_sigma}
shows the values of the mean age and metallicities calculated at 1.5
r$_{\rm eff}$ ($\sim$ 2.5 scale-lengths) as a function
of the  central velocity dispersion. The values of the central
velocity dispersion have been calculated using the high spectral
resolution 
version of the  CALIFA data and  will be presented in Falc\'on-Barroso et al. (in
preparation).
As can be seen, there is a relation for which galaxies with higher
velocity dispersion have older and more metal rich populations in the
disk. We have also plotted the central values of the galaxies for
comparison.  The relations between the age and metallicity and the
velocity dispersion follow by the bulge and the disc are similar, with
bulges
older, on average, and more metal rich. Note that the dispersion in
the  luminosity-weighted age at a given central velocity dispersion is larger 
for bulges than for the disk. 

\begin{figure}
\centering
\resizebox{0.5\textwidth}{!}{\includegraphics[angle=0]{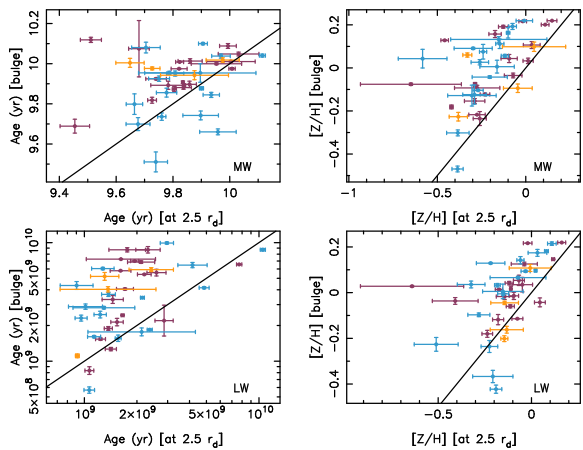}}
\caption{Relation between the mean, luminosity- and mass-weighted,
  stellar population models measured at 1.5 r$_{\rm eff}$ and in the center of
  the galaxies of our sample. The meaning of the colors is the same 
as in previous figures.\label{fig:disk_bulge}}
\end{figure}

Figure~\ref{fig:disk_bulge} shows the relation between the stellar
populations at 1.5 r$_{\rm eff}$ and that in the center for our sample of
galaxies.
Both parameters are highly correlated.
Correlations between the structural parameters and colors between the bulge and
the disk have been found by previous authors (Peletier \& Balcells
1996; Courteau, de Jong \&
Broeils 1996; de Jong 1996b;  Carollo
et al. 2007\nocite{2007ApJ...658..960C}; Gadotti \& Dos Anjos 2001)
 and have been interpreted as evidences of
internal secular evolution. However, the relation can also be driven
by a third, global  parameter that drives both the star formation
history of bulges and disks.
 The stellar populations 
in the bulge of this sample will be the subject of a future paper and
we  wont discuss this issue further.
\section{Discussion}
\label{sec:discussion}
In the present paper, we have analysed the spatially resolved stellar population properties
in the disc of galaxies with and without bars to test observationally  numerical
simulations predicting that the non-linear coupling between bars and spiral waves can
make the process of radial migration due to resonant scattering of resonances much
more efficient and fast than when a single perturber (e.g., spiral
arms alone) is considered
(Friedli \& Benz 1995; Friedli 1998; Minchev \& Famaey 2010; Minchev et al. 2011a; Brunetti et al. 2011;
Di Matteo et al. 2013).

In our analysis, we do {\it not} find any difference in the slope of the stellar-phase metallicity
gradients between barred and unbarred galaxies.
Does this mean that the non-linear coupling of bars and spirals is much weaker than what has been
proposed so far and, therefore, that numerical simulations are failing at predicting the influence
of bars on the evolution of the stellar disc?  Not necessarily. Here we discuss some possibilities that can
explain the lack of differences in our data.

(i)  Radial migration due to the exchange of angular momentum at
corotation is not efficient in hot discs (Brunetti et al. 2010\nocite{2011A&A...534A..75B}).
Brunetti et al. (2011) showed that not all barred galaxies experience
strong diffusion and that the efficiency of the diffusion   depends
on the bar strength and thus, ultimately, on the stability of the disk.
Hot discs do not respond to the perturbation created by the non-linear
coupling of bar and spiral arms.
If this is true we will expect to find the  differences between barred
and unbarred galaxies in those
discs with lower vertical dispersion  ($\sigma_z$).
However, we have compared the metallicity gradients for those galaxies with low $\sigma_Z$
(using the methodology described in Gerssen, Kuijken \& Merrifield 1997\nocite{1997MNRAS.288..618G})
without finding any difference.

(ii) The lack of differences between the gradients of barred and
unbarred galaxies could be related with the  longevity of bars. If bars are not long-lasting structures but  recurrent patterns
(Bournaud  \& Combes 2002) then, the fact that we do not find differences between barred and
unbarred galaxies would not necessarily imply that bars are not important for stellar migrations
but, simply, that unbarred galaxies could have been barred in the recent past.
However, most numerical simulations show that once they formed, bars are robust structures
(Shen \& Sellwood 2004\nocite{2004ApJ...604..614S}; Athanassoula,
Lambert \& Dehnen 2005\nocite{2005MNRAS.363..496A}; Debattista et al. 2006\nocite{2006ApJ...645..209D}; 
Berentzen et al.\ 2007\nocite{2007ApJ...666..189B}; Villa-Vargas et
al. 2010\nocite{2010ApJ...719.1470V}; 
Kraljic, Bournaud \& Martig 2012\nocite{2012ApJ...757...60K}; Athanassoula et al. 2013\nocite{2013MNRAS.429.1949A}).
Kraljic  et al. (2012)\nocite{2012ApJ...757...60K} showed, in a series of cosmological simulated discs that the majority
of bars formed at redshifts $z\sim 0.8 -1$ survive to redshift $z=0$
without changing its strength, although cosmological gas infall is necessary to maintain some of them.
They also show that some bars formed earlier, when the gas fraction of the galaxies was higher, get
destroyed and refurbish again but, again, these galaxies represent a minority of barred galaxies.

Furthermore, at least in massive disc galaxies, bars have old and
metal rich stellar
populations, 
older than the disc and similar in age and metallicity to that of the bulges 
(see S\'anchez-Bl\'azquez et al. 2011\nocite{2011MNRAS.415..709S} and
P\'erez, S\'anchez-Bl\'azquez \& Zurita 2009\nocite{2009A&A...495..775P})
which also support the idea that formed long time ago.
The longevity of bars is also suggested in studies
of bar fraction evolution with redshift (e.g., Sheth et al.\ 2008\nocite{2008ApJ...675.1141S}) that find a similar bar
fraction at z$\sim$0.8 than that at the present day for galaxies with
M$_{*} \geq $10$^{11}$M$_{\odot}$.
Lower mass galaxies, however, have a bar fraction that was much lower at z$\sim$0.8 than now,
which can be explained with a delay in the bar formation (Athanassoula
et al. 2013\nocite{2013MNRAS.429.1949A}; Sheth et al. 2012\nocite{2012ApJ...758..136S}).
However, the debate regarding the survability of bars is still not closed and, therefore, 
the longevity of the bars remains a possibility to explain the lack of differences
in the metallicity gradient.

(iii) In the theory of the corotation scattering mechanism described in  Sellwood \& Binney (2002), the
migration is only effective if the structure
is transient.  Bars then could not be effective at disc mixing once
they were formed, due to their long-lived nature.
However, recent N-body Tree-SPH simulations by Minchev et al. (2012ab) show that bars are the most effective drivers of
radial migration {\it through the galactic evolution} despite the fact that they are not transient,
but only slowly evolving.  This is also consistent with the findings
by Brunetti et al. (2011)\nocite{2011A&A...534A..75B}.

(iv) Another  possibility is that there is radial migration but the consequences
are not visible in the metallicity gradient. This could happen, for example,
if the metallicity gradient is not very steep (and it was also not very steep in the past).
We have found, indeed, fairly flat present-day metallicity gradients, which could support this possibility.
In our study of the evolution of the metallicity gradient, we have found steeper
slopes for older populations but still, not very steep (except in some galaxies). Furthermore,
we cannot discard a flattening of the gradients due to, precisely, stellar migrations.
The only  way to check this point is studying the gas-phase metallicity gradient at different redshifts.
In the last years, several authors have obtained radial abundance gradients at high redshifts
(Cresci et al. 2010\nocite{2010Natur.467..811C}; Jones et al. 2010\nocite{2010ApJ...725L.176J}; 
Queyrel et al. 2012\nocite{2012A&A...539A..93Q}; and Yuan et al. 2011\nocite{2011ApJ...732L..14Y}). Yuan et
 al.
(2011) show that at least on one Grand Design disc at redshift $z~1.5$, the metallicity
gradient is significantly steeper ($-0.16$~dex/kpc) than the typical gradient encounter today.
However, despite such advances, it is still very difficult to obtain high resolution data
for galaxies at high redshift and, therefore, a flat metallicity gradient throughout the 
evolution of galactic discs remains a possibility to explain the lack of differences in the 
metallicity gradients of barred and unbarred galaxies.

The results can also indicate that the mechanism proposed by Minchev \& Famaey (2010) is
not as important as producing radial migrations as predicted.
This does not mean that radial migration is not important.
Other mechanism not related with the presence of bars can be at play.
Sellwood \& Binney (2002) suggested a  mechanism based on resonant scattering of stars under
the effect of transient spiral waves (see Sec.~\ref{sec:introduction}). 
Haywood (2008) estimated upper
values for the migration rate from 1.5 to 3.7 kpc/Gyr, which agree with the values in L\'epine
et al. (2003)\nocite{2003ApJ...589..210L} for the radial wandering due to the scattering mechanism assumed by Sellwood \&
Binney (2002).
Observationally, claimed evidences of stellar migration, as upturns in the
age distribution of stars have been reported in galaxies
without bars (e.g., de Jong et al. 2007\nocite{2007ApJ...667L..49D}; Yoachim et al. 2010, 2012\nocite{2012ApJ...752...97Y}\nocite{2010ApJ...716L...4Y},
 Radburn-Smith 2012\nocite{2012ApJ...753..138R}).
However, it was shown in S\'anchez-Bl\'azquez et al. (2009)\nocite{2009MNRAS.398..591S} that an upturn
in the age distribution beyond the break is not an unequivocal sign of radial migrations and
that this can be produced by other reasons, for example,  a warp in the gaseous discs that
decrease the star formation density in the external parts.
Furthermore, there are many galaxies with breaks where the upturn in the stellar population has not been
observed (e.g., Gorgarten et al. 2010\nocite{2010ApJ...712..858G}; Yoachim et al. 2012\nocite{2012ApJ...752...97Y}).
The stellar migration models should explain why this is the case.
Another $"$evidence$"$ of the occurrence of radial migration is the slope and scatter of the age-metallicity 
relation in the Solar neighborhood (ie., Ro{\v s}kar et al. 2008b). However, this can be consequence of 
the errors in the determination of ages and abundances in individual
stars (Anguiano et al., in preparation).
However, considerably work remains to be done and all of these
conclusions remain speculative.

To finish we want to stress here that our results do not contradict
previous observational results in the field. Studies of stellar
populations in disc galaxies comparing galaxies with and without bars
have only been done using colors (Gadotti \& Dos Anjos 2001\nocite{2001AJ....122.1298G}).
This study found that  there was an excess of barred galaxies among the objects with null or 
positive (bluish inward) color gradients. However, the gradients were calculated  
in the whole galaxy, not just the disc (as in our study), and measure
differences between the stellar populations in the bulge and the disc. 

On the other hand, previous studies comparing the gas-phase
metallicity gradient of galaxies with and without bars had found
flatter 
gradients in barred galaxies compared with unbarred ones (e.g.,
Martin \& Roy 1994; Zaritsky et al. 1994). 
However, gas and stars suffer from very different evolutionary processes; the gas is 
mainly dominated by the gravitational torque of the non-axisymmetric mass component, while the stars 
are mainly affected by different orbital mixing.
Furthermore, the above cited studies suffer from the poor
number statistics (only 5 barred galaxies were analysed in Zaritski et al.).
Using a much larger sample of 306 galaxies from the CALIFA survey, 
S\'anchez et al. (2014)\nocite{2014A&A...563A..49S} found no difference between the gas-phase metallicity 
gradient of barred and unbarred galaxies.

\section{Conclusions}
\label{sec:conclusions}
We have presented a spatially resolved analysis of the star formation histories and metallicity
evolution for a sample of disc galaxies using data from the CALIFA survey. The main motivation
of this study is to compare the stellar population gradients in galaxies with and without bars,
to analyse the influence of these axisymmetric structures in producing radial migrations.
However, we also present generalities related with the relation of the stellar population
gradients and other properties of the galaxies and the evolution of the metallicity gradients
with time.

The main conclusions of the present study are:
\begin{itemize}
\item The mass-weighted age in disc galaxies reflect the fact that nearby galaxies in the mass range studied
by us are dominated, at all radii, by  old  stars.
\item The luminosity-weighted age profile show a shallow, but negative slope for the majority  of our galaxies.
The mean values are $-0.116\pm 0.200$ and $-0.014\pm 0.135$
dex/r$_{\rm eff}$ for the light- and the mass-weighted values.
This implies that
there is a larger proportion of young {\it vs.} old stars in the external parts of the disc with respect to the
inner parts.
\item The mean luminosity-weighted metallicity gradient is also shallow when normalized
to the effective radius of the disc. In most cases, the luminosity-weighted metallicity
gradient is steeper than the mass-weighted one. The
mean values are $-0.051\pm 0.126$ dex/$r_{\rm eff}$ and $-0.089\pm
0.151$ dex/$r_{\rm eff}$ for the mass and the light-weighted gradientes
respectively, where the errors represent the RMS dispersion.
\item The analysis of the metallicity gradients at different stellar ages
reflects steeper gradients for older stars.
\item The stellar-metallicity gradient (normalized to the effective radius of the disc) is not correlated
with other global galactic properties, as the morphological type the mass or the luminosity.
\item We compare the stellar-metallicity gradients of galaxies with
  and without bars and we did not find any 
significant difference. This is contrary to the predictions of
some numerical simulations indicating a flattening of the metallicity
gradient due to the bar presence.
\end{itemize}
\begin{acknowledgements}
P.S-B acknowledges support from the Ram\'on y Cajal program, grant ATA2010-21322-C03-02 from 
the Spanish Ministry of Economy and Competitiveness (MINECO). 
F.F.R.O. acknowledges the Mexican National Council for Science and Technology (CONACYT) for financial support
under the programme Estancias Posdoctorales y Sab\'aticas al Extranjero para la Consolidaci\'on de Grupos de
Investigaci\'on, 2010-2012.
We acknowledge financial support from the research projects 
under grants AYA2010-21887-C04-03, AYA2010-21322-C03-03 and AYA2010-15081 by the Spanish Ministerio 
de Ciencia e Innovaci\'on.
J. F.-B. acknowledges support from the Ramon y Cajal Program, grants AYA2010-21322-C03-02 from the Spanish
Ministry of Economy and Competitiveness (MINECO). 
R.A.M is funded by the Spanish program of International Campues of Excellence Moncloa (CEI).
C.J.W acknowledges support through the Marie Curie Career Integration Grant 303912.
I.M. acknowledges financial support from the Junta de Andaluc\'{\i}a
through projects PO8-TIC-03531 and TIC114, and the Spanish Ministry of
Economy and Competitiveness (MINECO) through the project AYA2010-15169.
We acknowledge the usage of the HyperLeda database (http://leda.univ-lyon1.fr).
Based on observations collected at the
Centro Astronmico Hispano-Alean (CAHA) at Calar Alto,
operated jointly by the Max-Planck Institut ur Astronomie
and the Instituto de Astrofsica de Andaluca (CSIC).
FCT/MCTES (Portugal) and POPH/FSE (EC). He acknowledges support by
the Funda\c{c}\~{a}o para a Ci\^{e}ncia
e a Tecnologia (FCT) under project FCOMP-01-0124-FEDER-029170 (Reference FCT
PTDC/FIS-AST/3214/2012), funded by FCT-MEC (PIDDAC) and FEDER (COMPETE).
We also thanks the referee por his/her suggestions that have improved 
the final version of this paper.
\end{acknowledgements}
\bibliographystyle{aa}
\bibliography{references}

\label{lastpage}

\appendix
\section{Derivation of the age-metallicity relation}
\label{appendix1}
One of the advantages of deriving star formation histories with {\tt STECKMAP} is that, in principle, one can 
also derive the metallicity of the stellar population with different ages. However, due to the existent degeneracies
in the determination of these parameters, one has to test if this is really possible. In S\'anchez-Bl\'azquez et al. (2011) 
-- see their appendix B -- we already probed that we were able to recover the evolution of metallicity with time for the 
data with a similar signal-to-noise, resolution and wavelength coverage than that used in paper. In this appendix we 
repeat the test but using synthetic data with  the characteristics of the CALIFA spectra. We have generated synthetic 
spectra for an exponentially declining star formation history $e^{(-t/\tau)}$ with different values of $\tau$, 
$\tau=$1, 5, 10 and 20 Gyr and 
with two different chemical evolution, one with constant metallicity and one where the metallicity 
increase with age with the law $Z=0.2-(age (Gyr) -1.0)/(16.0)$. 

\begin{figure*}
\centering
\resizebox{0.4\textwidth}{!}{\includegraphics[angle=0]{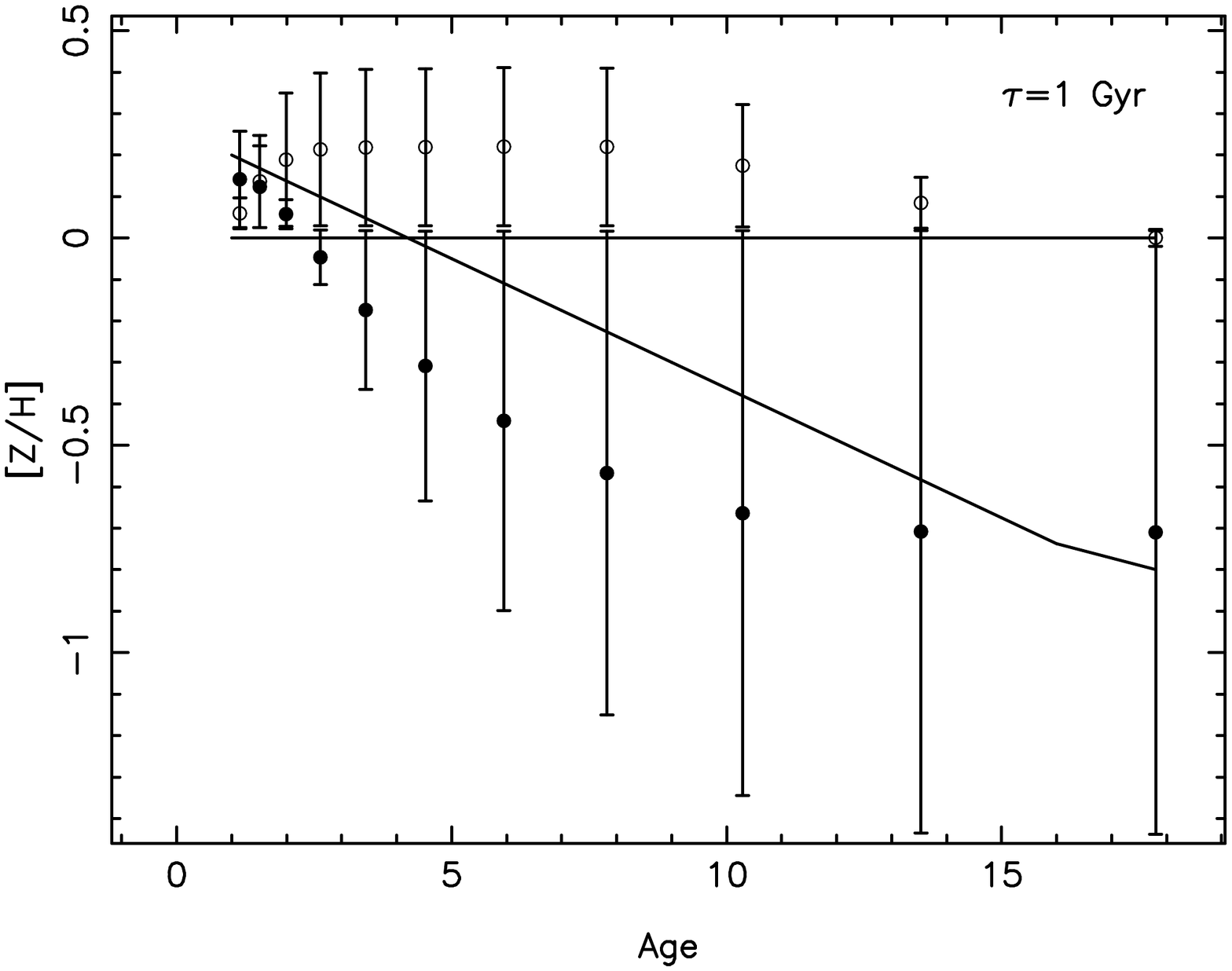}}
\resizebox{0.4\textwidth}{!}{\includegraphics[angle=0]{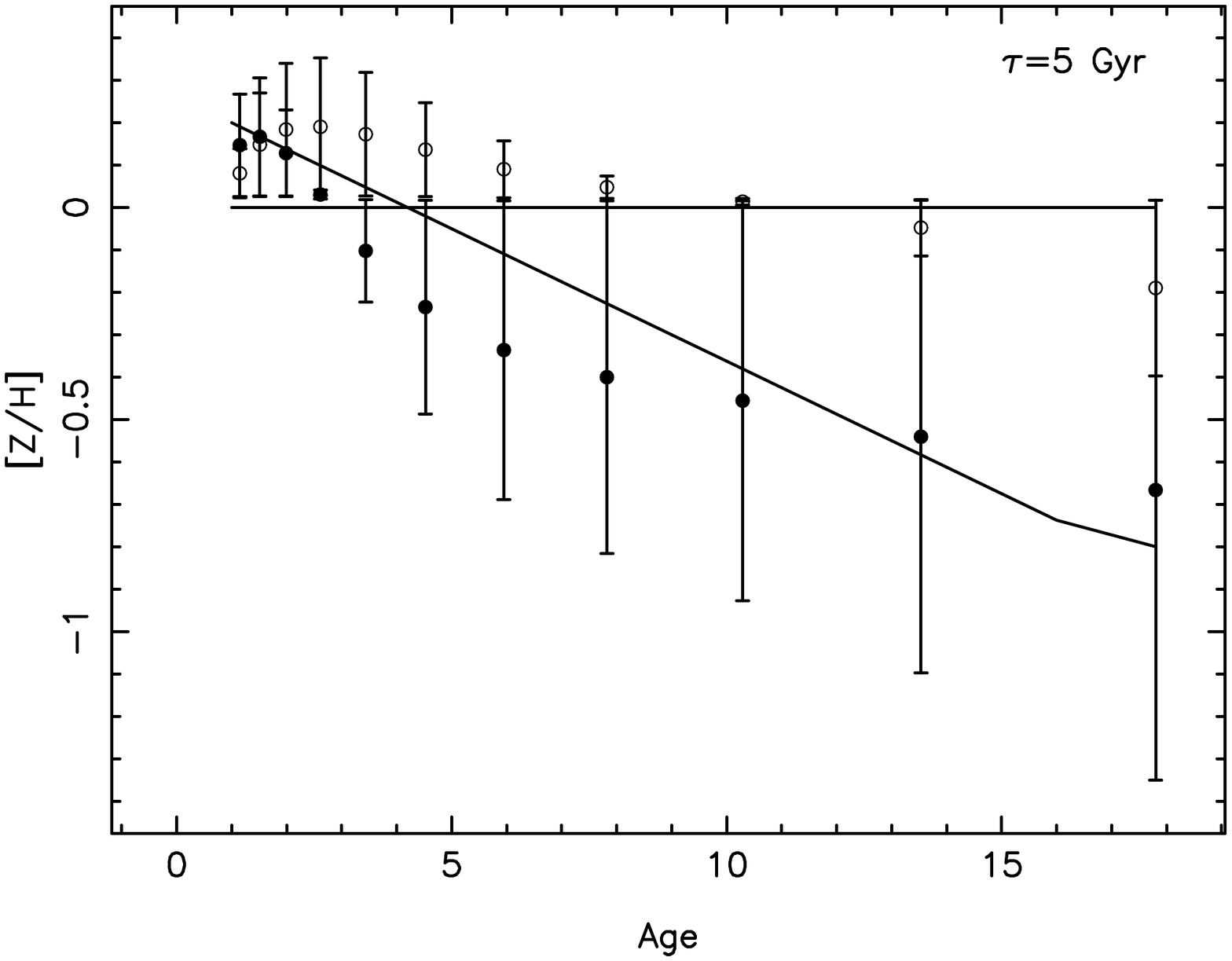}}
\resizebox{0.4\textwidth}{!}{\includegraphics[angle=0]{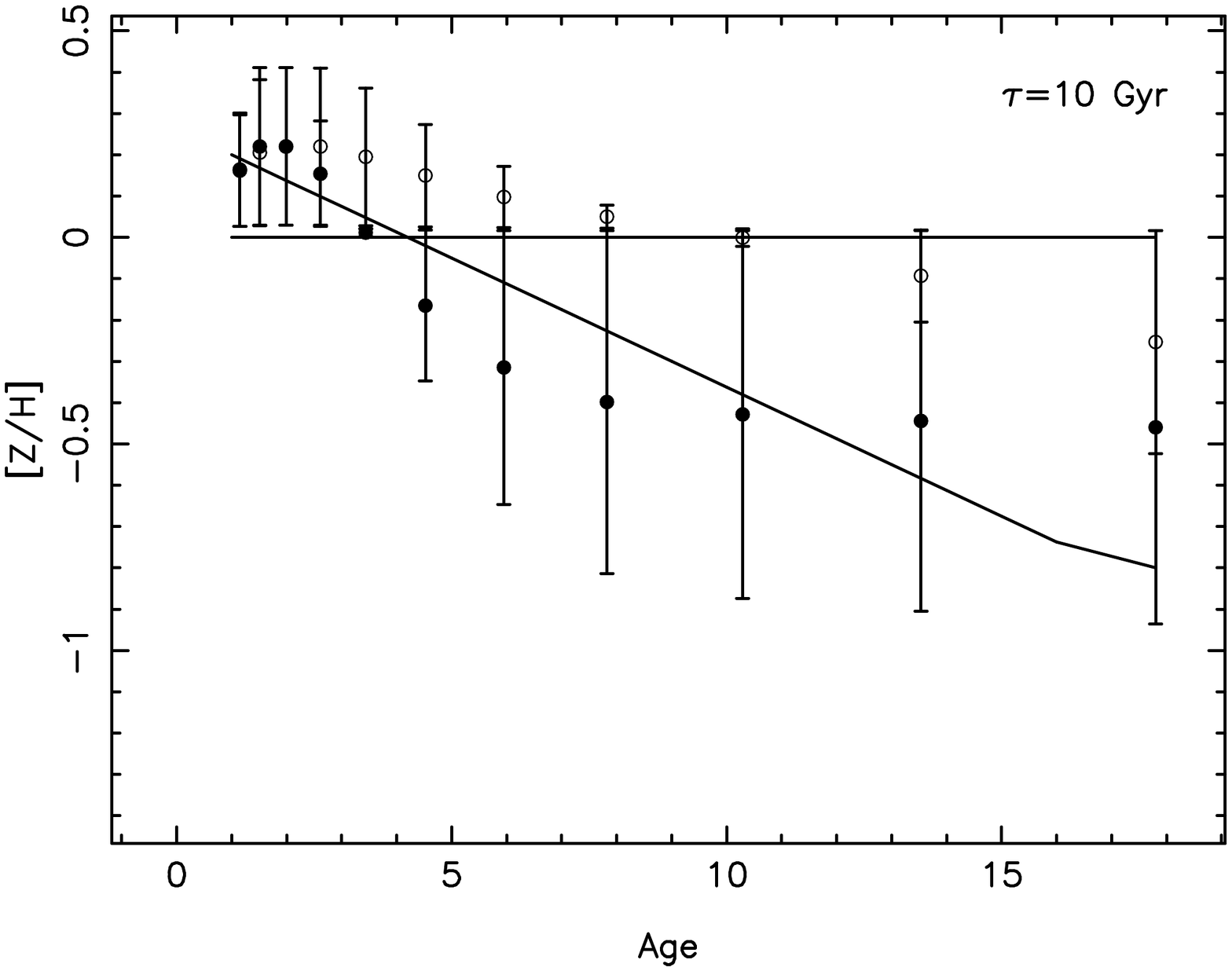}}
\resizebox{0.4\textwidth}{!}{\includegraphics[angle=0]{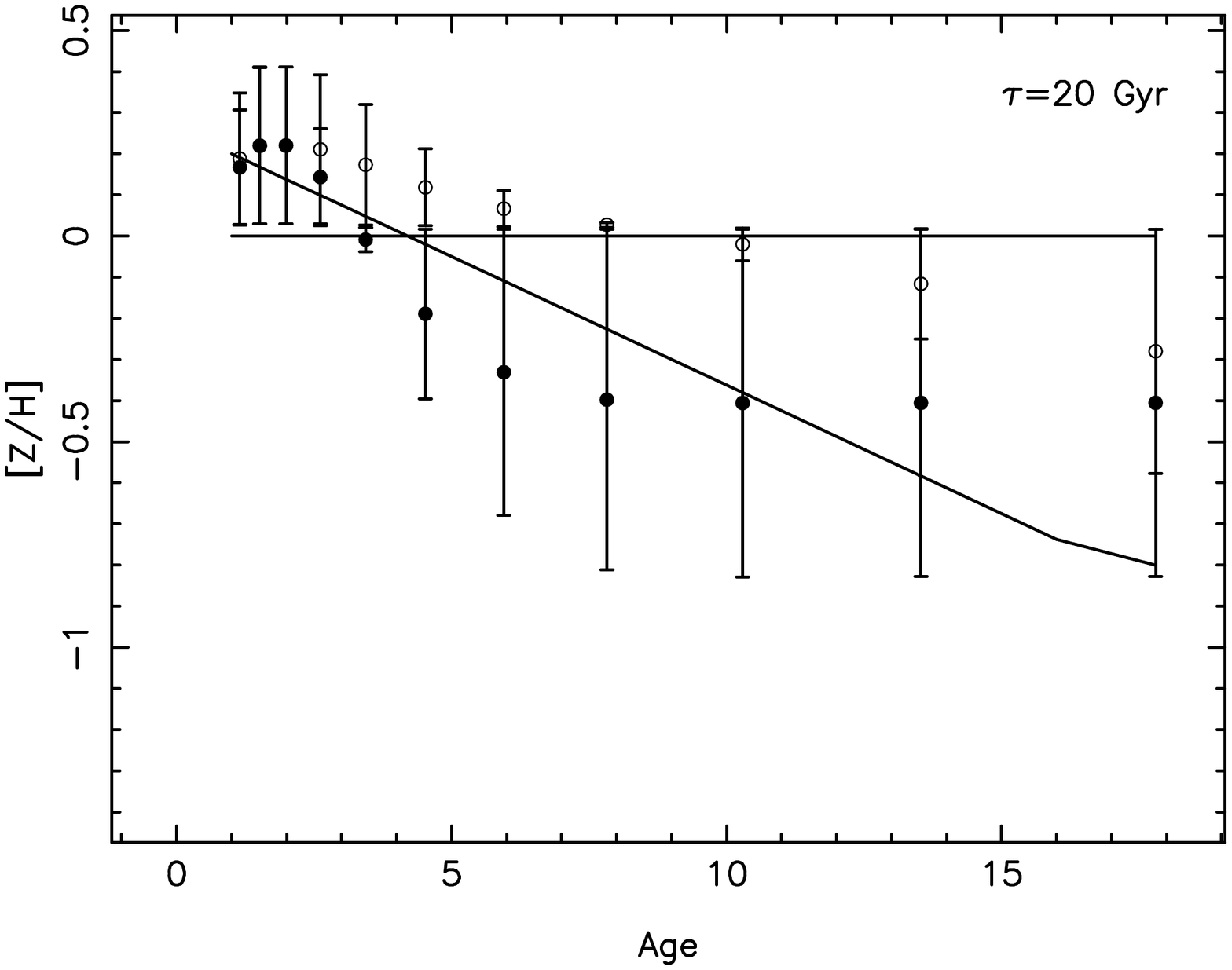}}
\caption{Age metallicity relation for a simulated exponential star formation history with different $\tau$
and with two different chemical evolution models. Solid lines indicate the simulated age-metallicity relation 
while the points show the recovered values. Open symbols shows the recovered age-metallicity relation in the 
case of constant metallicity with age while the filled symbols show the recovered age-metallicity relation in the 
case where the metallicity increase with time. Error bars represent the RMS dispersion of a set of 50 
Monte Carlo simulations
where each pixel is moved according to a Gaussian distribution of width given by the errors. The synthetic spectra
have been degraded to the resolution of the data and noise have been added to simulate a signal-to-noise per $\AA$ of 40.
\label{test_simus}}
\end{figure*}

We have used the models of V10 broadened to 150 km/s and 
we have added noise to simulate a spectrum of S/N=40. We then run {\tt STECKMAP} on these spectra.
Figure~\ref{test_simus} shows the simulated and the recovered age
metallicity relation for the different synthetic spectra. The error bars in the figure are computed as the RMS dispersion of the
values obtained in 50 MonteCarlo simulations in which each pixel of the synthetic spectrum was modified randomly following a
Gaussian distribution with a width given by the noise spectrum. 
As can be seen, although at all ages the metallicity is not very well constrained (the scatter from the different simulations 
is large), we can recover the age-metallicity relation in the synthetic spectra.

\section{Age gradients}
\label{appendix2}
\begin{figure*}
\centering
\resizebox{0.22\textwidth}{!}{\includegraphics[angle=0]{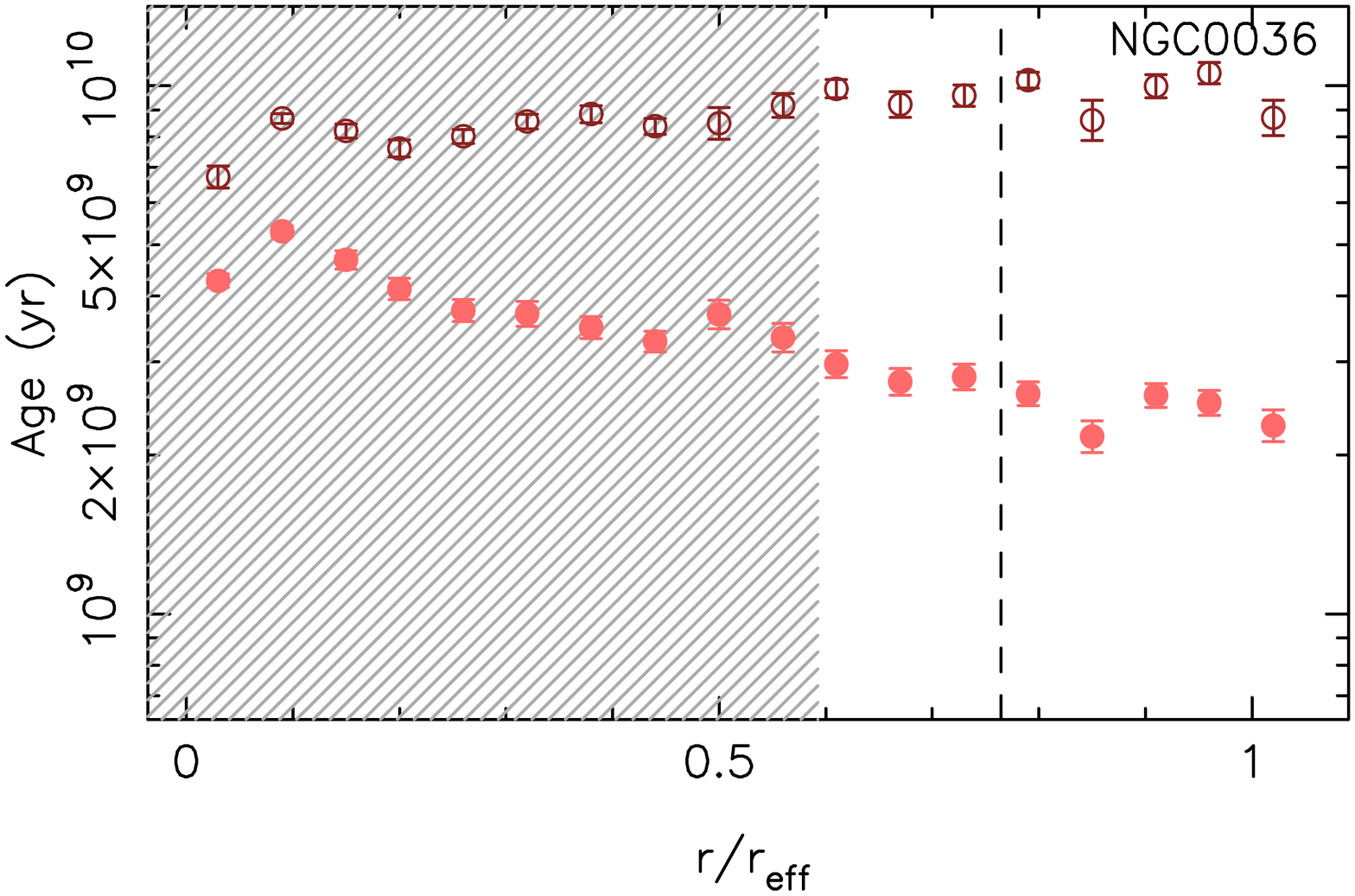}}
\resizebox{0.22\textwidth}{!}{\includegraphics[angle=0]{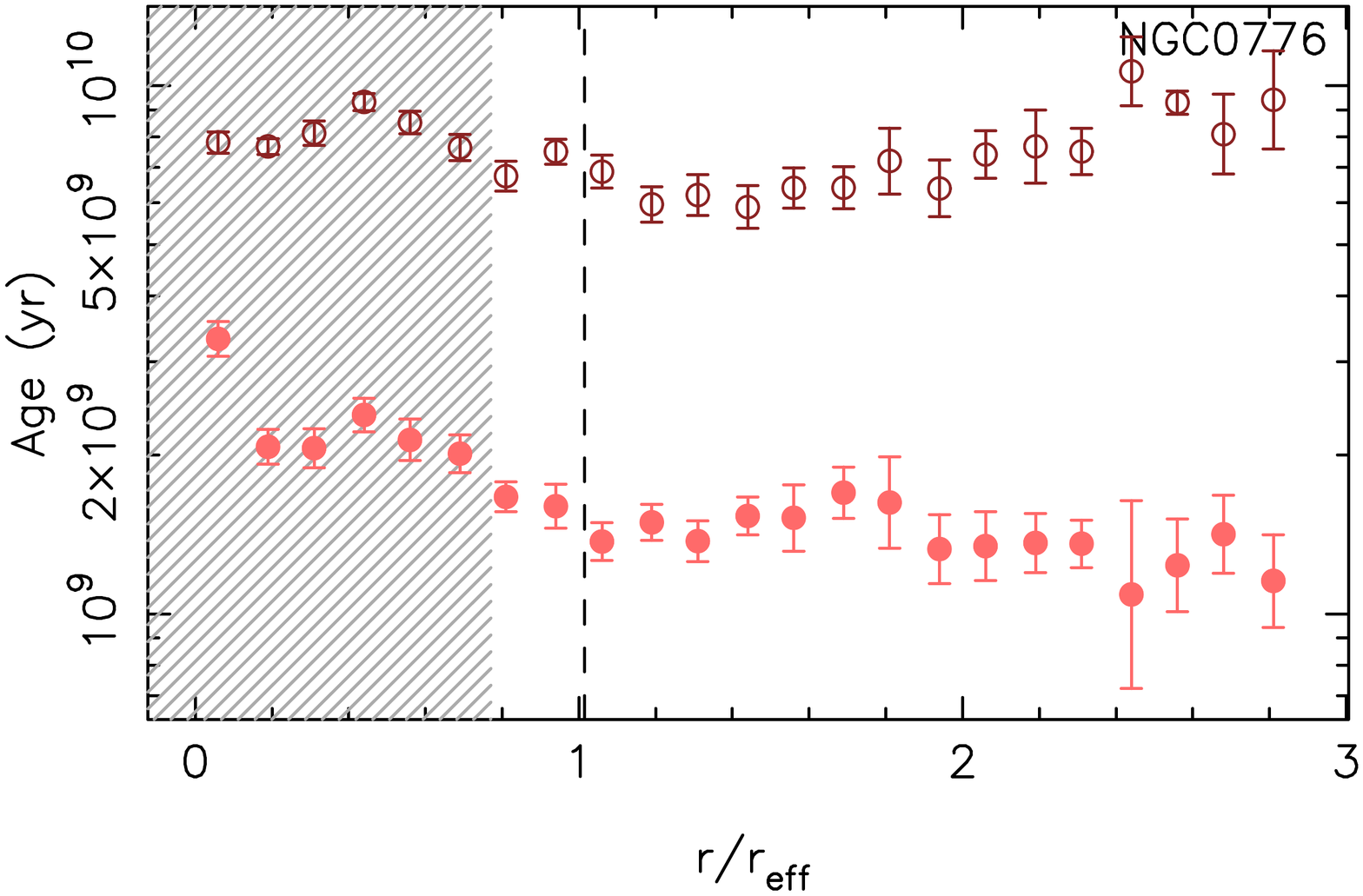}}
\resizebox{0.22\textwidth}{!}{\includegraphics[angle=0]{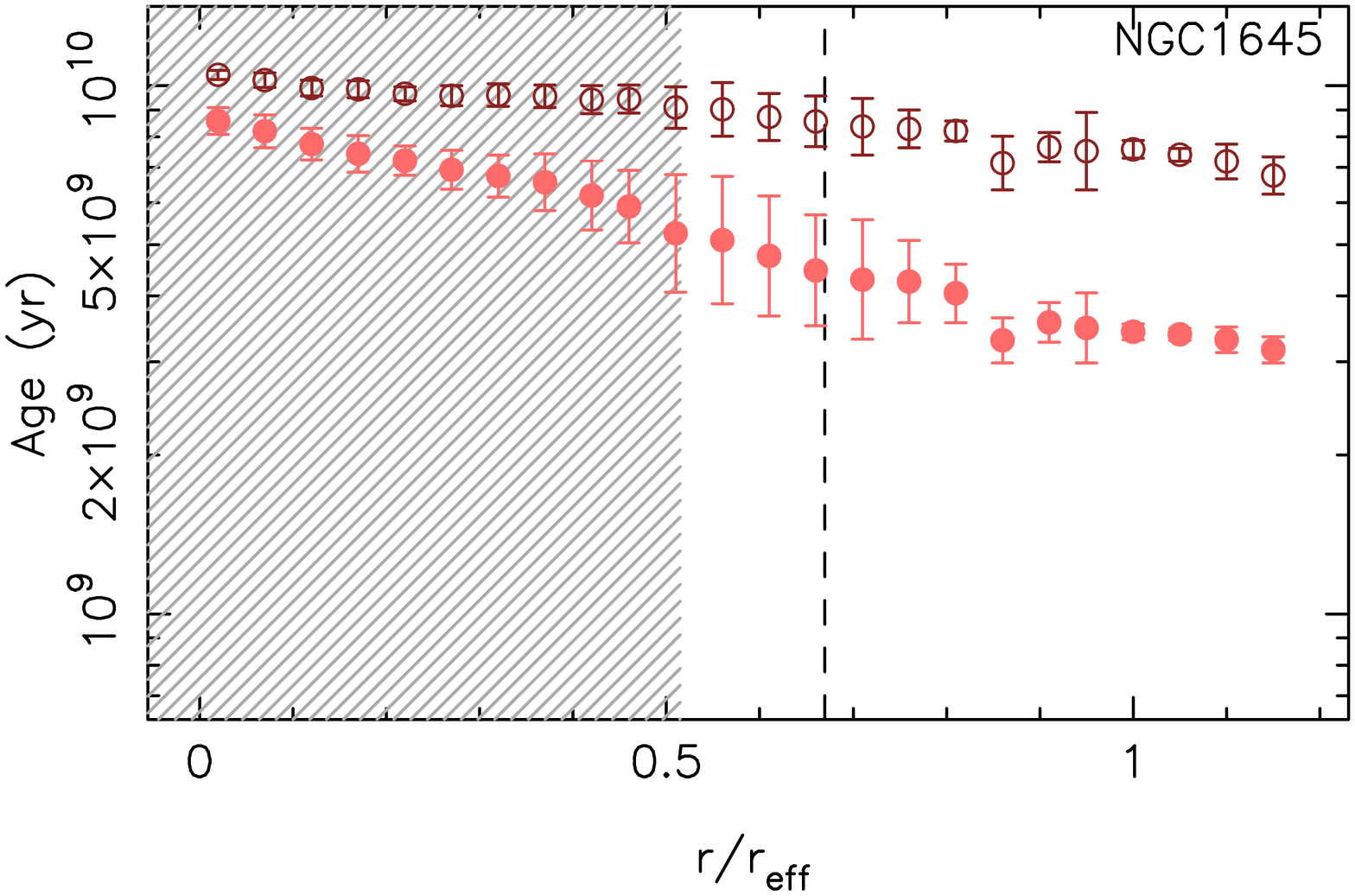}}
\resizebox{0.25\textwidth}{!}{\includegraphics[angle=0]{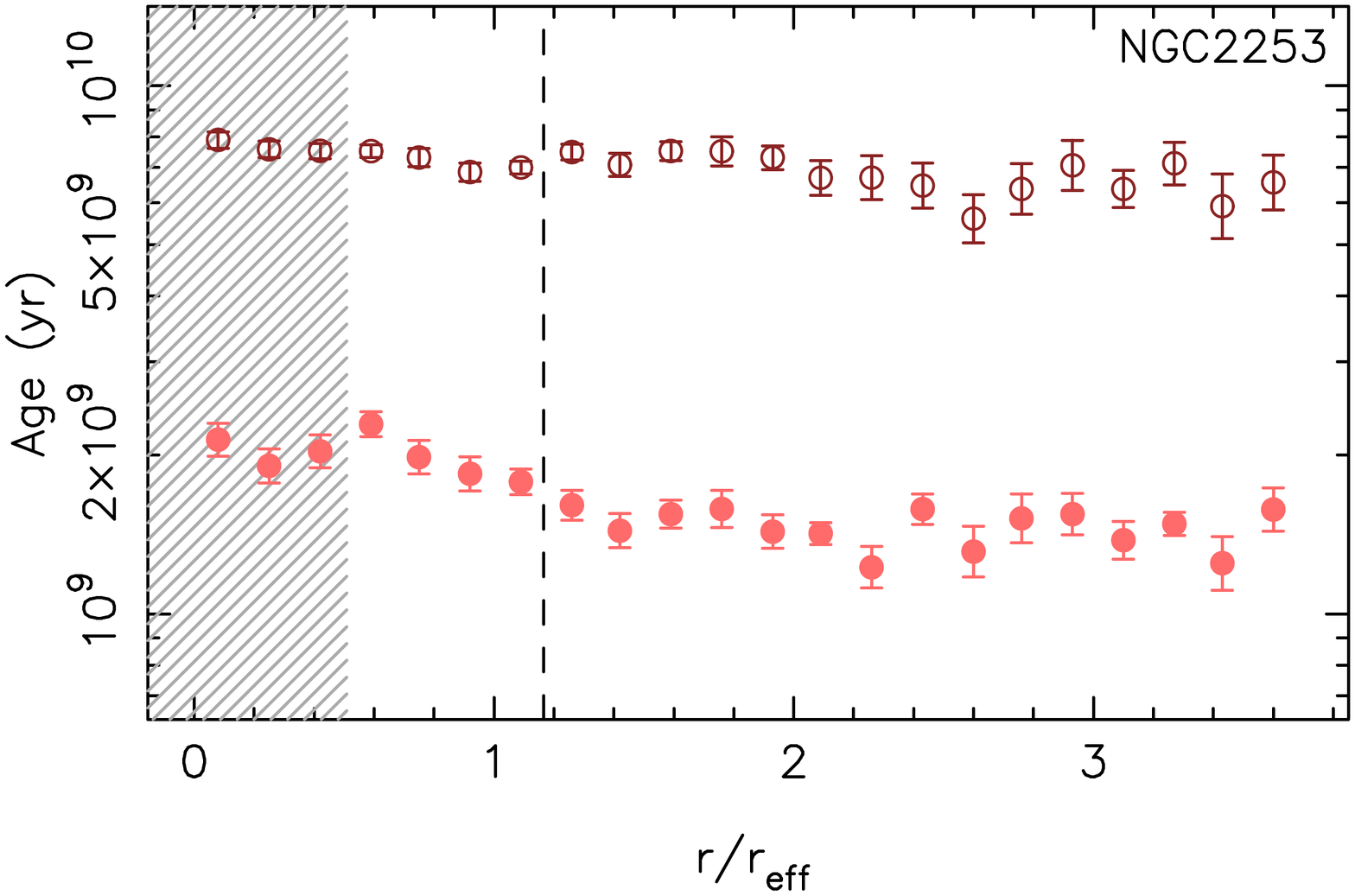}}
\resizebox{0.22\textwidth}{!}{\includegraphics[angle=0]{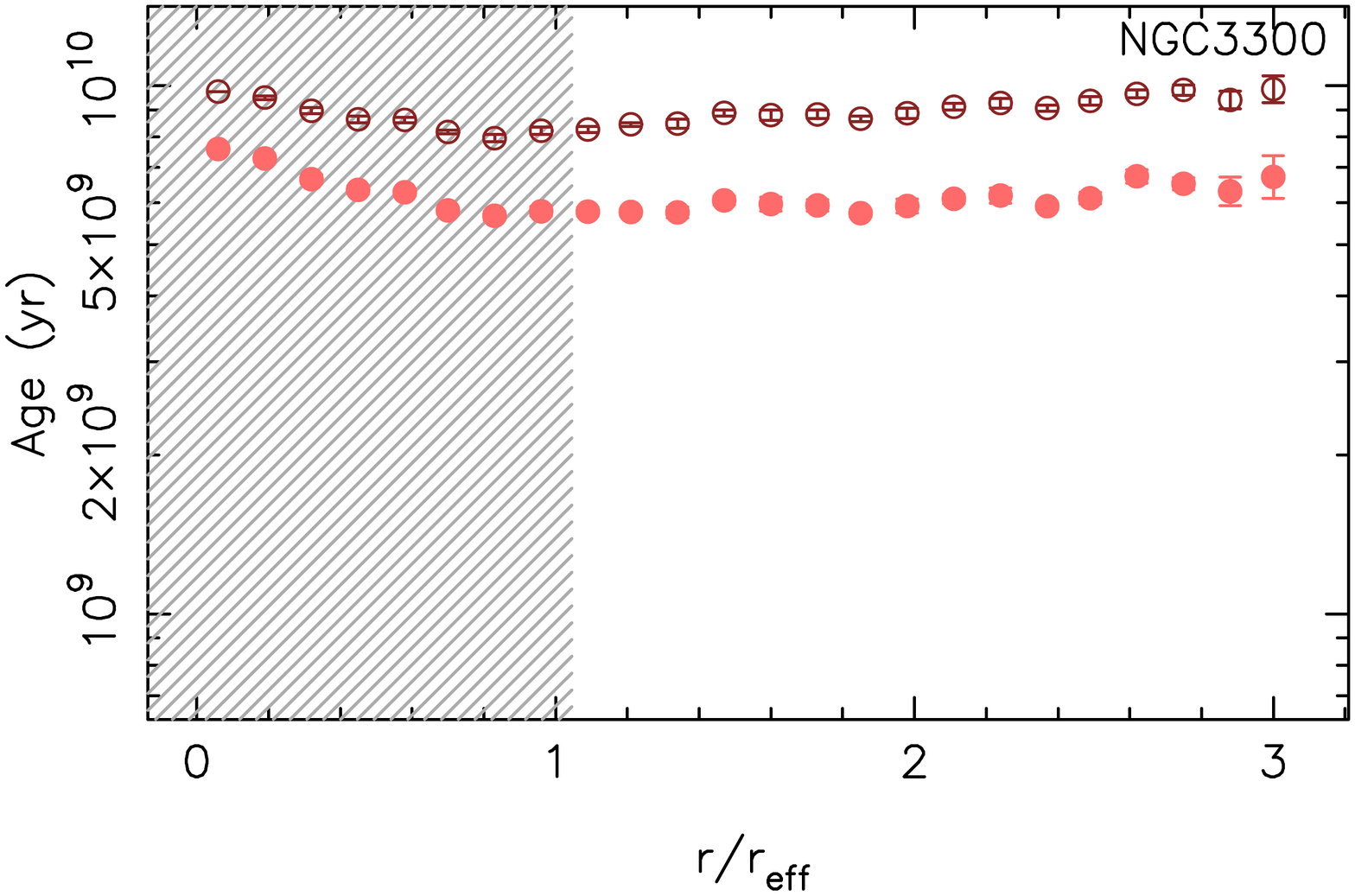}}
\resizebox{0.22\textwidth}{!}{\includegraphics[angle=0]{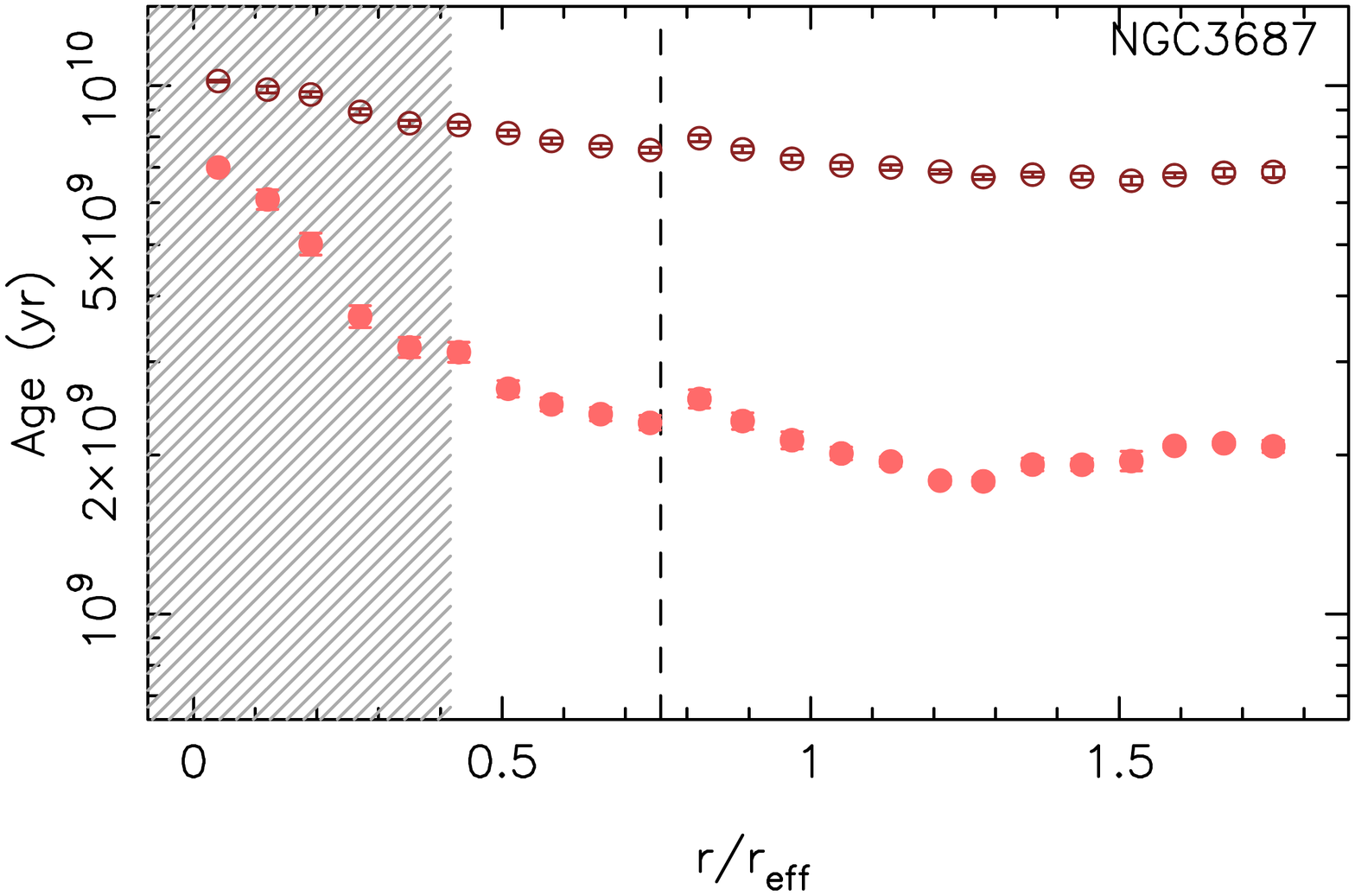}}
\resizebox{0.22\textwidth}{!}{\includegraphics[angle=0]{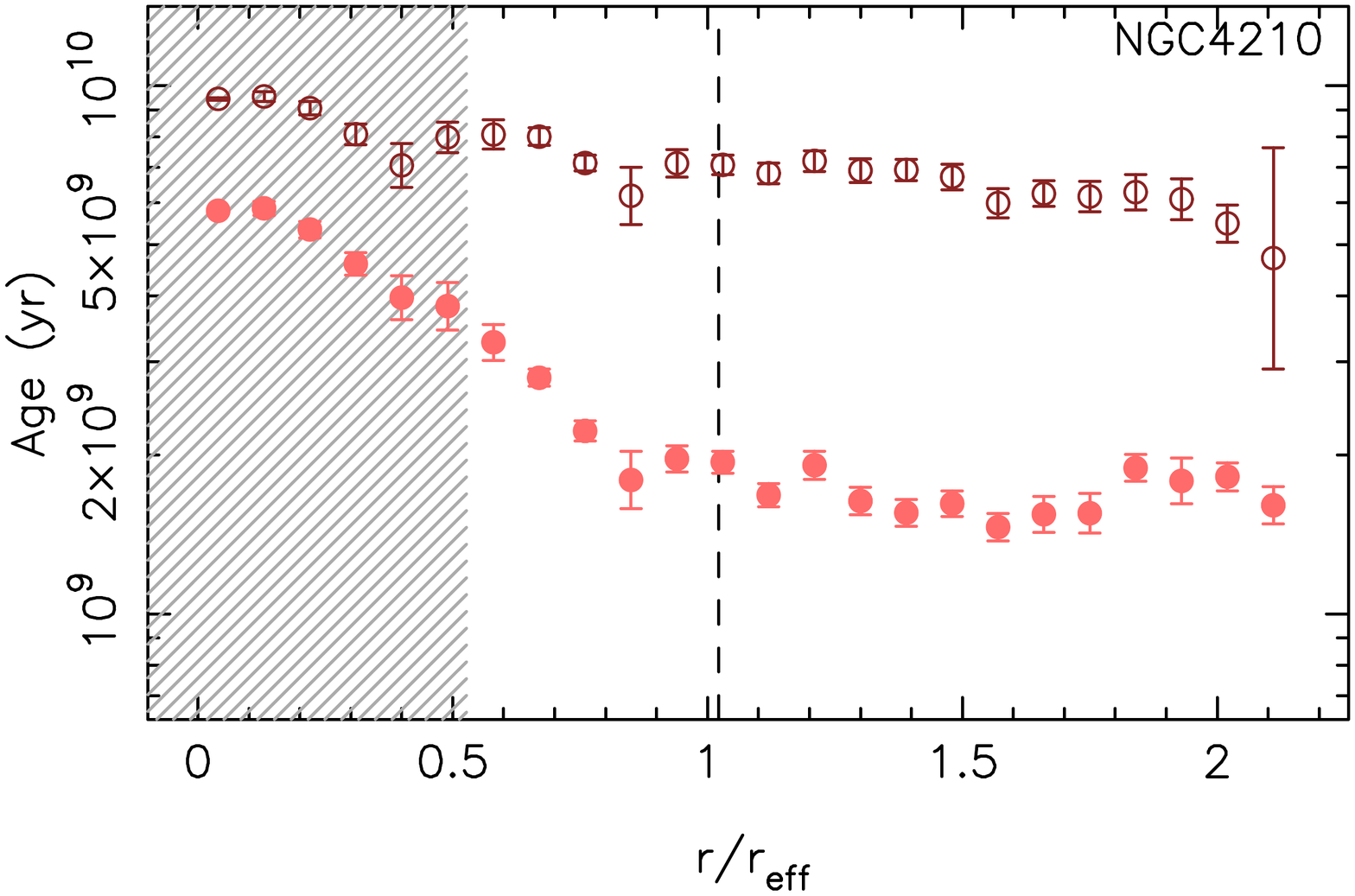}}
\resizebox{0.22\textwidth}{!}{\includegraphics[angle=0]{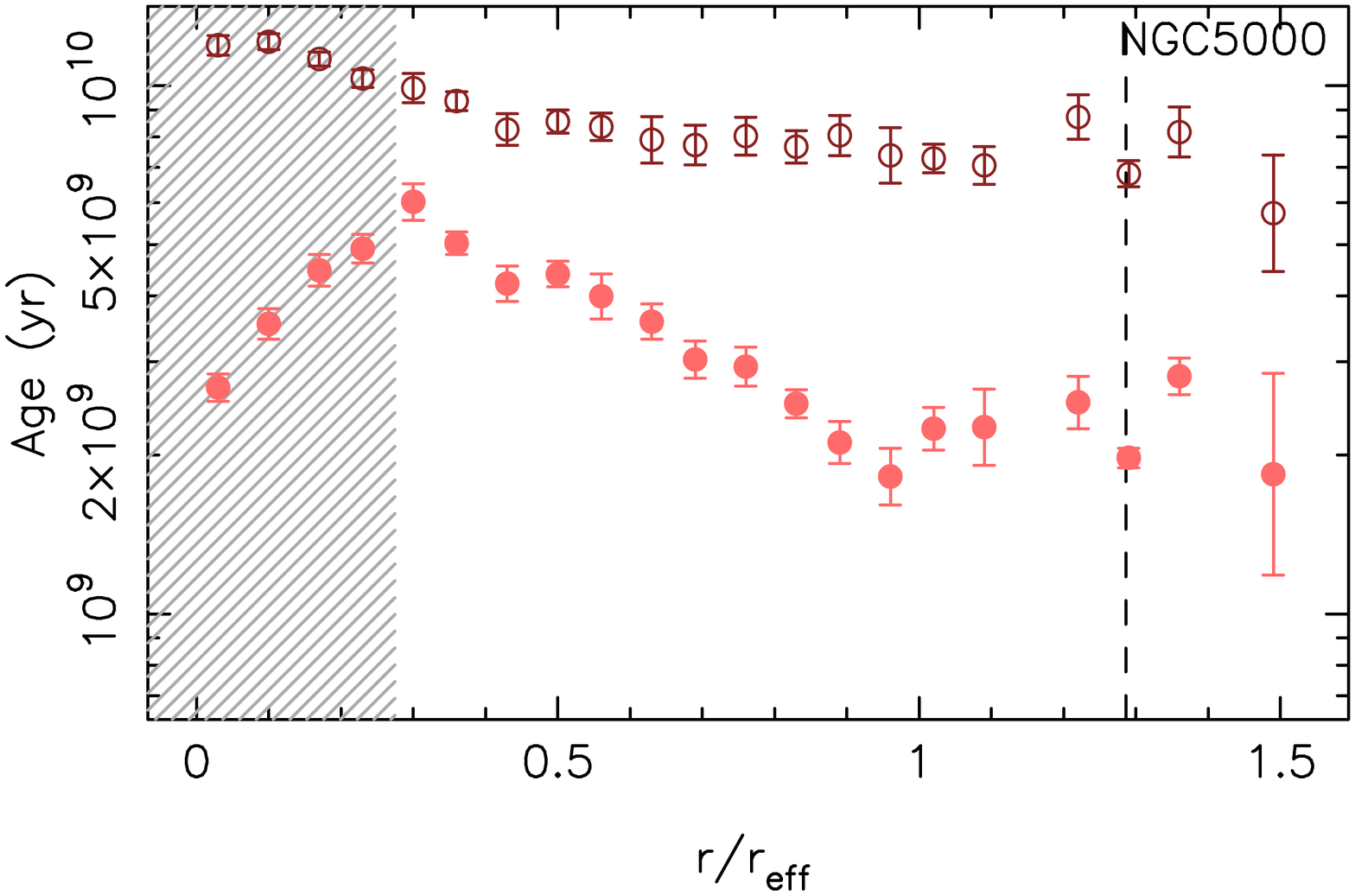}}
\resizebox{0.22\textwidth}{!}{\includegraphics[angle=0]{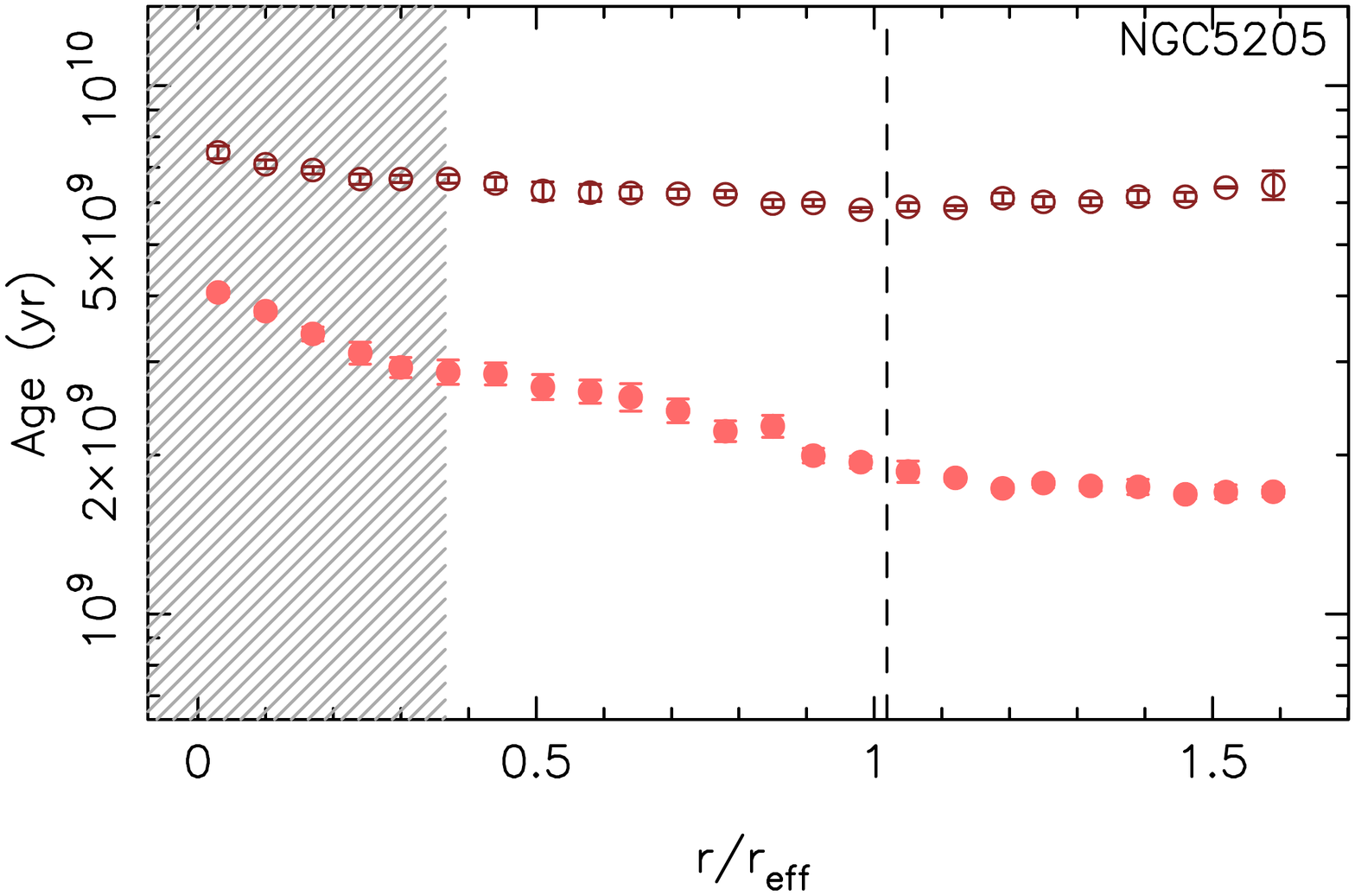}}
\resizebox{0.22\textwidth}{!}{\includegraphics[angle=0]{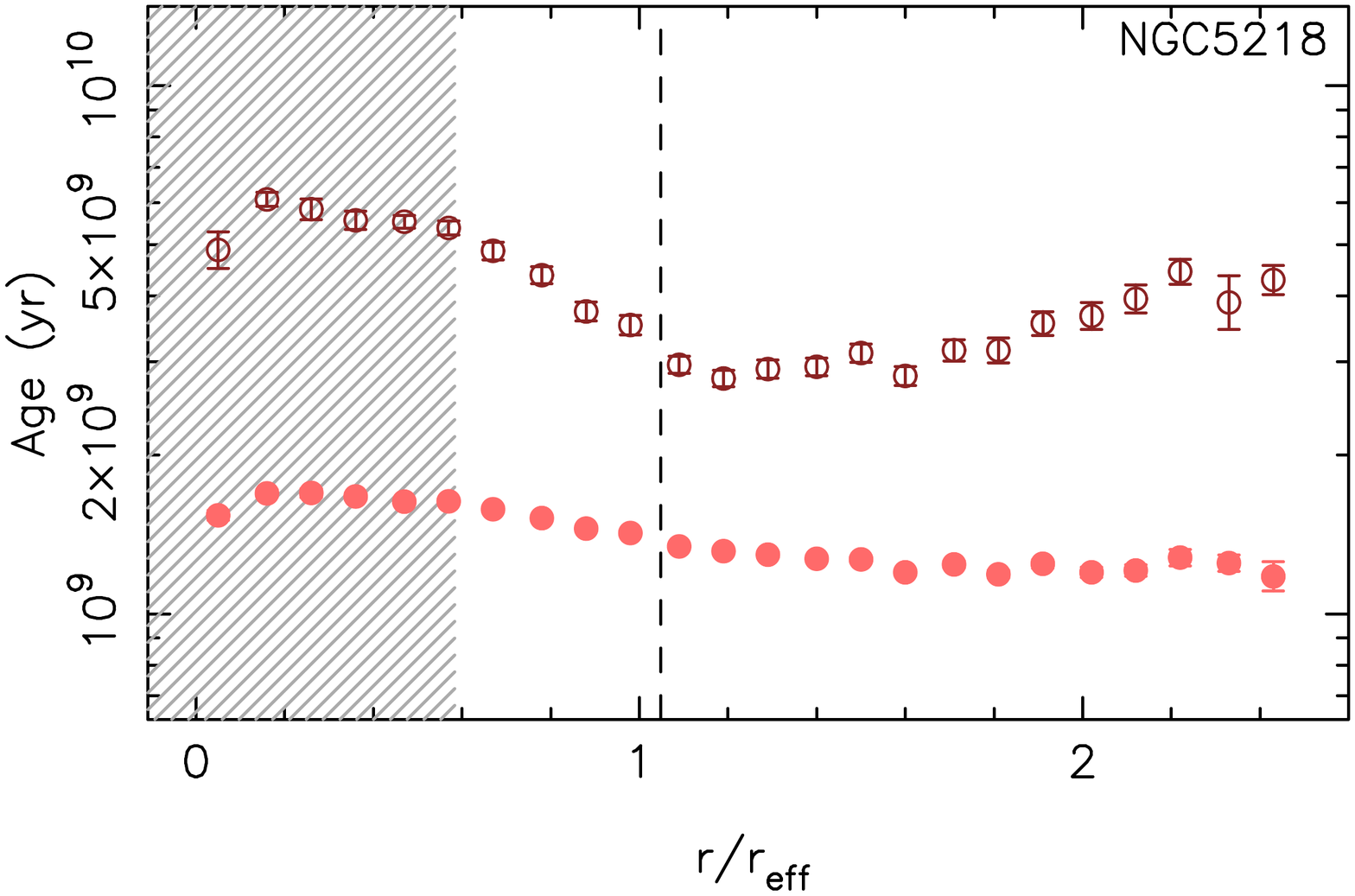}}
\resizebox{0.22\textwidth}{!}{\includegraphics[angle=0]{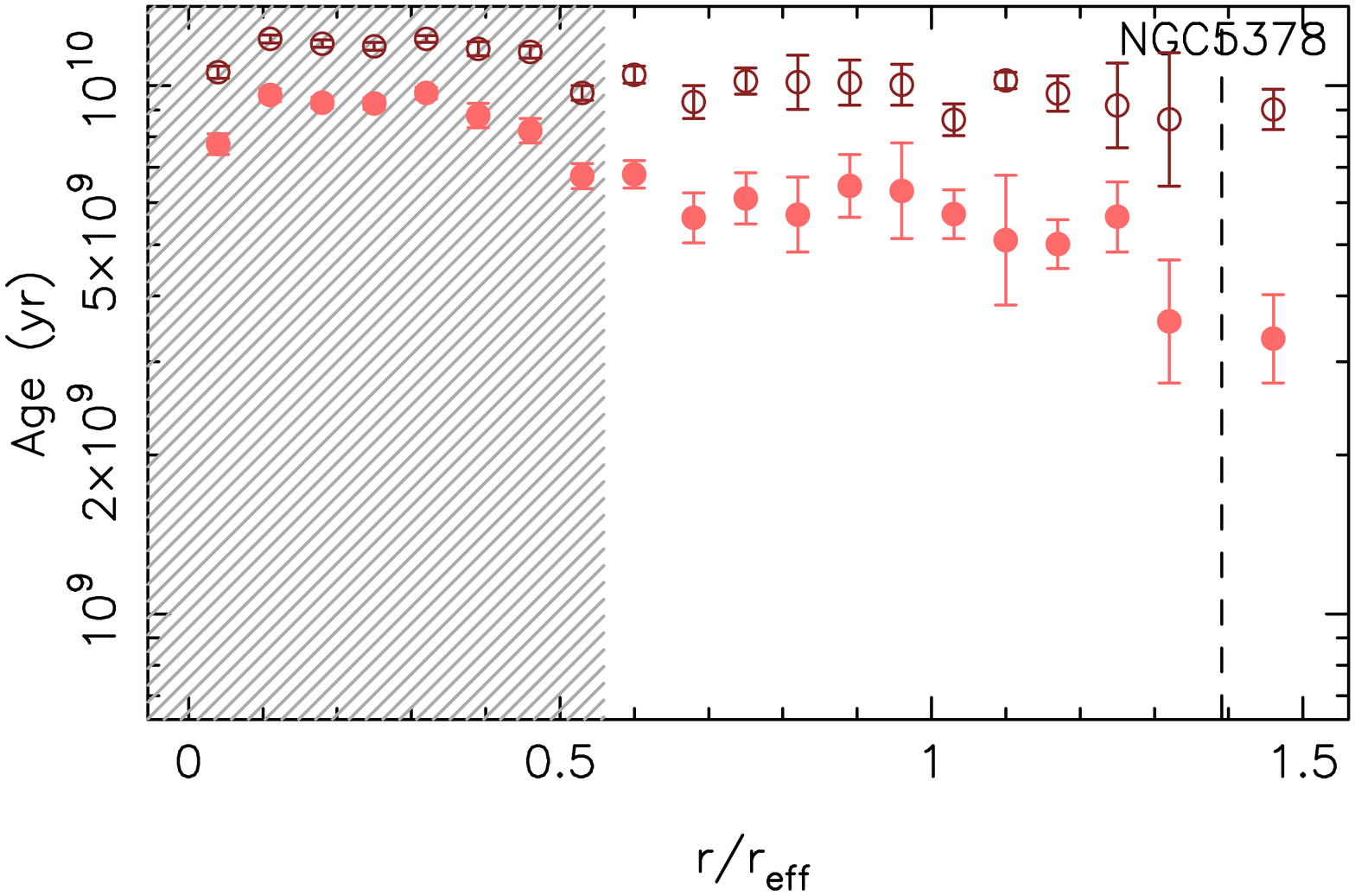}}
\resizebox{0.22\textwidth}{!}{\includegraphics[angle=0]{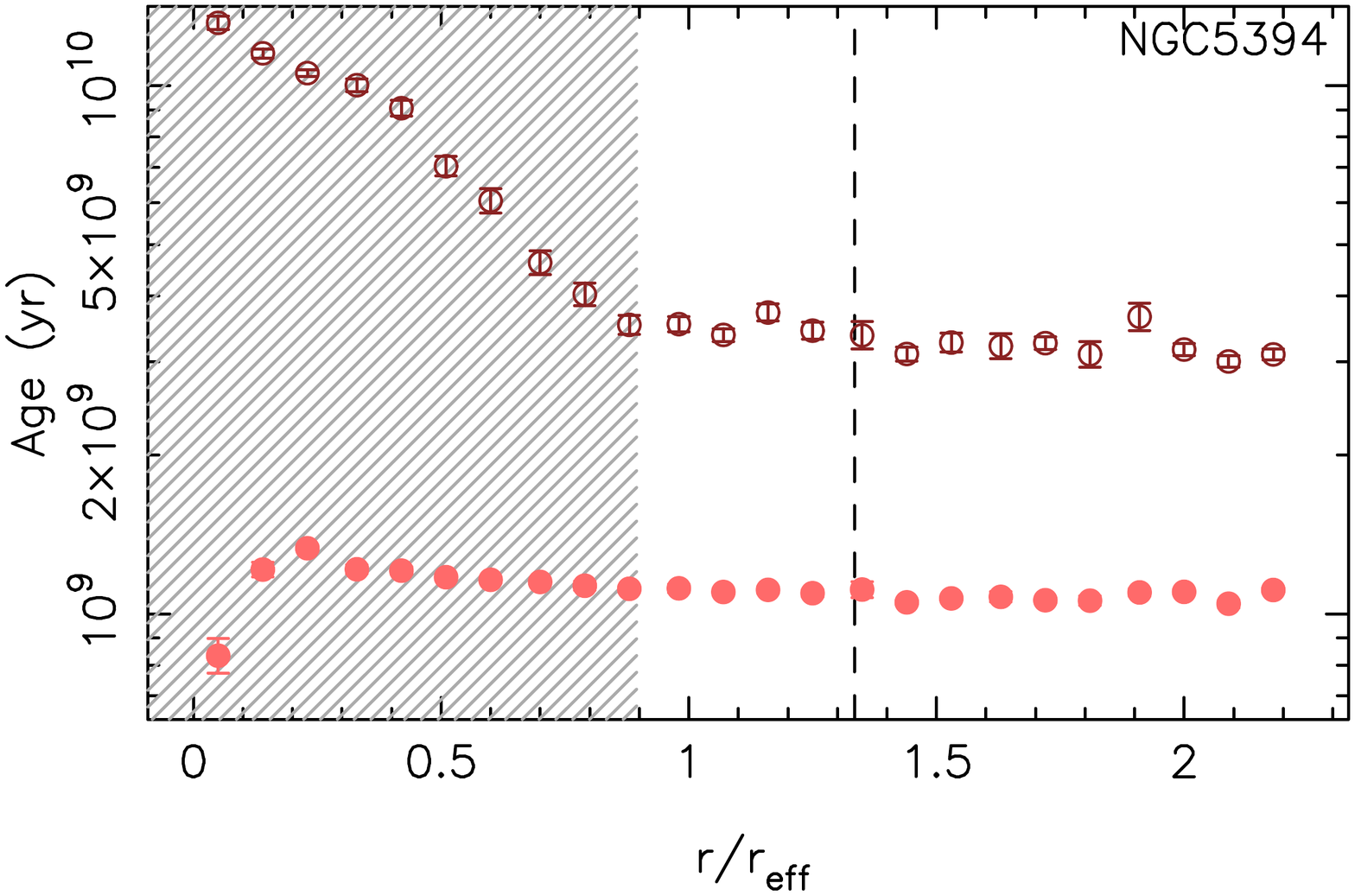}}
\resizebox{0.22\textwidth}{!}{\includegraphics[angle=0]{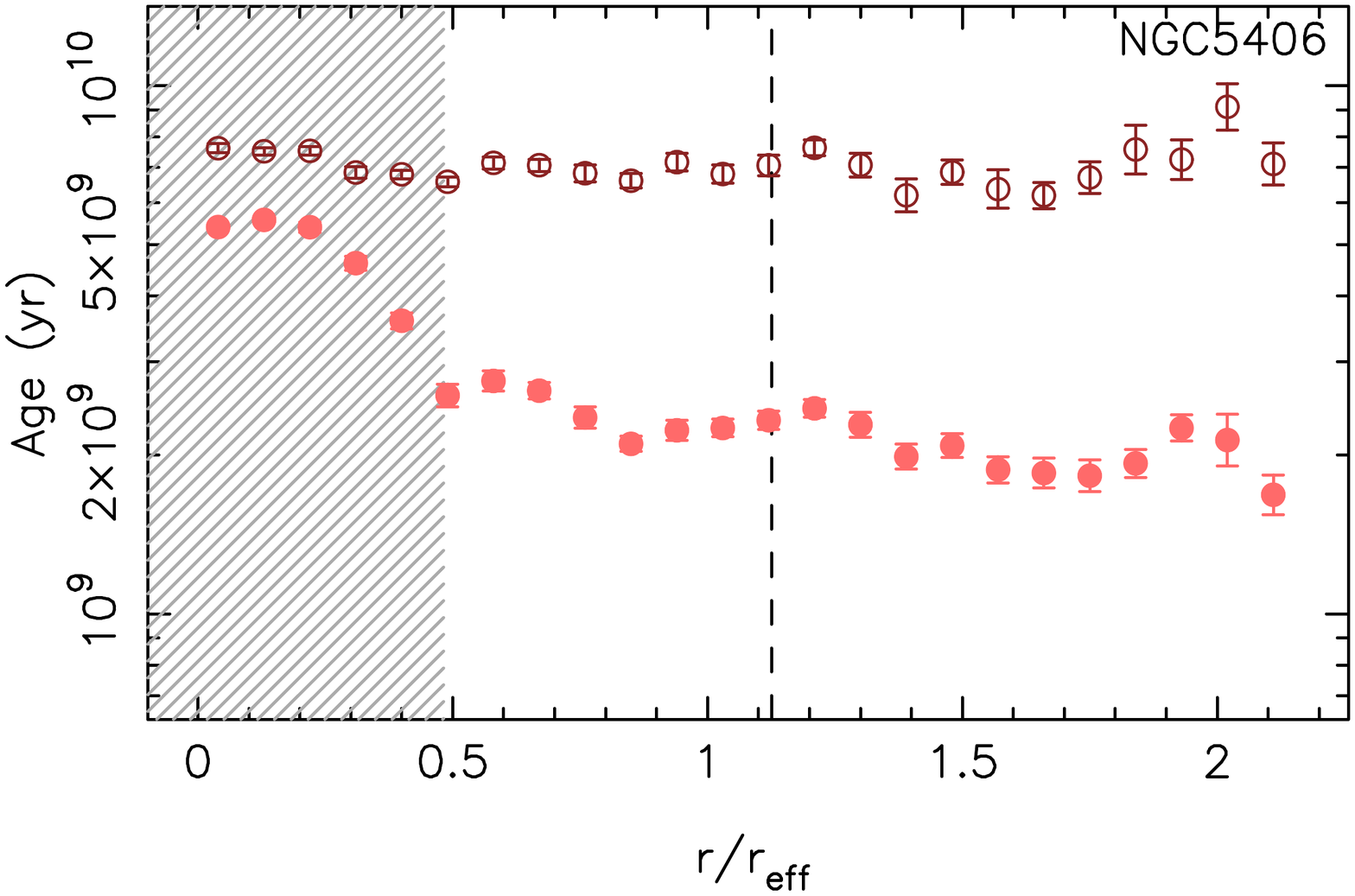}}
\resizebox{0.22\textwidth}{!}{\includegraphics[angle=0]{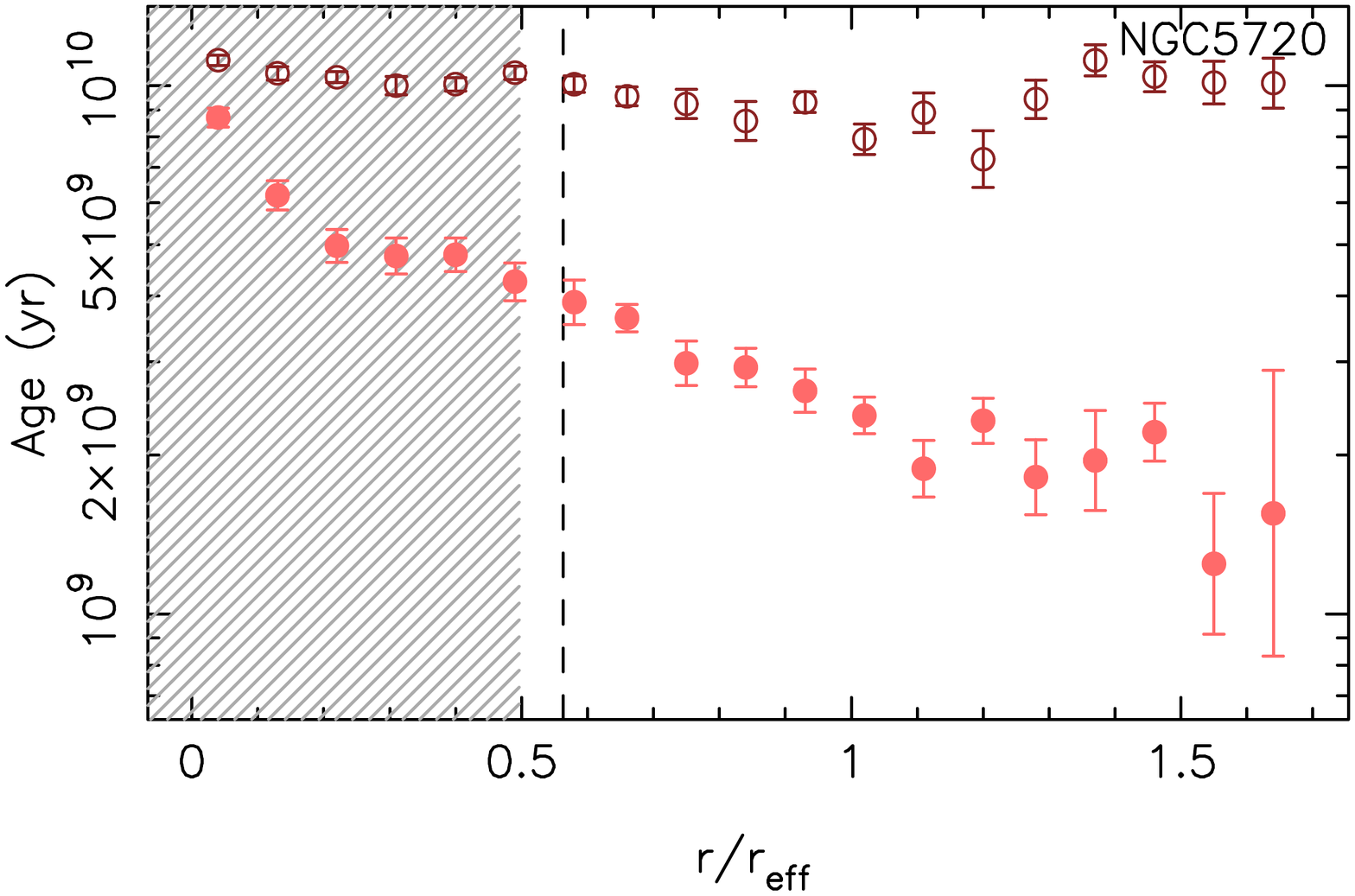}}
\resizebox{0.22\textwidth}{!}{\includegraphics[angle=0]{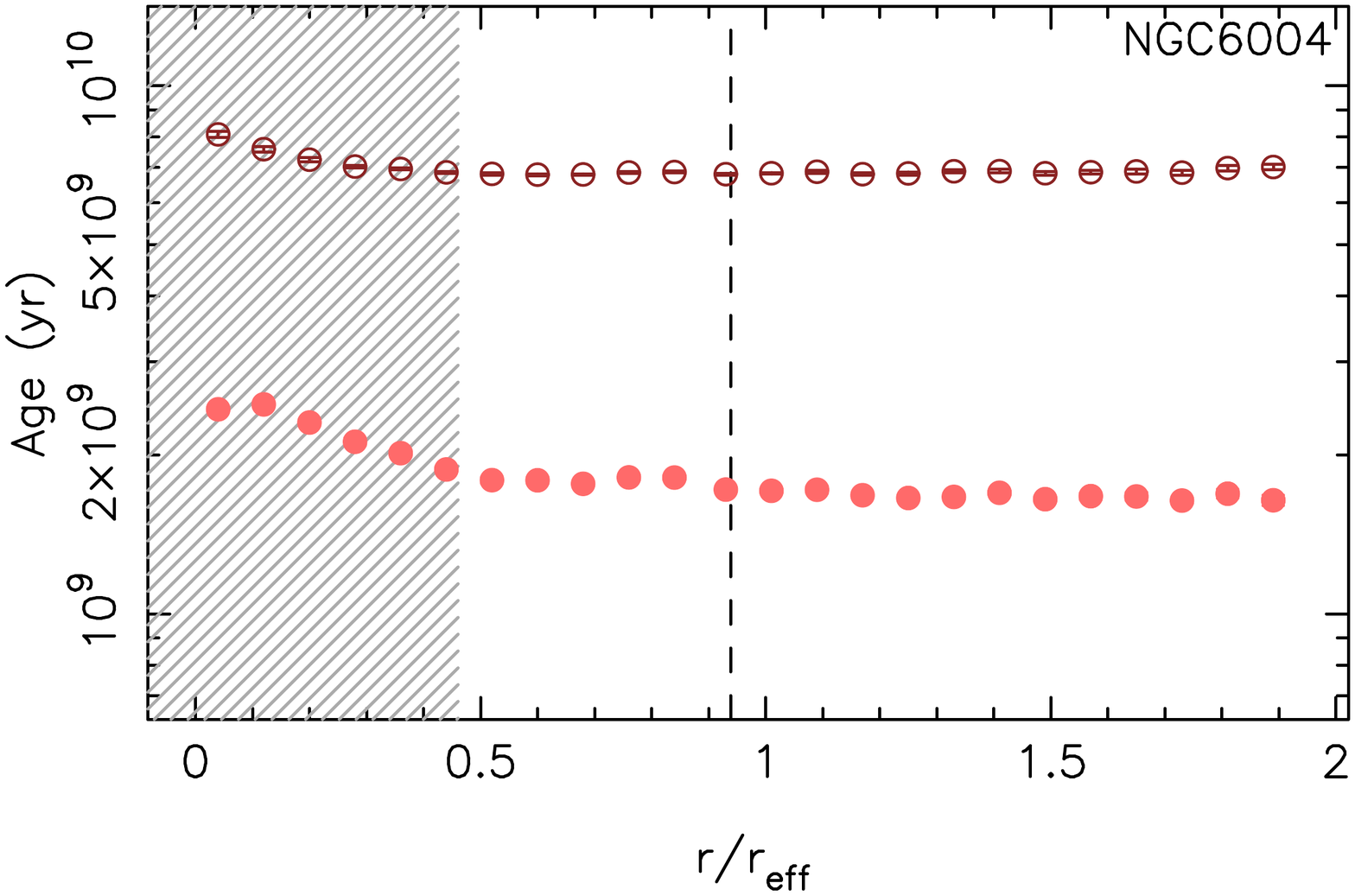}}
\resizebox{0.22\textwidth}{!}{\includegraphics[angle=0]{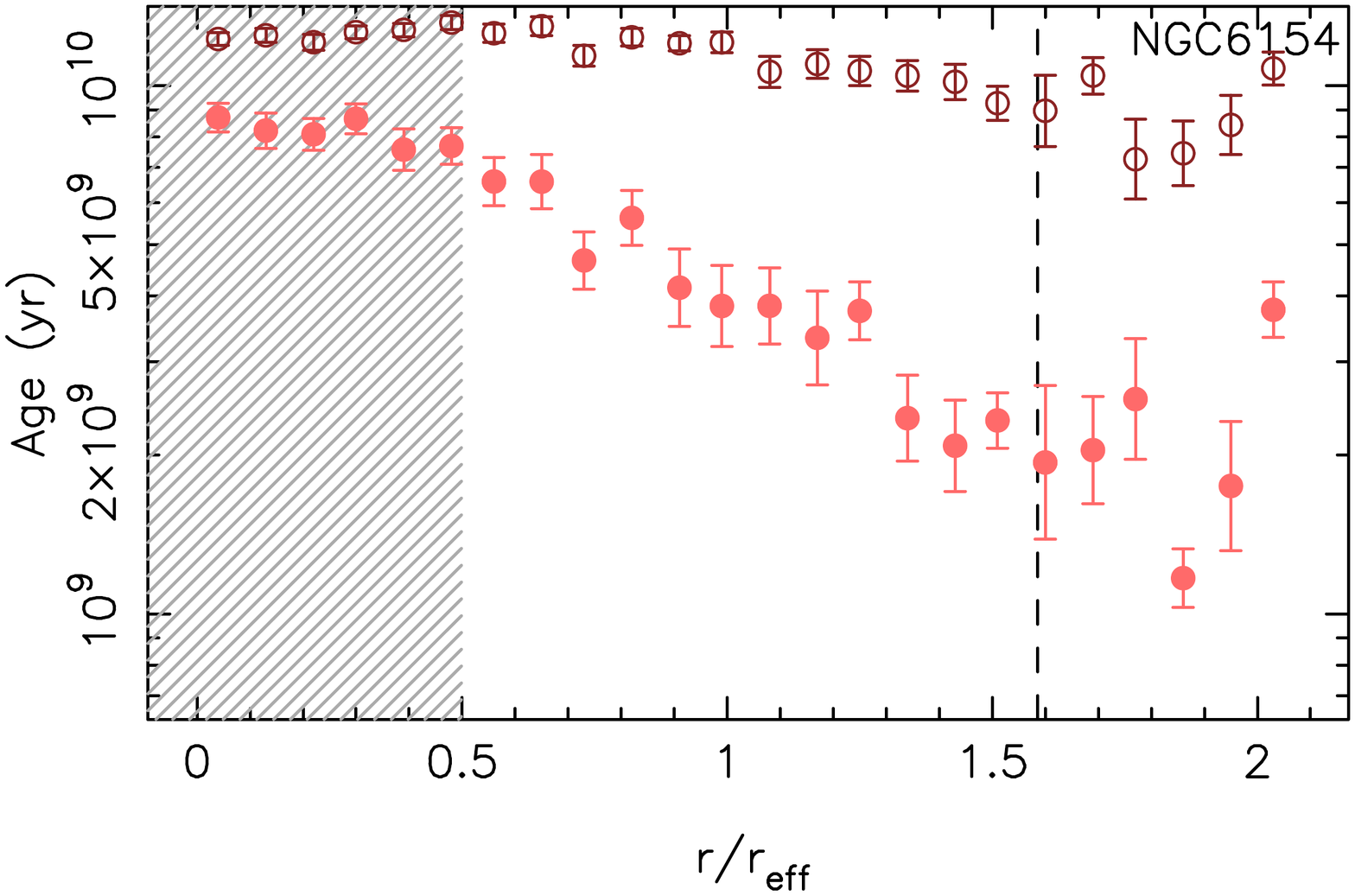}}
\resizebox{0.22\textwidth}{!}{\includegraphics[angle=0]{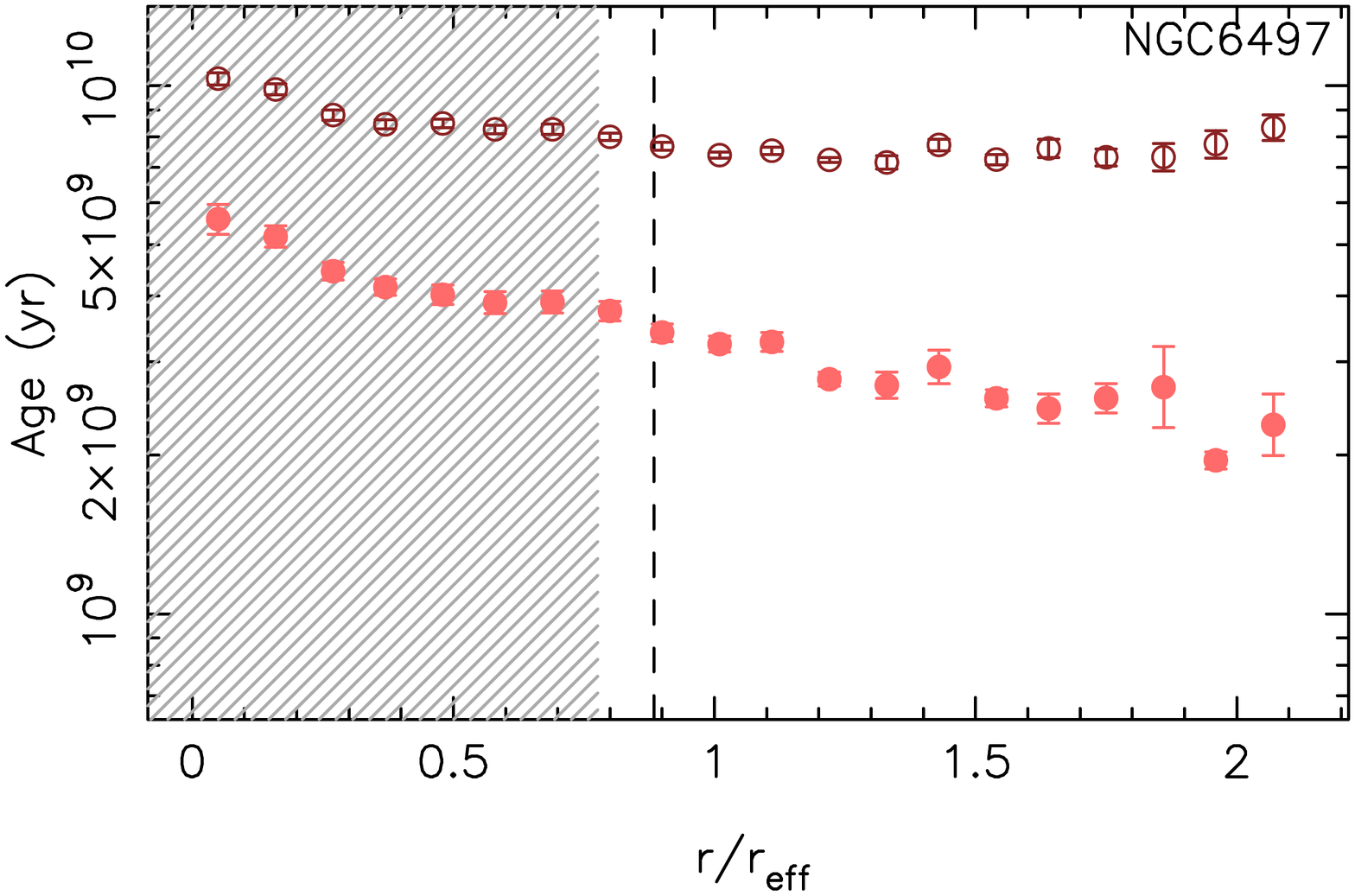}}
\resizebox{0.22\textwidth}{!}{\includegraphics[angle=0]{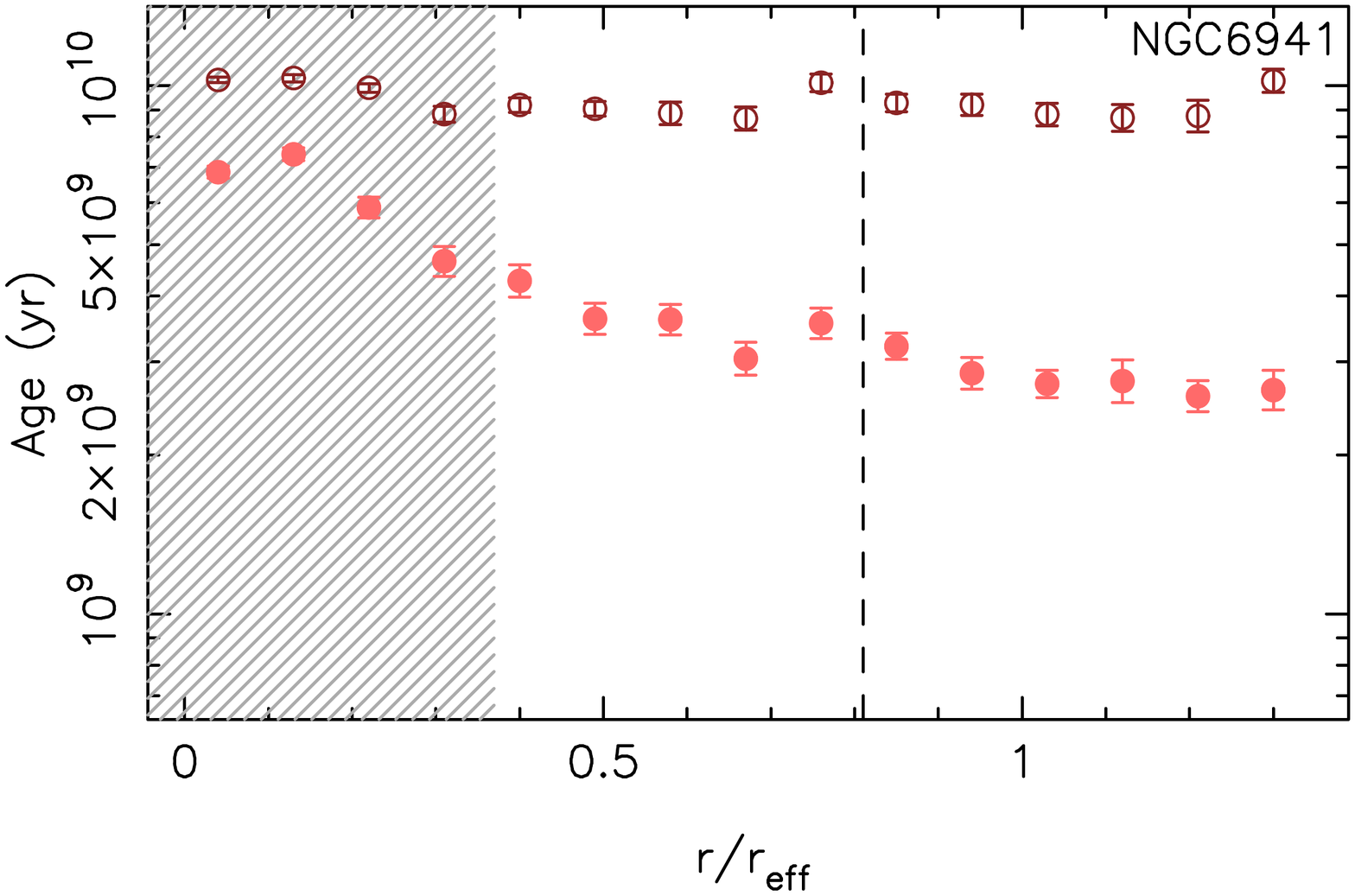}}
\resizebox{0.22\textwidth}{!}{\includegraphics[angle=0]{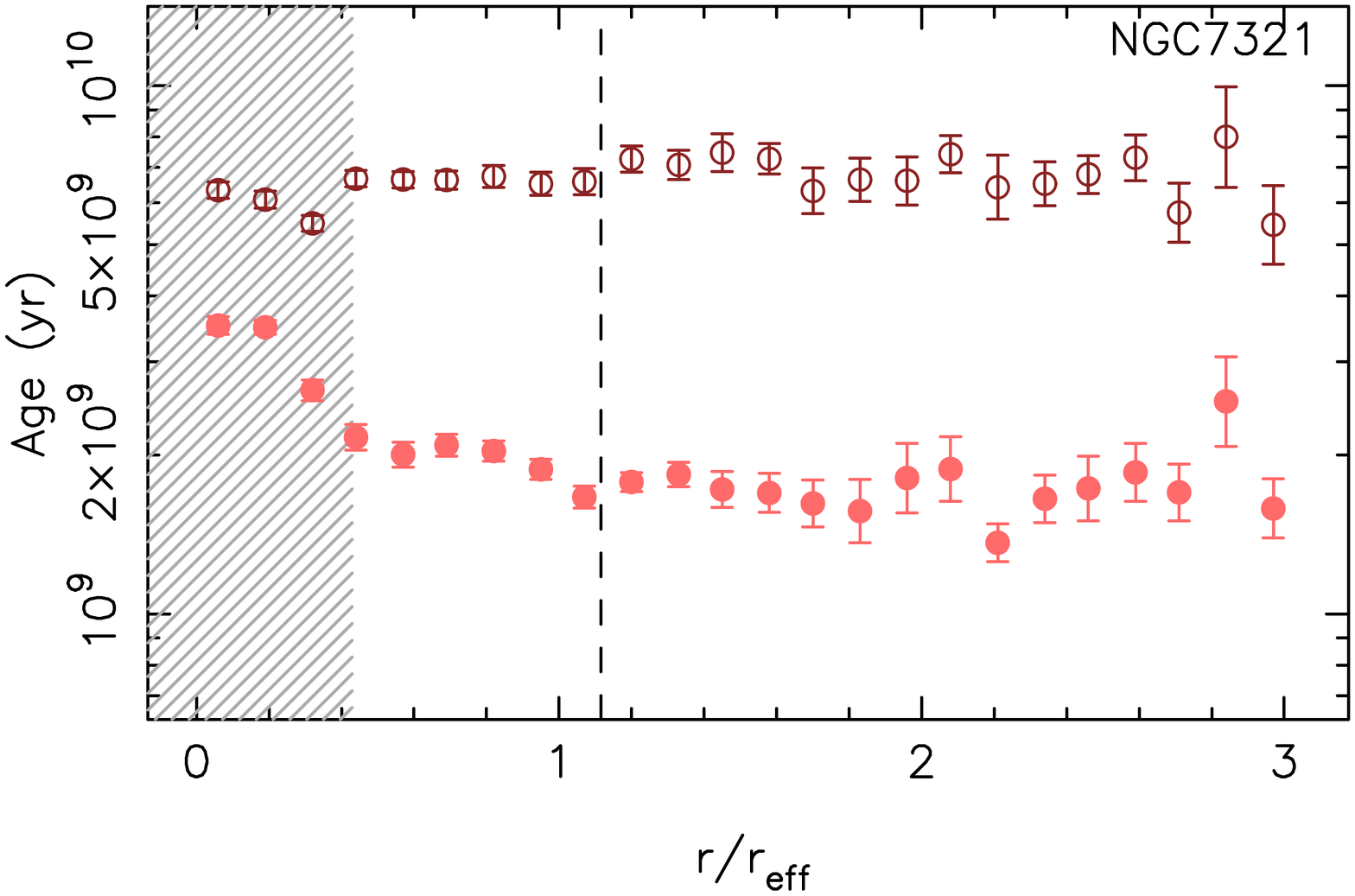}}
\resizebox{0.22\textwidth}{!}{\includegraphics[angle=0]{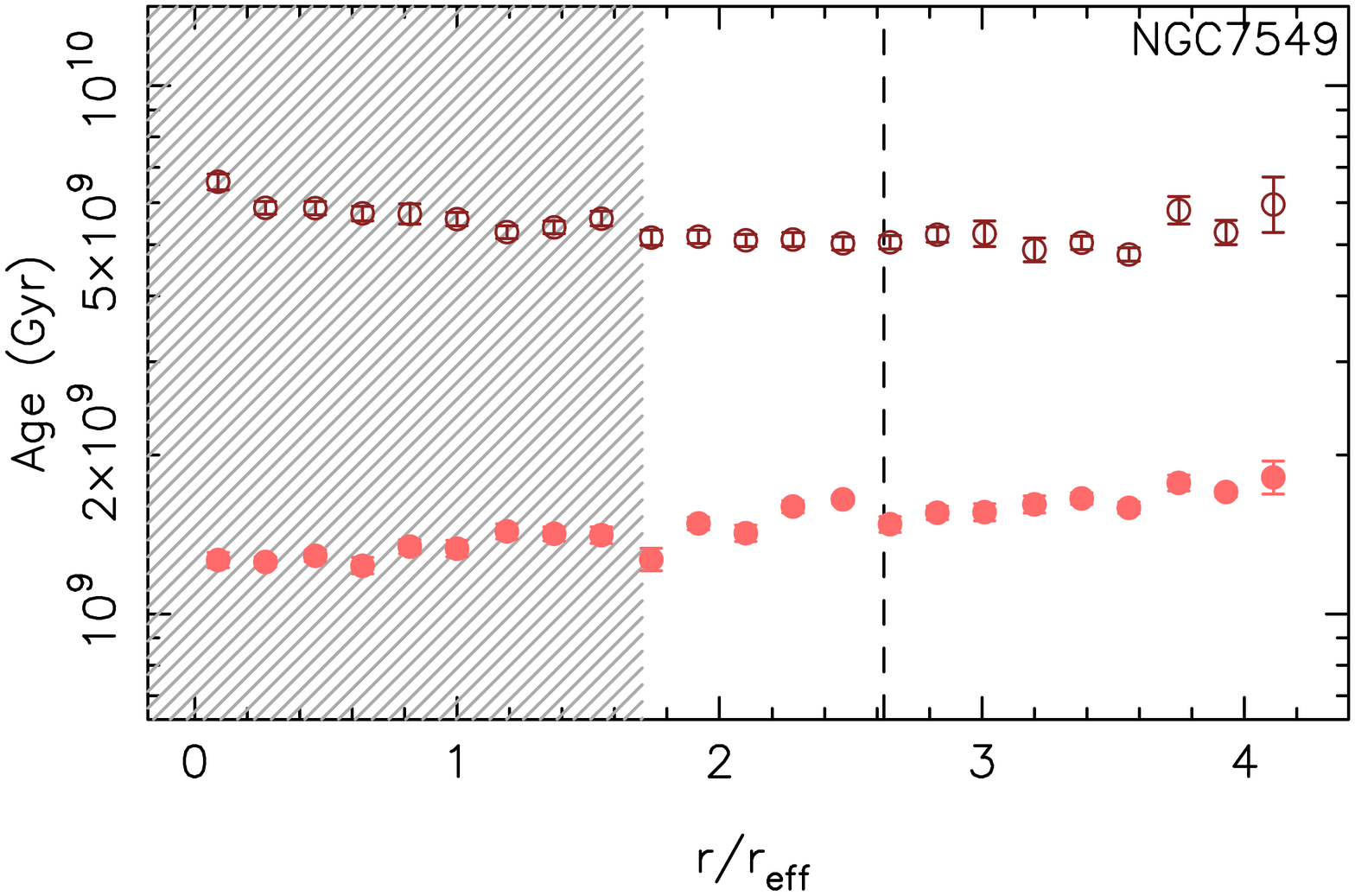}}
\resizebox{0.22\textwidth}{!}{\includegraphics[angle=0]{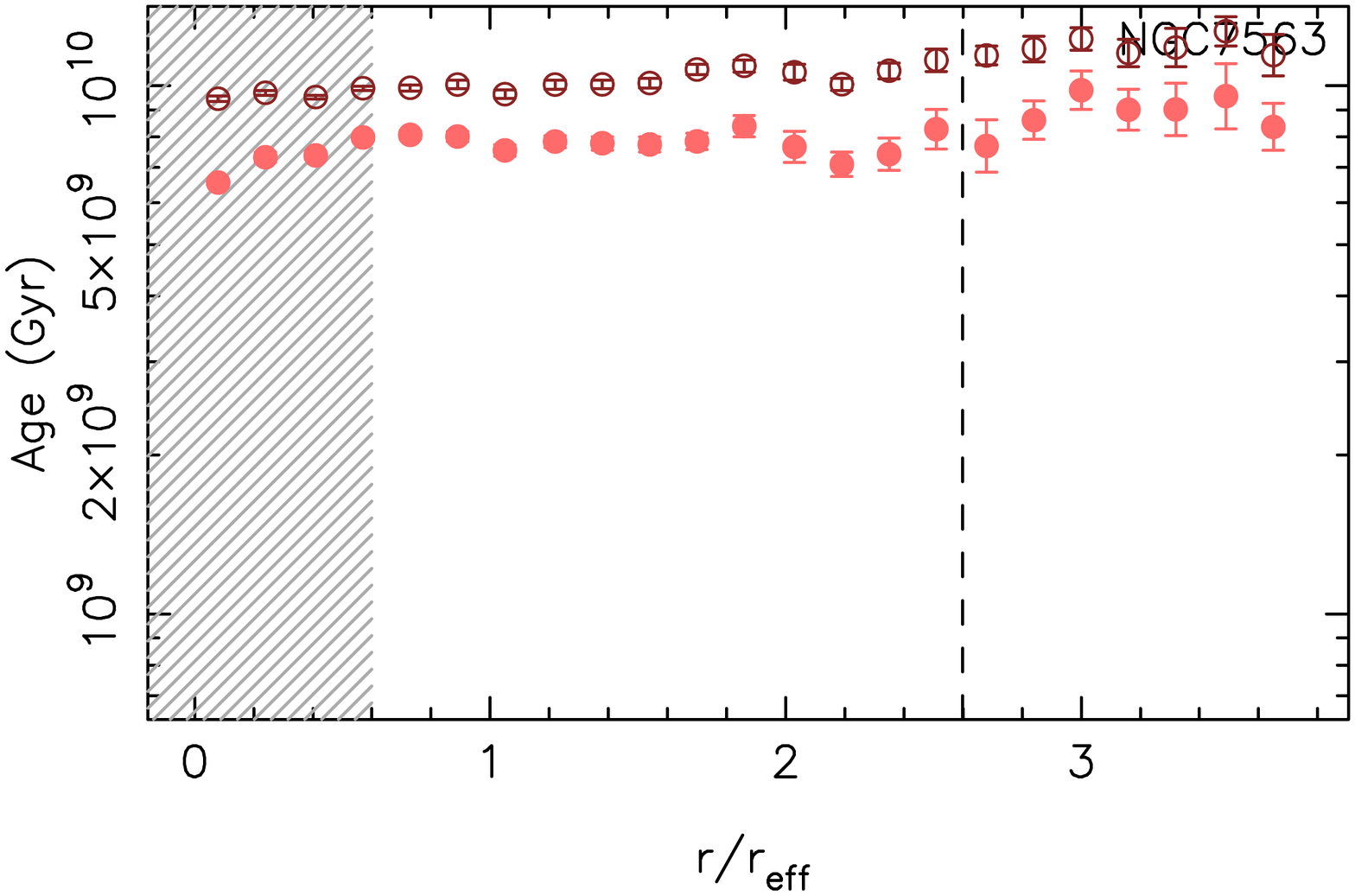}}
\resizebox{0.22\textwidth}{!}{\includegraphics[angle=0]{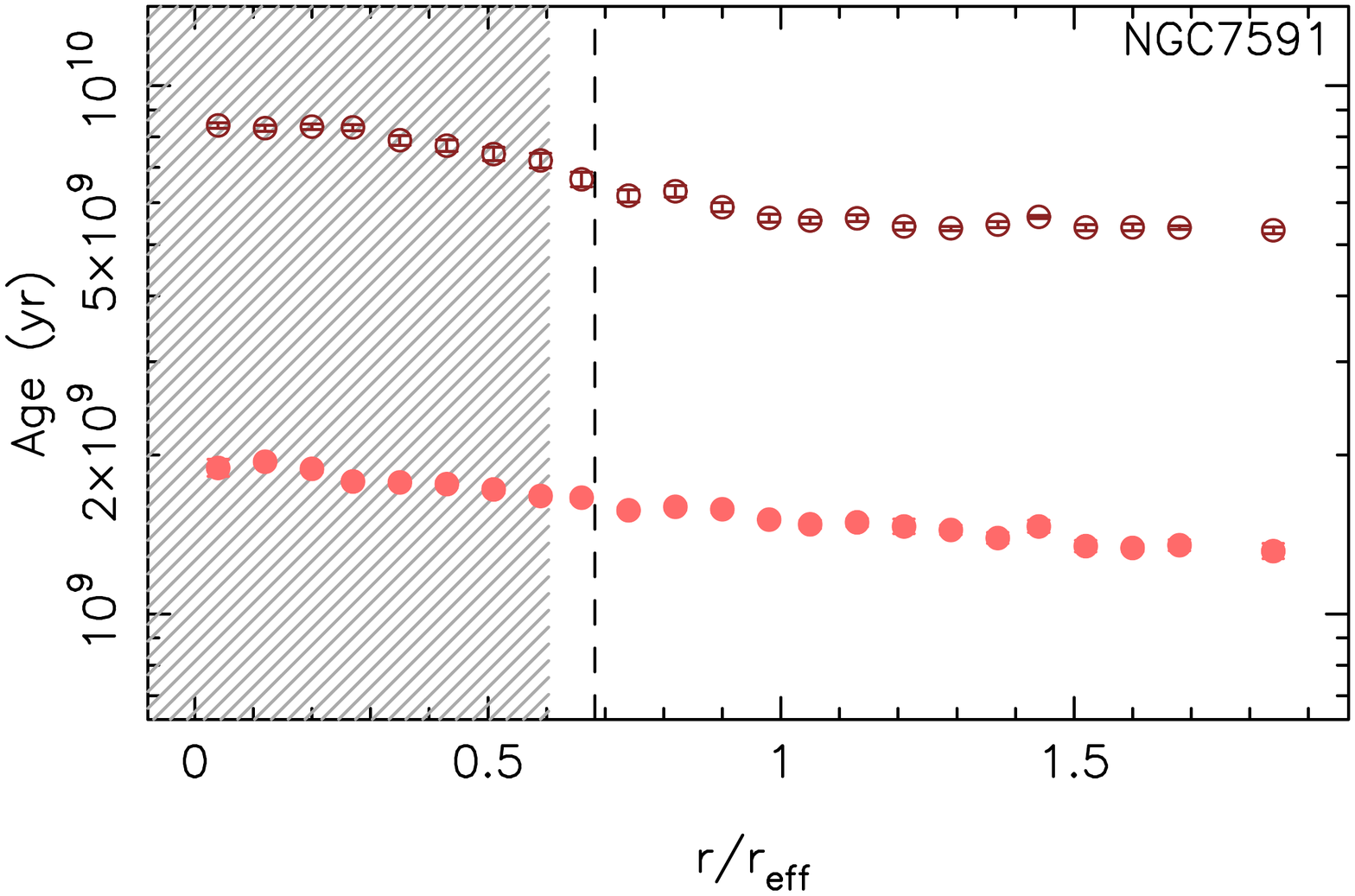}}
\resizebox{0.22\textwidth}{!}{\includegraphics[angle=0]{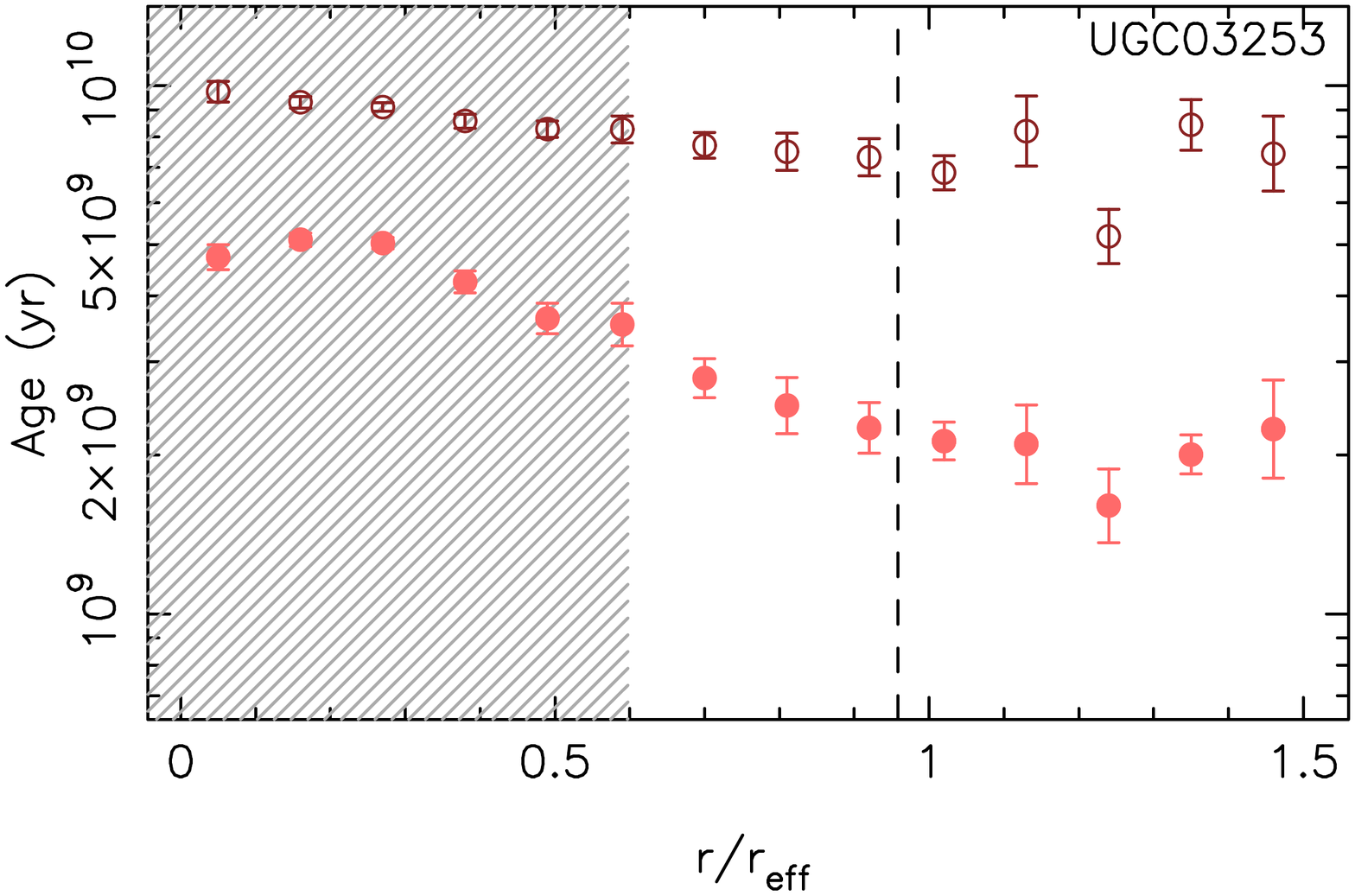}}
\resizebox{0.22\textwidth}{!}{\includegraphics[angle=0]{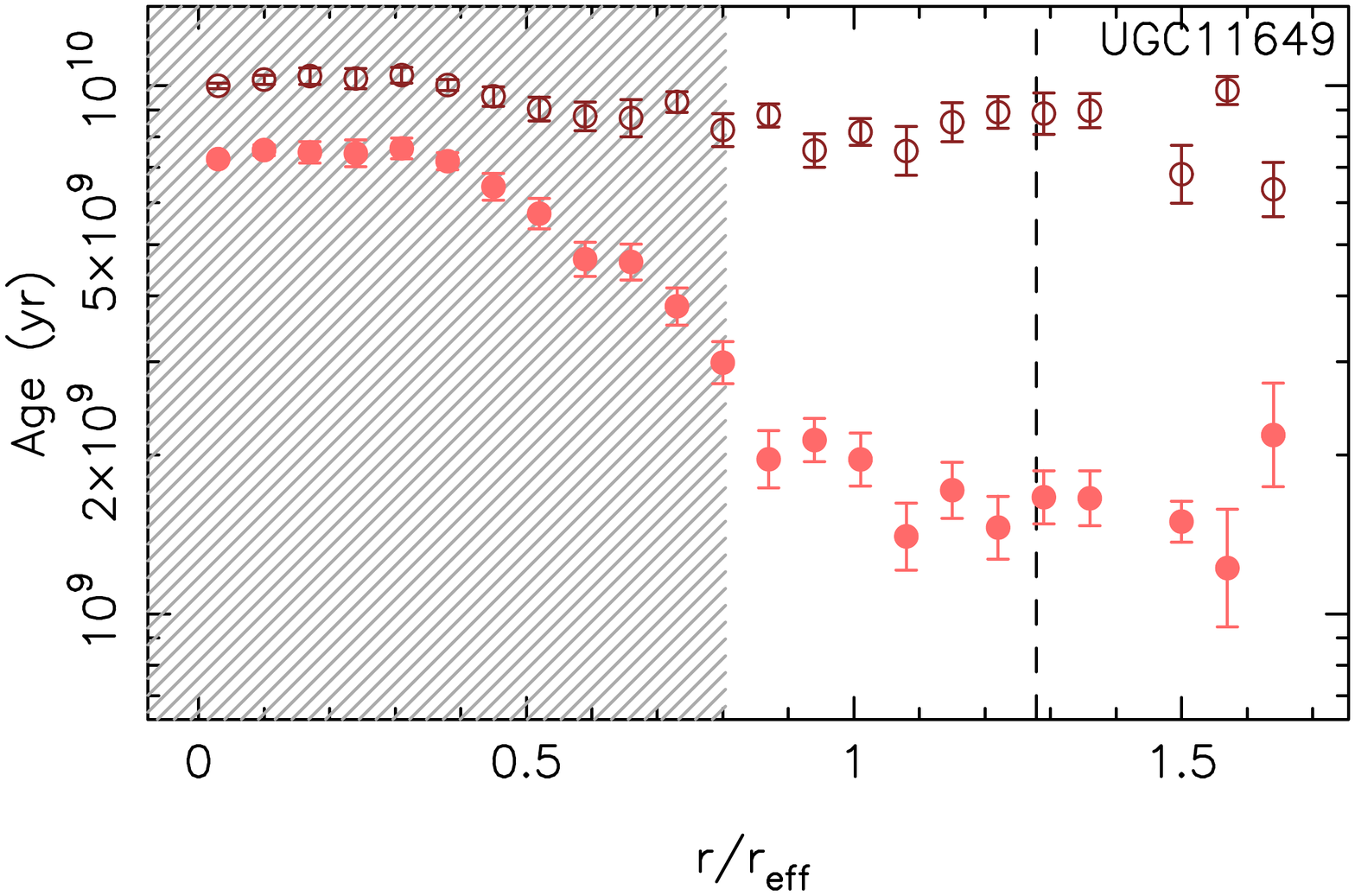}}
\resizebox{0.22\textwidth}{!}{\includegraphics[angle=0]{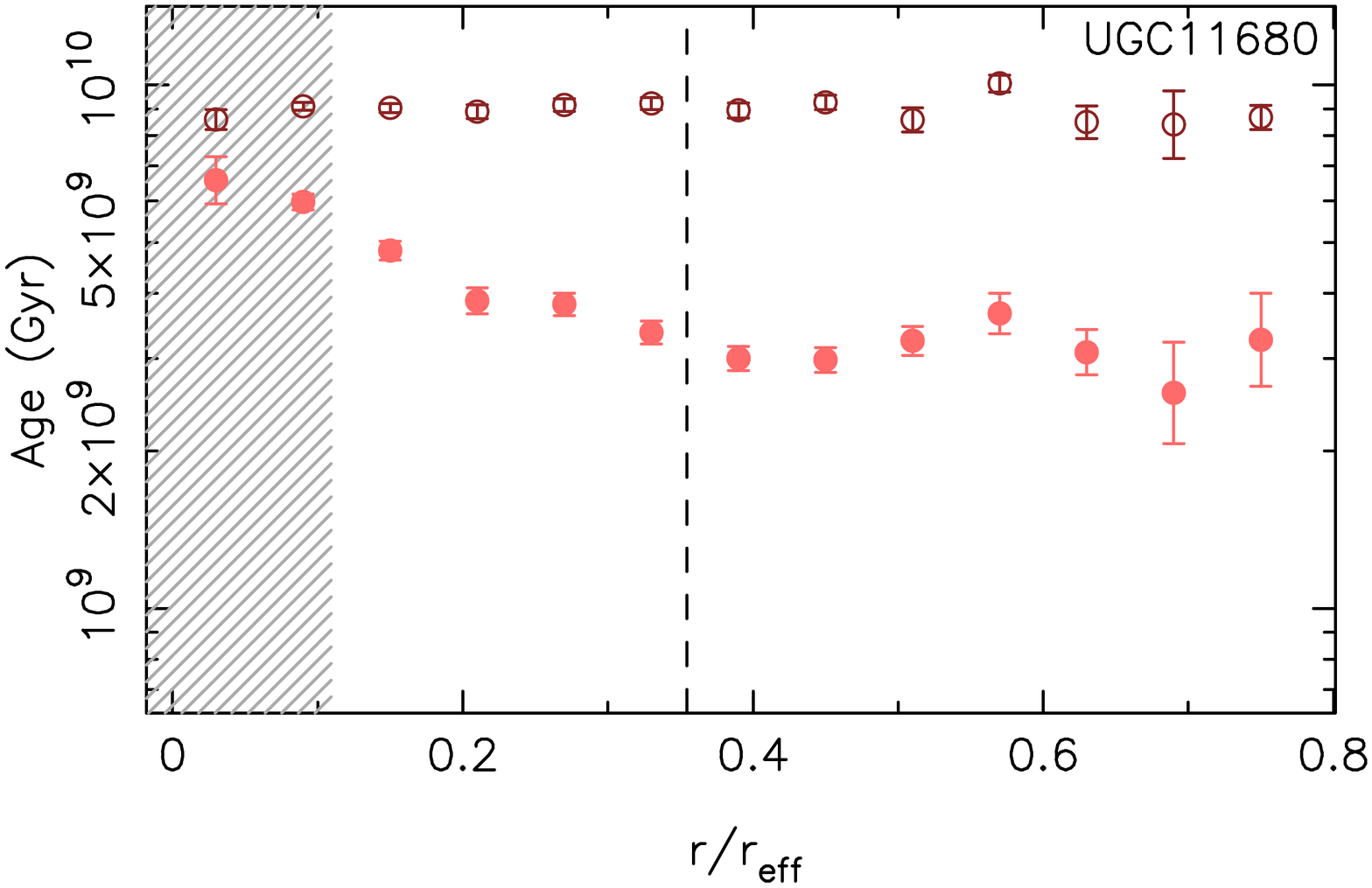}}
\caption{Mean luminosity- (light red) and mass- (dark red) age as a function of radius, normalized
to the effective radius of the disc. Shaded area indicate the radial range of the
surface brightness profile dominated by the bulge. The dashed line shows the
bar length measured as indicated in the text. Solid lines shows the linear
fit performed on the disc region. \label{fig:grad_age1}}
\end{figure*}

Figure~\ref{fig:grad_age1} shows the age gradients for our sample of galaxies.
\begin{figure*}
\centering
\resizebox{0.22\textwidth}{!}{\includegraphics[angle=0]{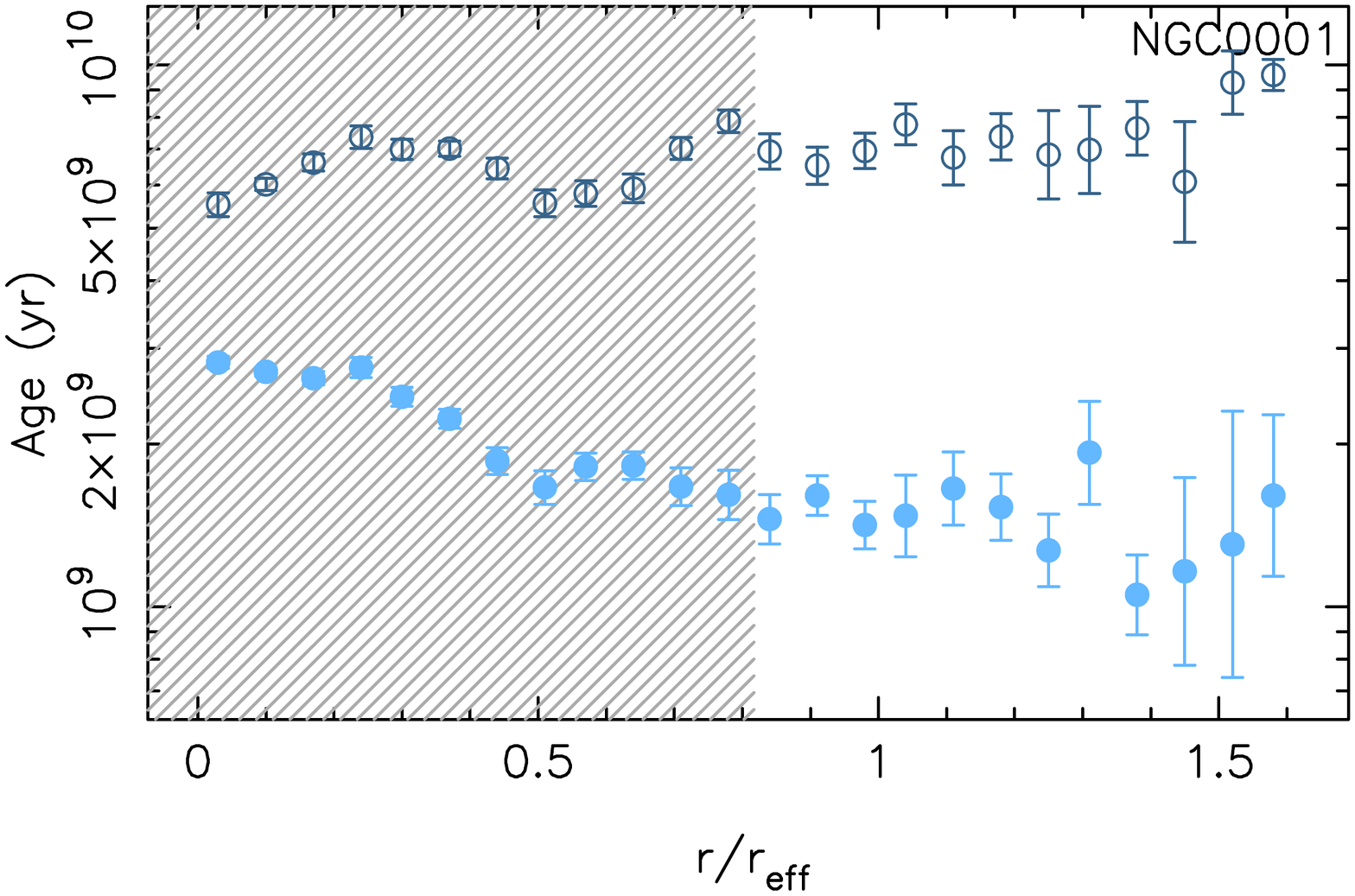}}
\resizebox{0.22\textwidth}{!}{\includegraphics[angle=0]{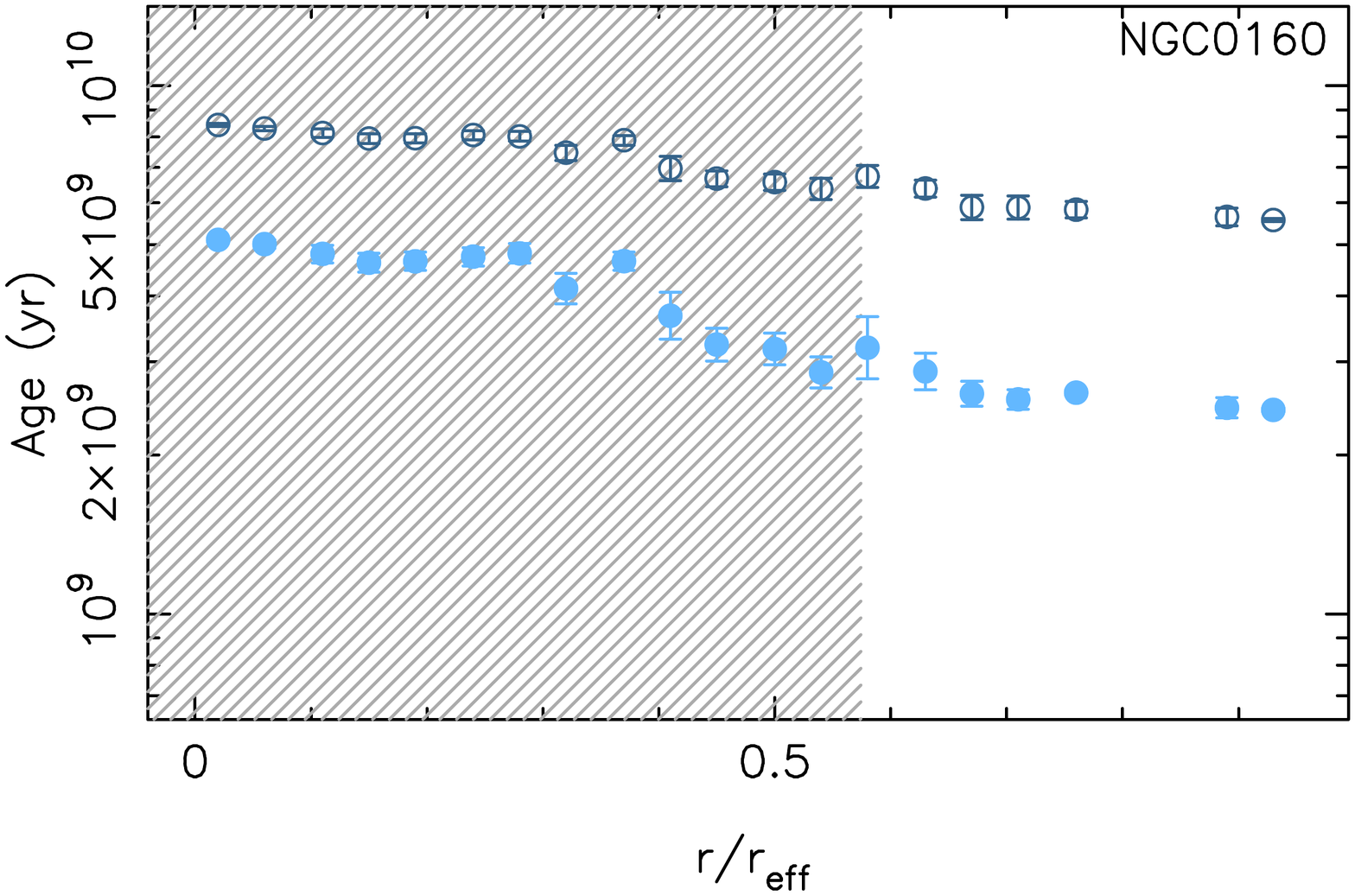}}
\resizebox{0.22\textwidth}{!}{\includegraphics[angle=0]{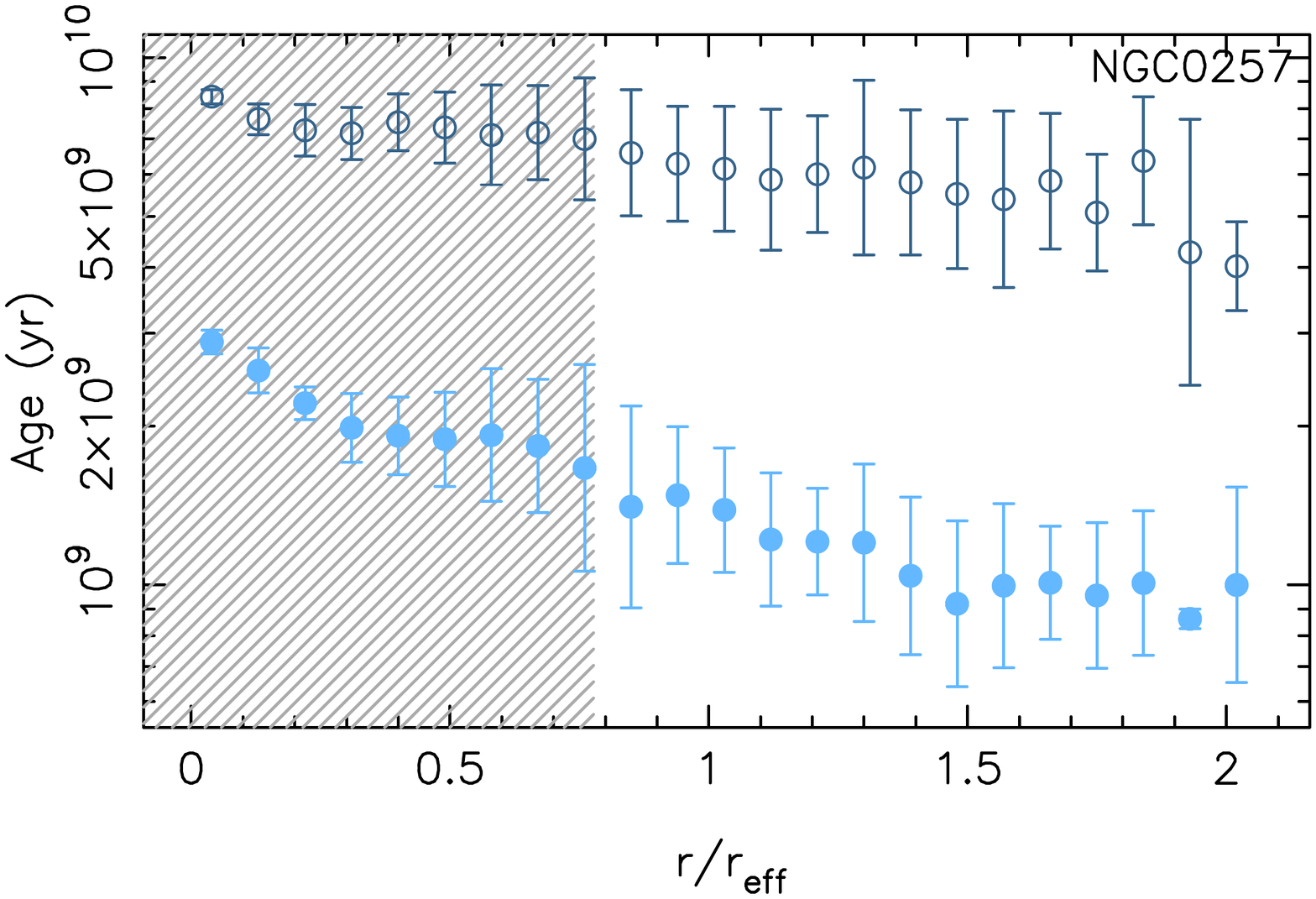}}
\resizebox{0.22\textwidth}{!}{\includegraphics[angle=0]{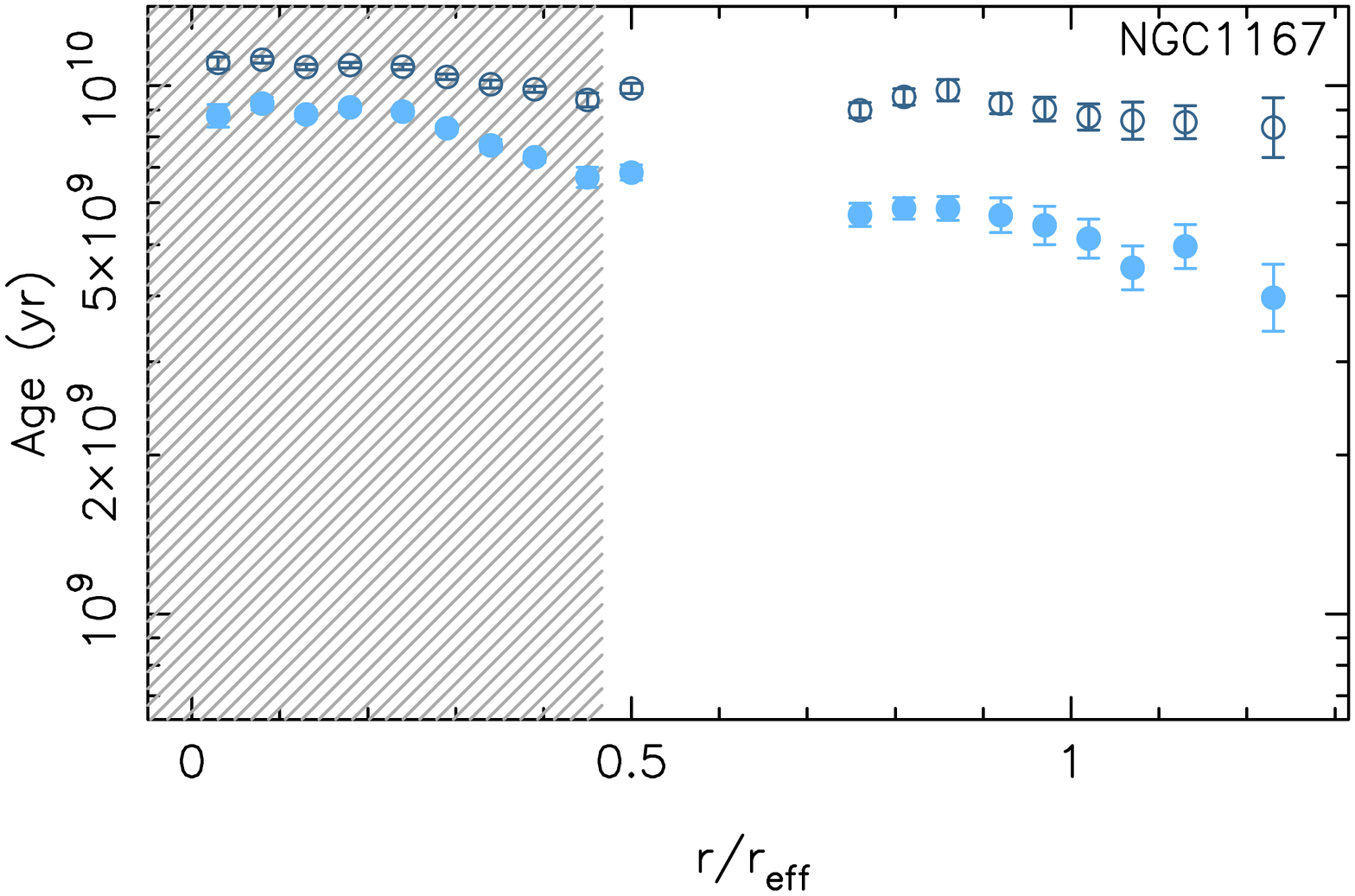}}
\resizebox{0.22\textwidth}{!}{\includegraphics[angle=0]{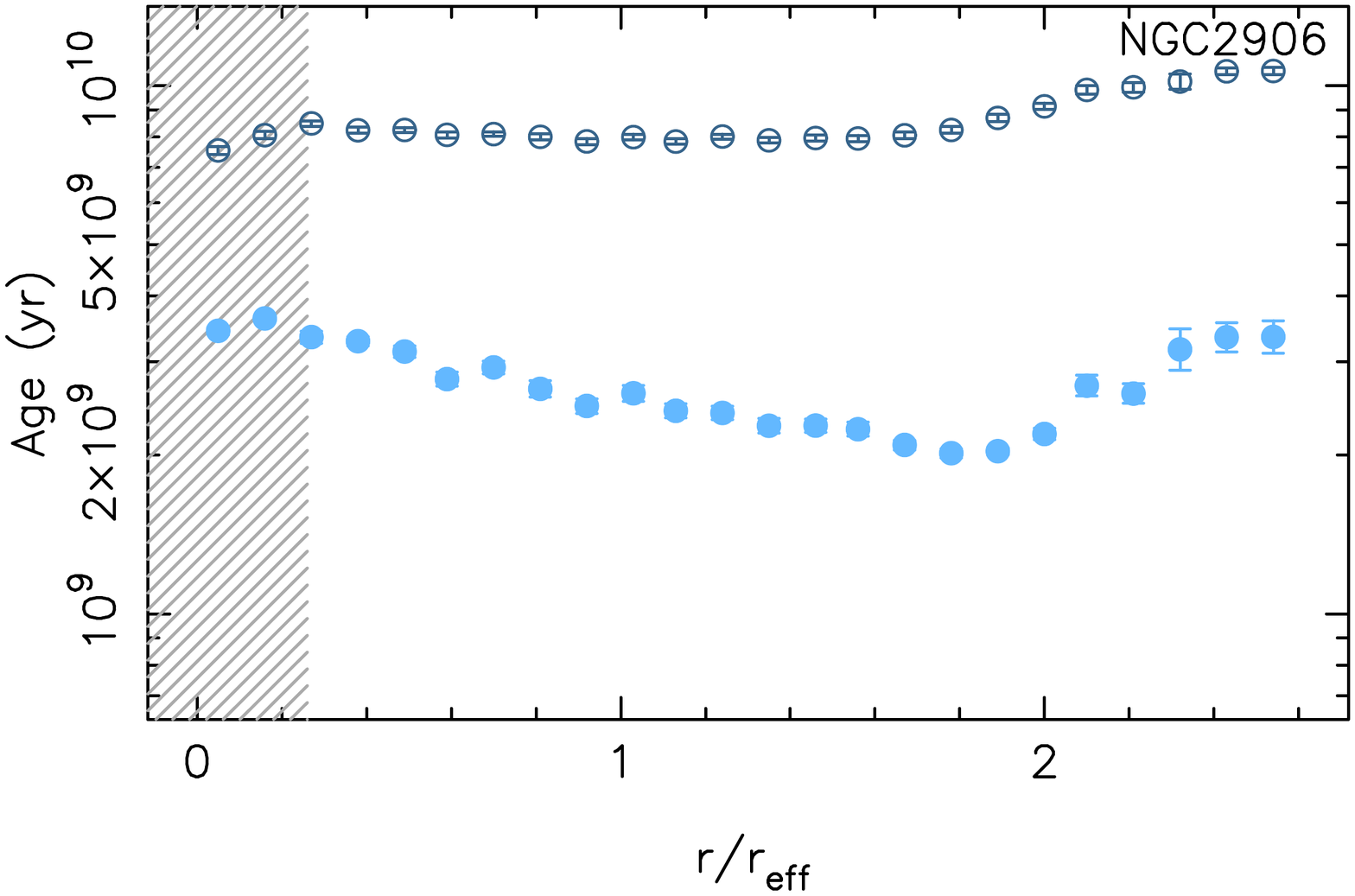}}
\resizebox{0.22\textwidth}{!}{\includegraphics[angle=0]{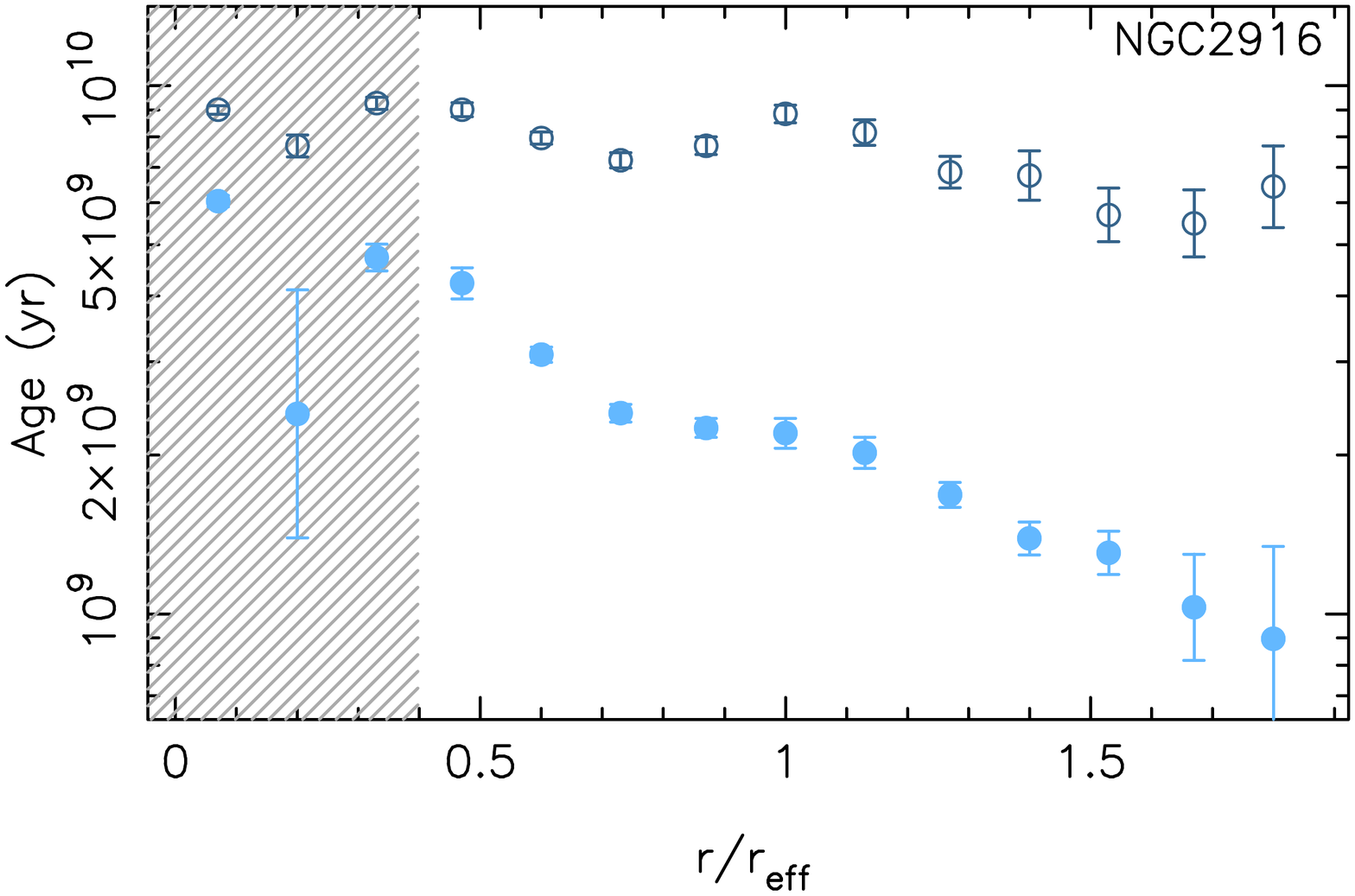}}
\resizebox{0.22\textwidth}{!}{\includegraphics[angle=0]{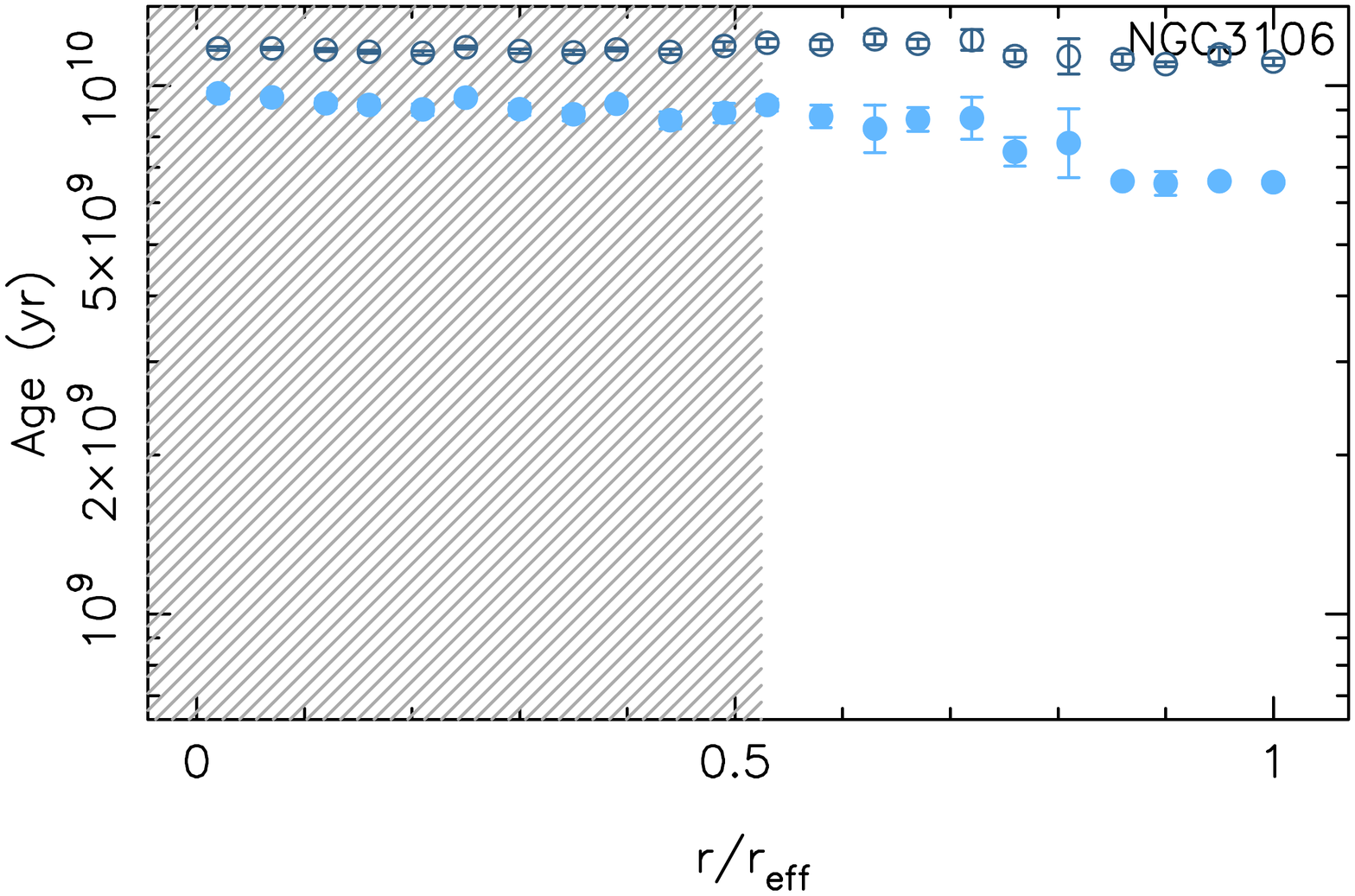}}
\resizebox{0.22\textwidth}{!}{\includegraphics[angle=0]{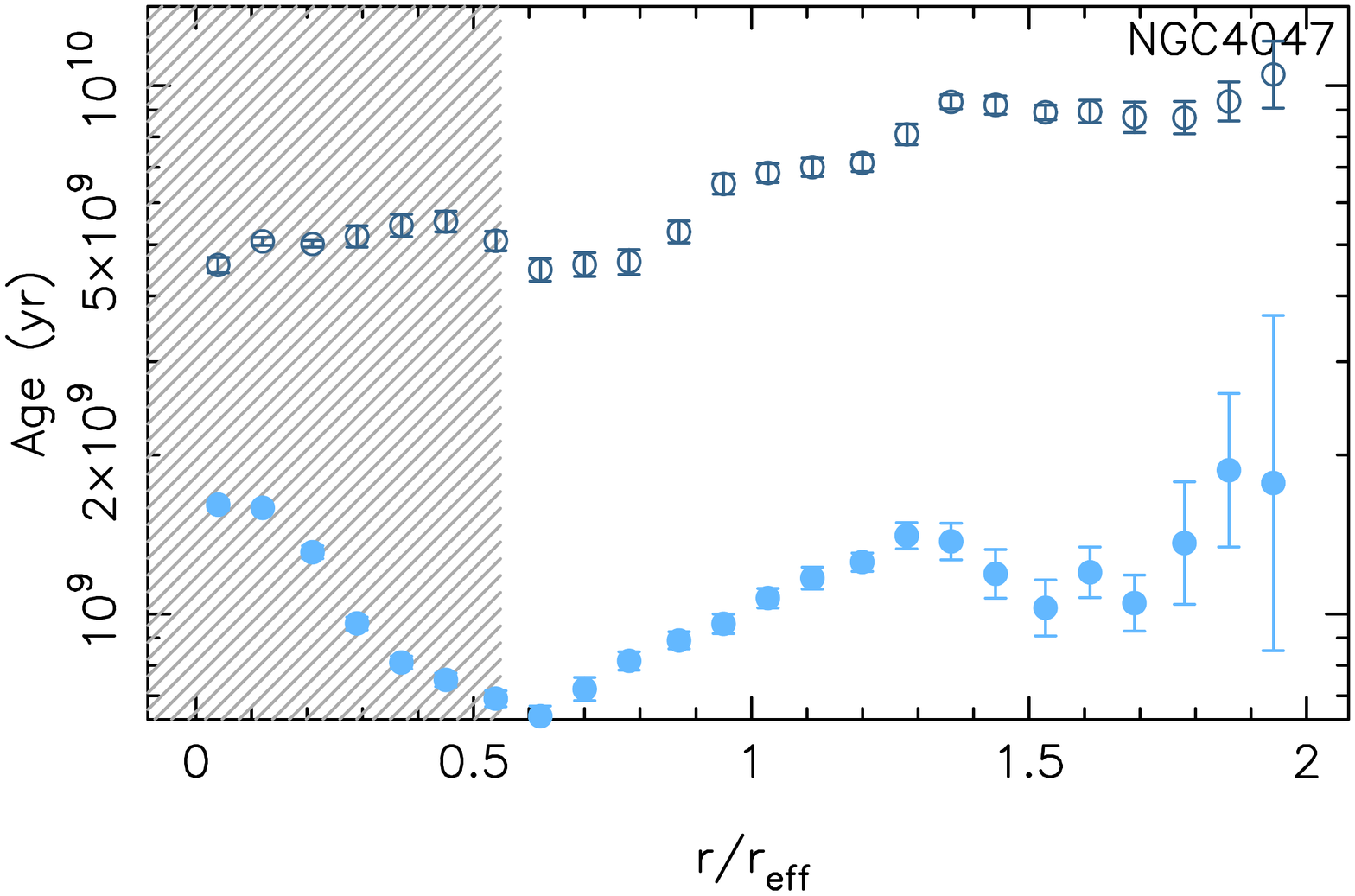}}
\resizebox{0.22\textwidth}{!}{\includegraphics[angle=0]{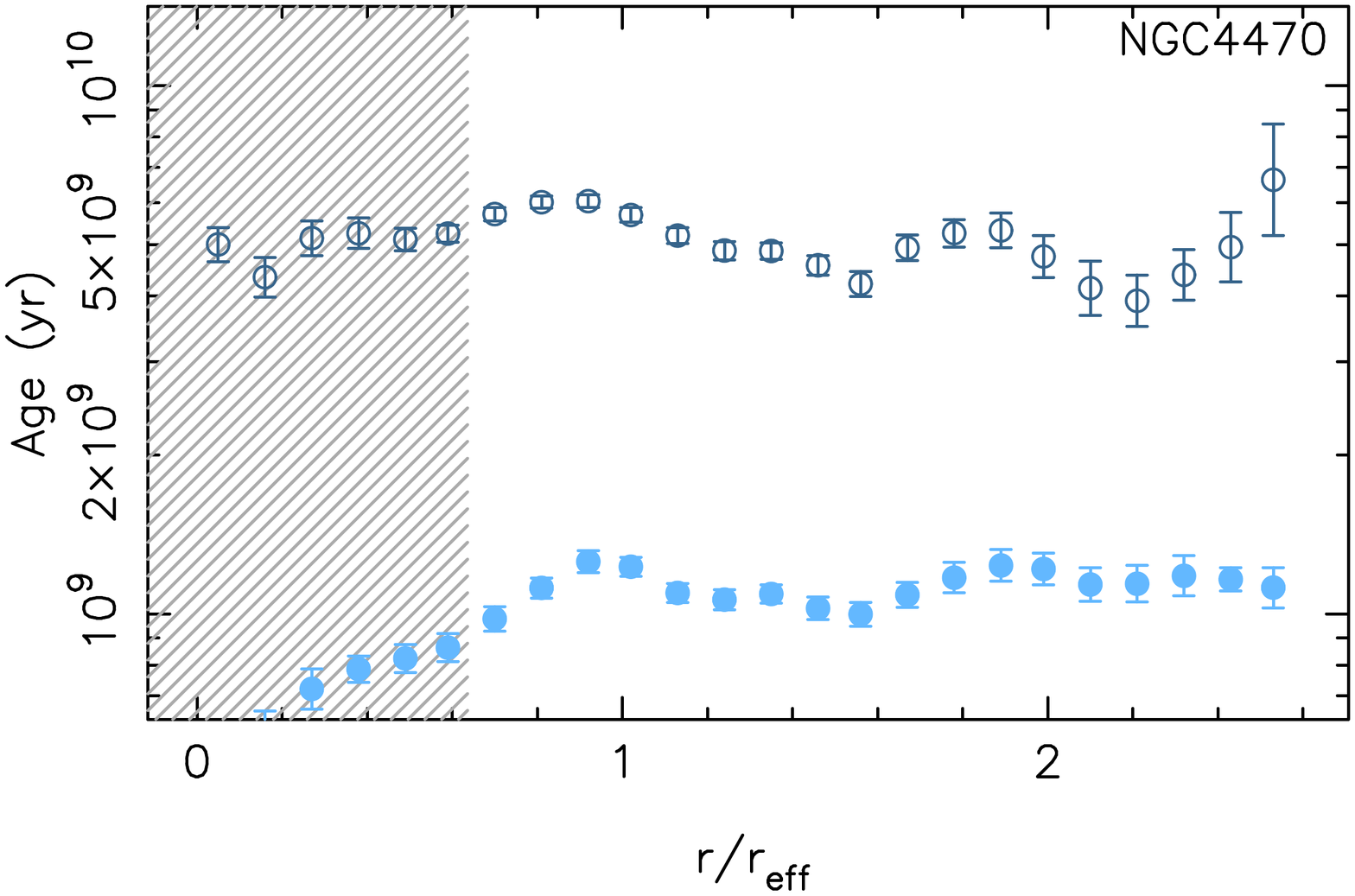}}
\resizebox{0.22\textwidth}{!}{\includegraphics[angle=0]{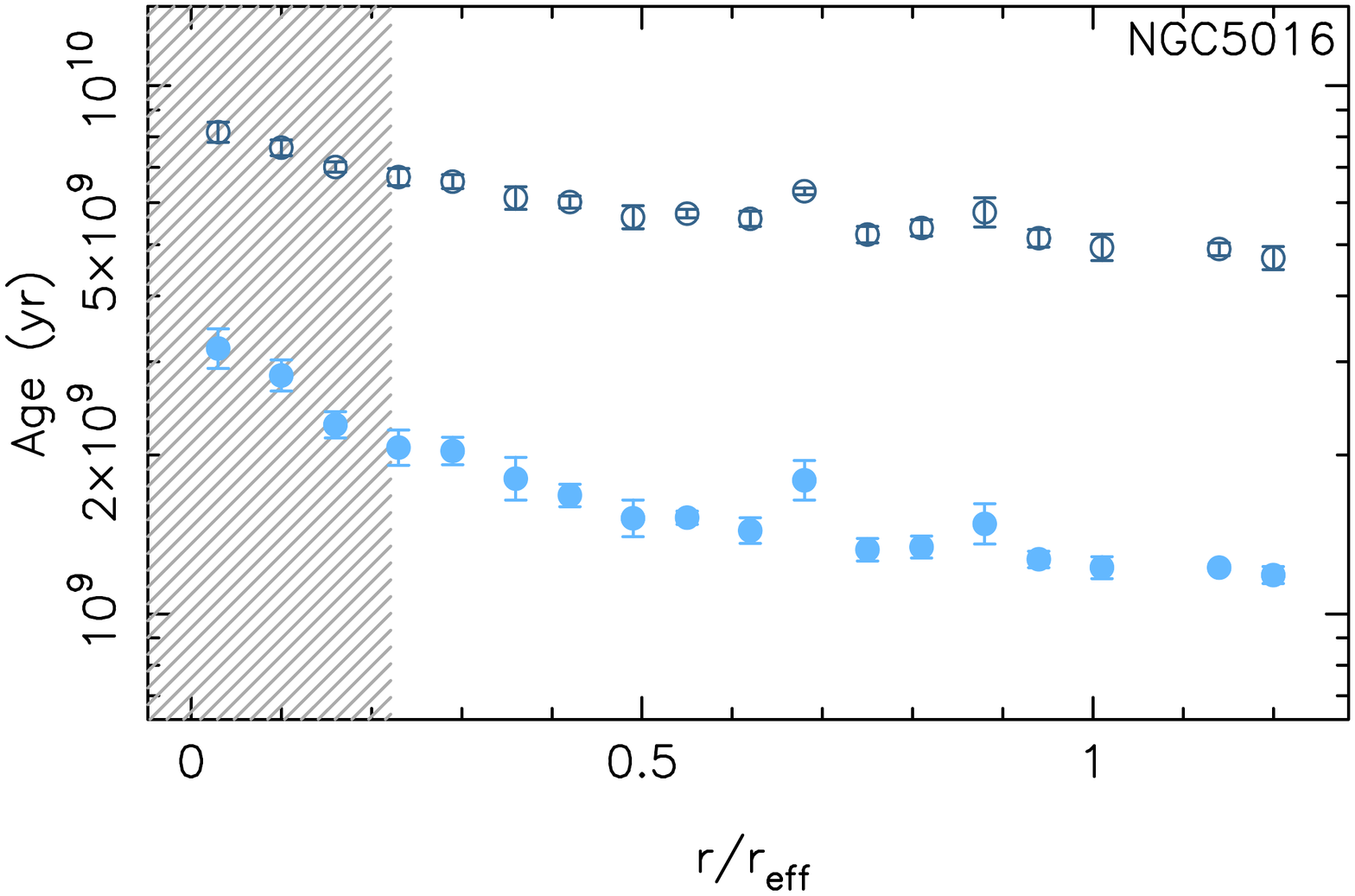}}
\resizebox{0.22\textwidth}{!}{\includegraphics[angle=0]{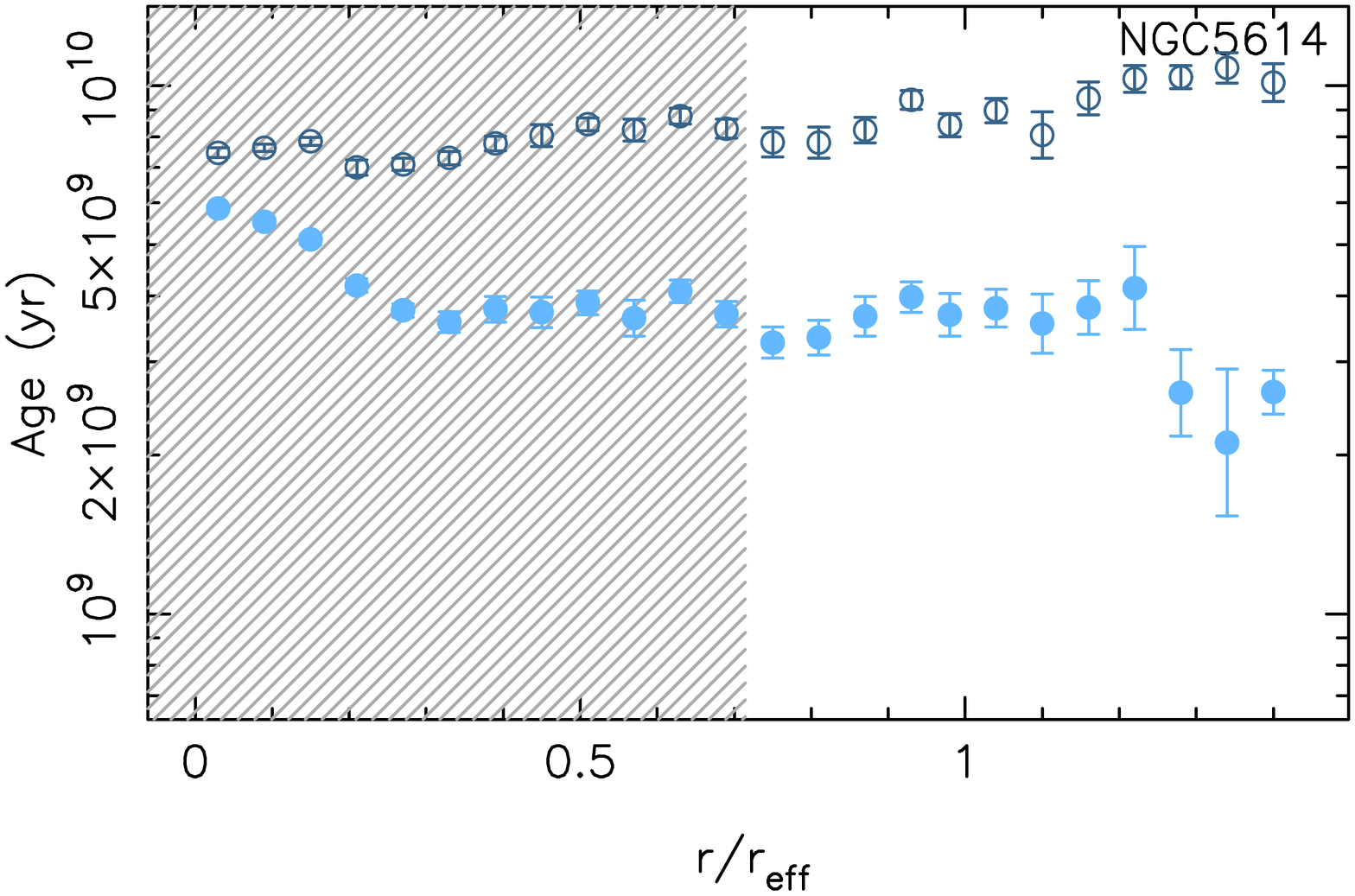}}
\resizebox{0.22\textwidth}{!}{\includegraphics[angle=0]{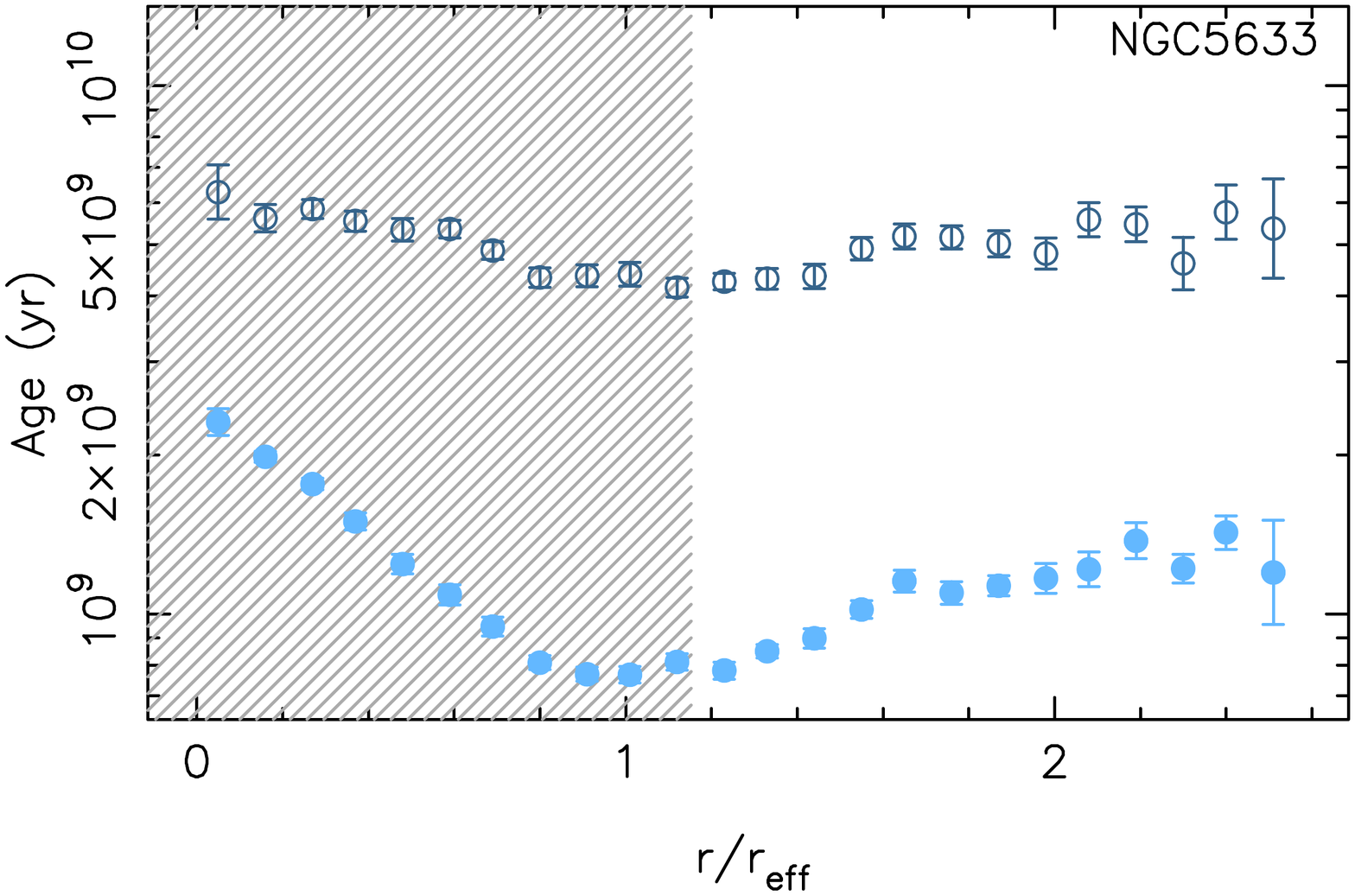}}
\resizebox{0.22\textwidth}{!}{\includegraphics[angle=0]{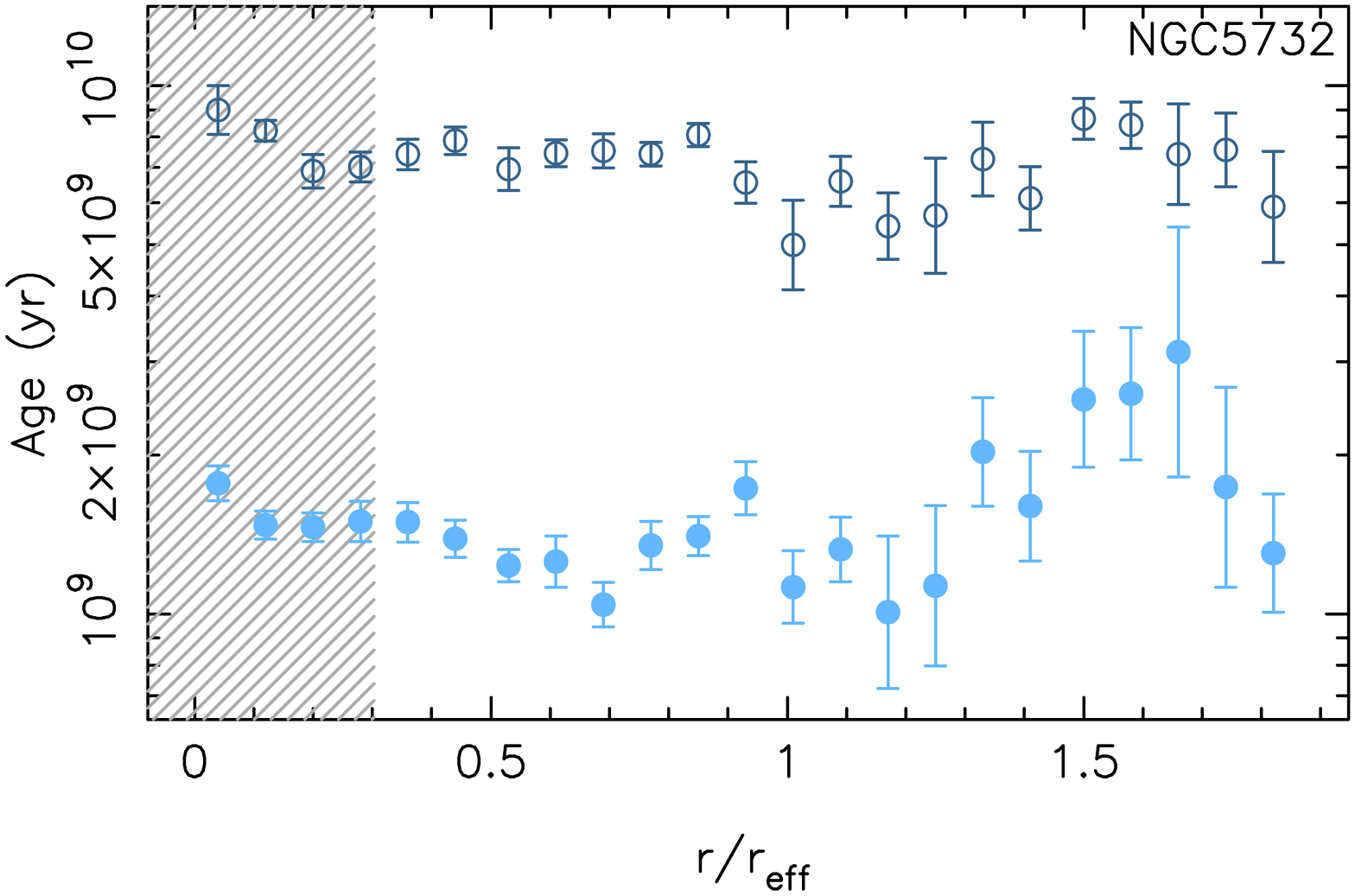}}
\resizebox{0.22\textwidth}{!}{\includegraphics[angle=0]{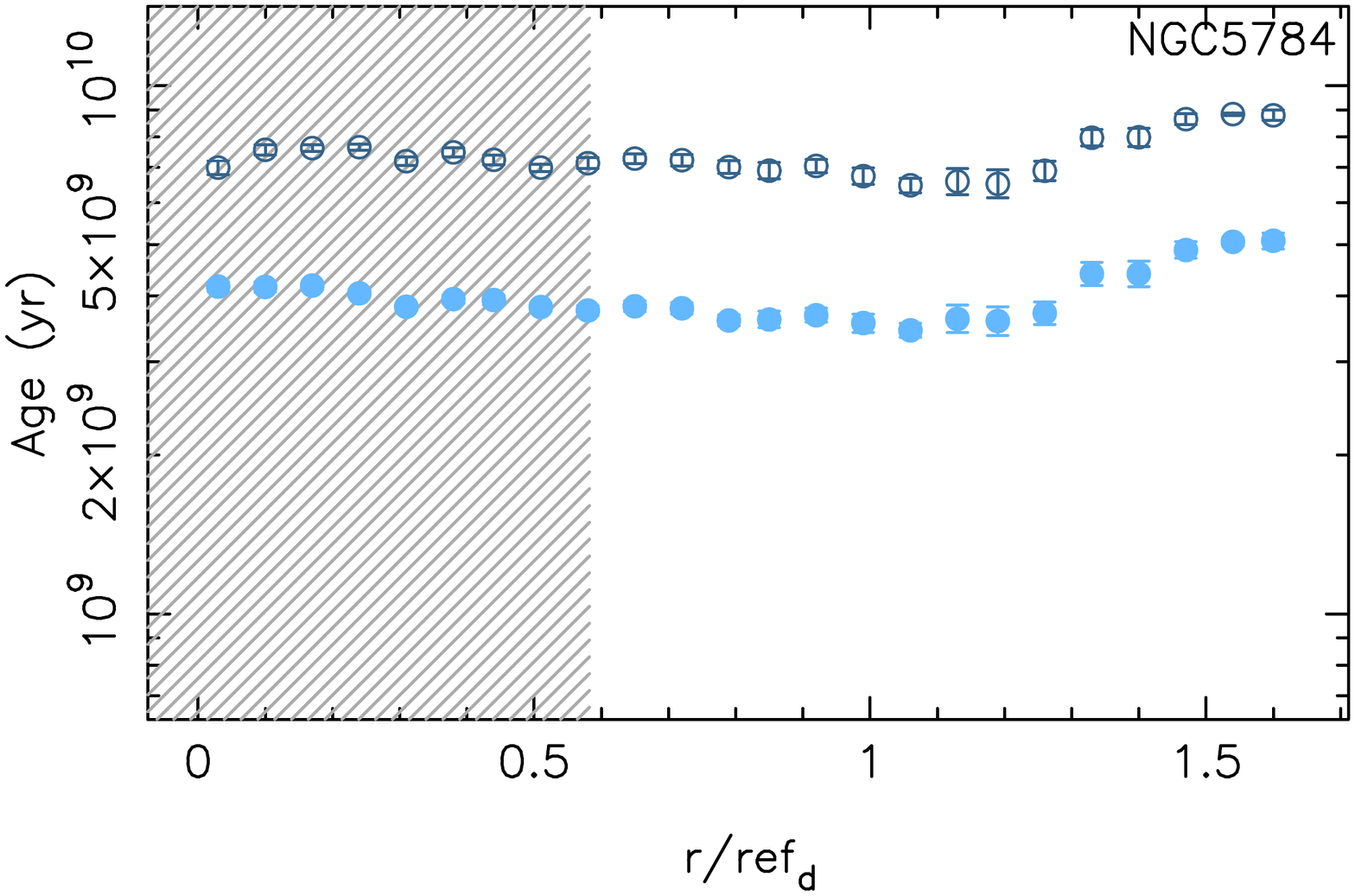}}
\resizebox{0.22\textwidth}{!}{\includegraphics[angle=0]{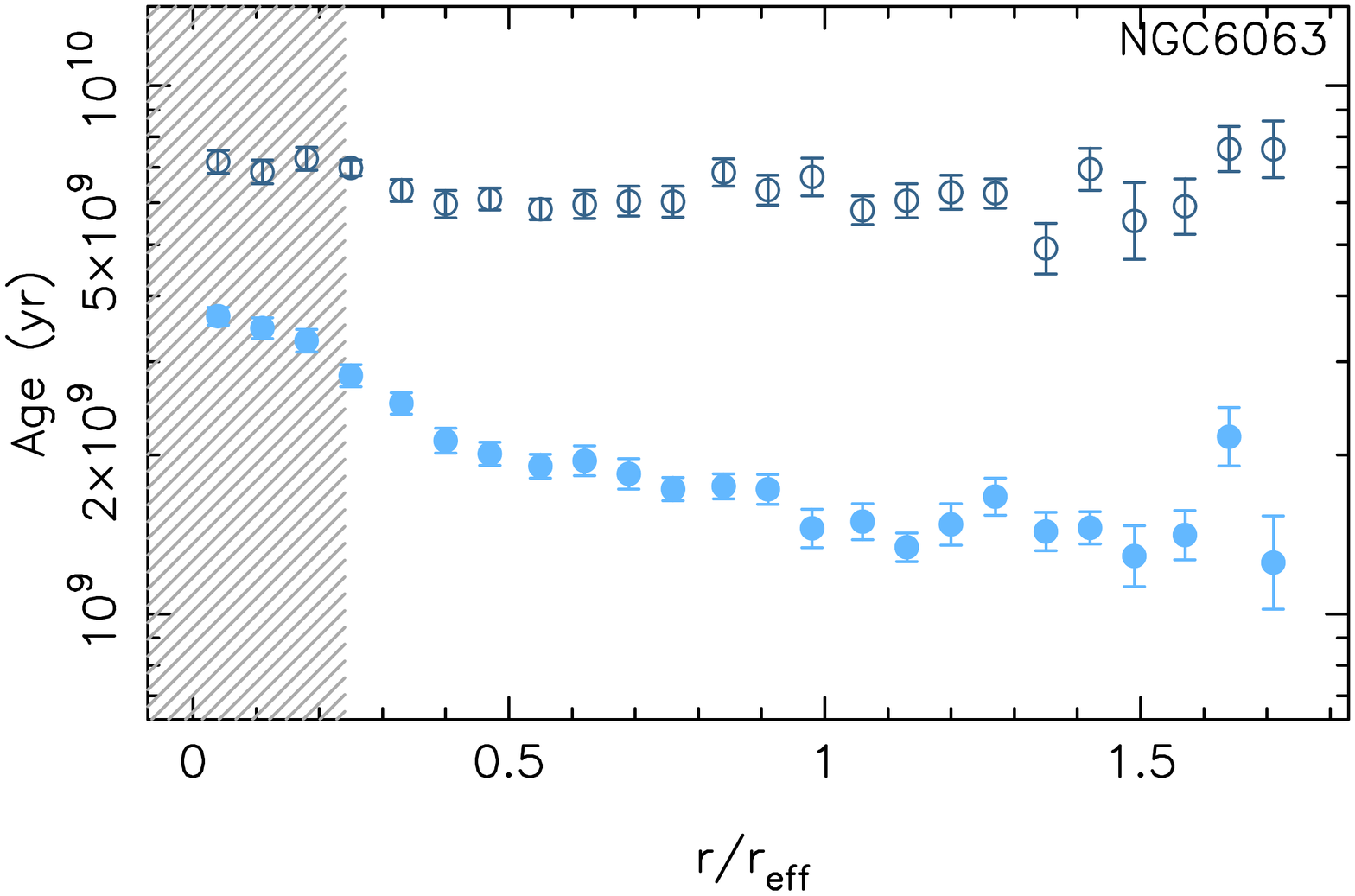}}
\resizebox{0.22\textwidth}{!}{\includegraphics[angle=0]{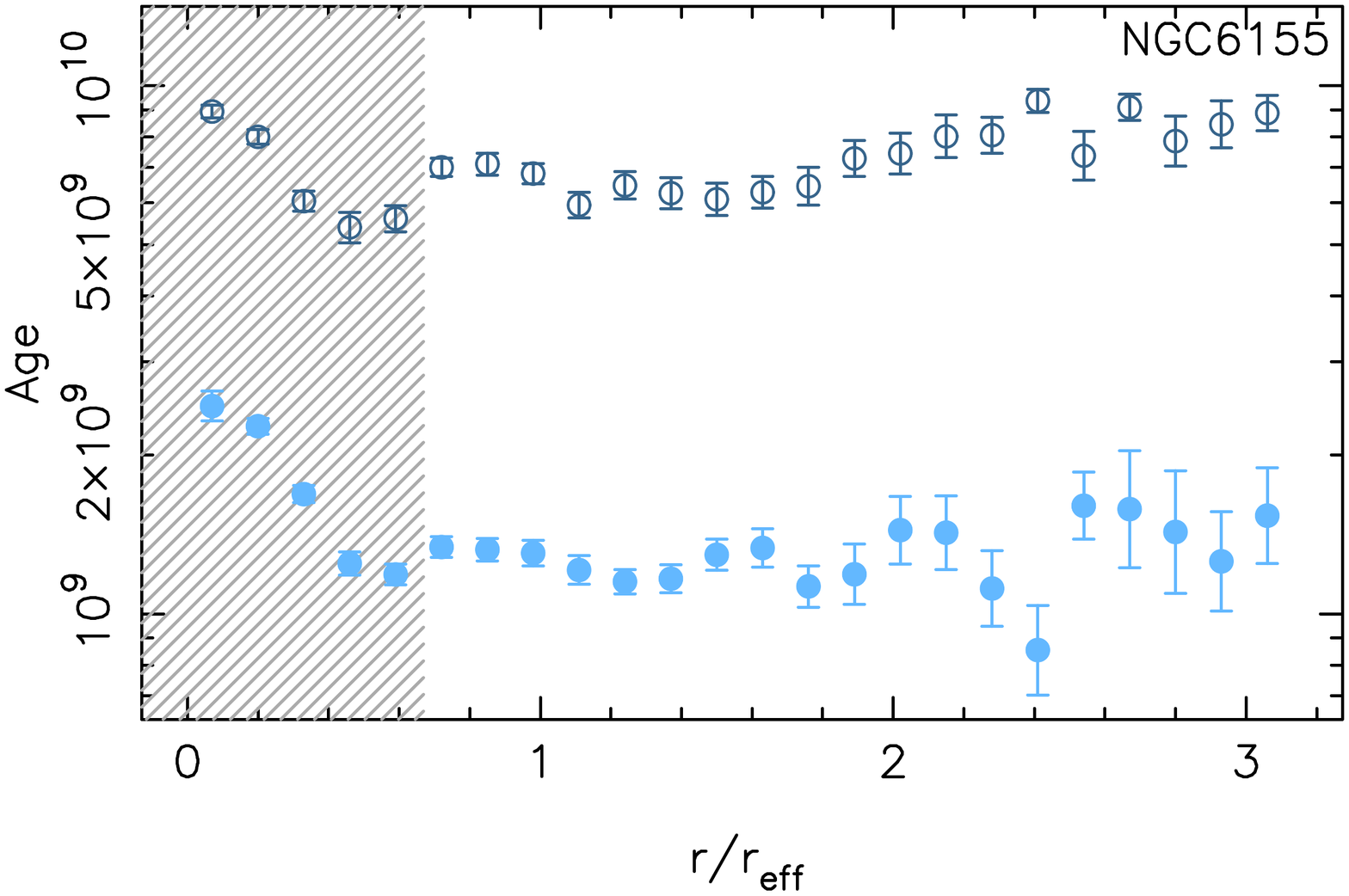}}
\resizebox{0.22\textwidth}{!}{\includegraphics[angle=0]{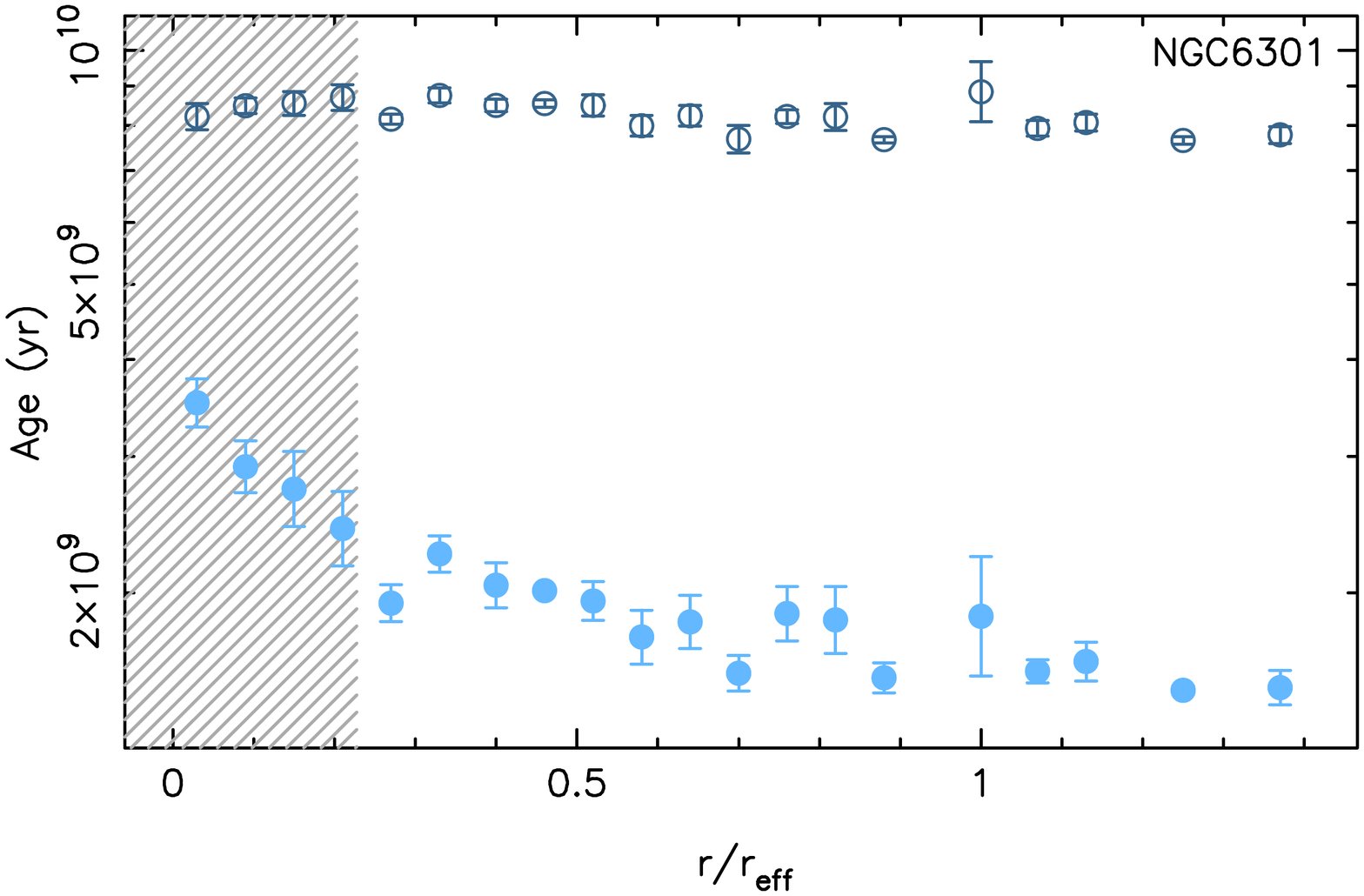}}
\resizebox{0.22\textwidth}{!}{\includegraphics[angle=0]{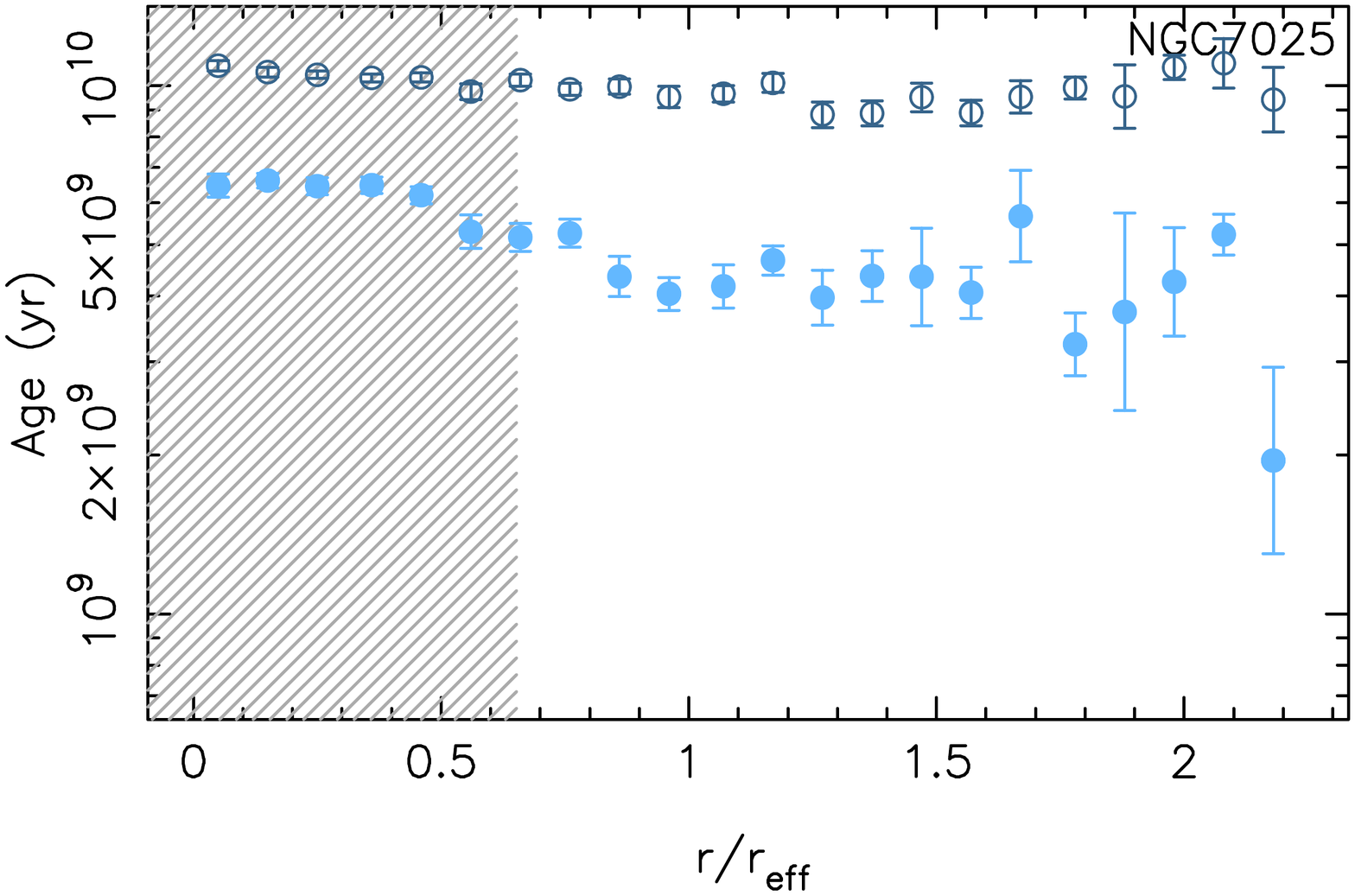}}
\resizebox{0.22\textwidth}{!}{\includegraphics[angle=0]{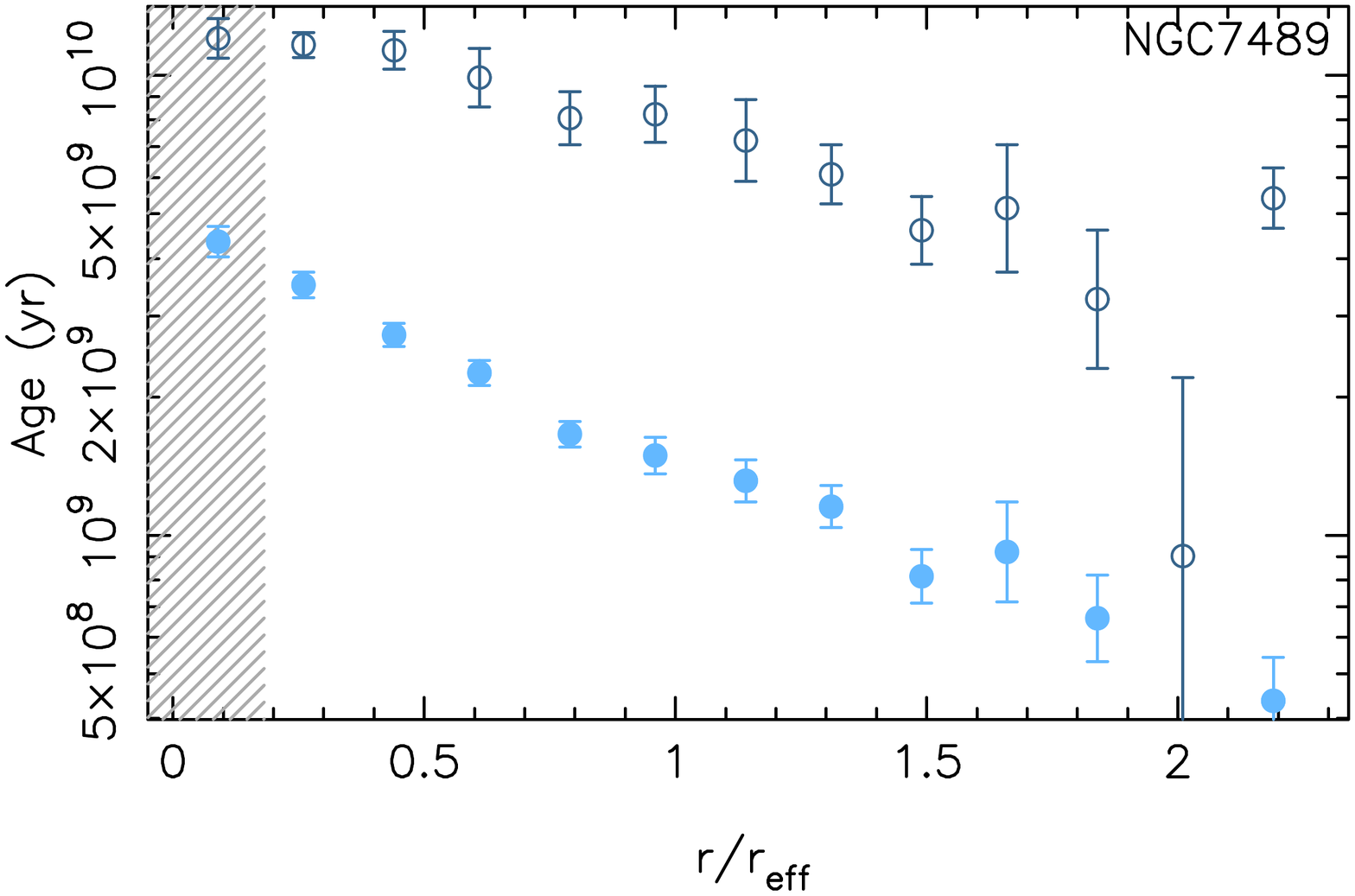}}
\resizebox{0.22\textwidth}{!}{\includegraphics[angle=0]{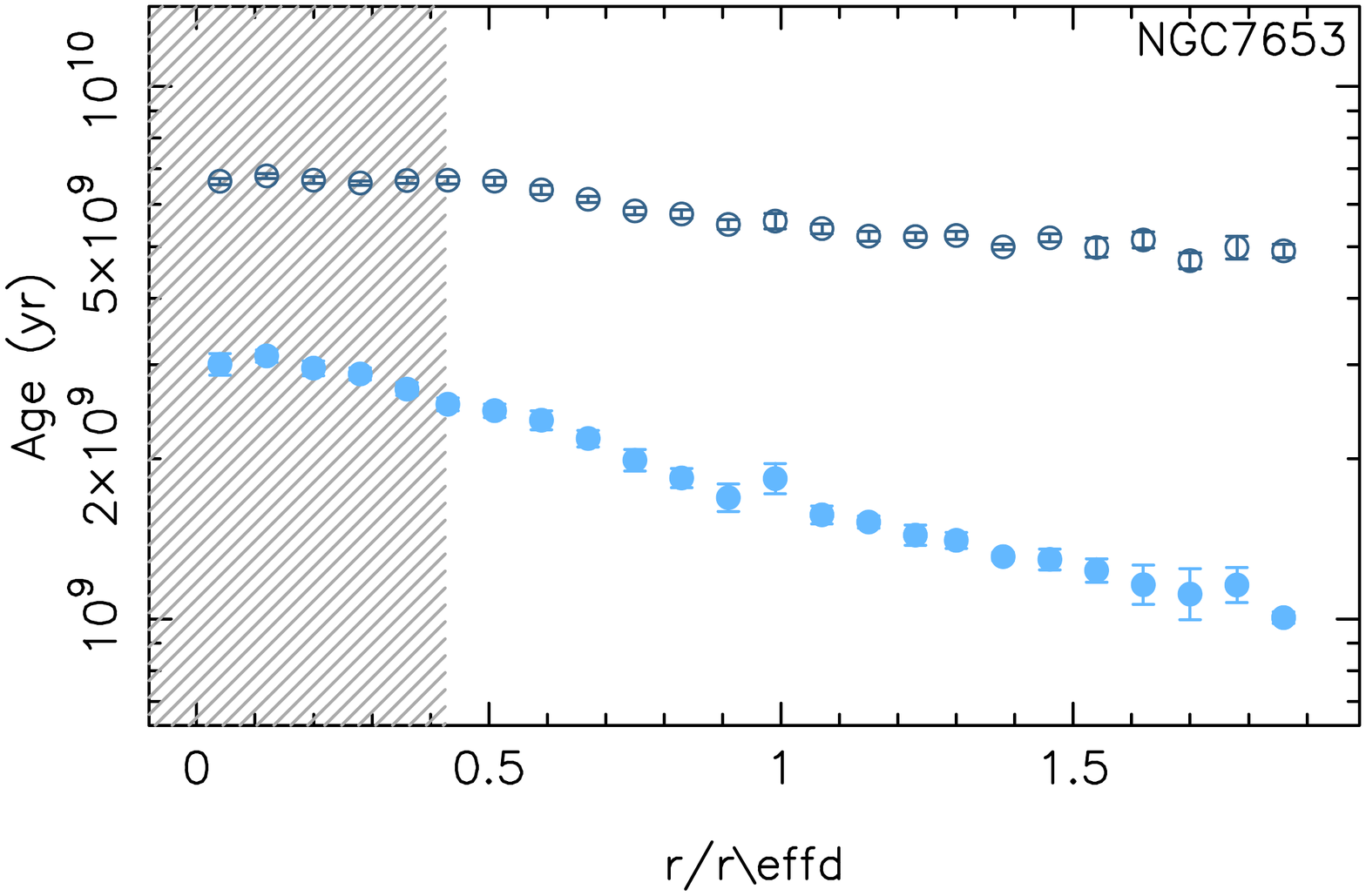}}
\resizebox{0.22\textwidth}{!}{\includegraphics[angle=0]{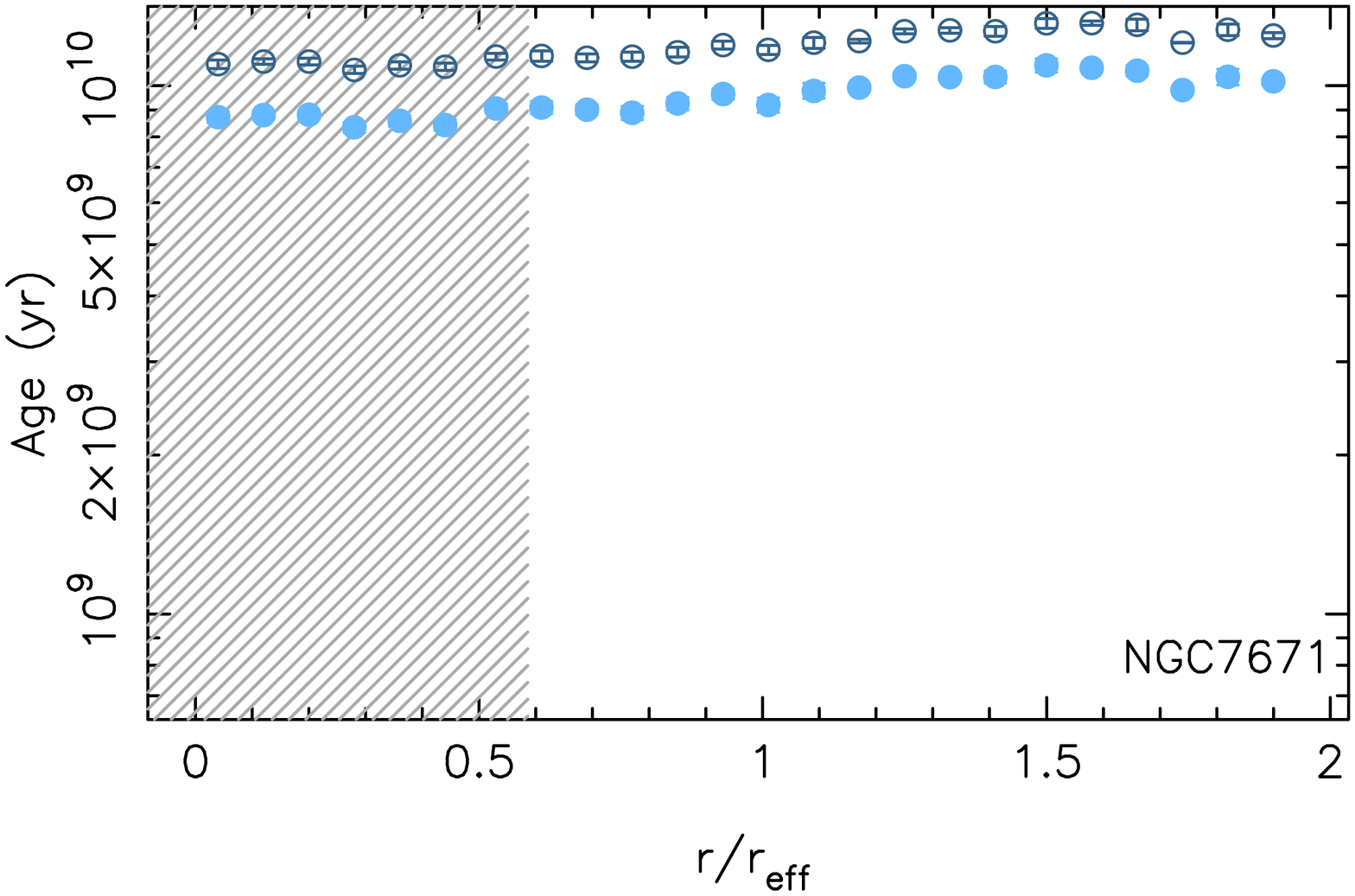}}
\resizebox{0.22\textwidth}{!}{\includegraphics[angle=0]{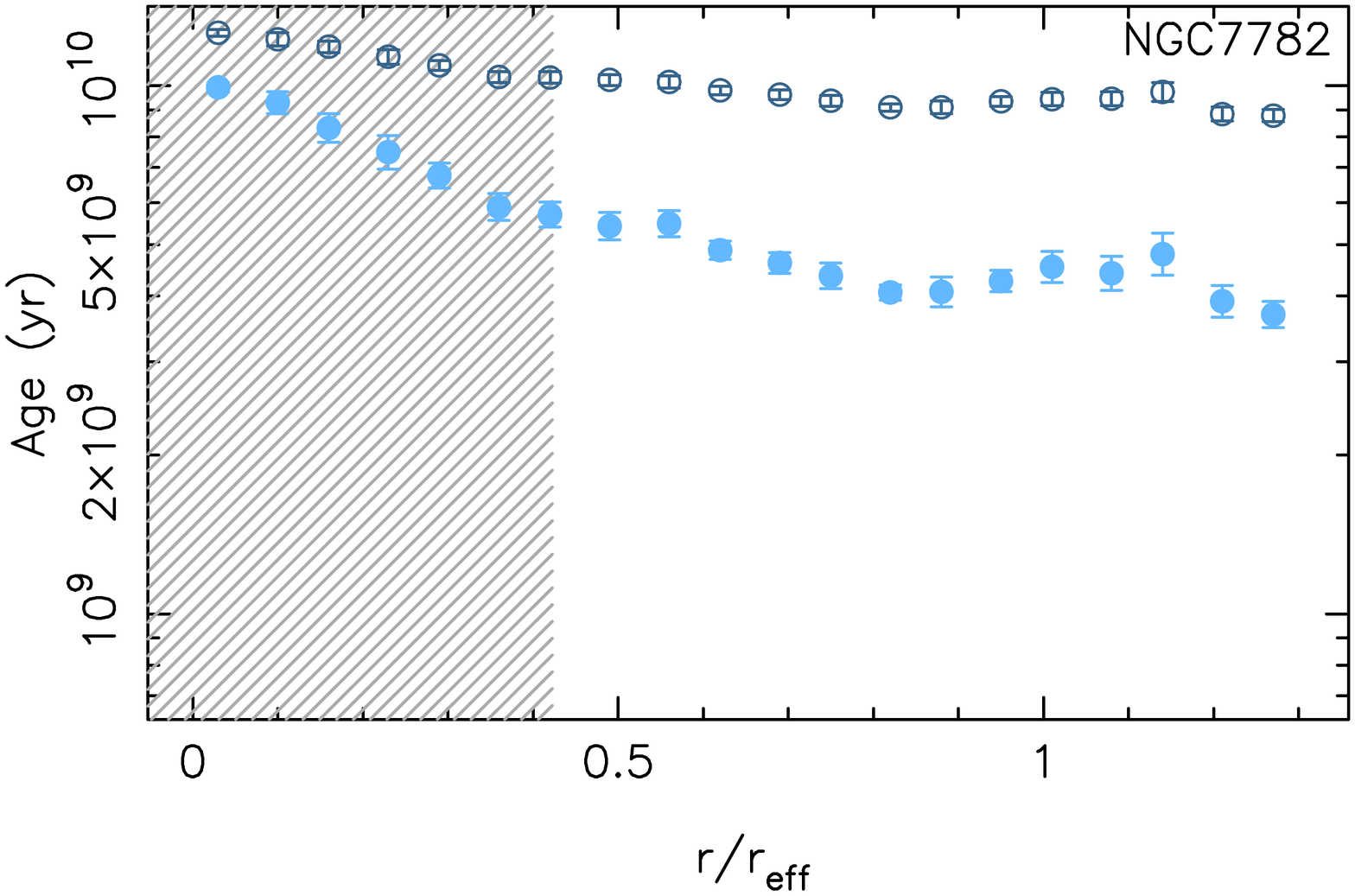}}
\resizebox{0.22\textwidth}{!}{\includegraphics[angle=0]{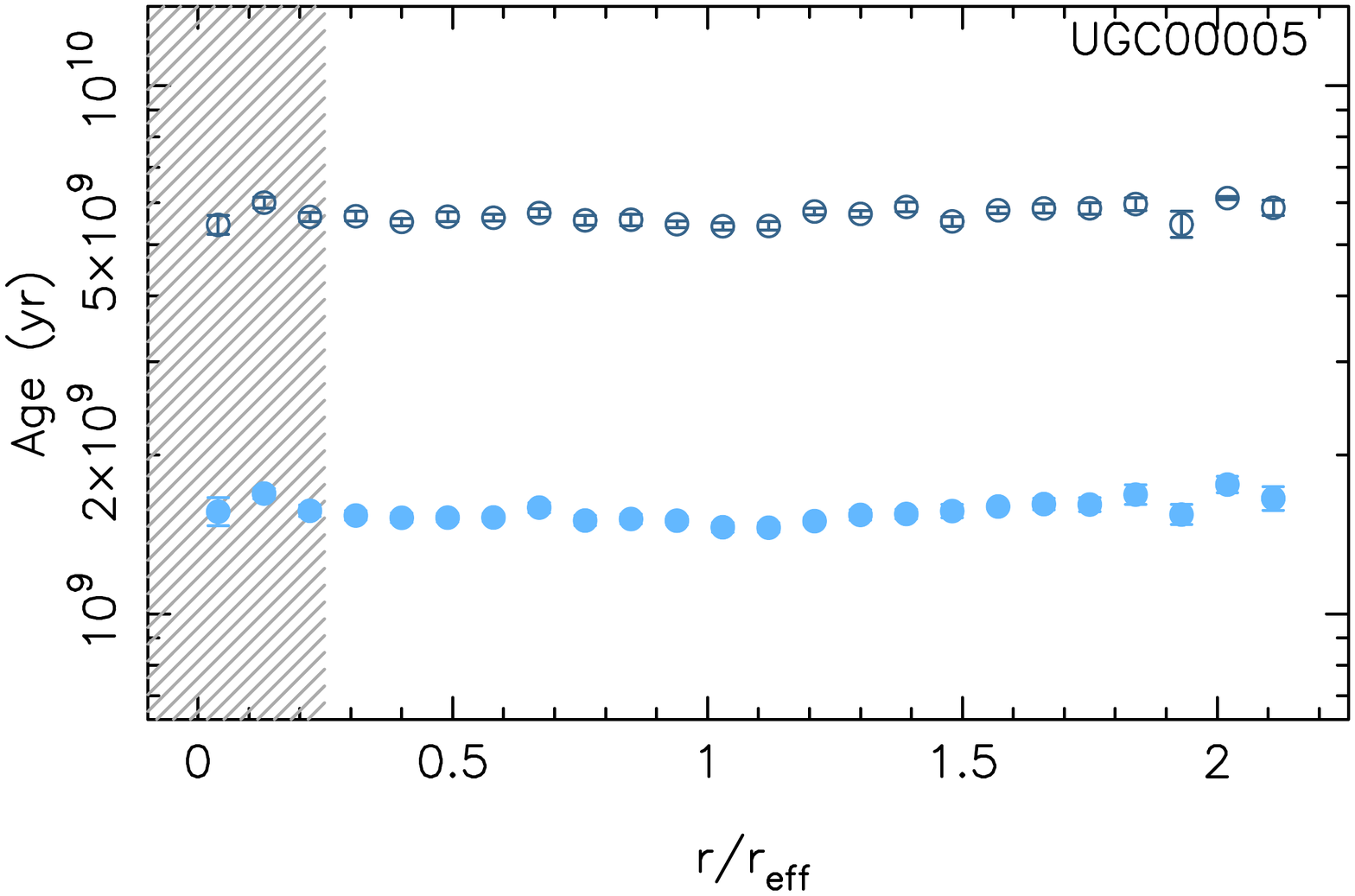}}
\resizebox{0.22\textwidth}{!}{\includegraphics[angle=0]{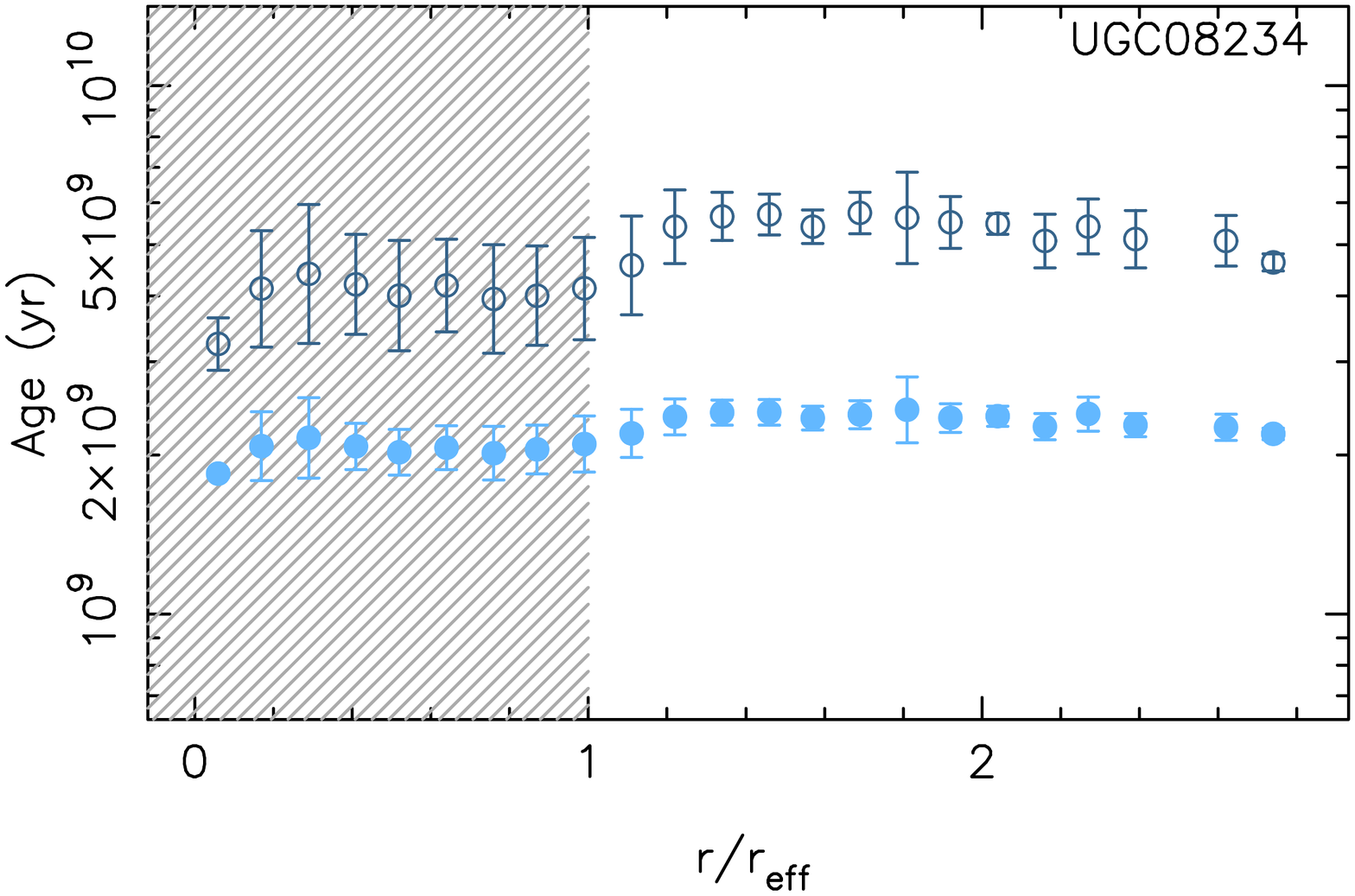}}
\resizebox{0.22\textwidth}{!}{\includegraphics[angle=0]{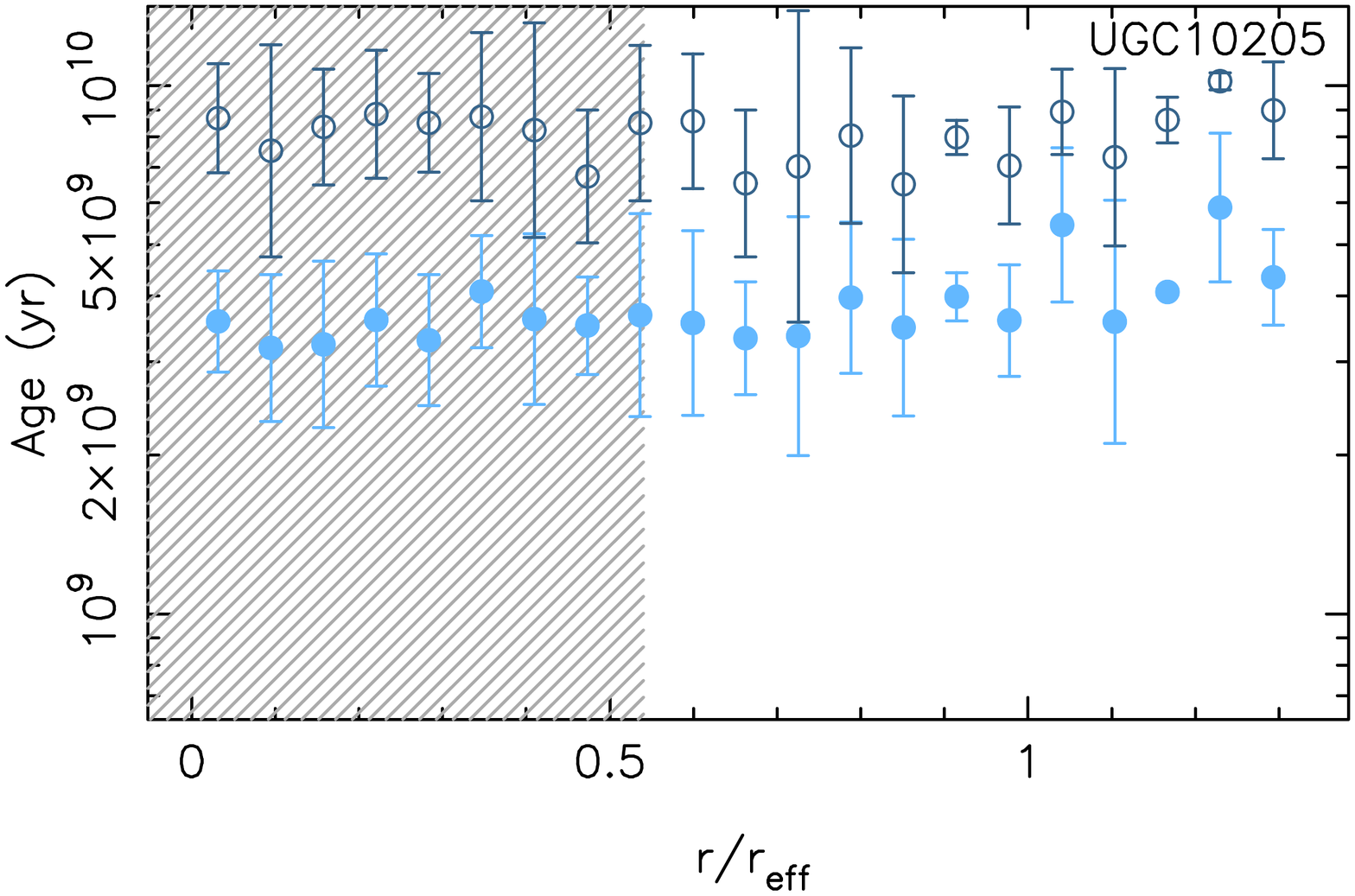}}
\resizebox{0.22\textwidth}{!}{\includegraphics[angle=0]{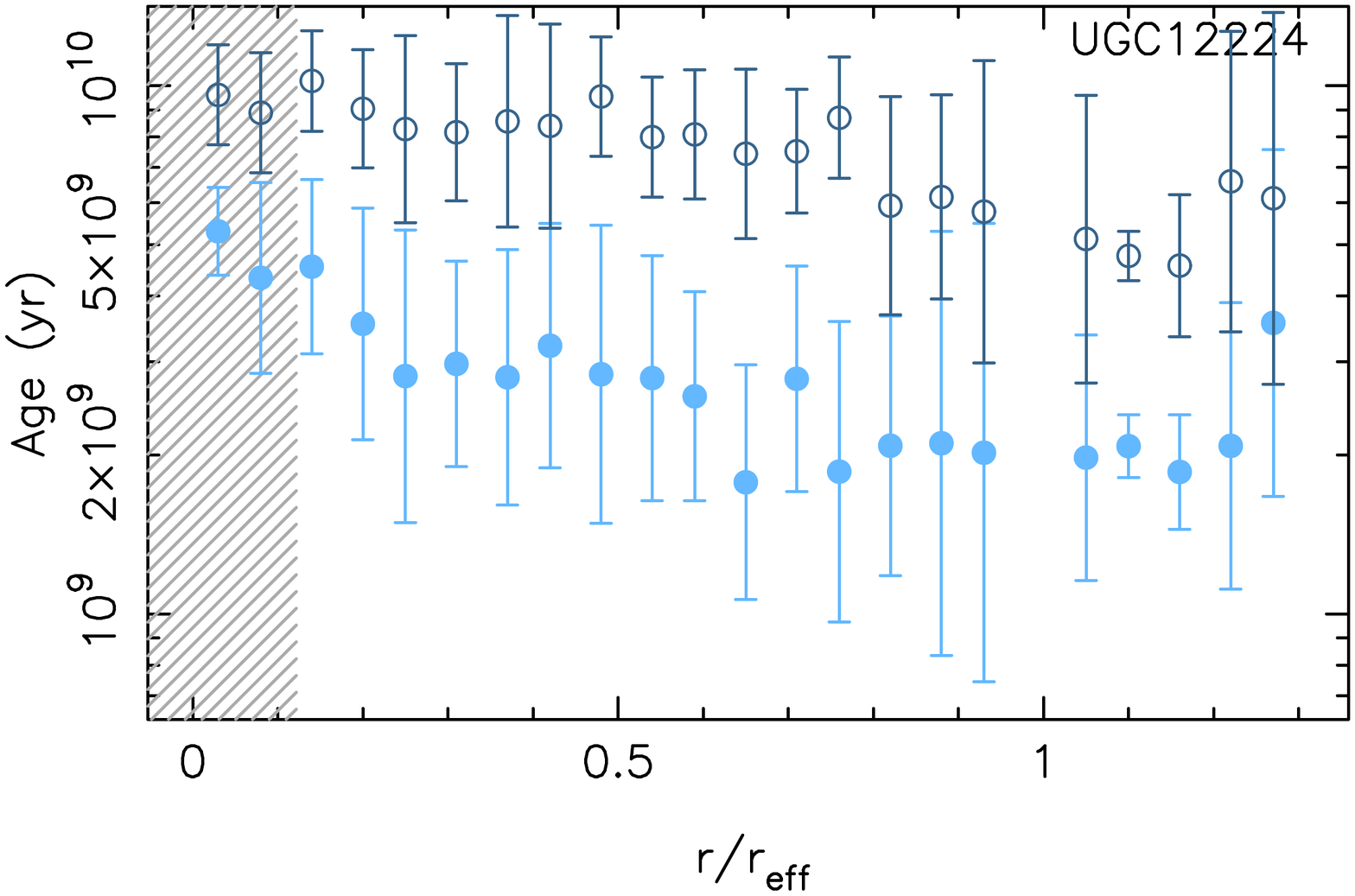}}
\resizebox{0.22\textwidth}{!}{\includegraphics[angle=0]{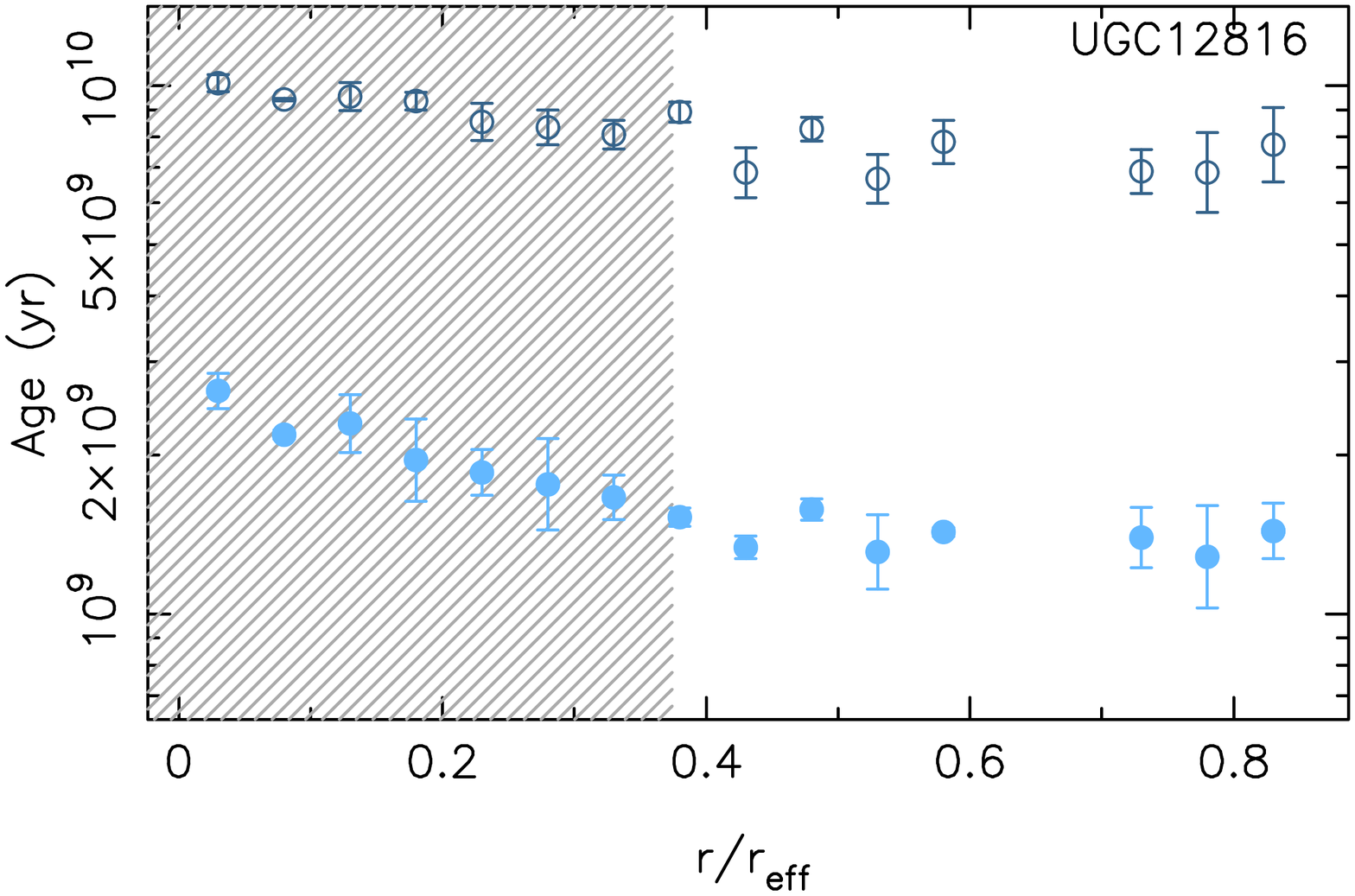}}
\caption{Age gradients of the sample of unbarred  galaxies\label{fig:grad_age2}}
\end{figure*}
\begin{figure*}
\centering
\resizebox{0.22\textwidth}{!}{\includegraphics[angle=0]{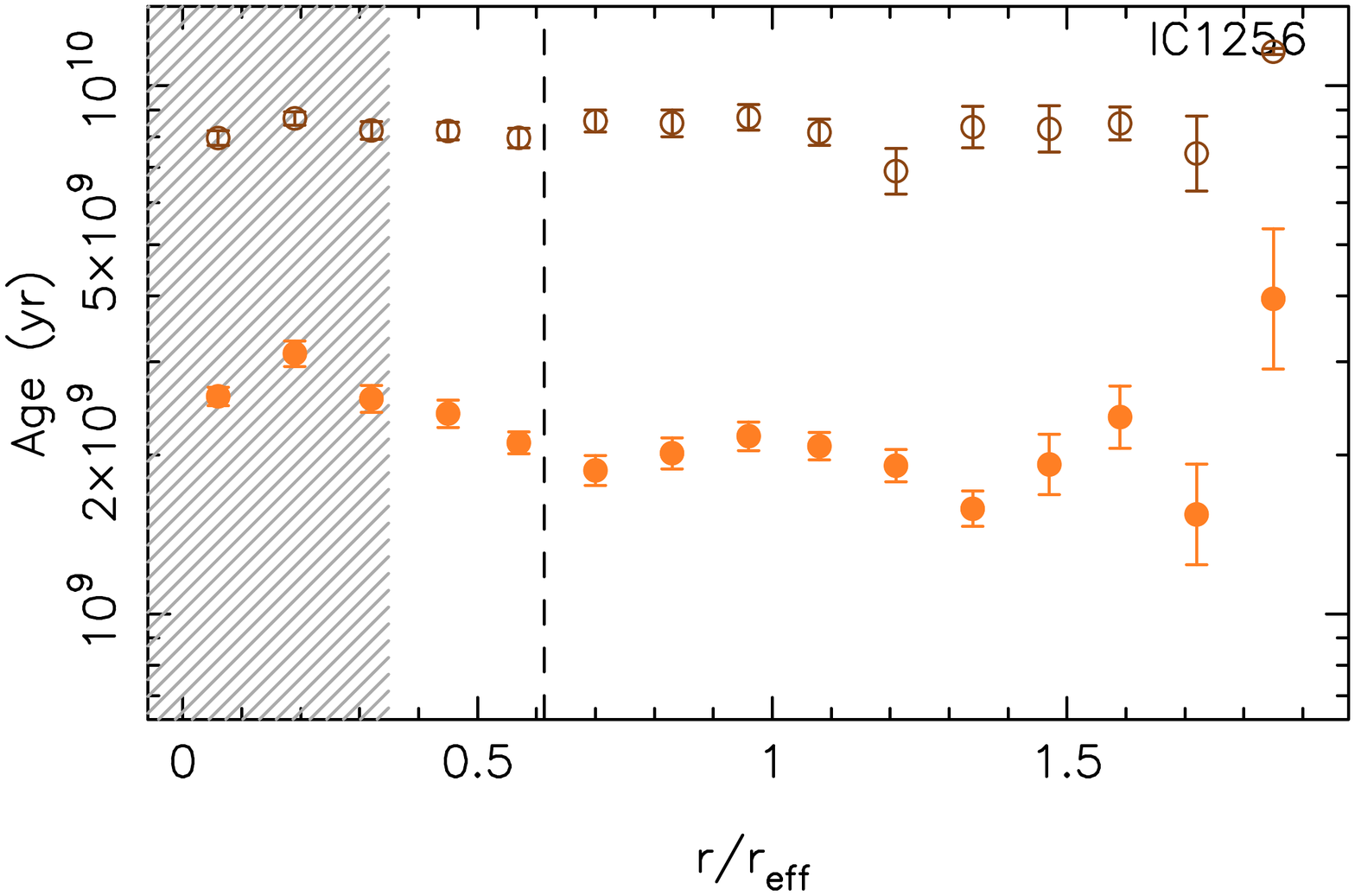}}
\resizebox{0.22\textwidth}{!}{\includegraphics[angle=0]{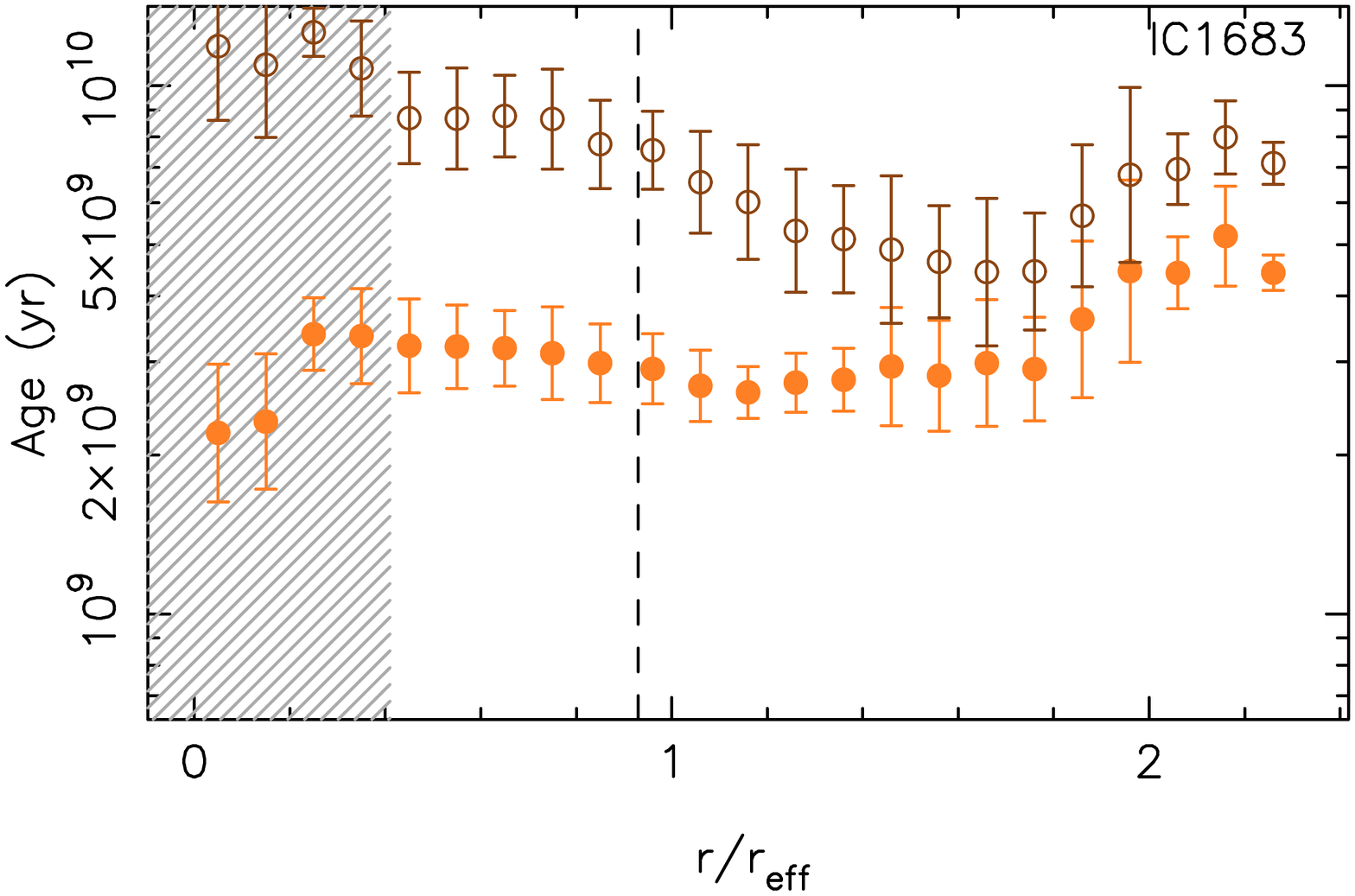}}
\resizebox{0.22\textwidth}{!}{\includegraphics[angle=0]{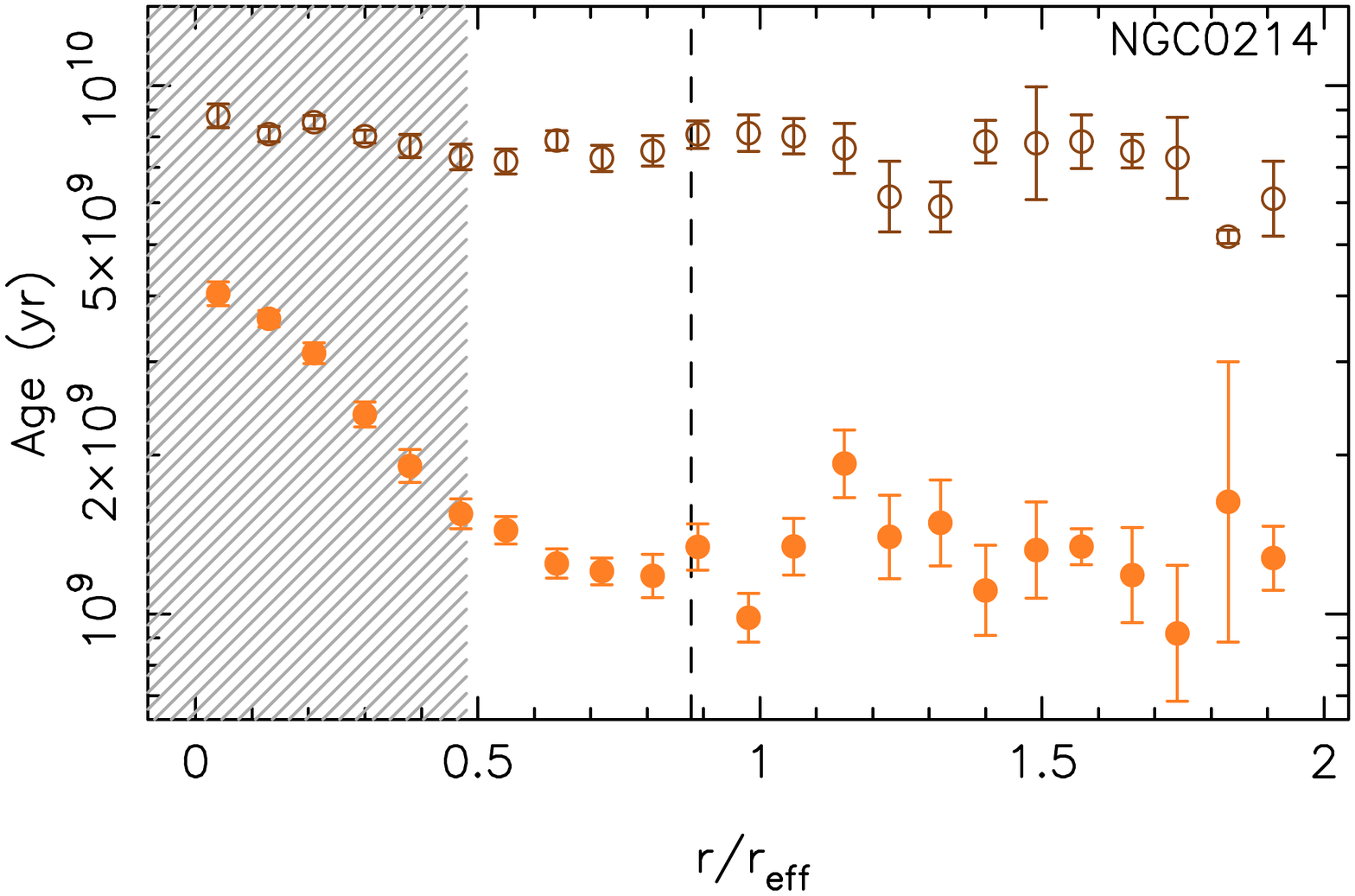}}
\resizebox{0.22\textwidth}{!}{\includegraphics[angle=0]{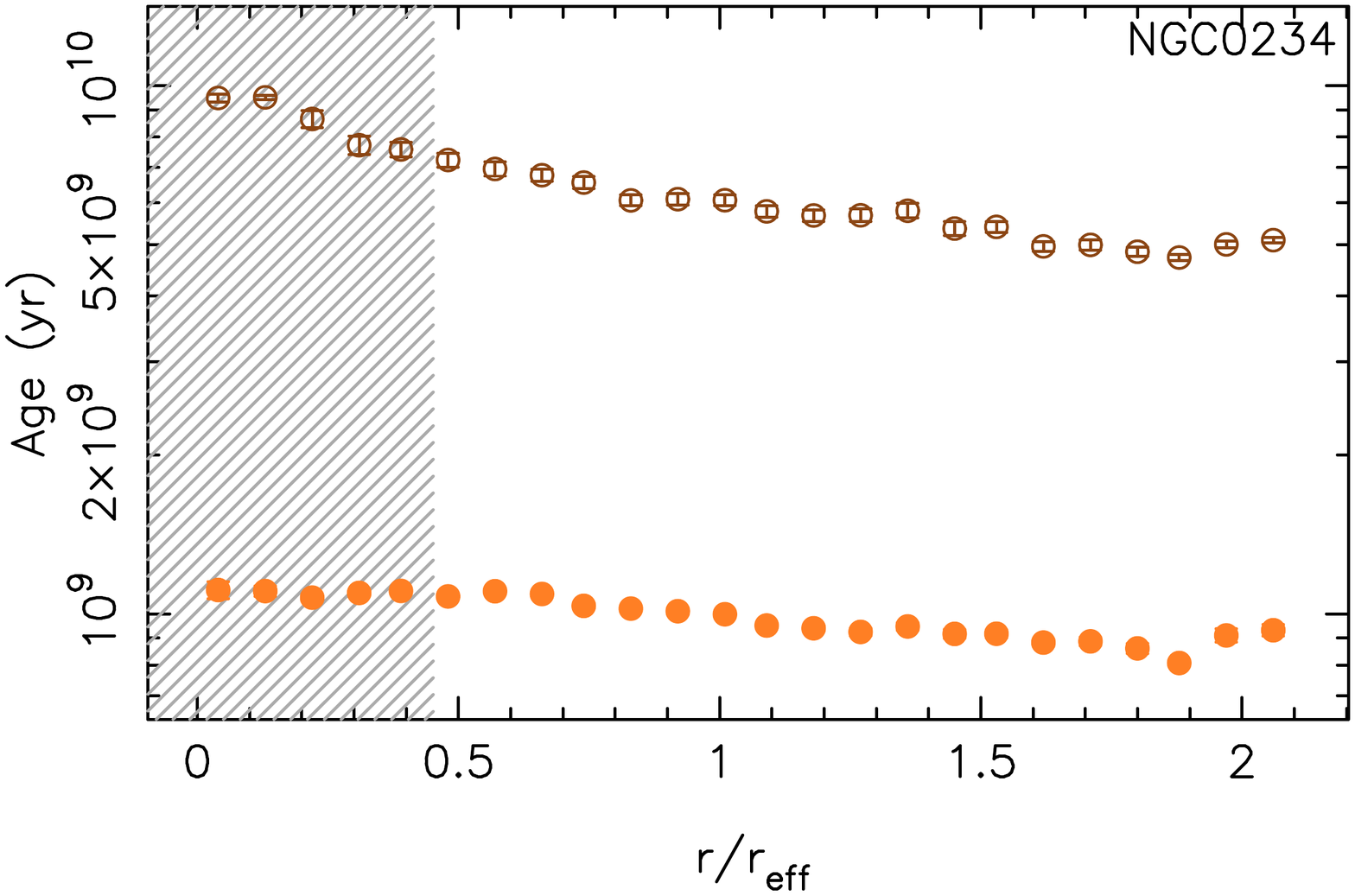}}
\resizebox{0.22\textwidth}{!}{\includegraphics[angle=0]{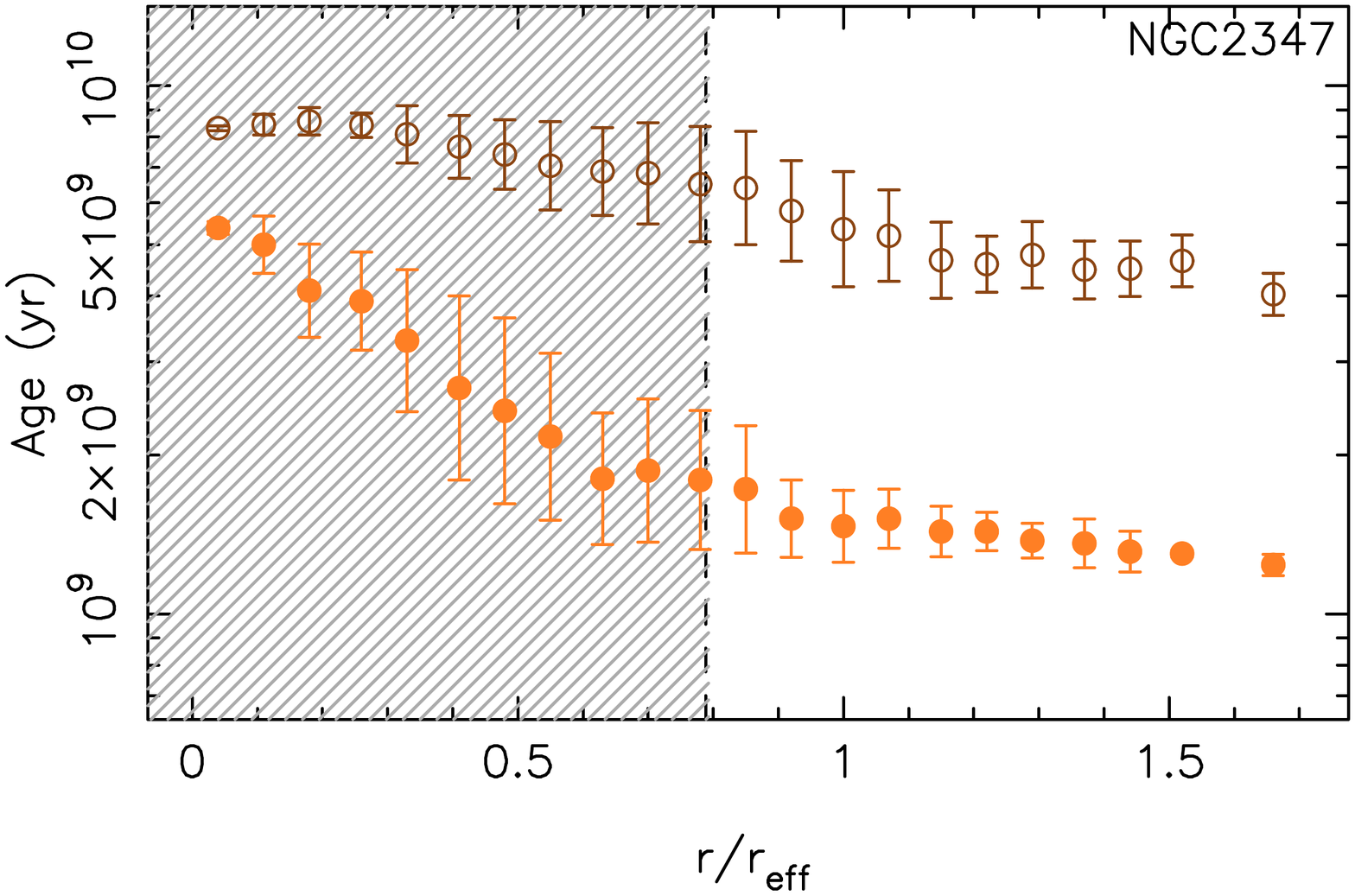}}
\resizebox{0.22\textwidth}{!}{\includegraphics[angle=0]{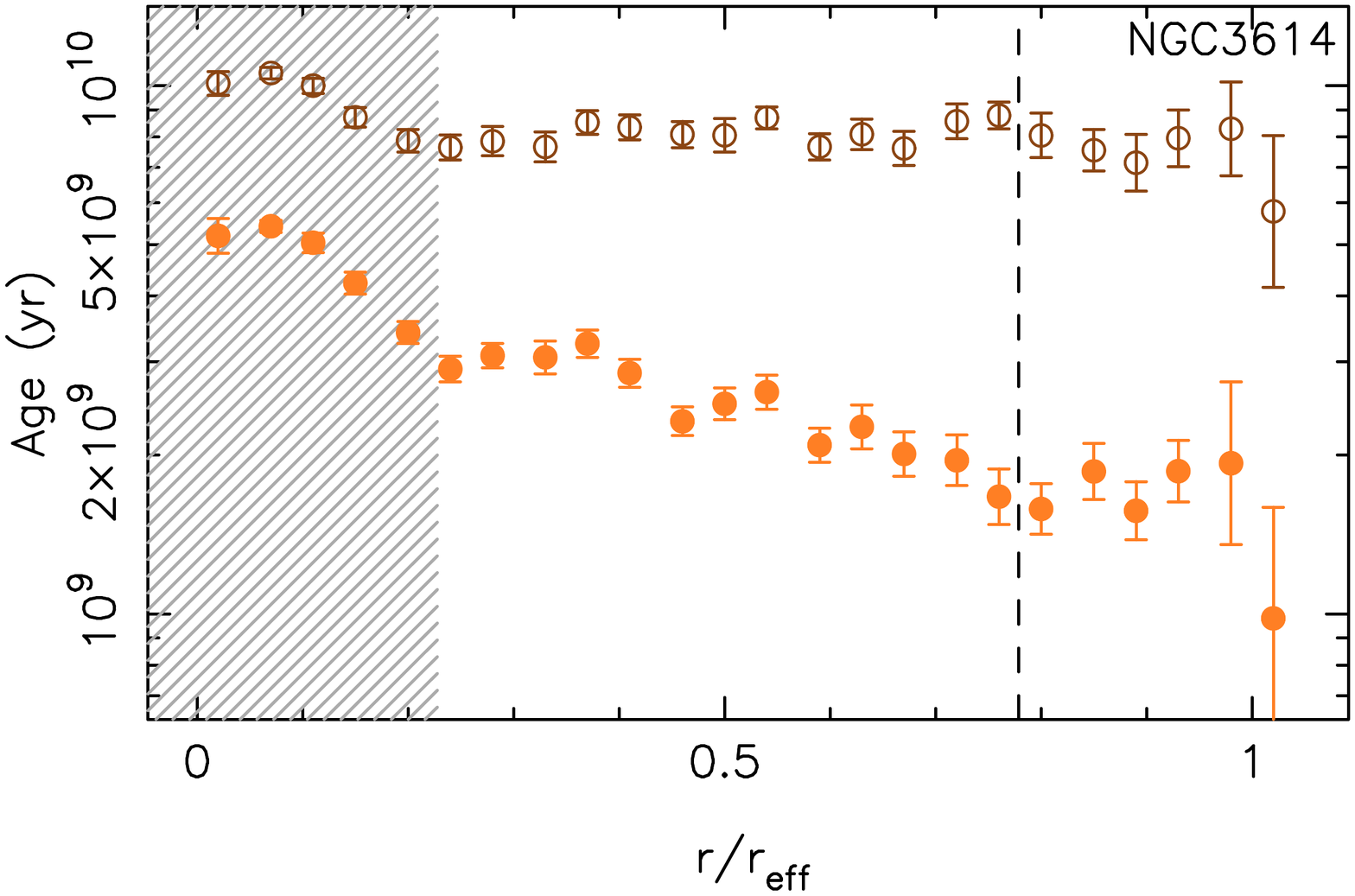}}
\resizebox{0.22\textwidth}{!}{\includegraphics[angle=0]{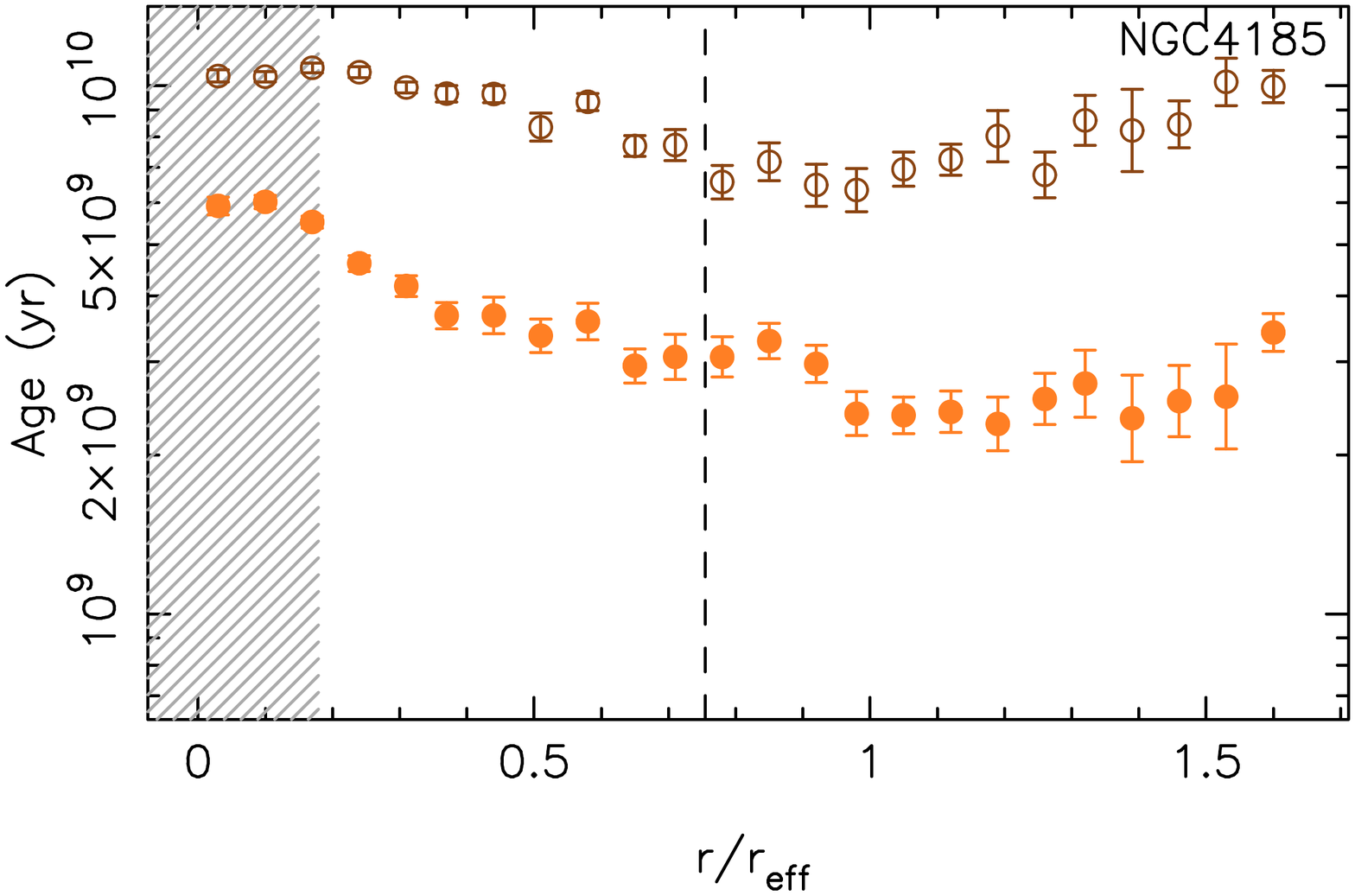}}
\resizebox{0.22\textwidth}{!}{\includegraphics[angle=0]{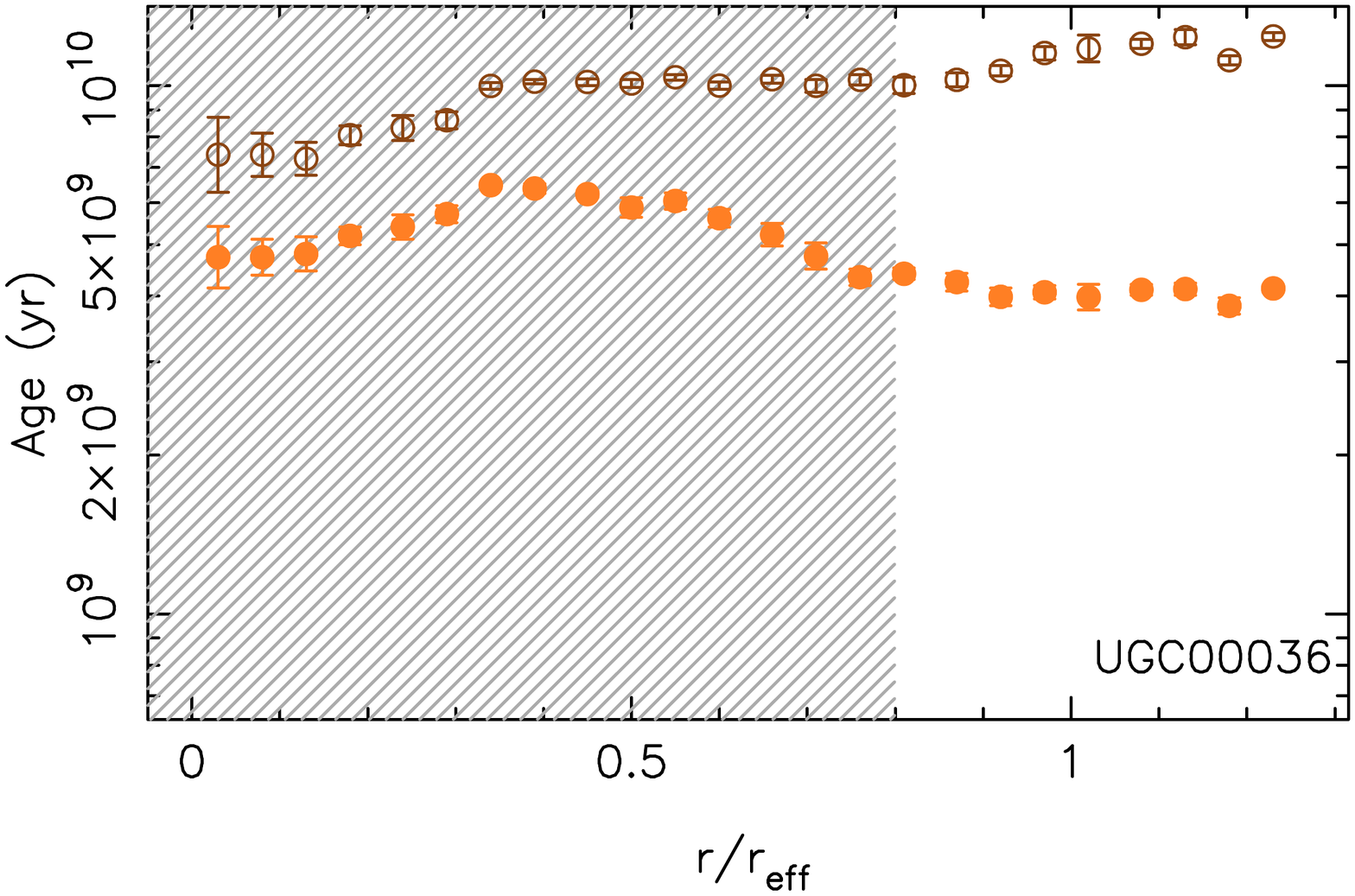}}
\resizebox{0.22\textwidth}{!}{\includegraphics[angle=0]{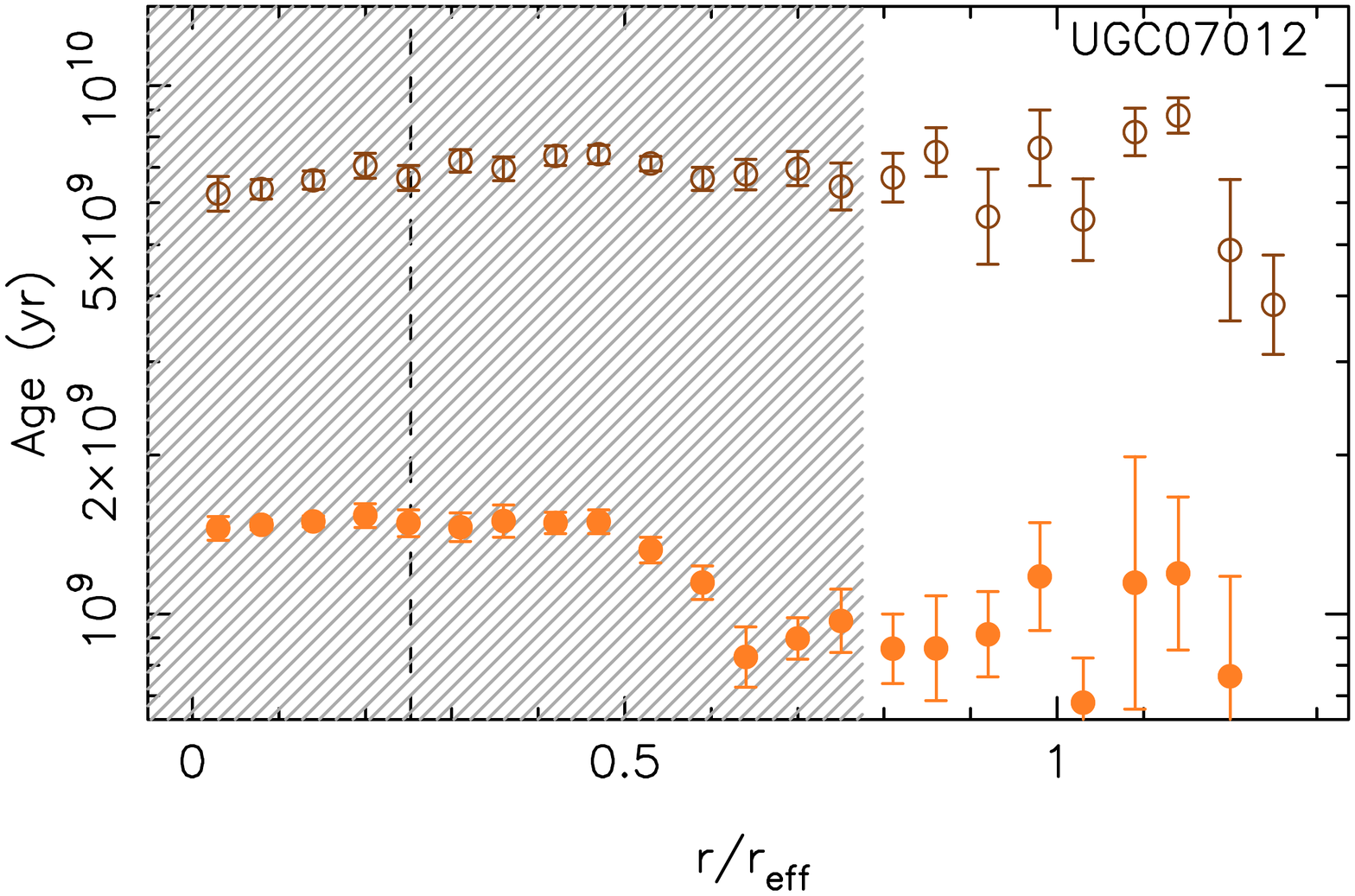}}
\caption{Age gradients of the sample of weakly barred  galaxies.\label{fig:grad_age3}}
\end{figure*}

\section{Metallicity gradients}
\begin{figure*}
\centering
\resizebox{0.22\textwidth}{!}{\includegraphics[angle=0]{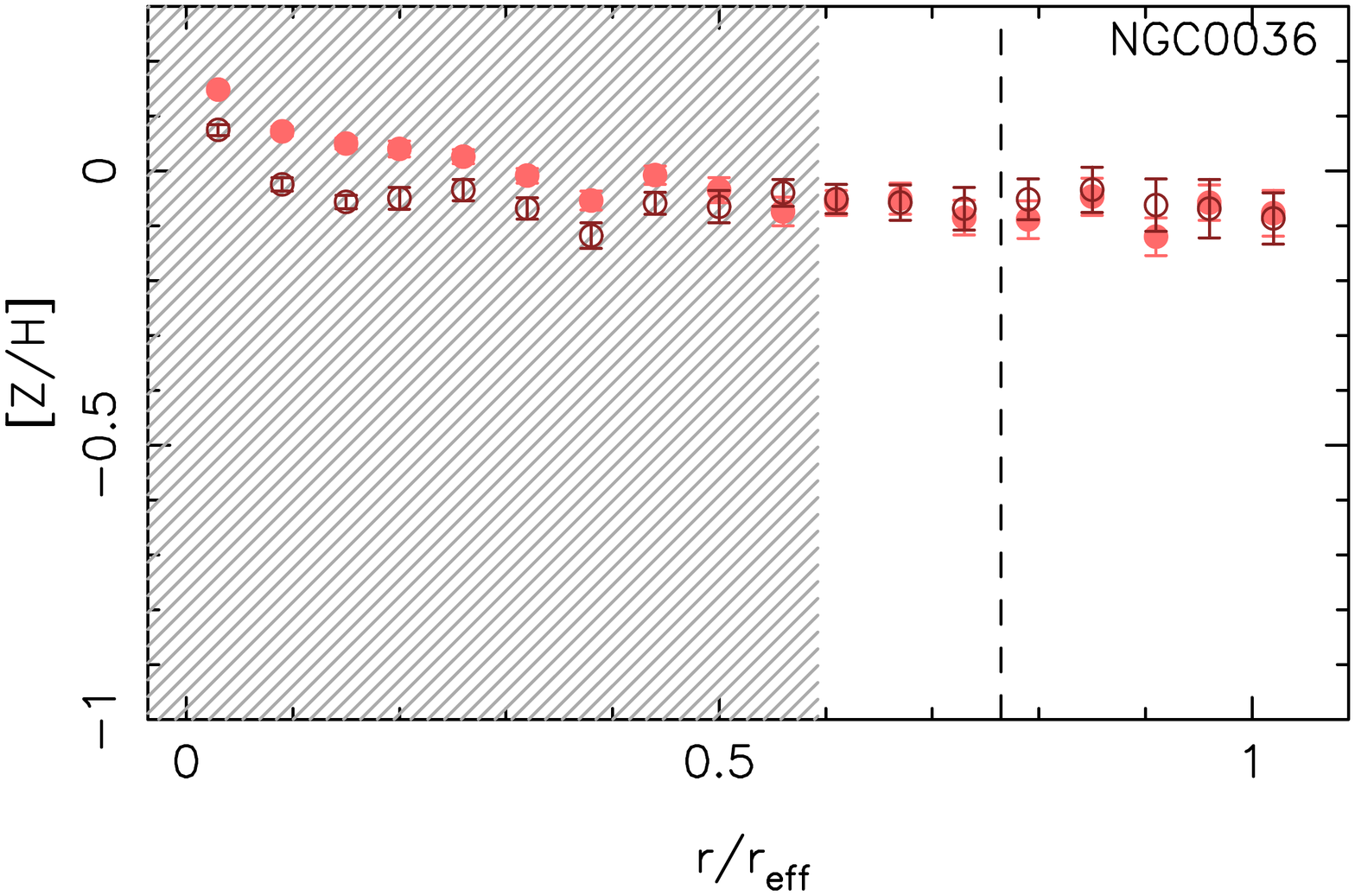}}
\resizebox{0.22\textwidth}{!}{\includegraphics[angle=0]{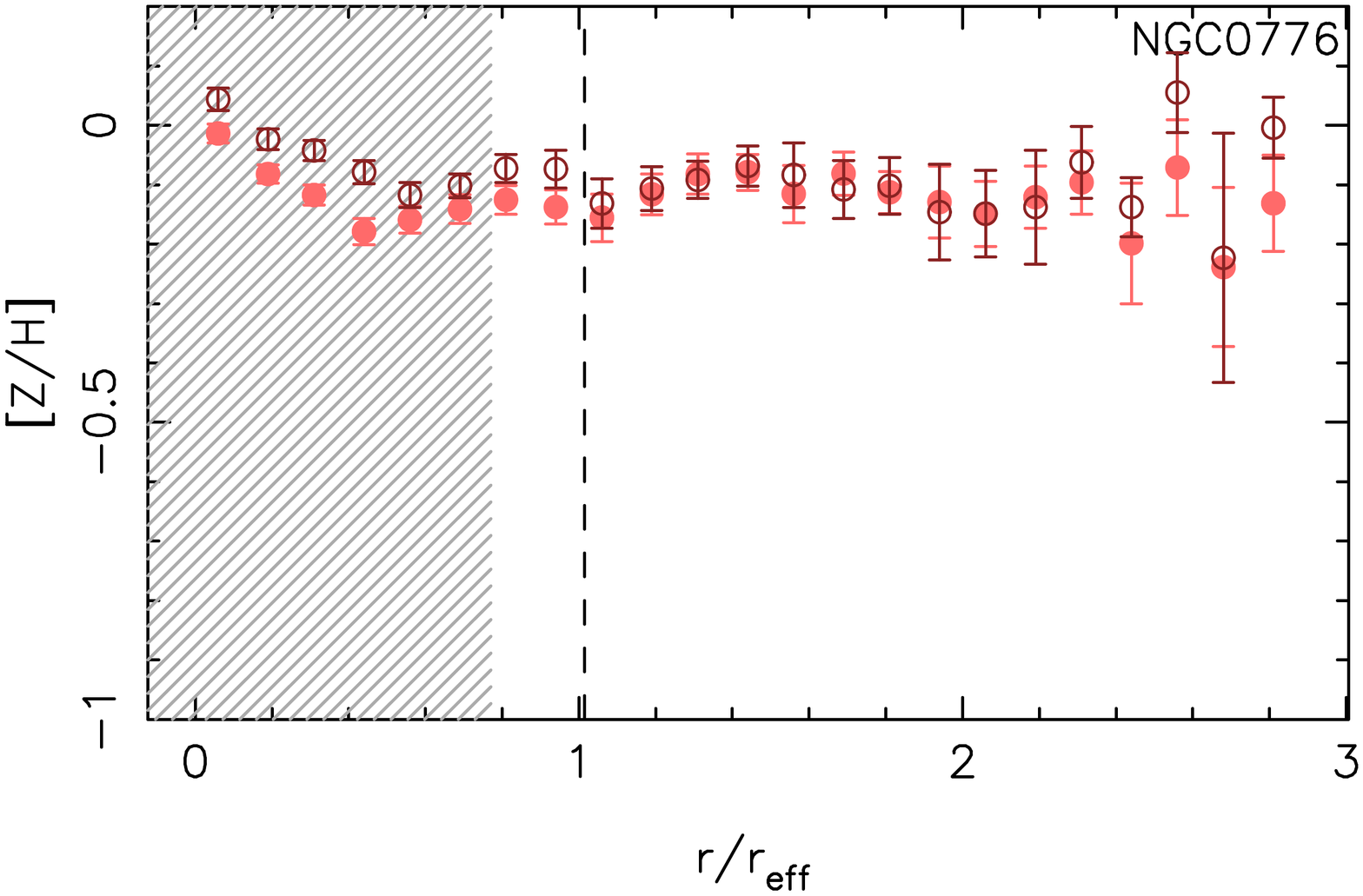}}
\resizebox{0.22\textwidth}{!}{\includegraphics[angle=0]{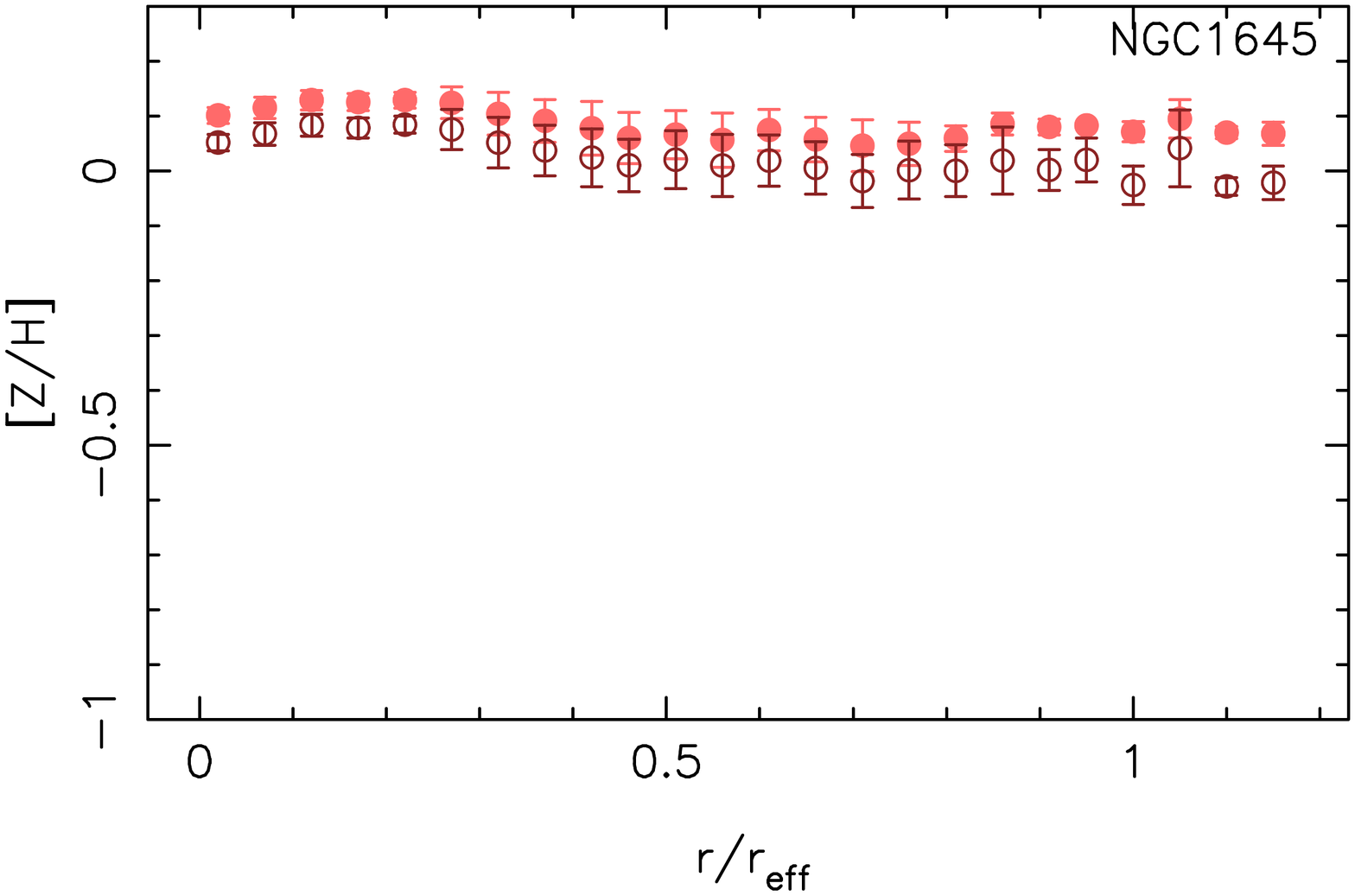}}
\resizebox{0.22\textwidth}{!}{\includegraphics[angle=0]{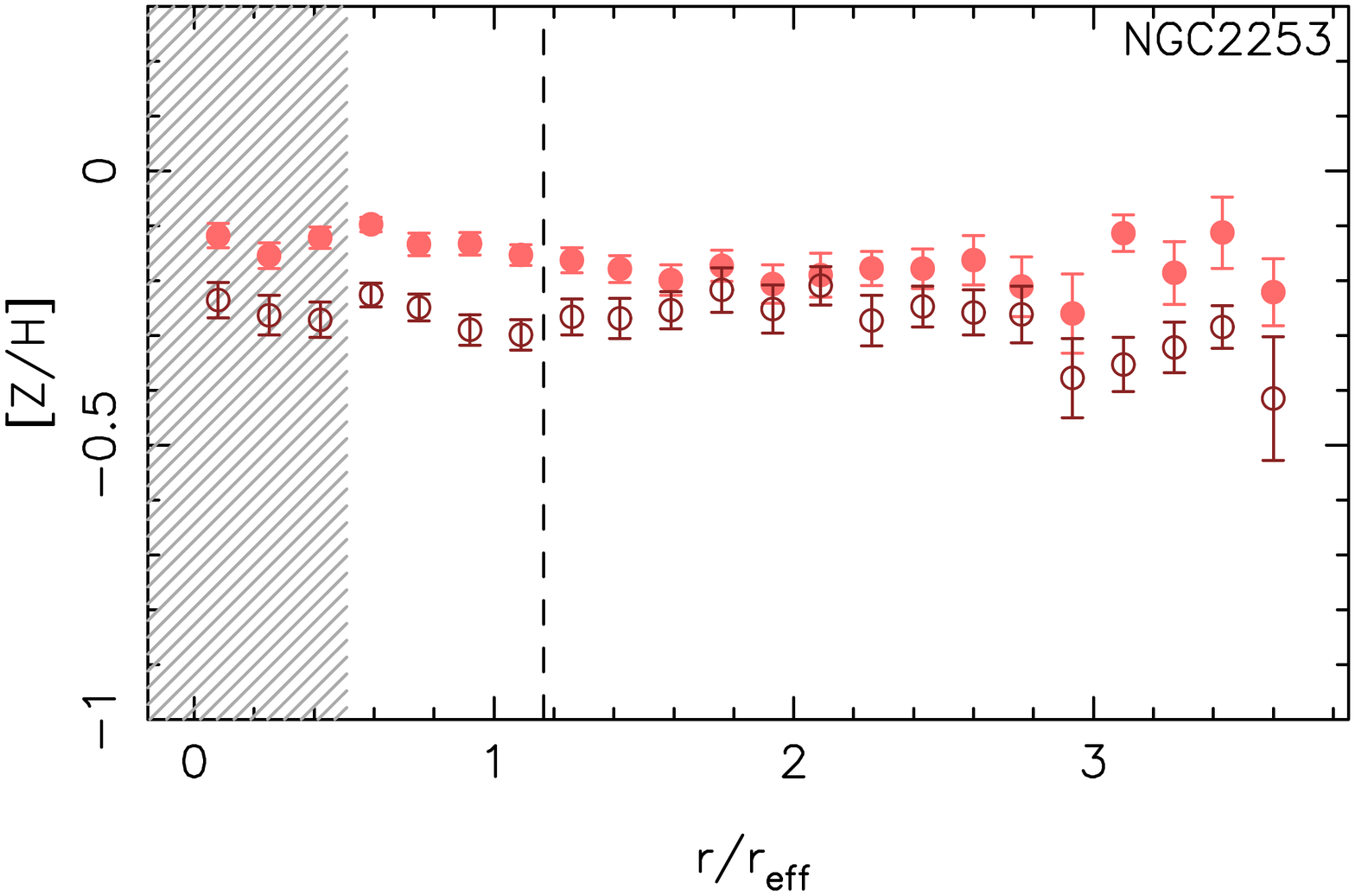}}
\resizebox{0.22\textwidth}{!}{\includegraphics[angle=0]{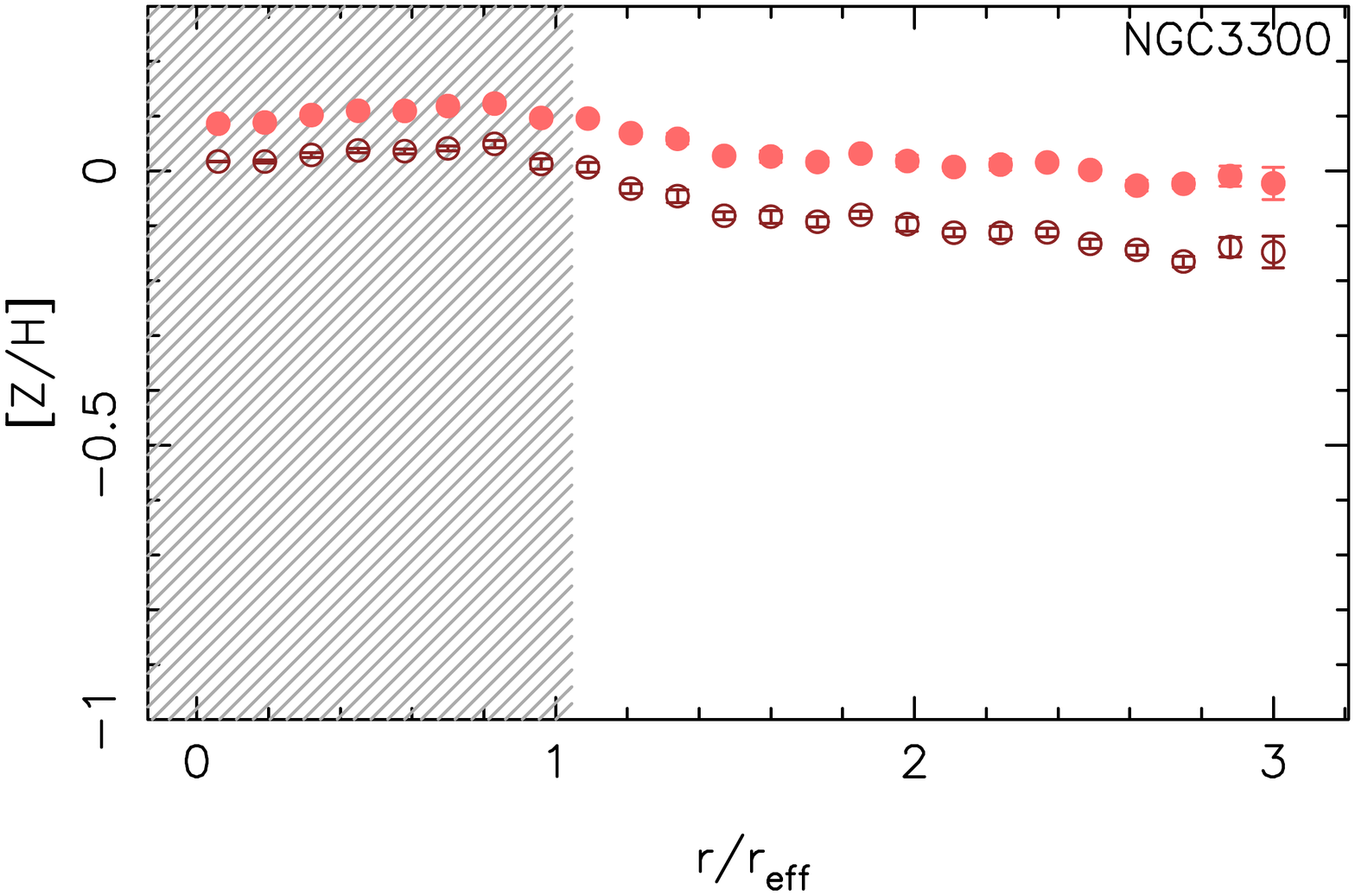}}
\resizebox{0.22\textwidth}{!}{\includegraphics[angle=0]{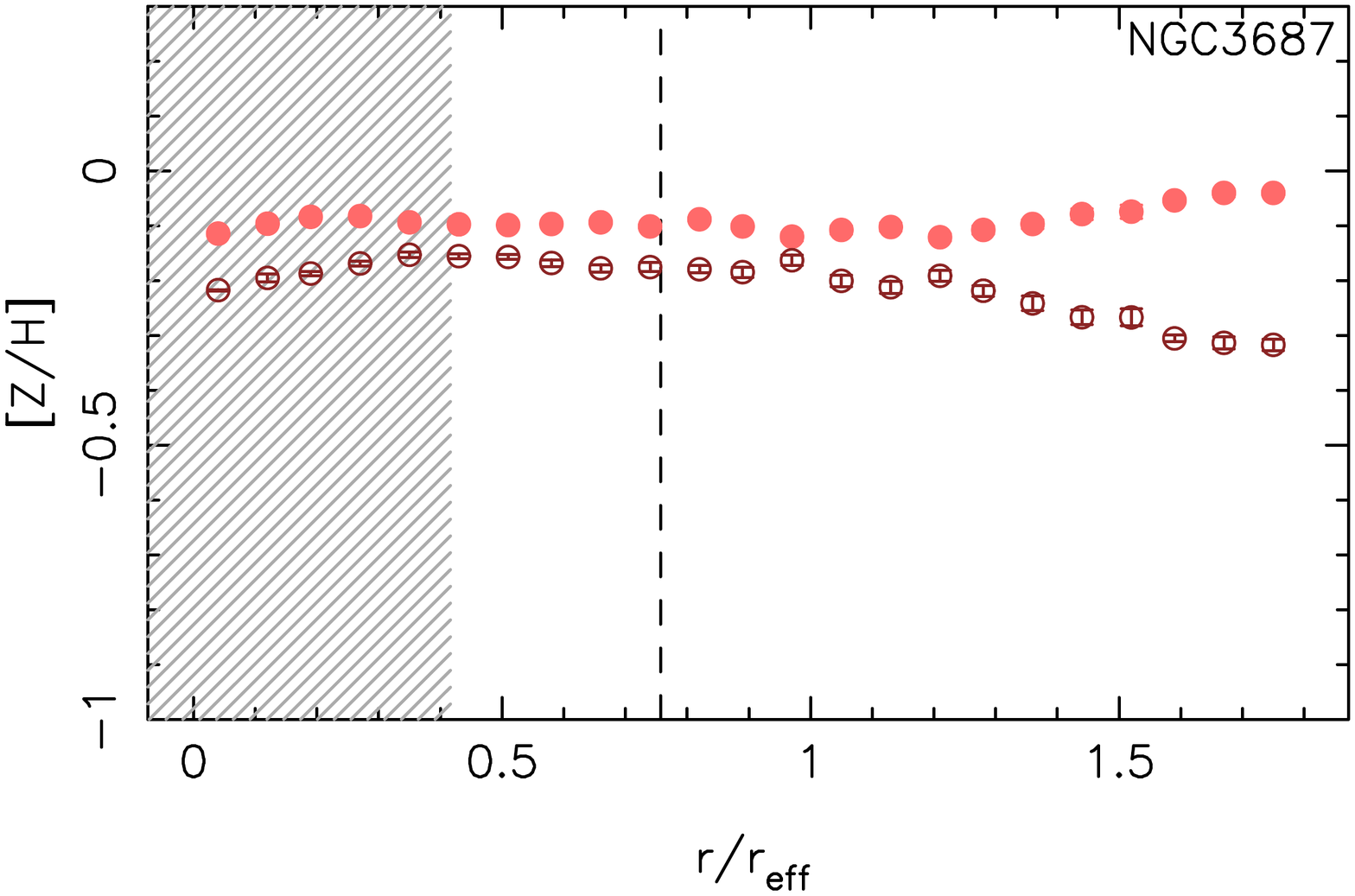}}
\resizebox{0.22\textwidth}{!}{\includegraphics[angle=0]{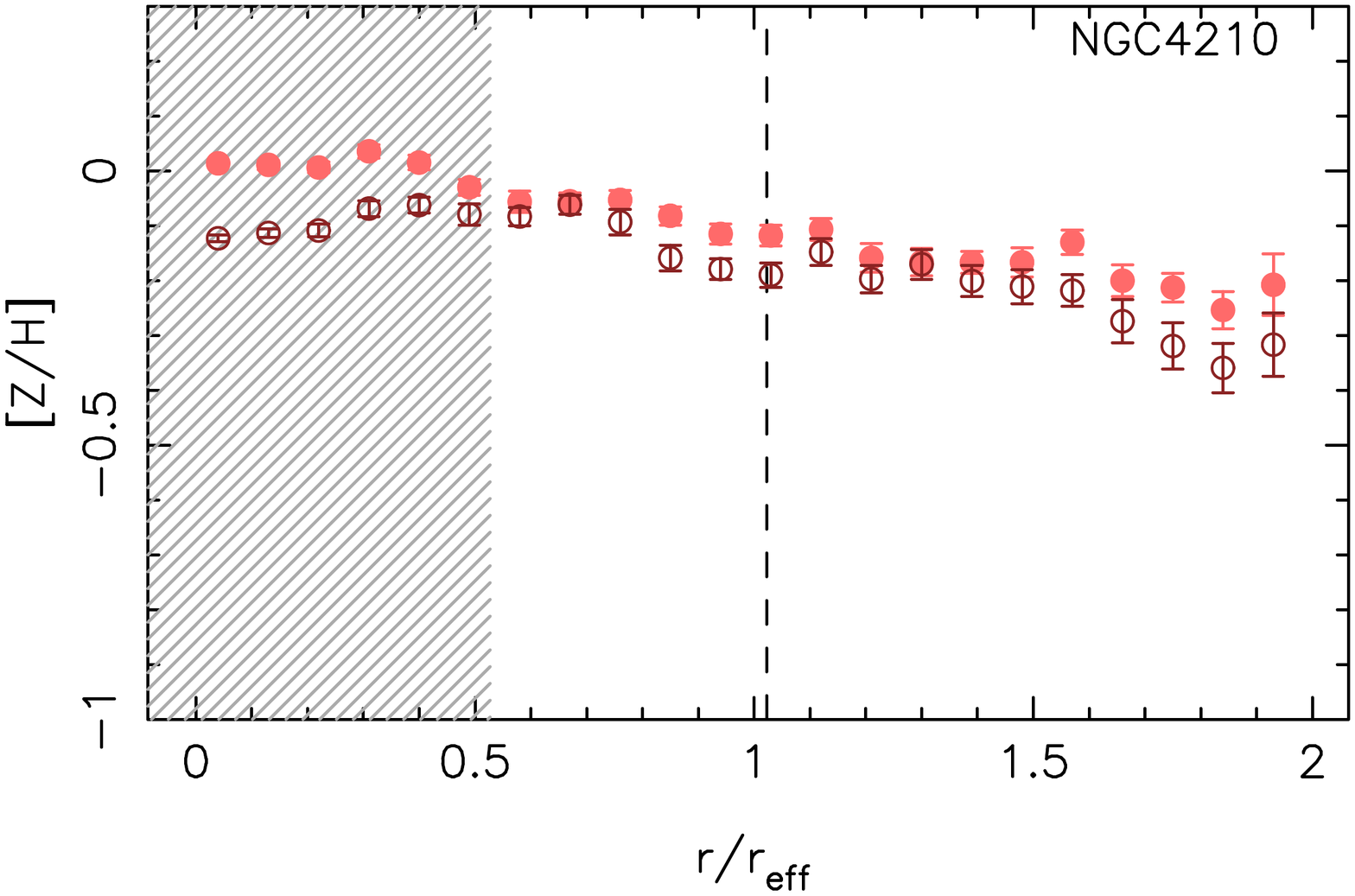}}
\resizebox{0.22\textwidth}{!}{\includegraphics[angle=0]{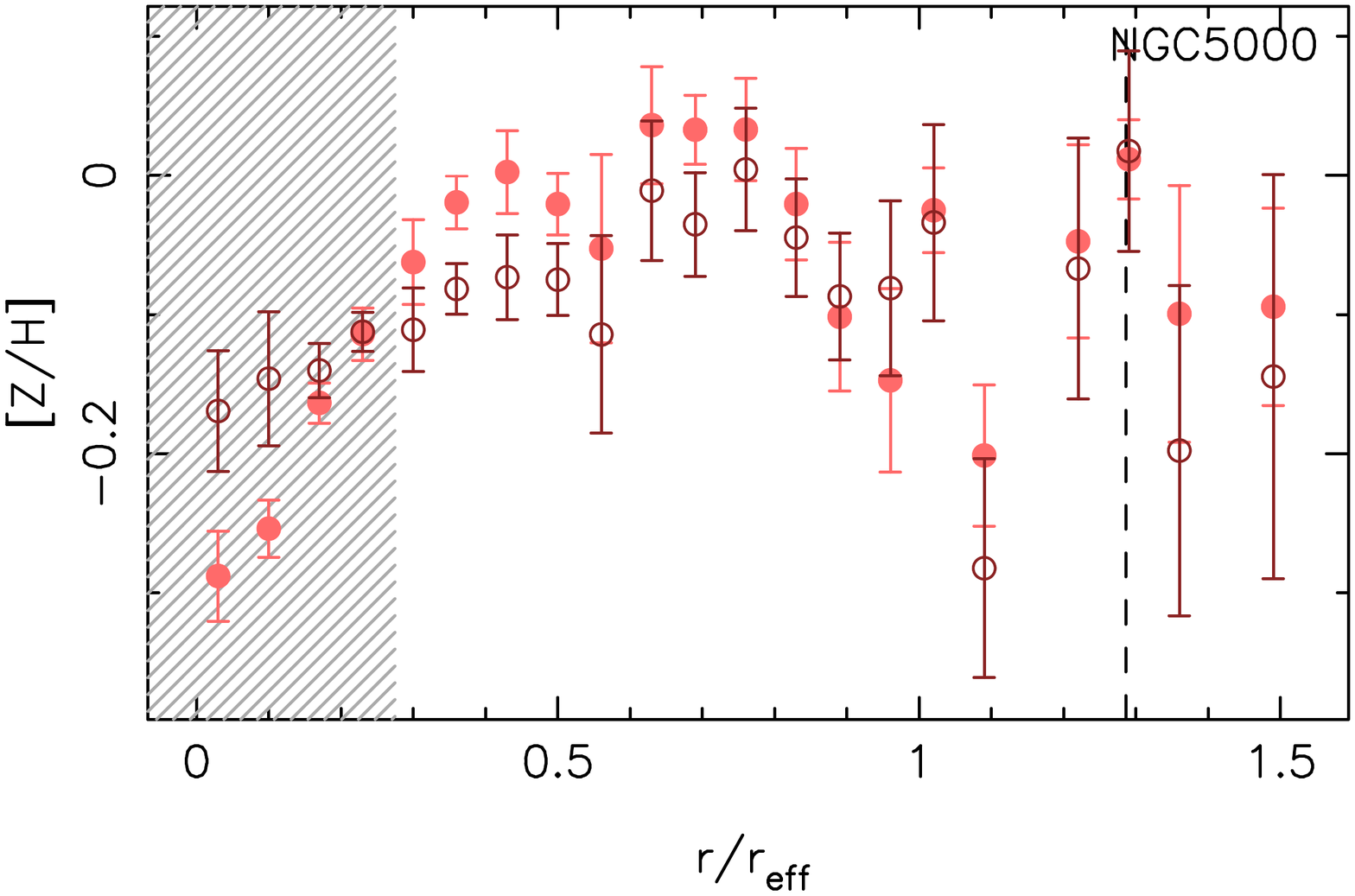}}
\resizebox{0.22\textwidth}{!}{\includegraphics[angle=0]{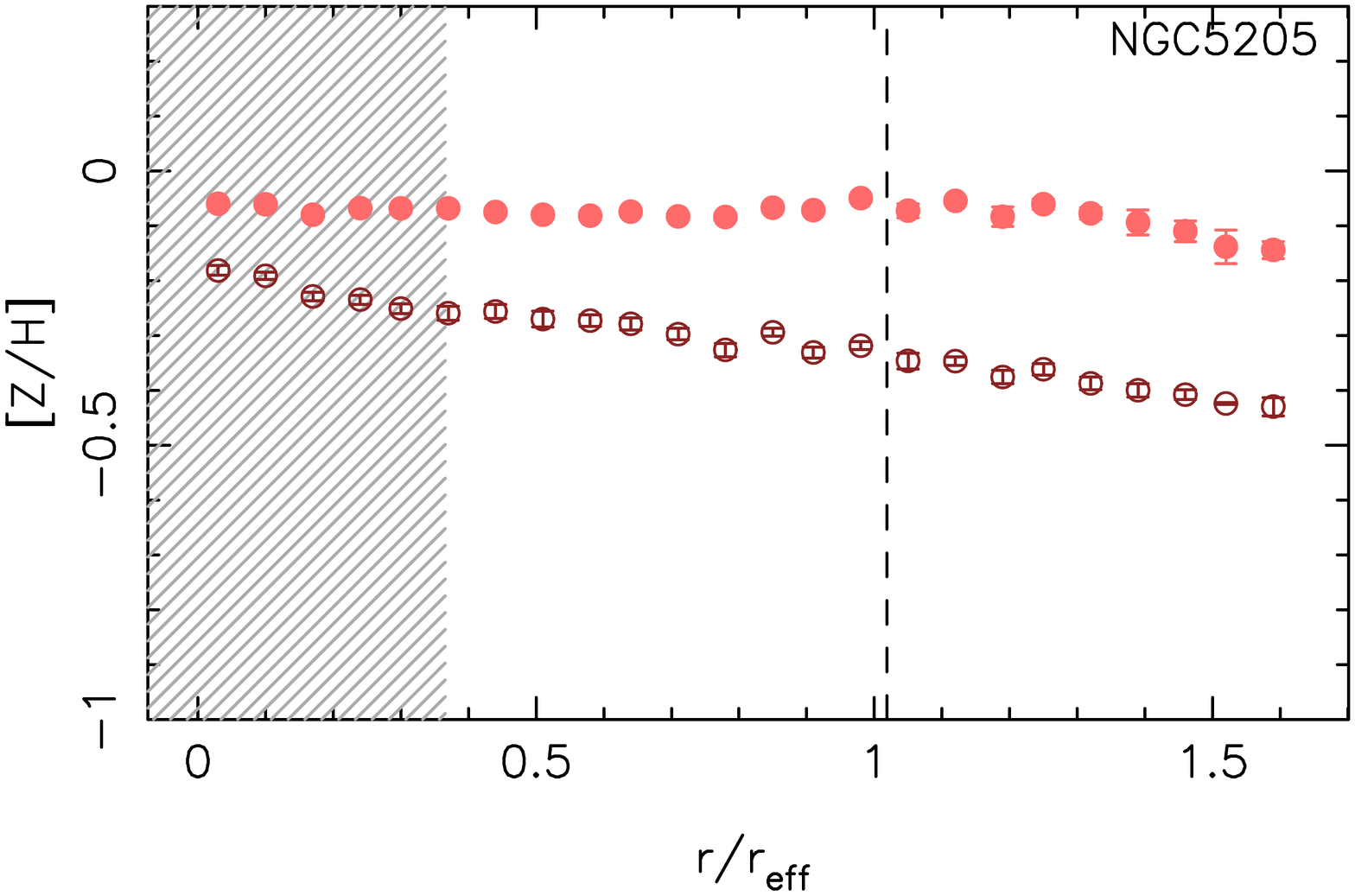}}
\resizebox{0.22\textwidth}{!}{\includegraphics[angle=0]{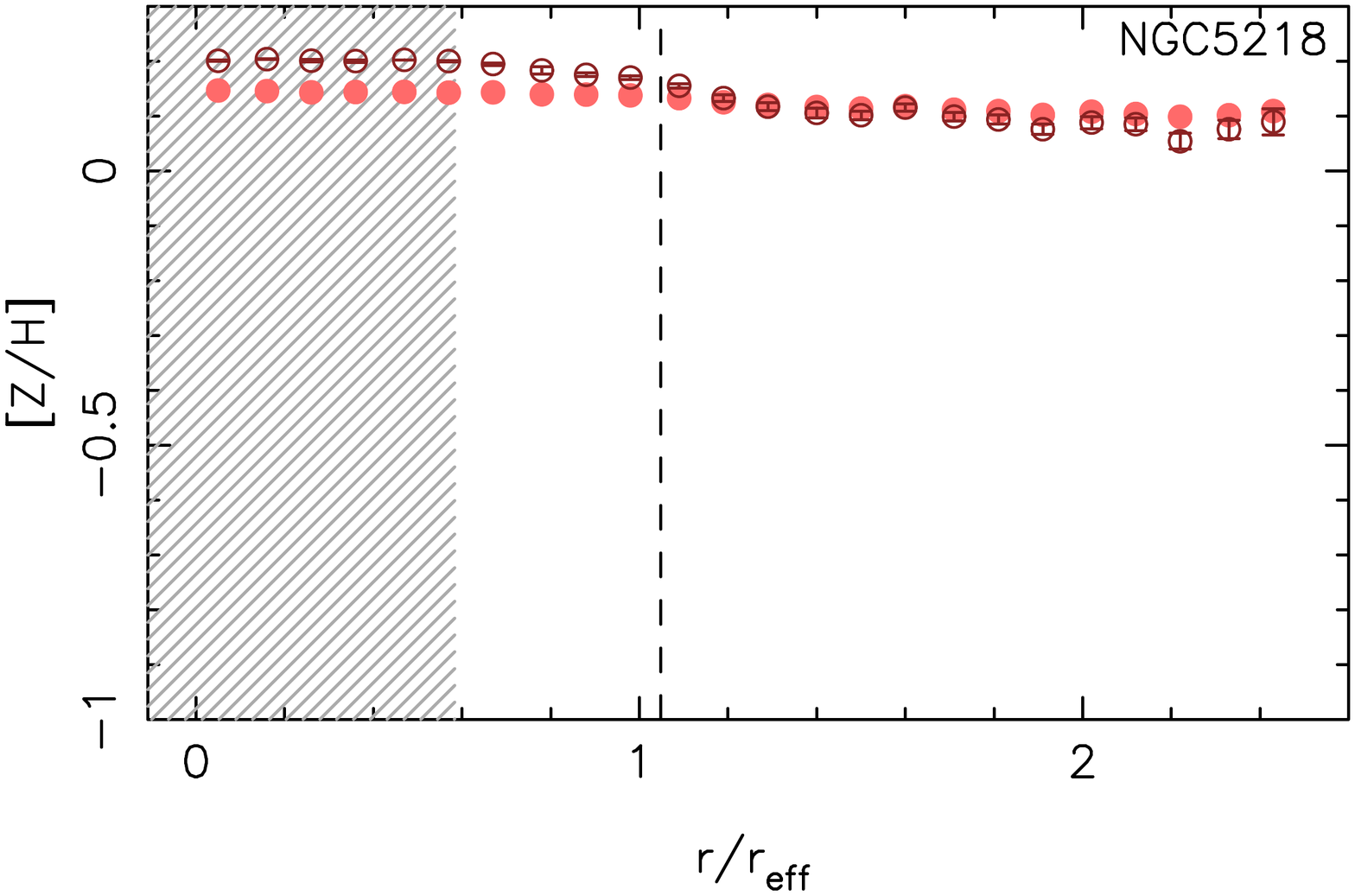}}
\resizebox{0.22\textwidth}{!}{\includegraphics[angle=0]{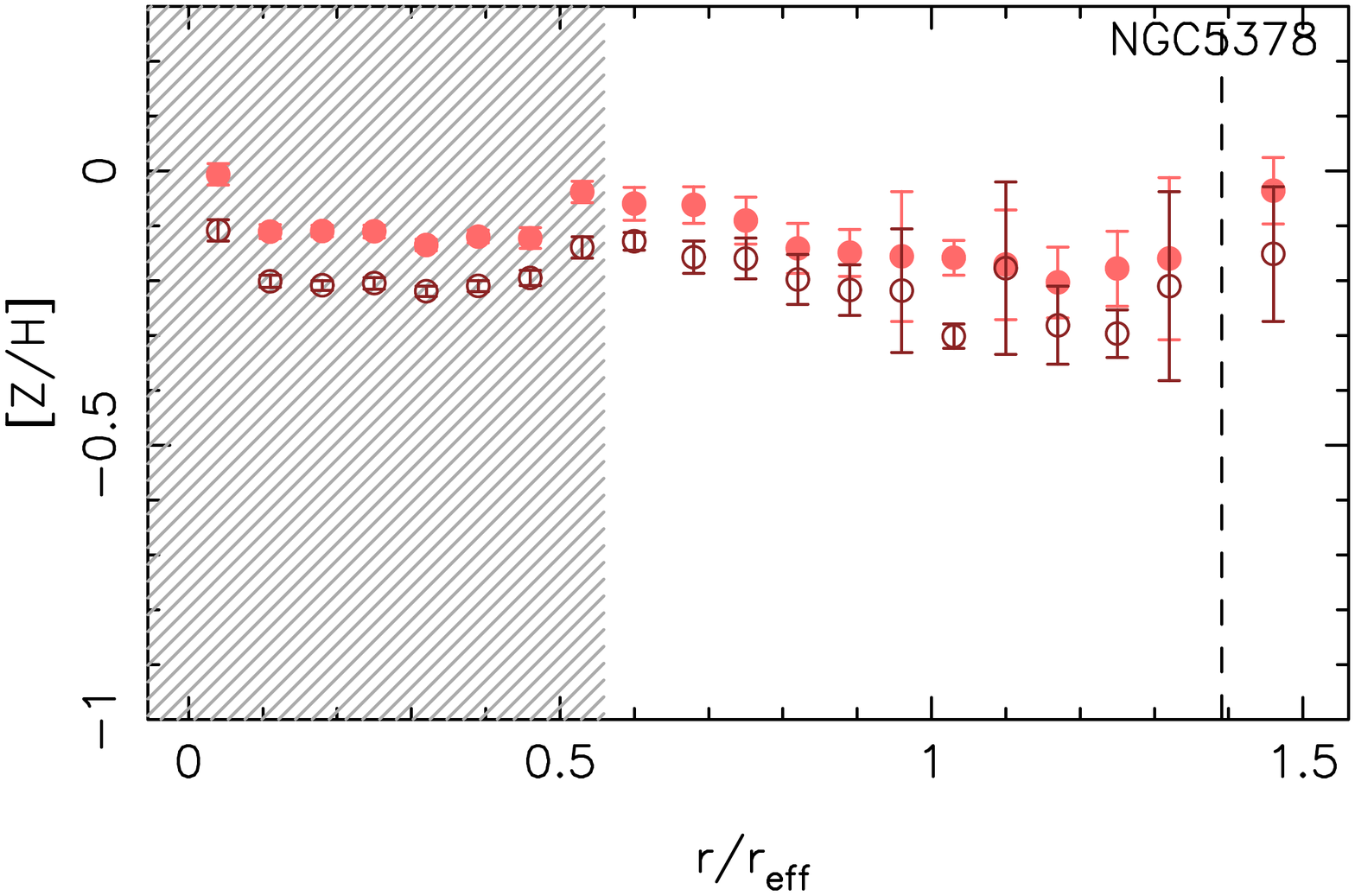}}
\resizebox{0.22\textwidth}{!}{\includegraphics[angle=0]{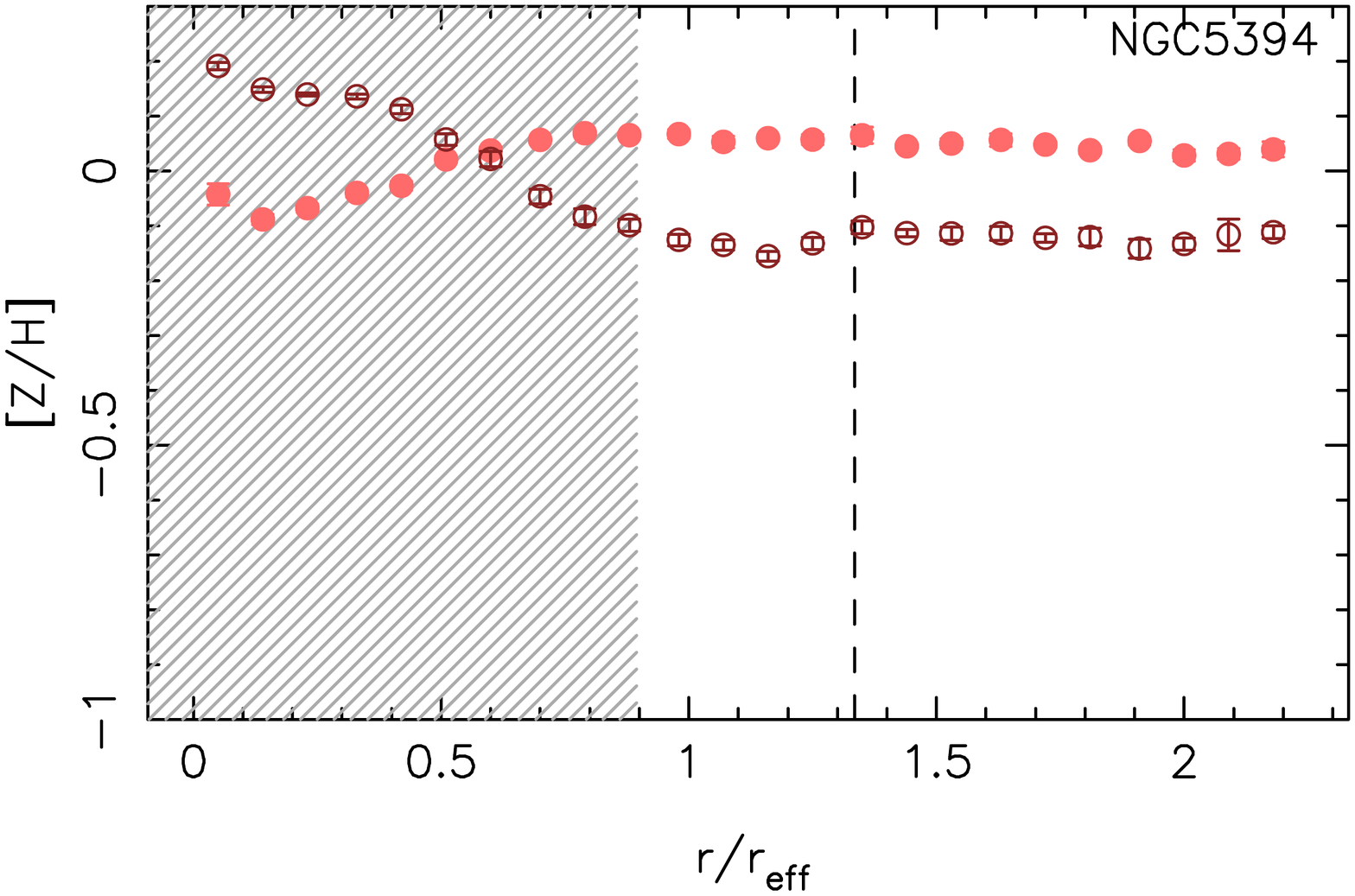}}
\resizebox{0.22\textwidth}{!}{\includegraphics[angle=0]{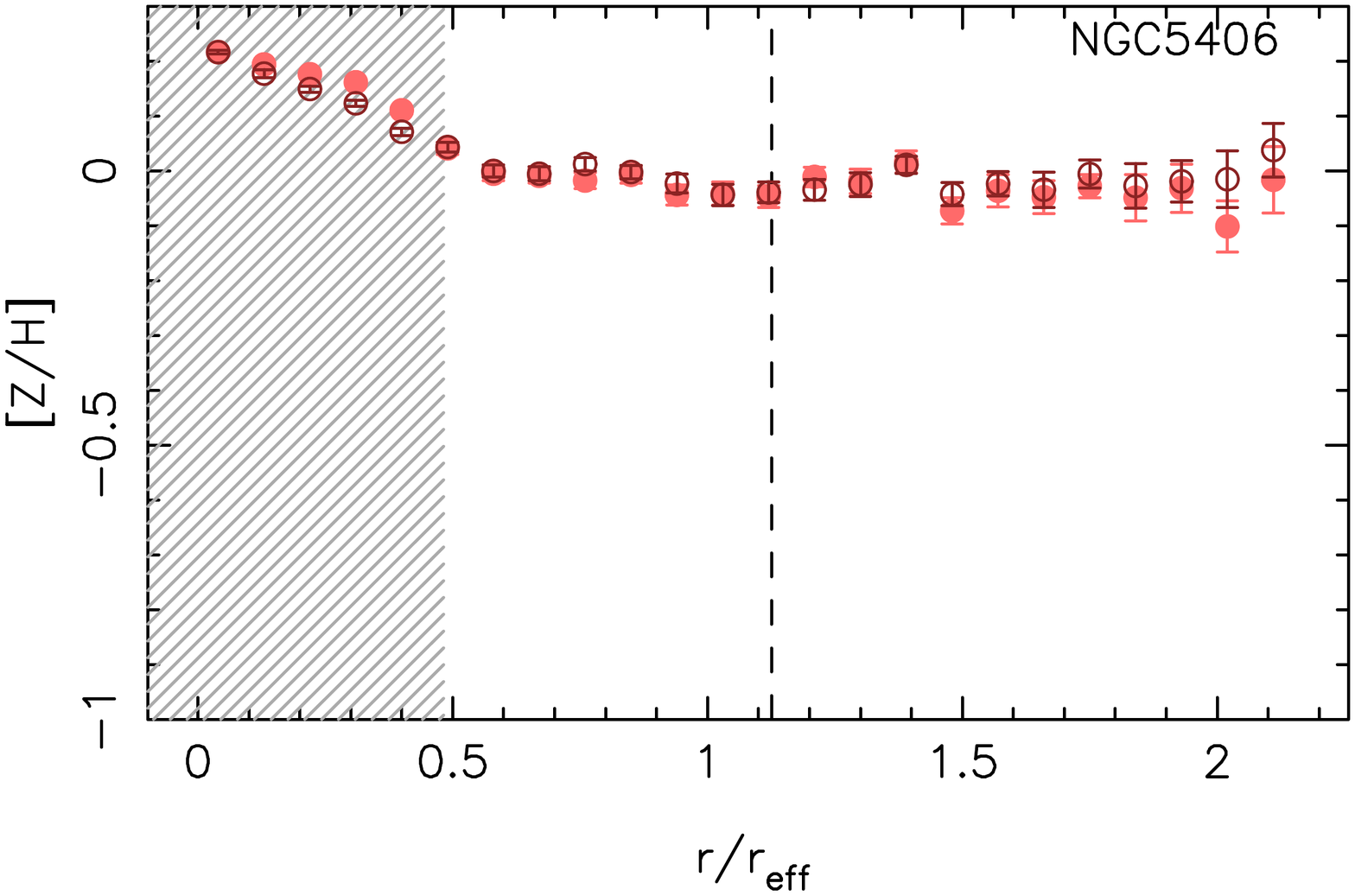}}
\resizebox{0.22\textwidth}{!}{\includegraphics[angle=0]{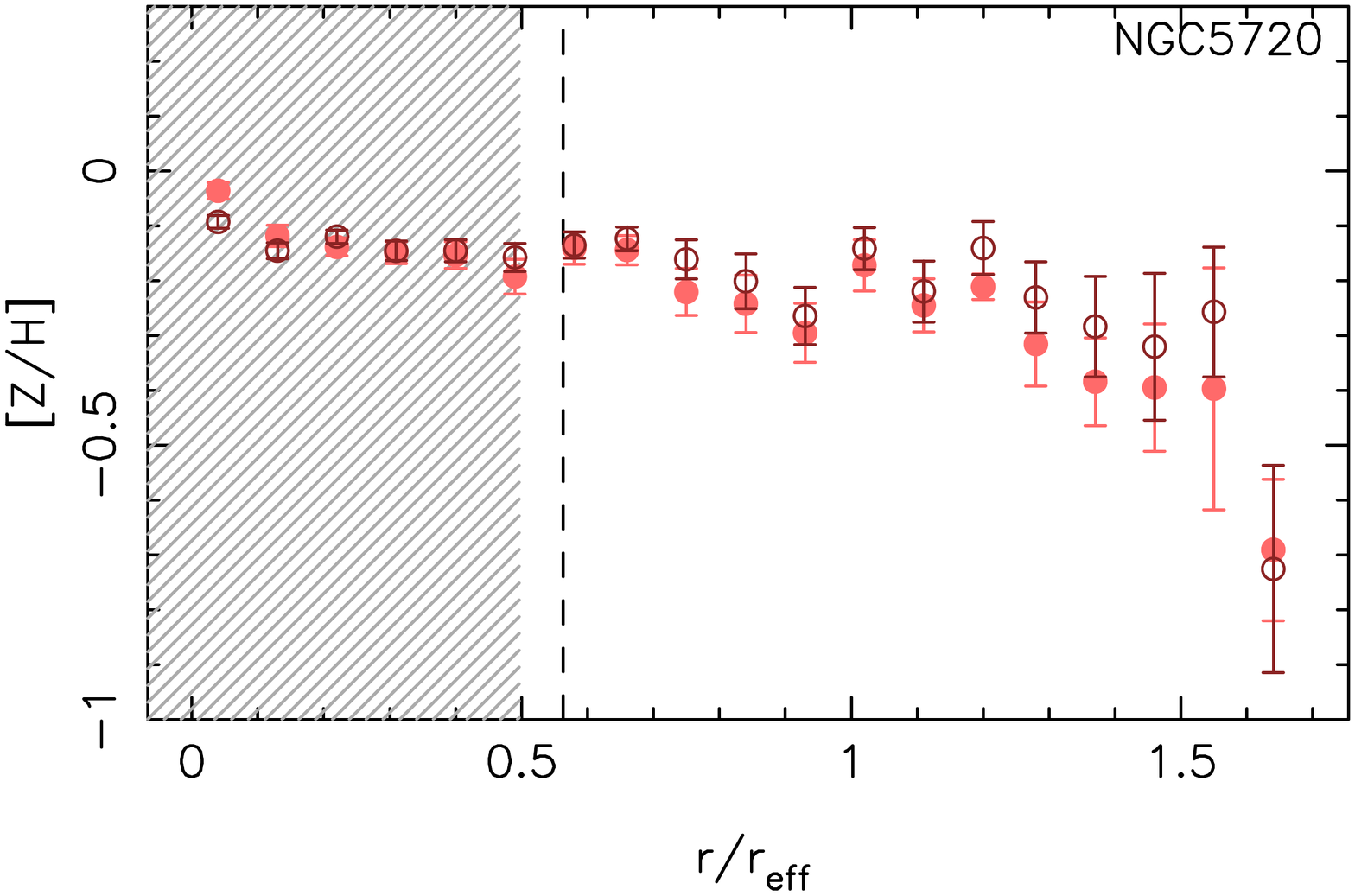}}
\resizebox{0.22\textwidth}{!}{\includegraphics[angle=0]{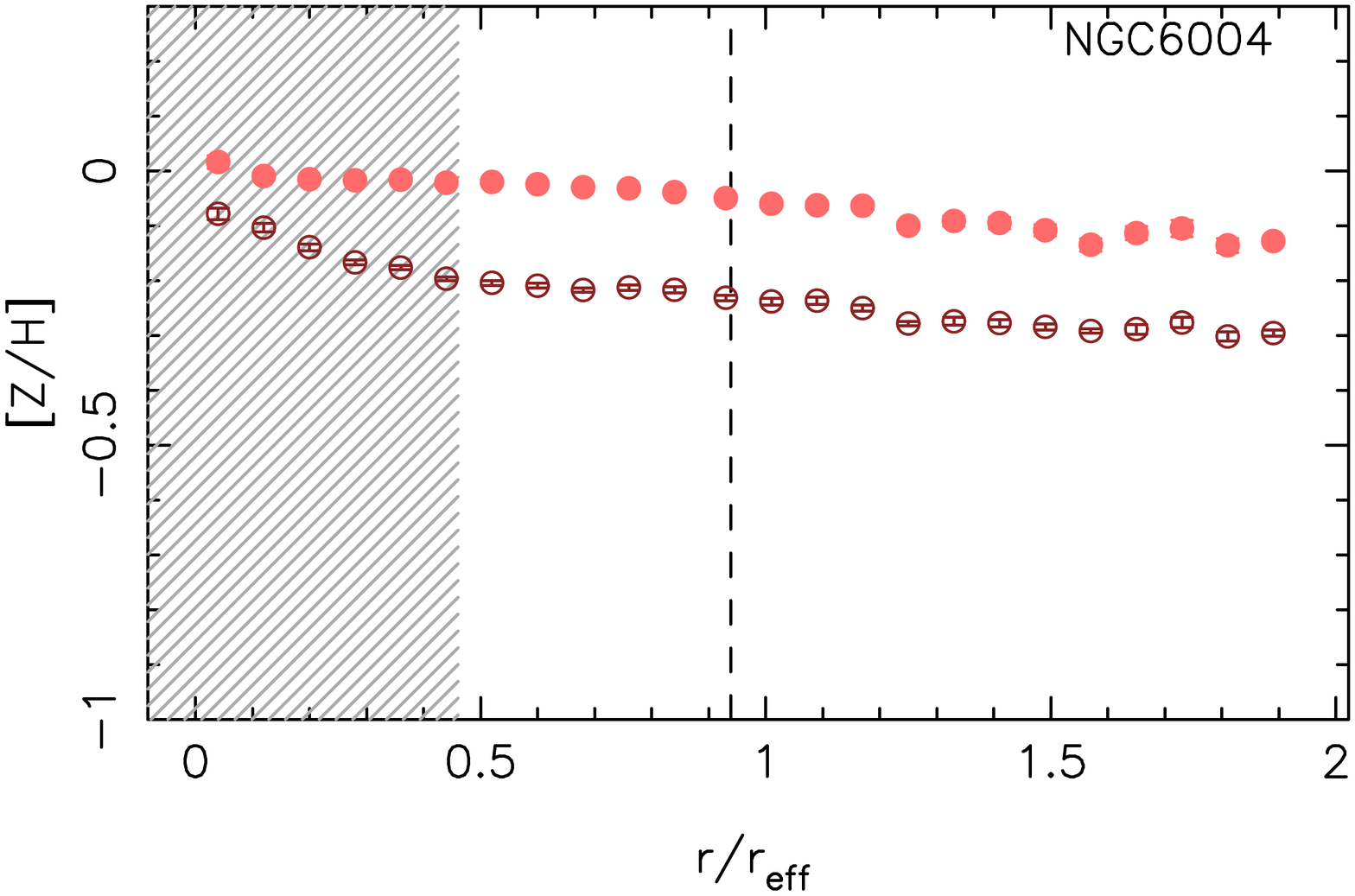}}
\resizebox{0.22\textwidth}{!}{\includegraphics[angle=0]{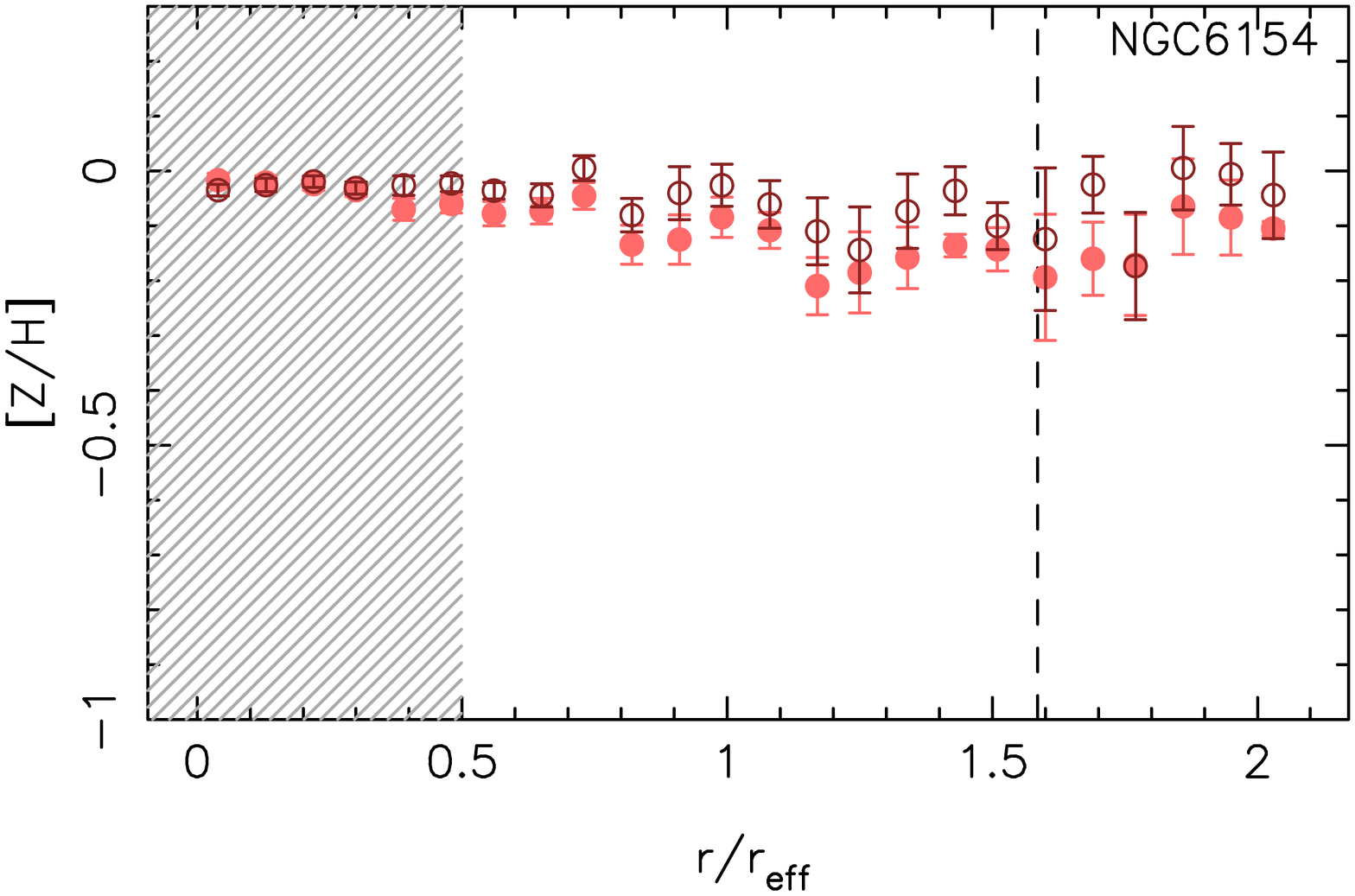}}
\resizebox{0.22\textwidth}{!}{\includegraphics[angle=0]{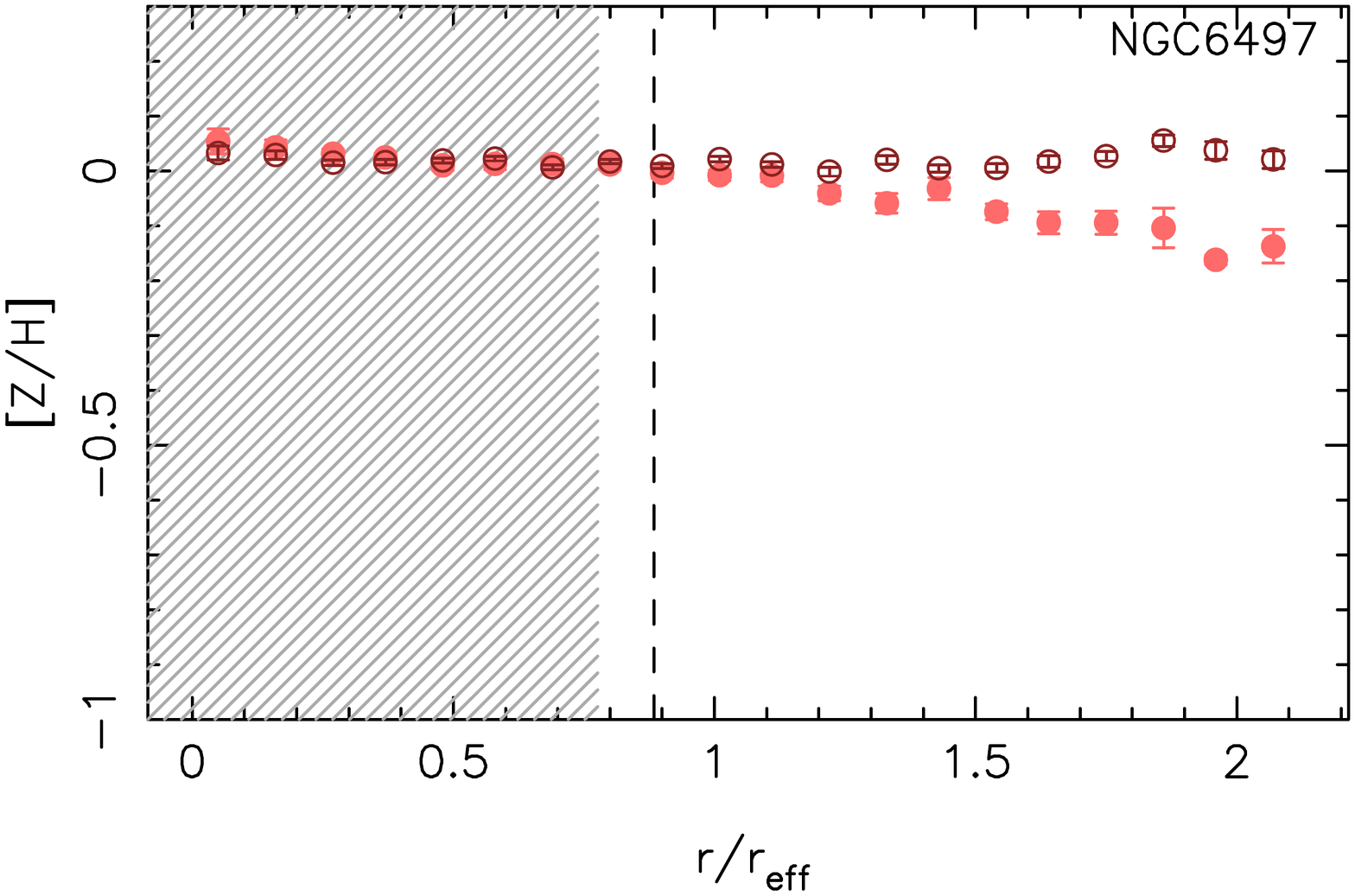}}
\resizebox{0.22\textwidth}{!}{\includegraphics[angle=0]{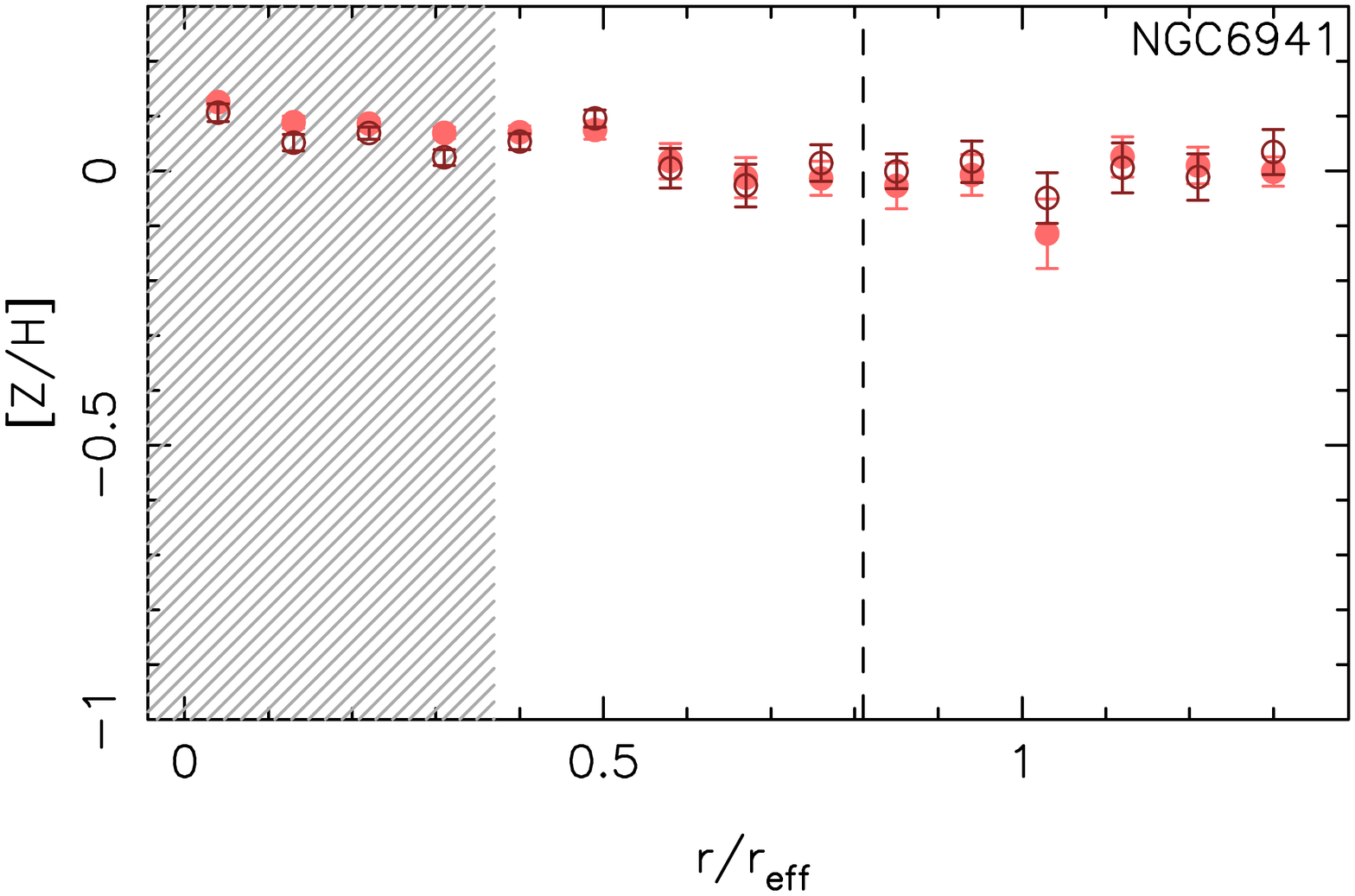}}
\resizebox{0.22\textwidth}{!}{\includegraphics[angle=0]{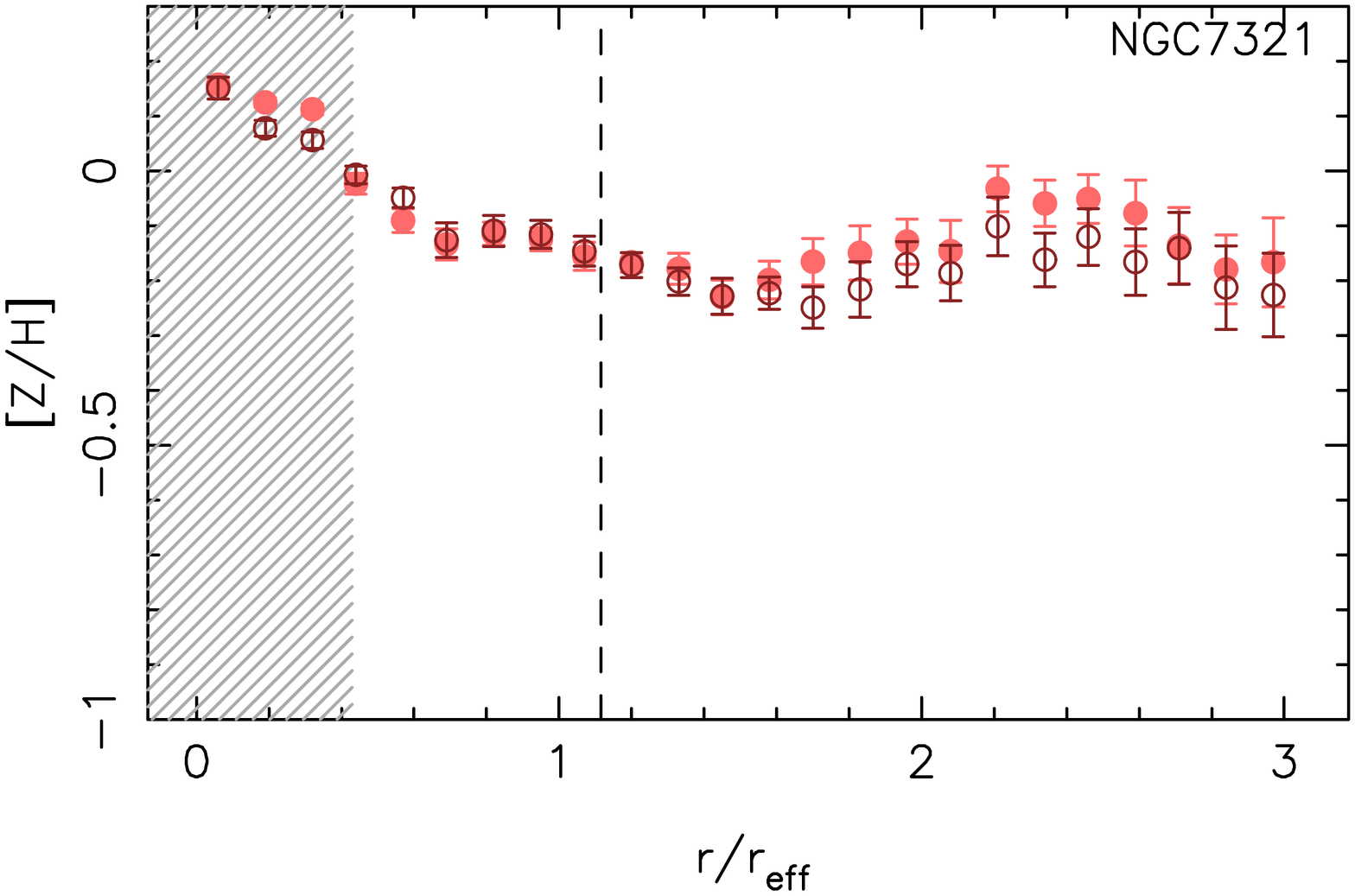}}
\resizebox{0.22\textwidth}{!}{\includegraphics[angle=0]{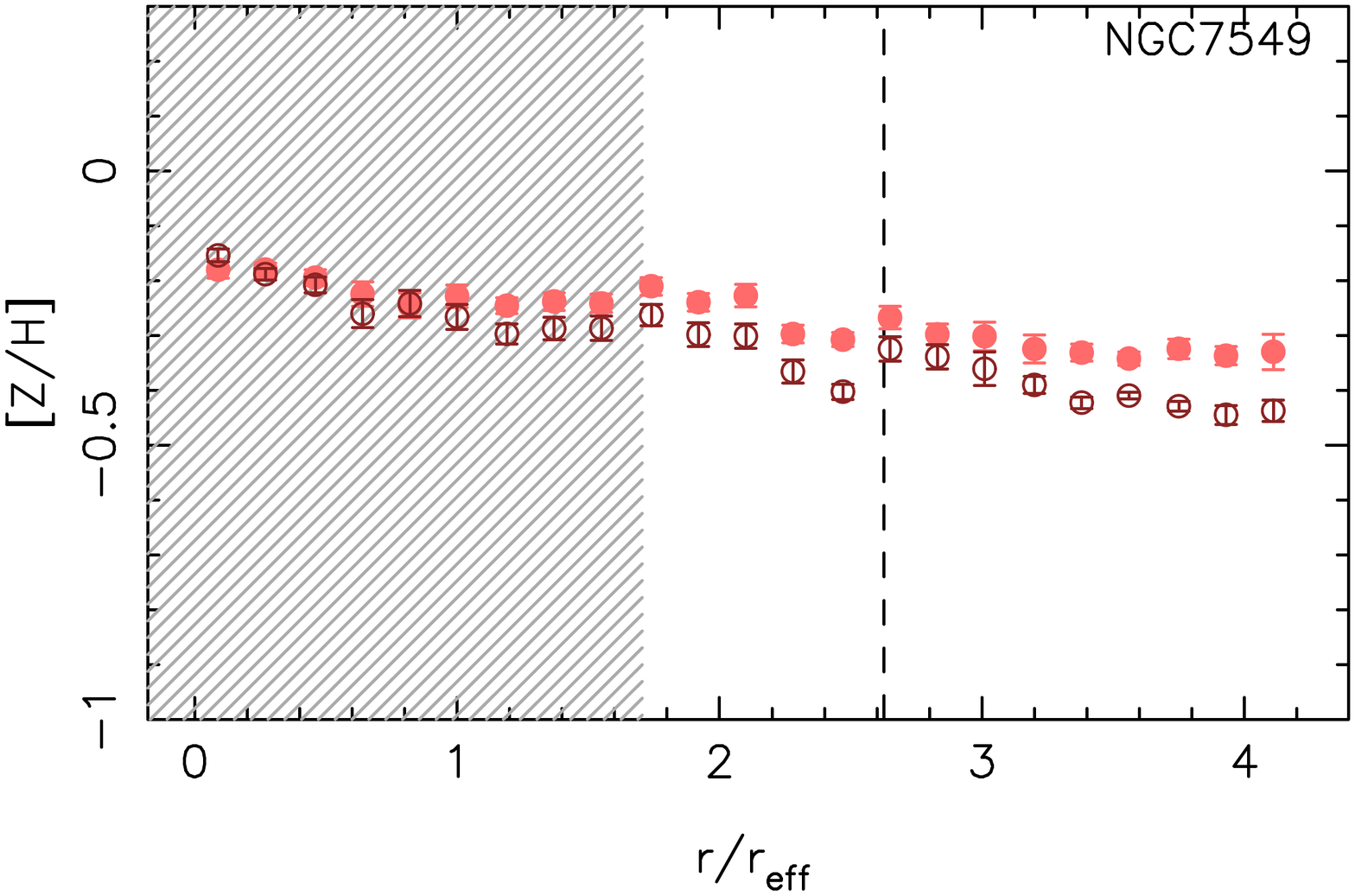}}
\resizebox{0.22\textwidth}{!}{\includegraphics[angle=0]{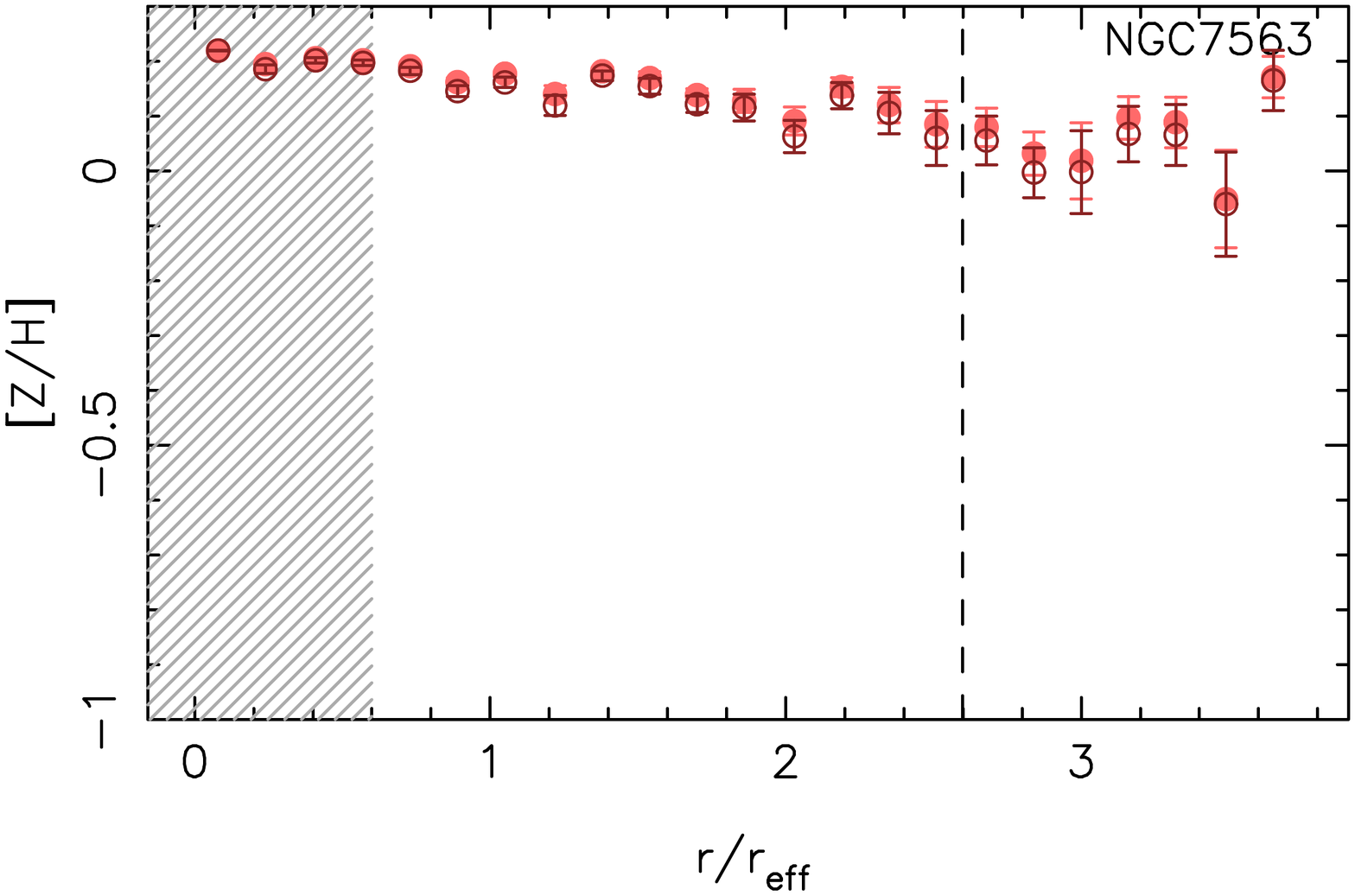}}
\resizebox{0.22\textwidth}{!}{\includegraphics[angle=0]{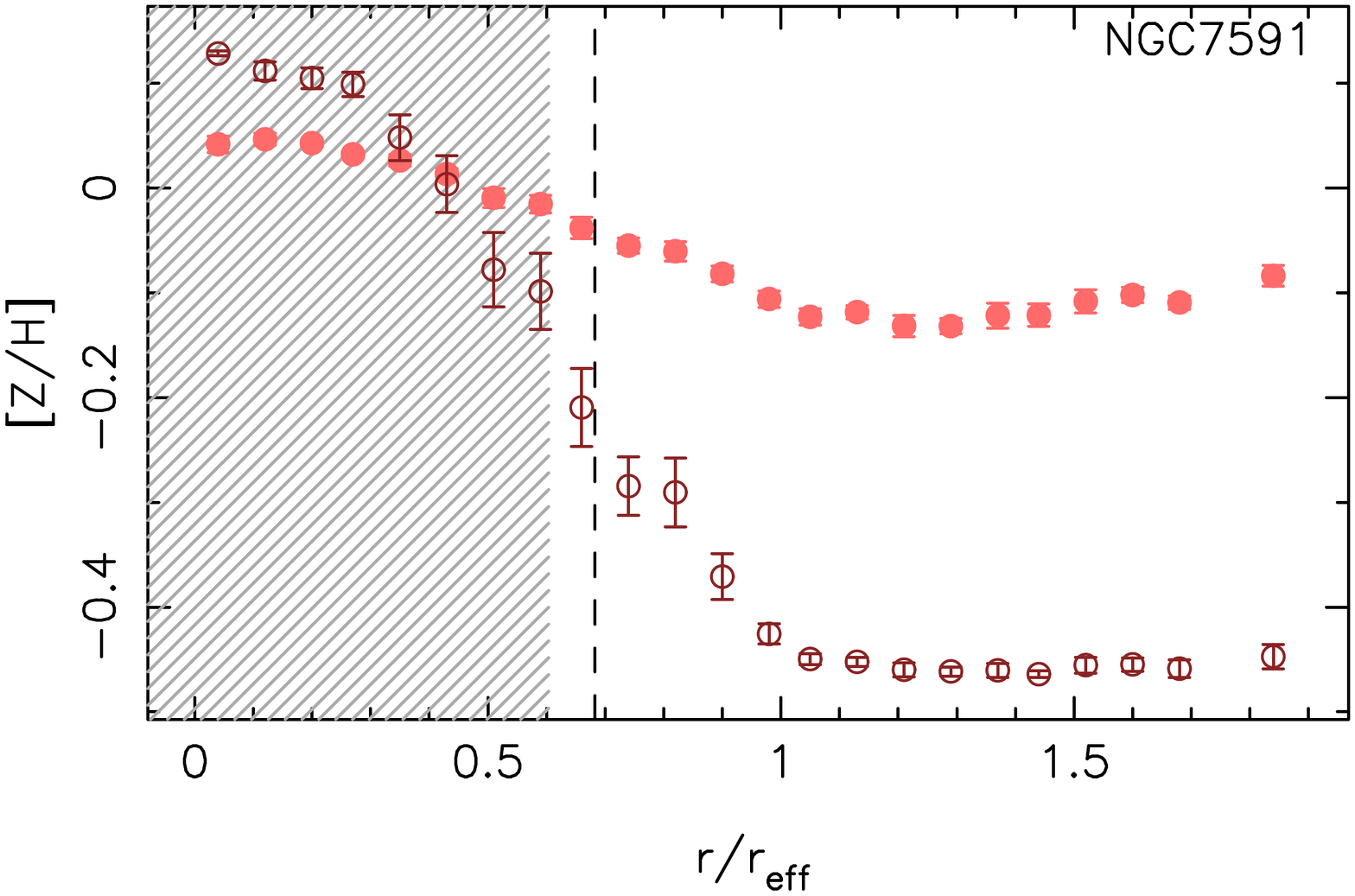}}
\resizebox{0.22\textwidth}{!}{\includegraphics[angle=0]{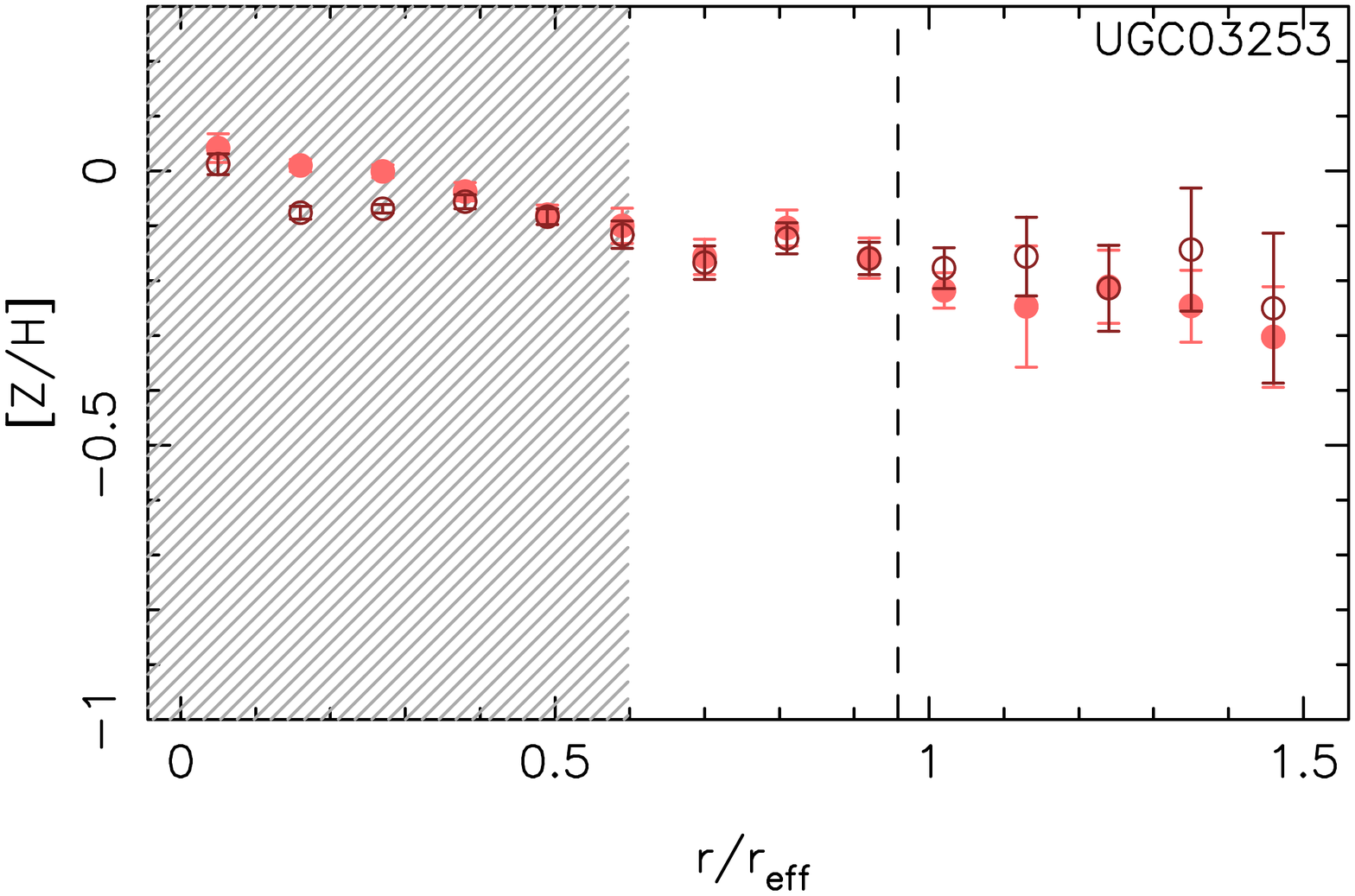}}
\resizebox{0.22\textwidth}{!}{\includegraphics[angle=0]{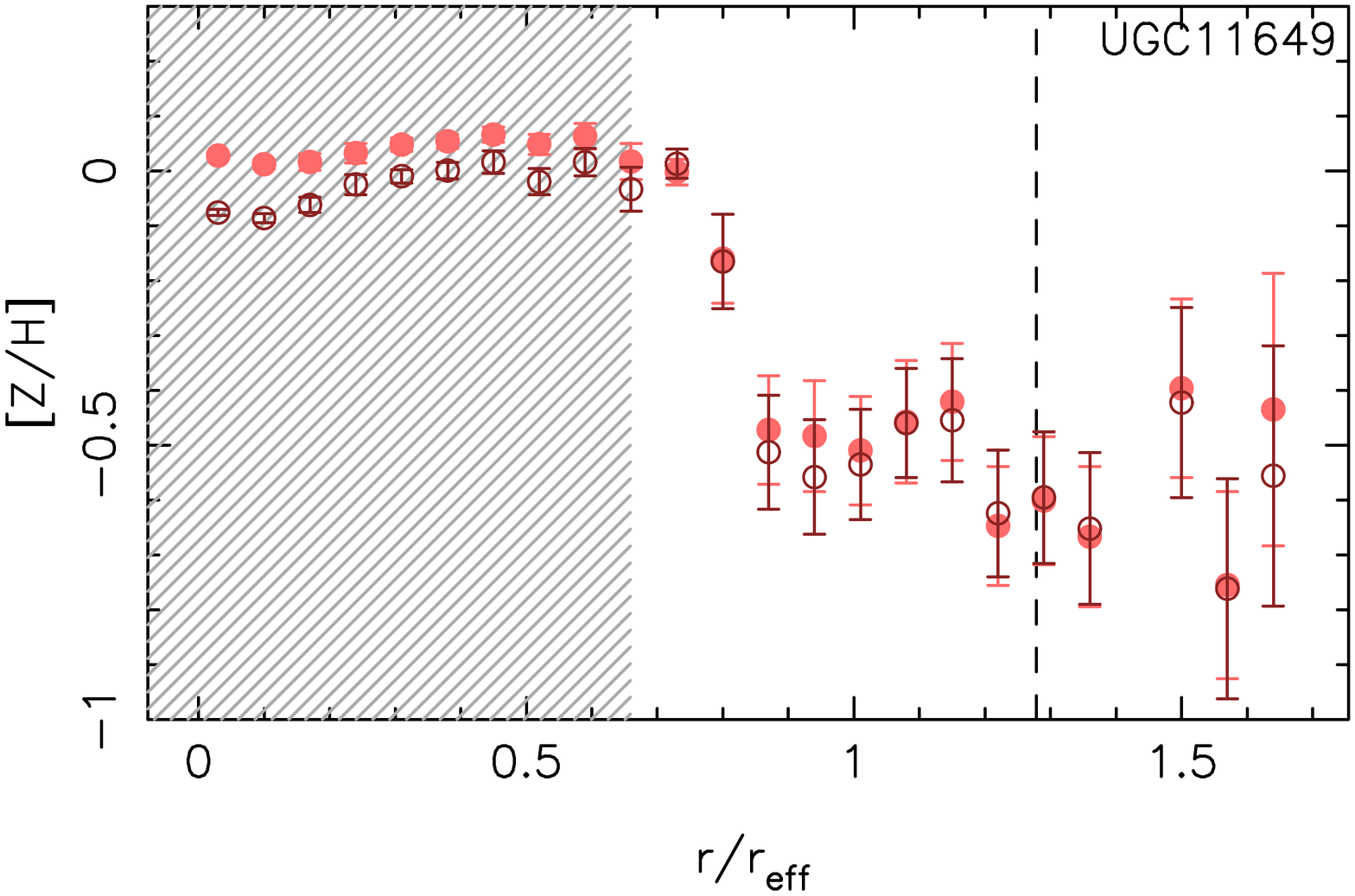}}
\resizebox{0.22\textwidth}{!}{\includegraphics[angle=0]{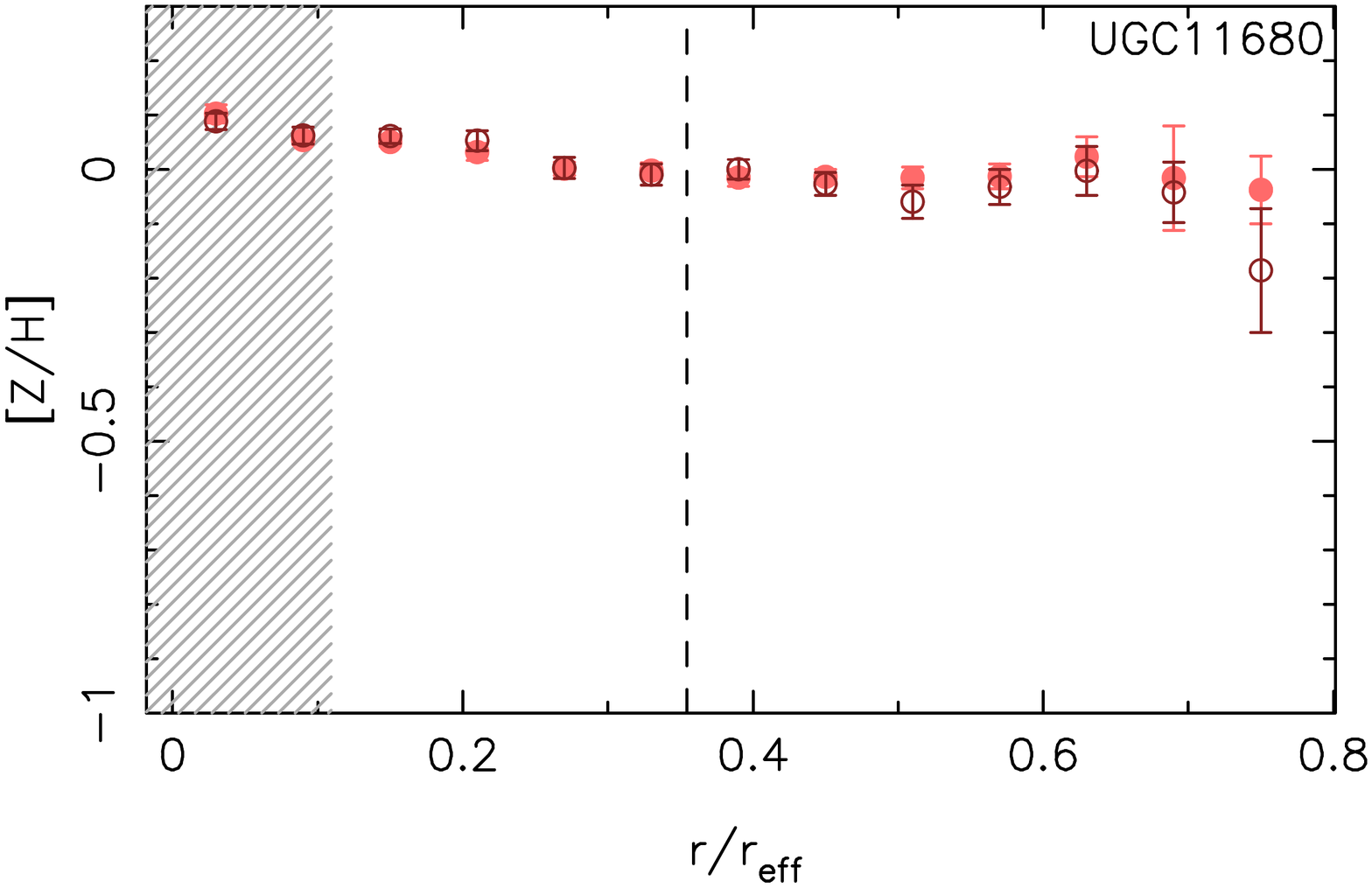}}
\caption{Luminosity- (light red) and Mass- (dark red) metallicity gradients
for our sample of barred galaxies. \label{fig:metgrads1}}
\end{figure*}

Figure~\ref{fig:metgrads1} shows the metallicity gradients for our
sample of galaxies.

\begin{figure*}
\centering
\resizebox{0.22\textwidth}{!}{\includegraphics[angle=0]{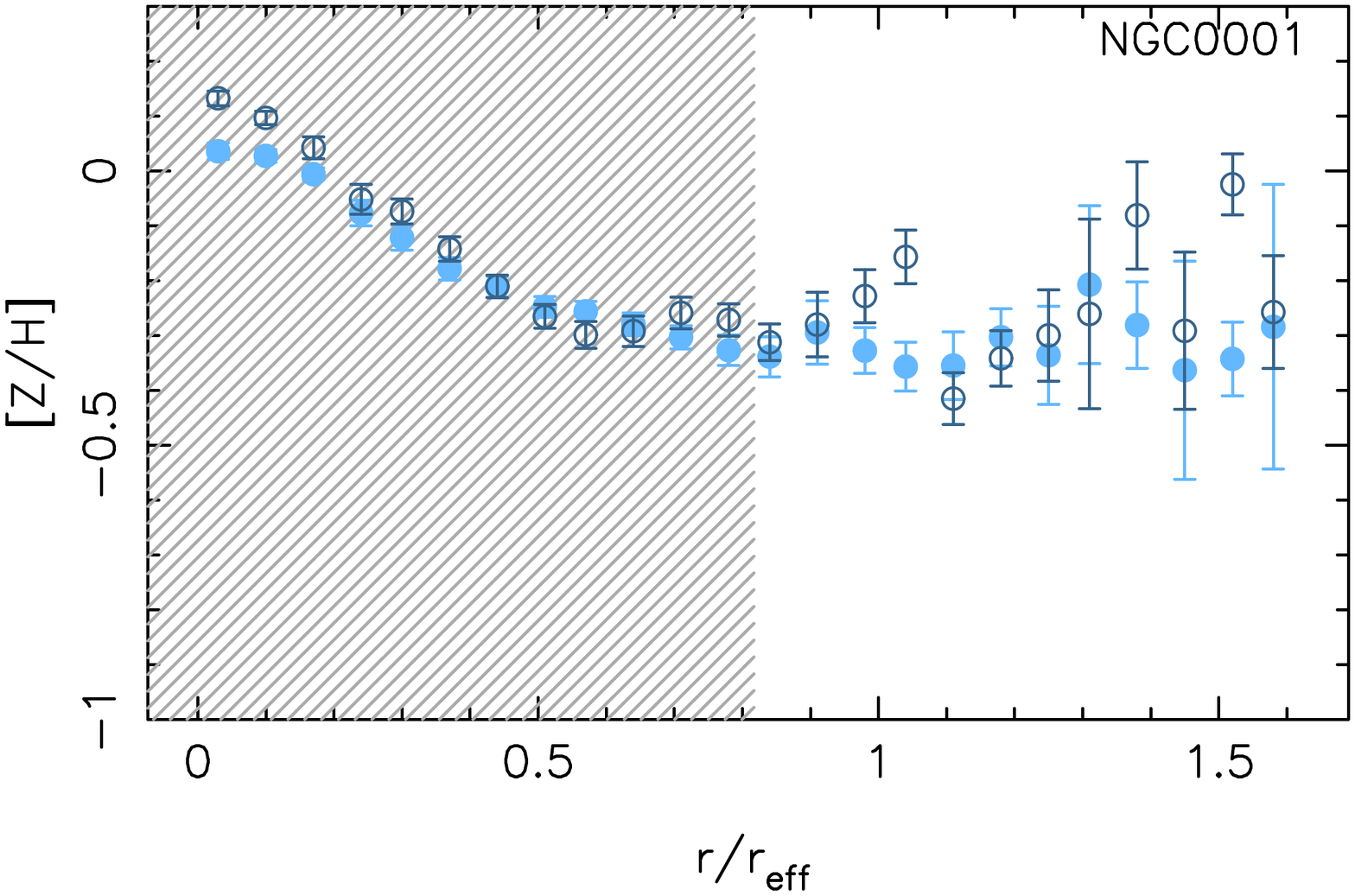}}
\resizebox{0.22\textwidth}{!}{\includegraphics[angle=0]{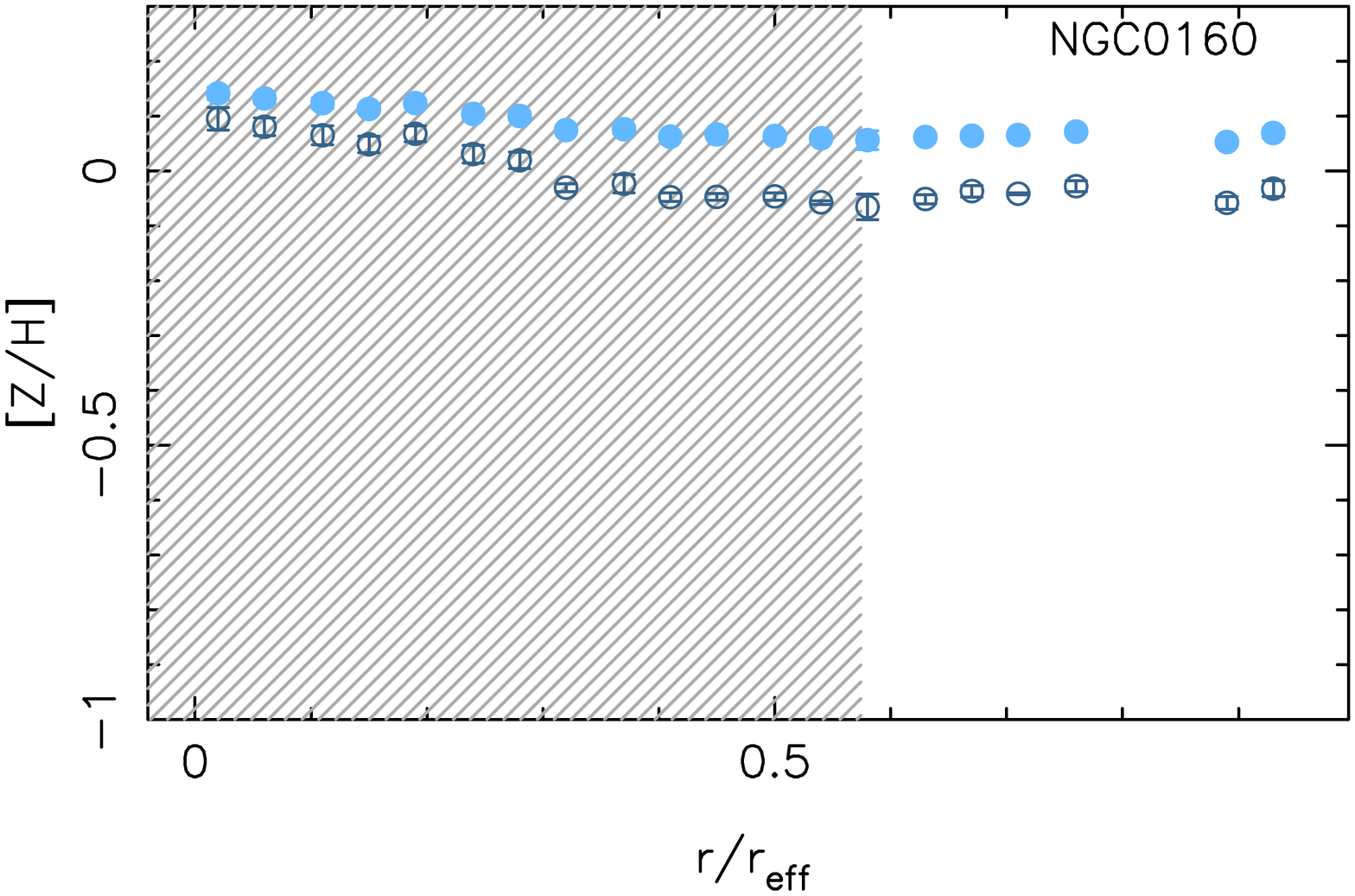}}
\resizebox{0.22\textwidth}{!}{\includegraphics[angle=0]{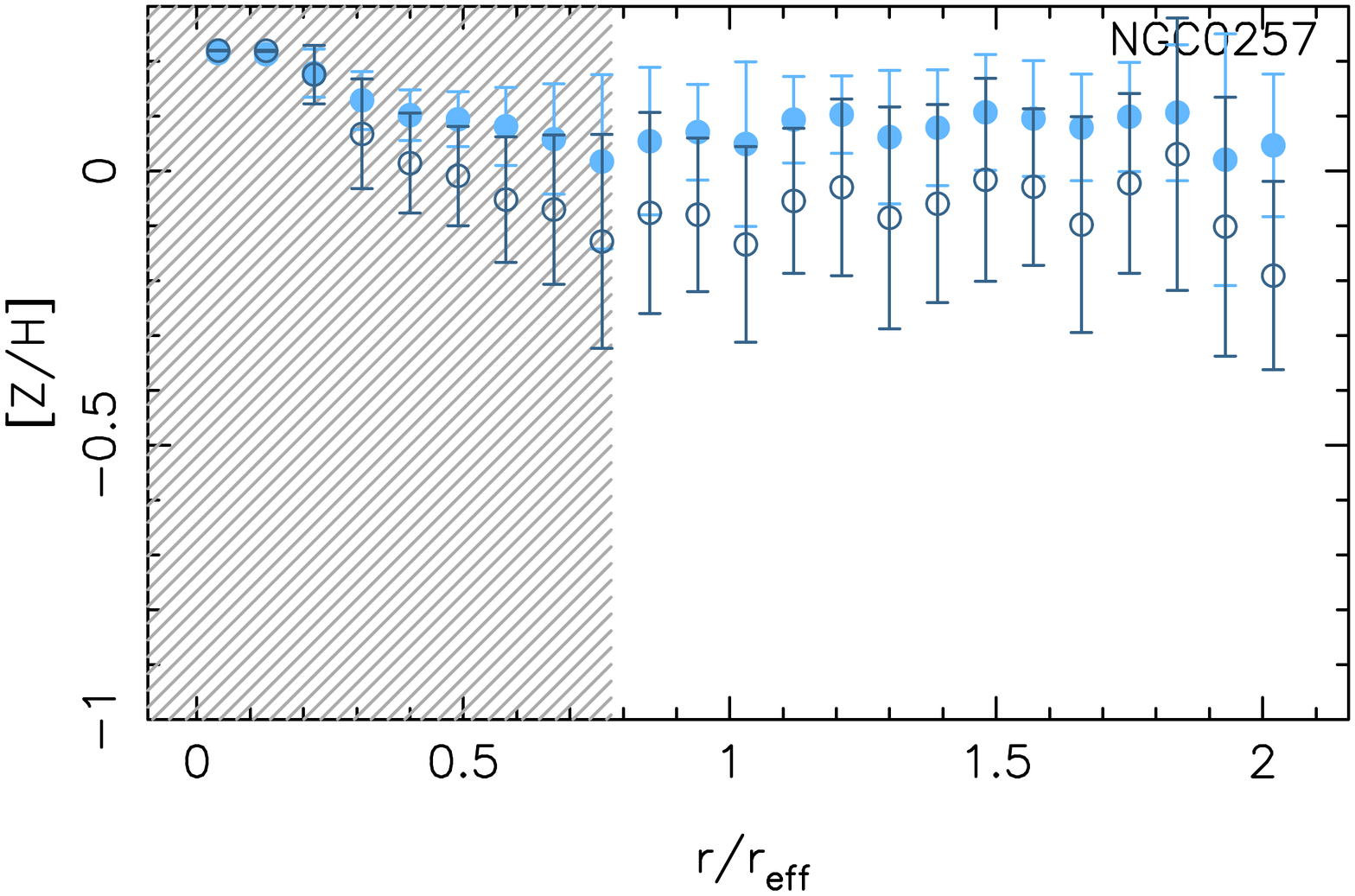}}
\resizebox{0.22\textwidth}{!}{\includegraphics[angle=0]{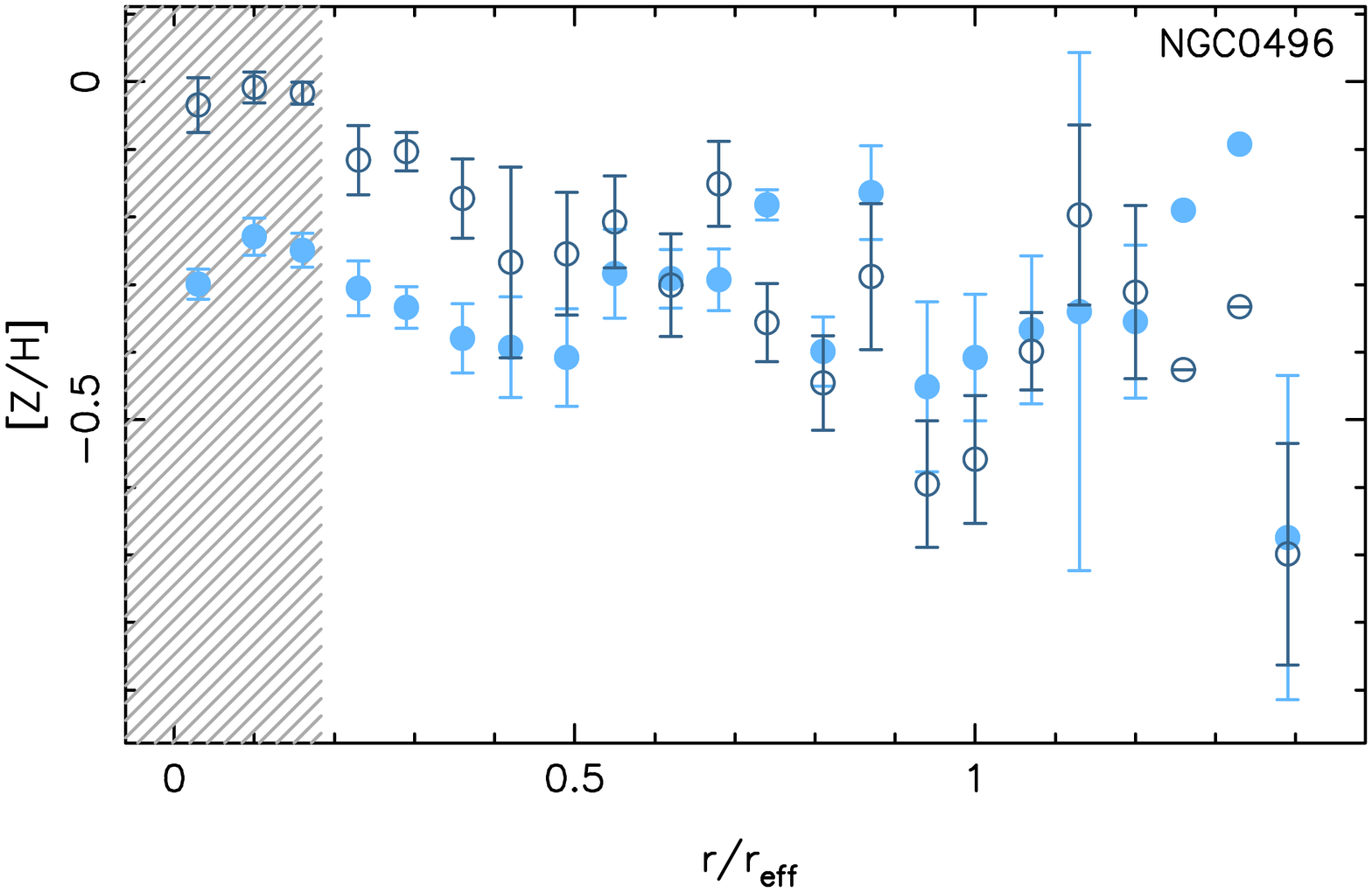}}
\resizebox{0.22\textwidth}{!}{\includegraphics[angle=0]{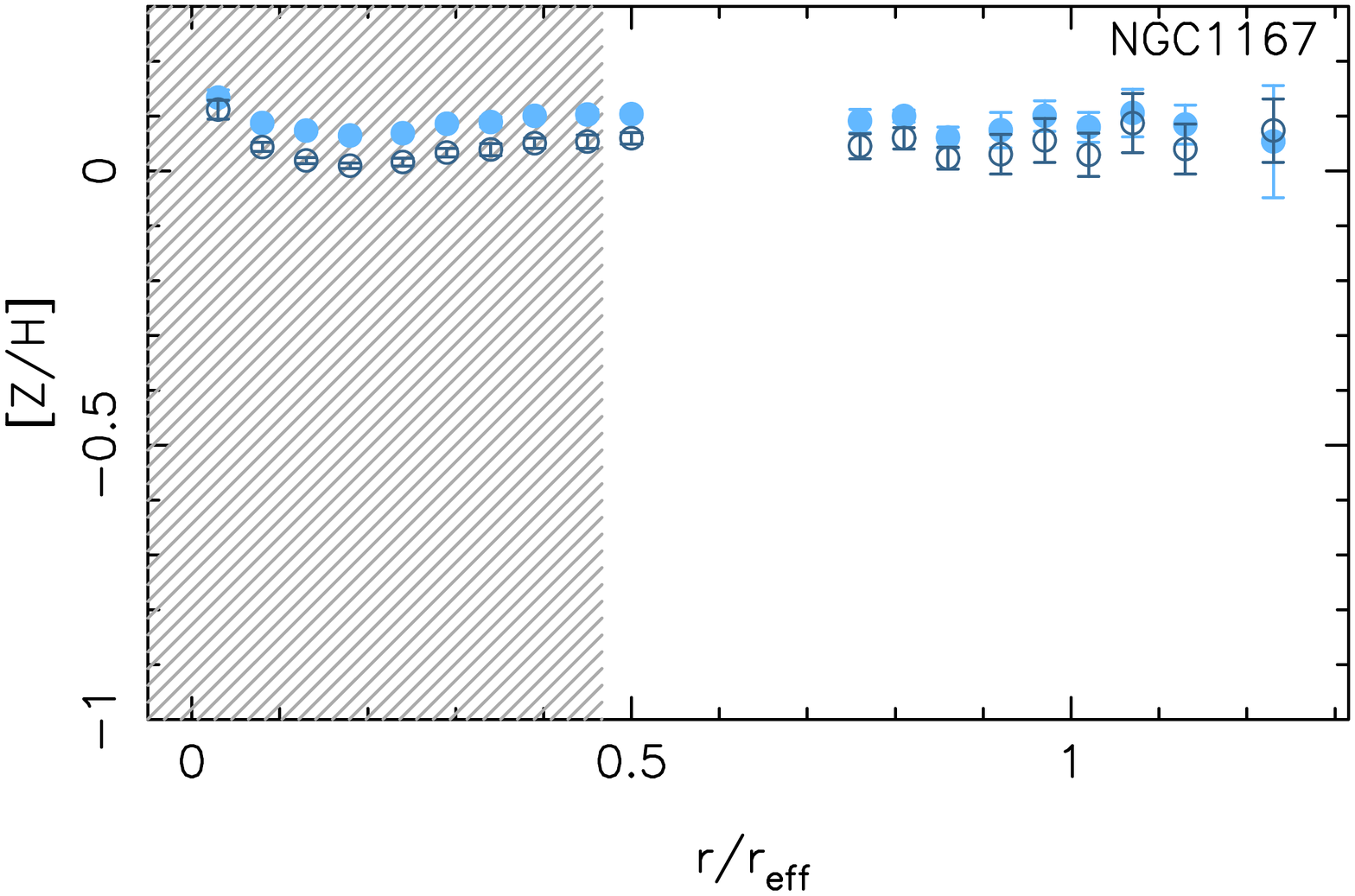}}
\resizebox{0.22\textwidth}{!}{\includegraphics[angle=0]{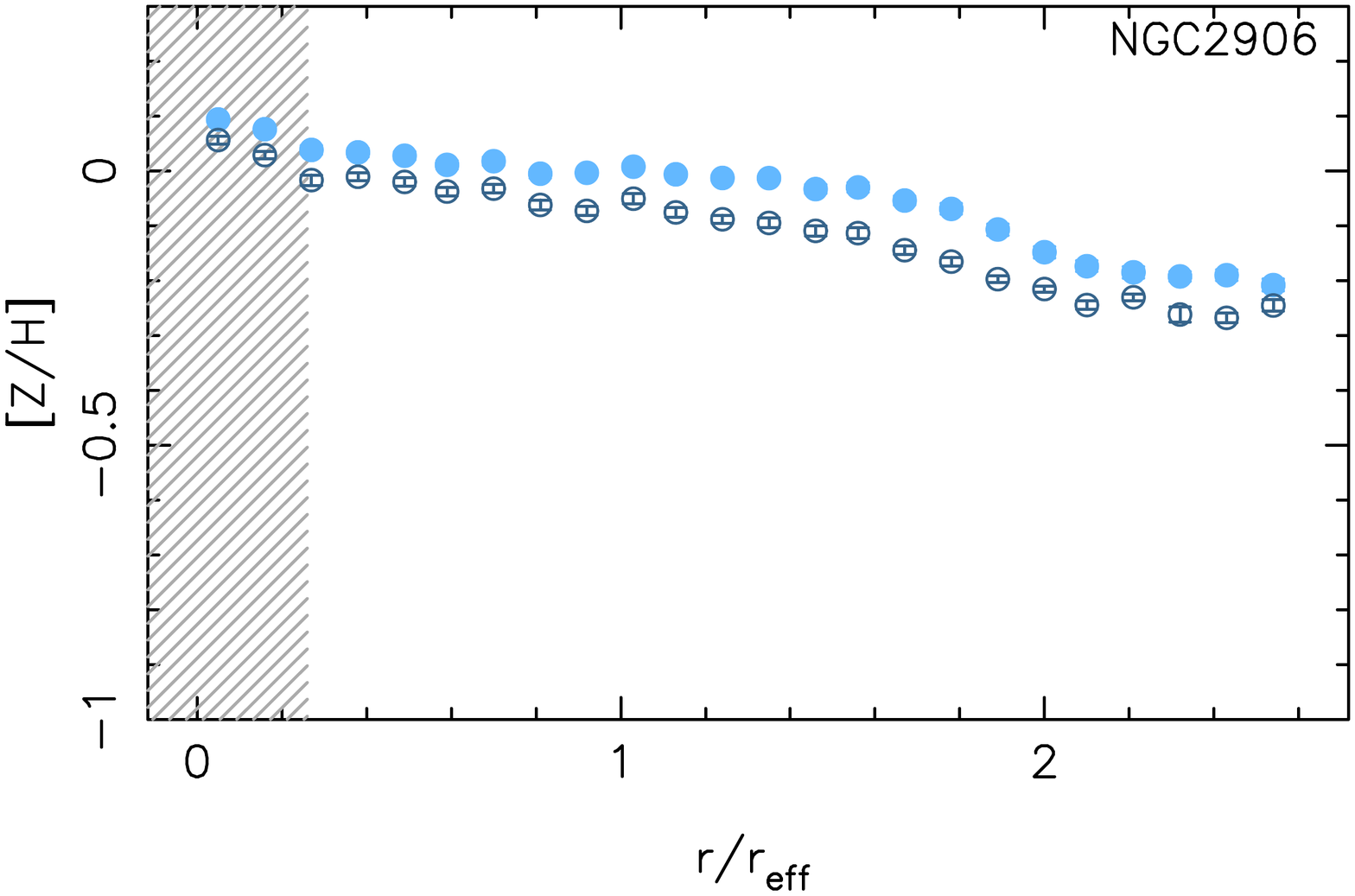}}
\resizebox{0.22\textwidth}{!}{\includegraphics[angle=0]{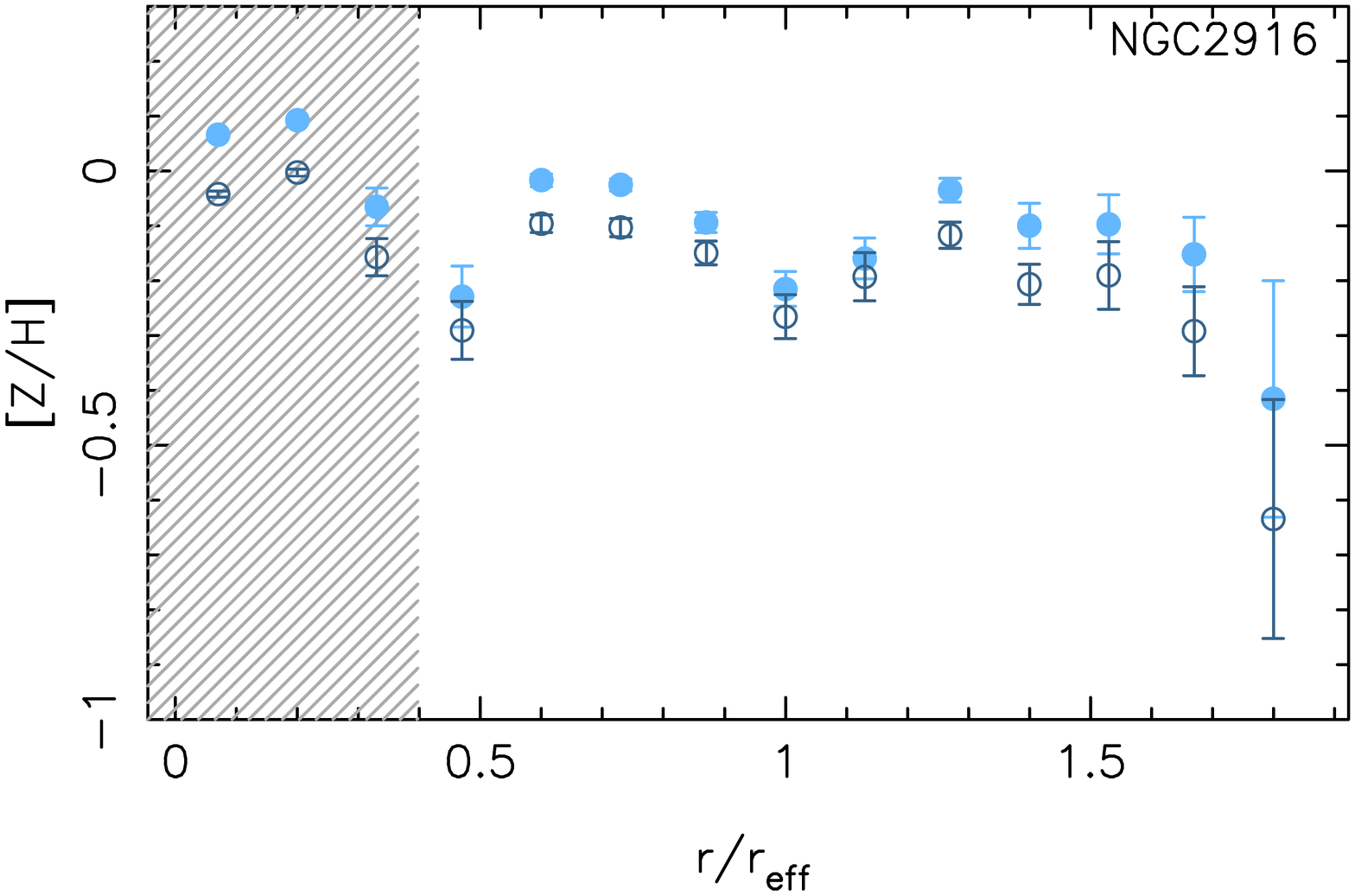}}
\resizebox{0.22\textwidth}{!}{\includegraphics[angle=0]{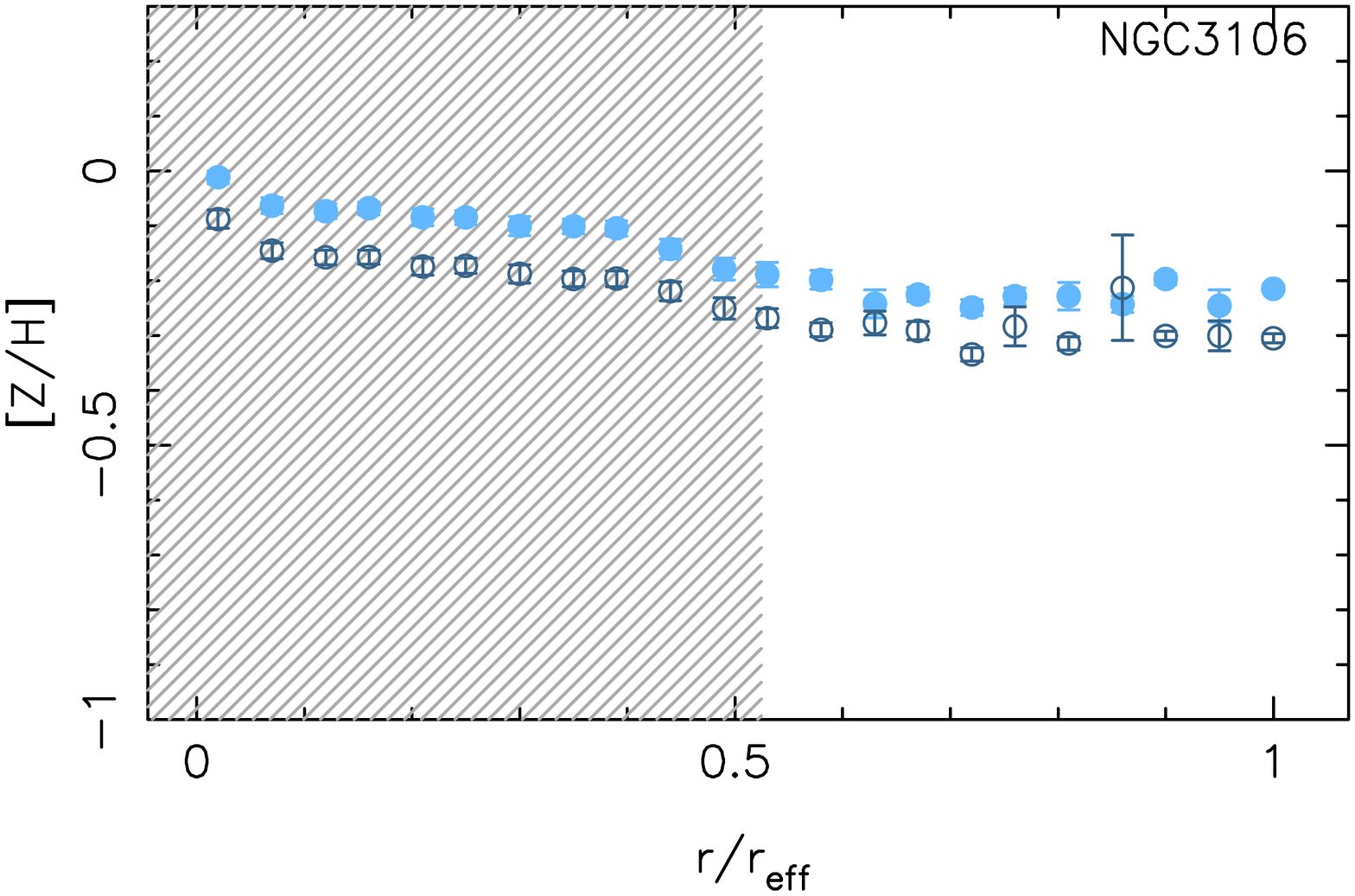}}
\resizebox{0.22\textwidth}{!}{\includegraphics[angle=0]{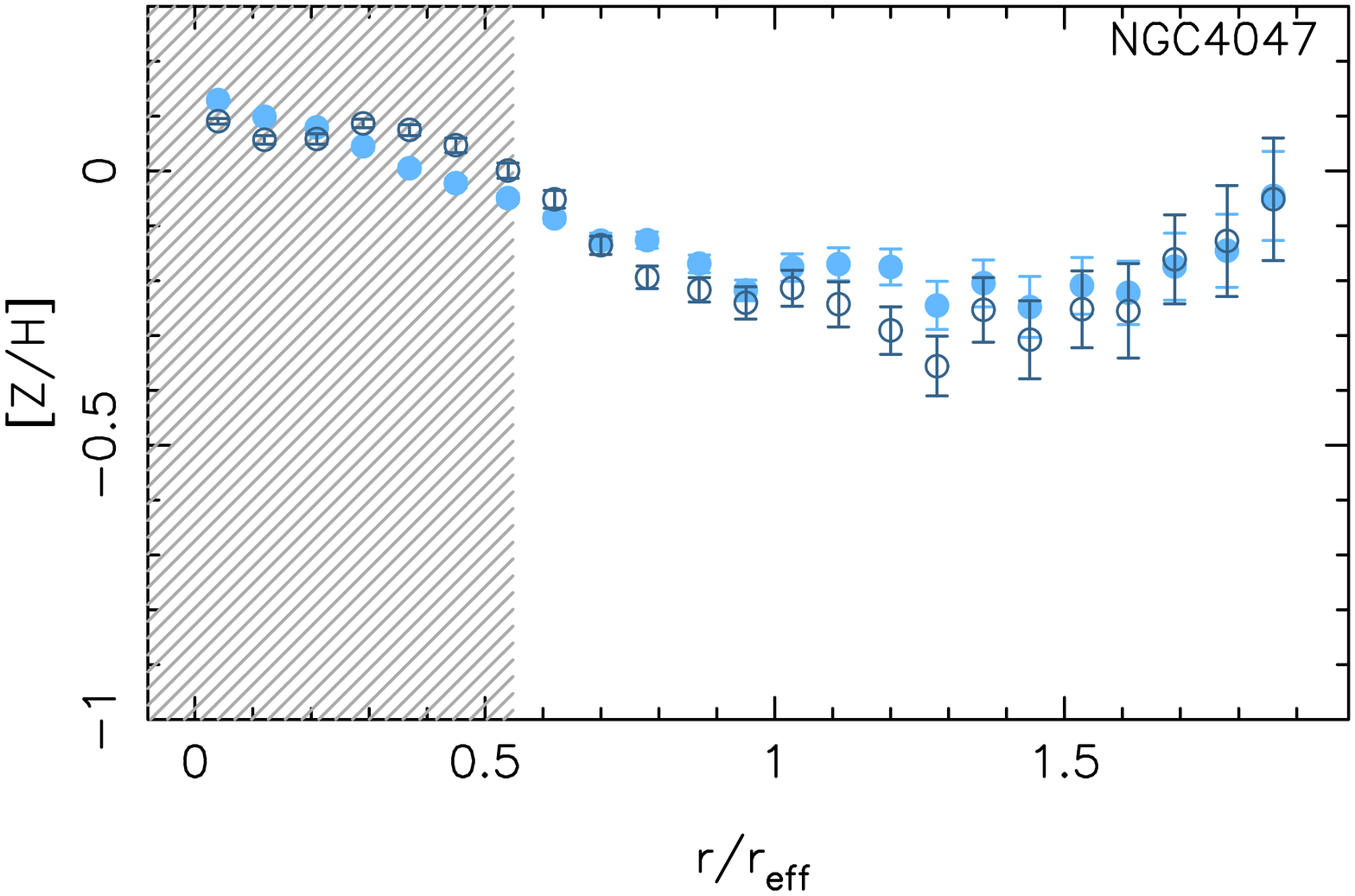}}
\resizebox{0.22\textwidth}{!}{\includegraphics[angle=0]{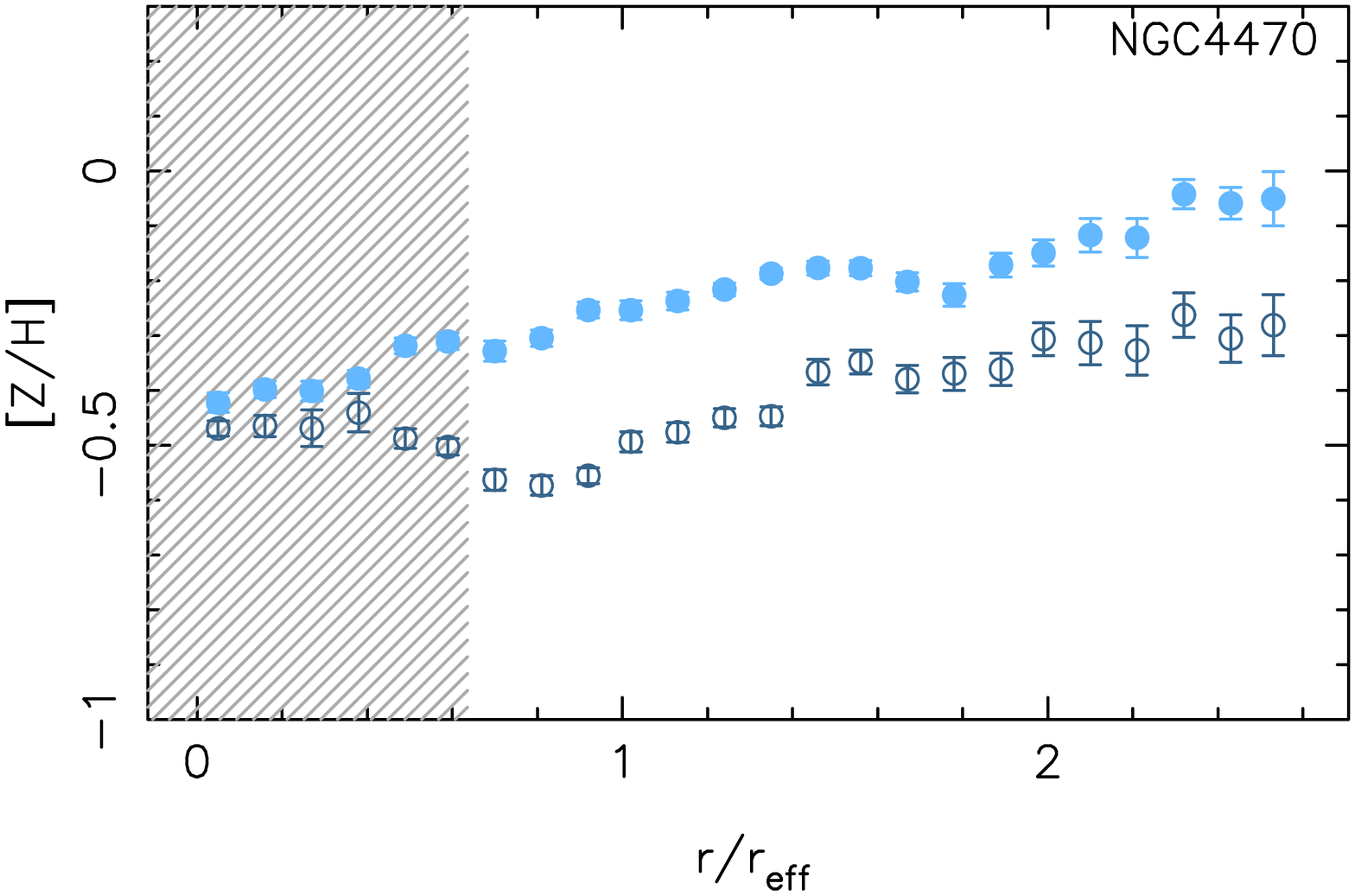}}
\resizebox{0.22\textwidth}{!}{\includegraphics[angle=0]{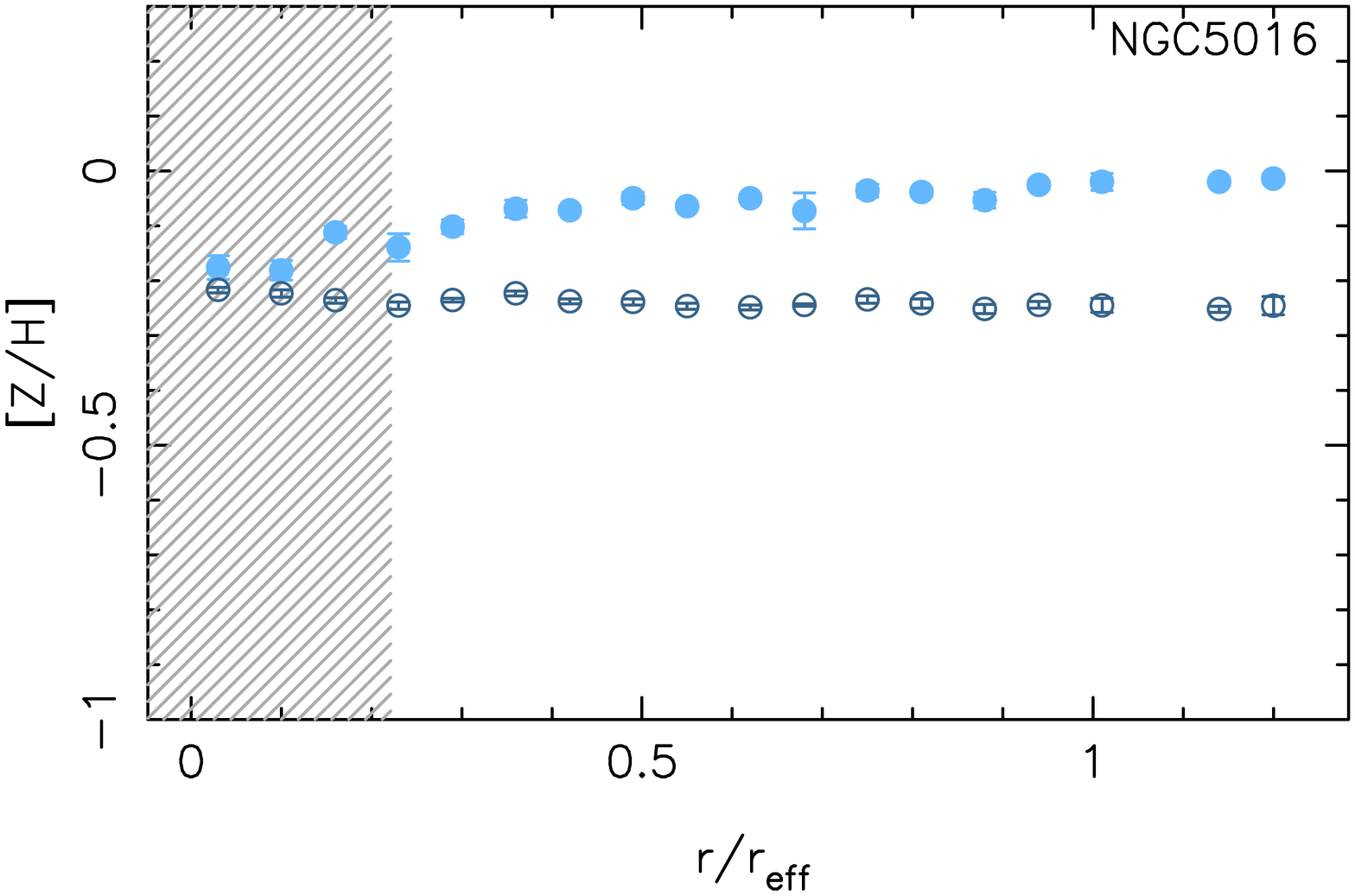}}
\resizebox{0.22\textwidth}{!}{\includegraphics[angle=0]{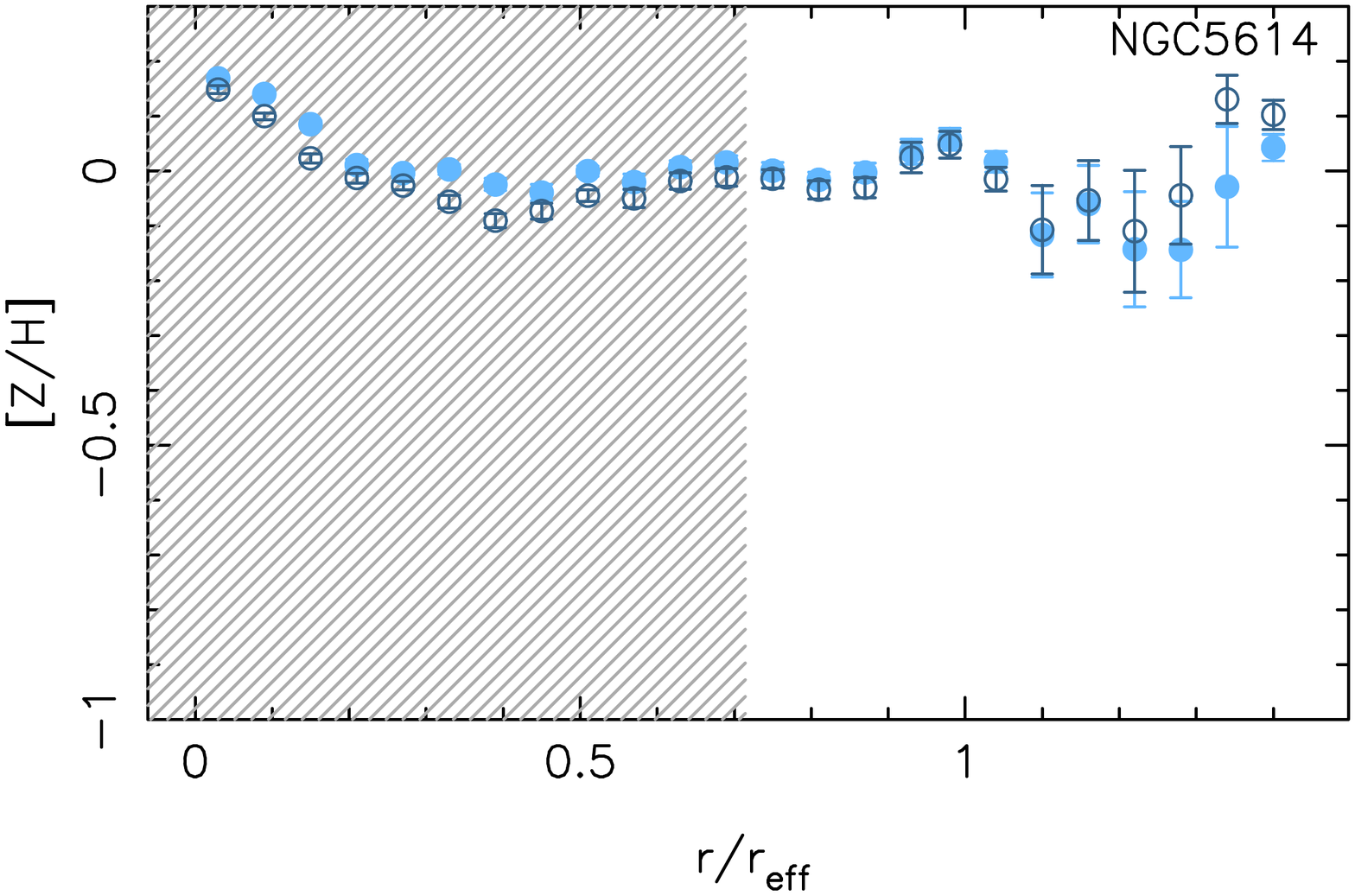}}
\resizebox{0.22\textwidth}{!}{\includegraphics[angle=0]{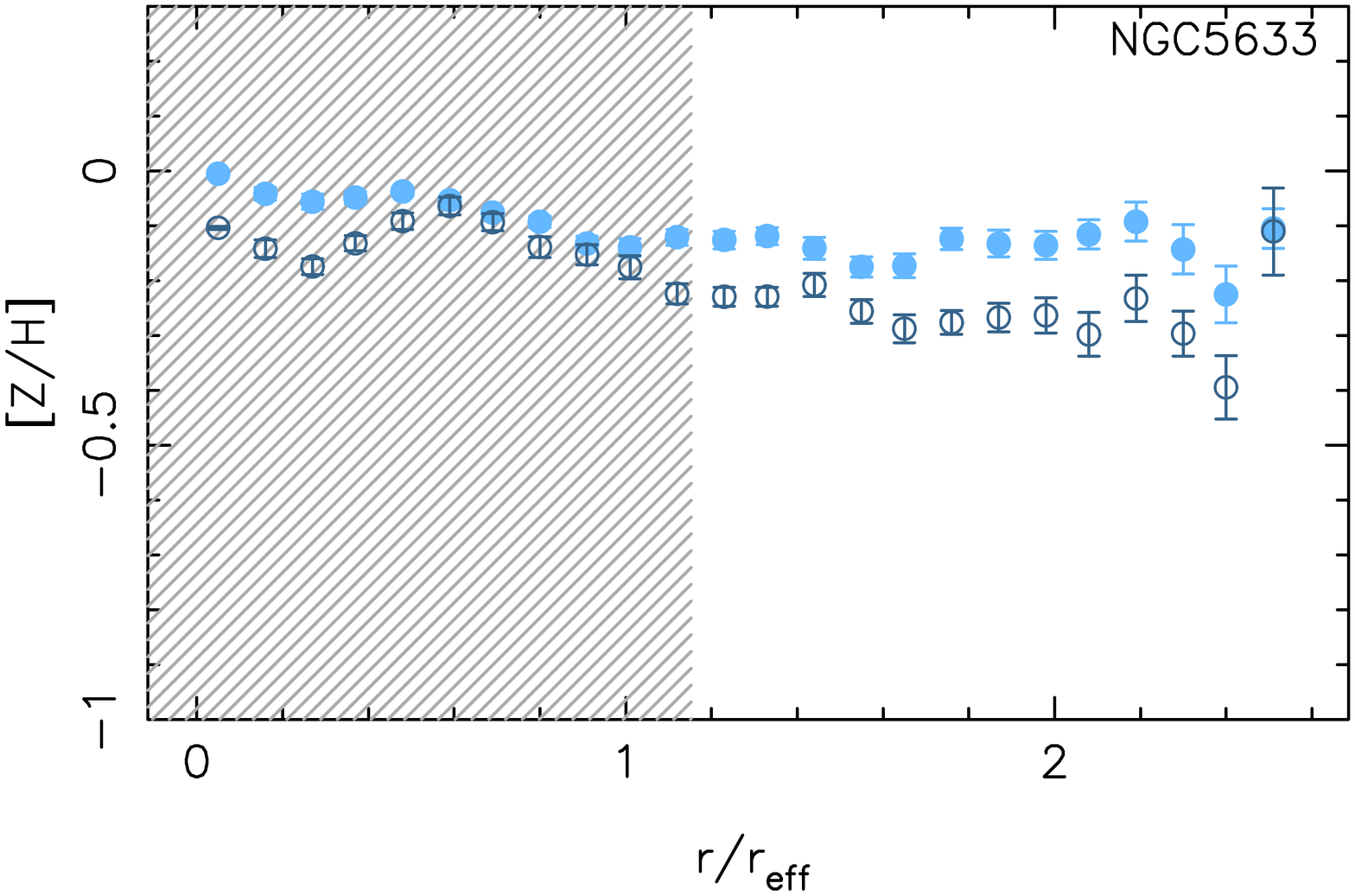}}
\resizebox{0.22\textwidth}{!}{\includegraphics[angle=0]{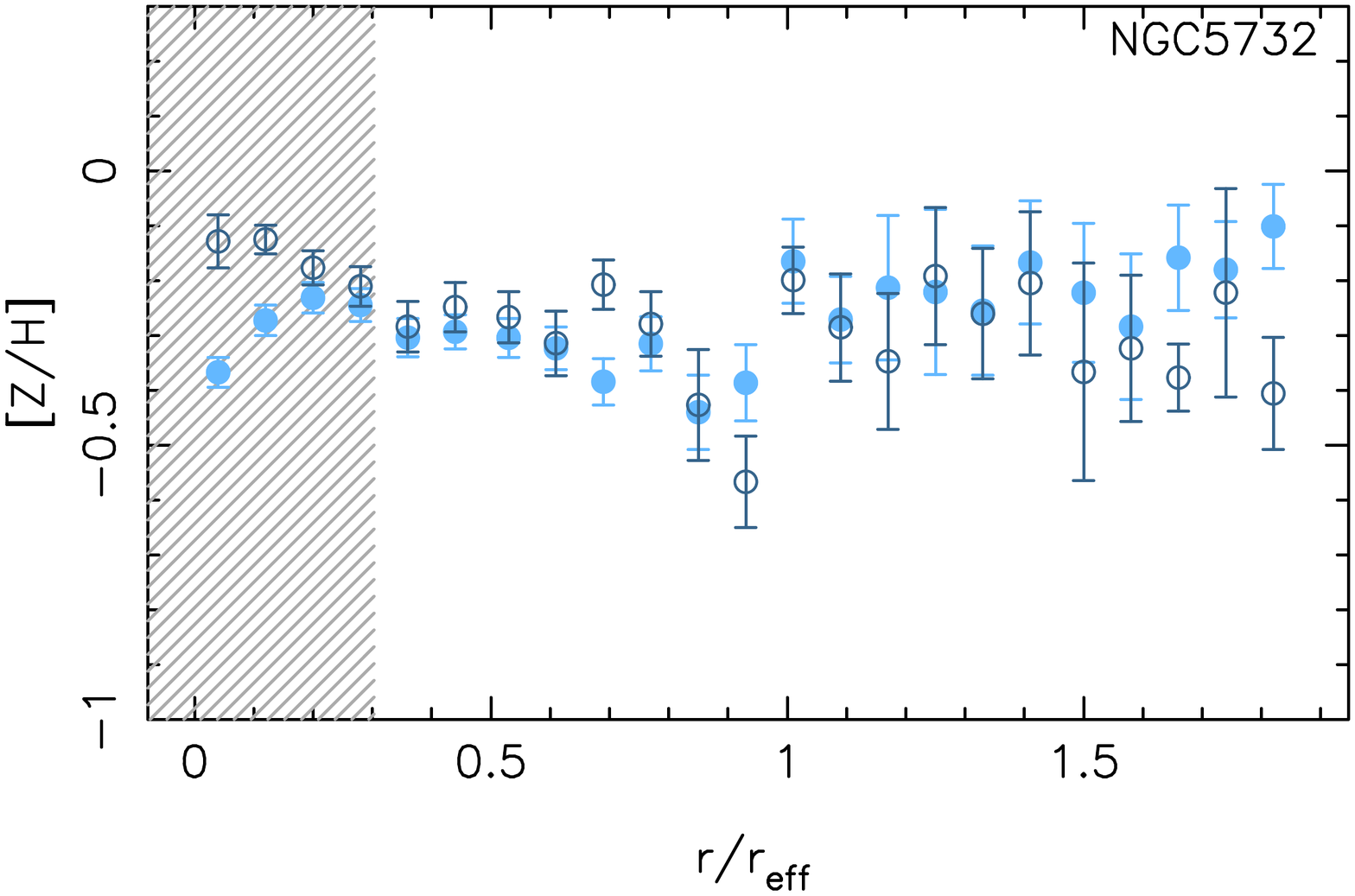}}
\resizebox{0.22\textwidth}{!}{\includegraphics[angle=0]{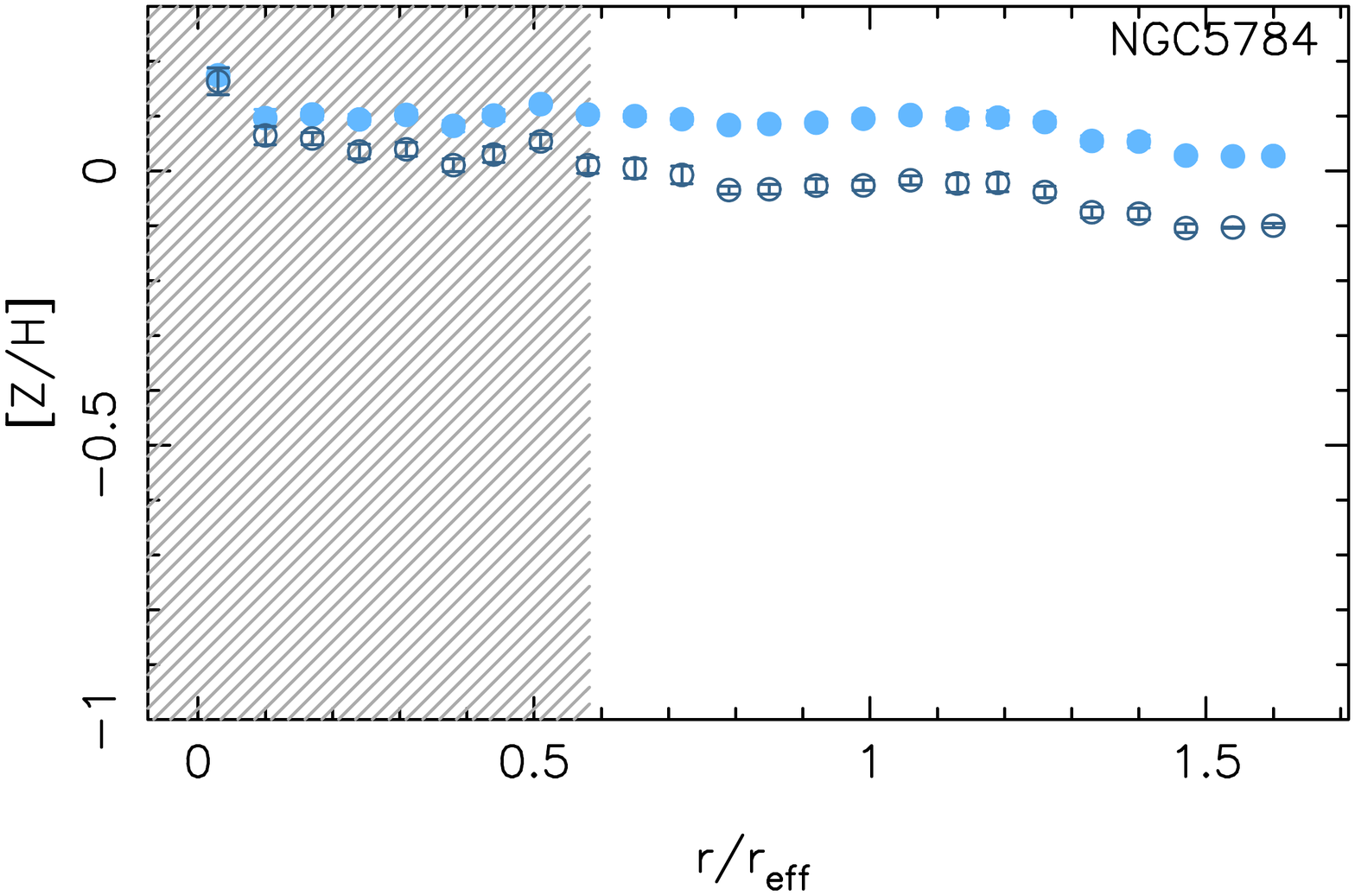}}
\resizebox{0.22\textwidth}{!}{\includegraphics[angle=0]{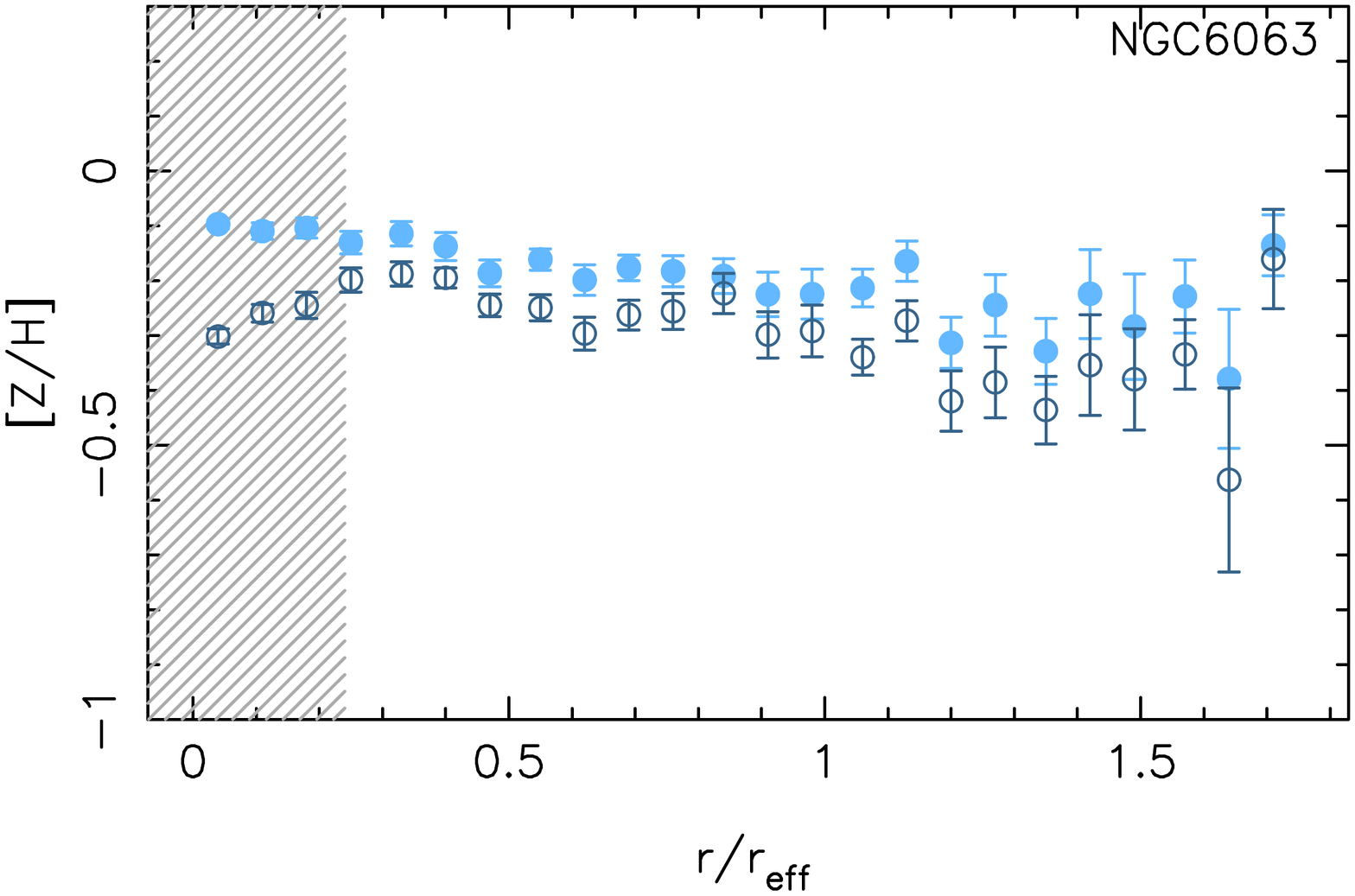}}
\resizebox{0.22\textwidth}{!}{\includegraphics[angle=0]{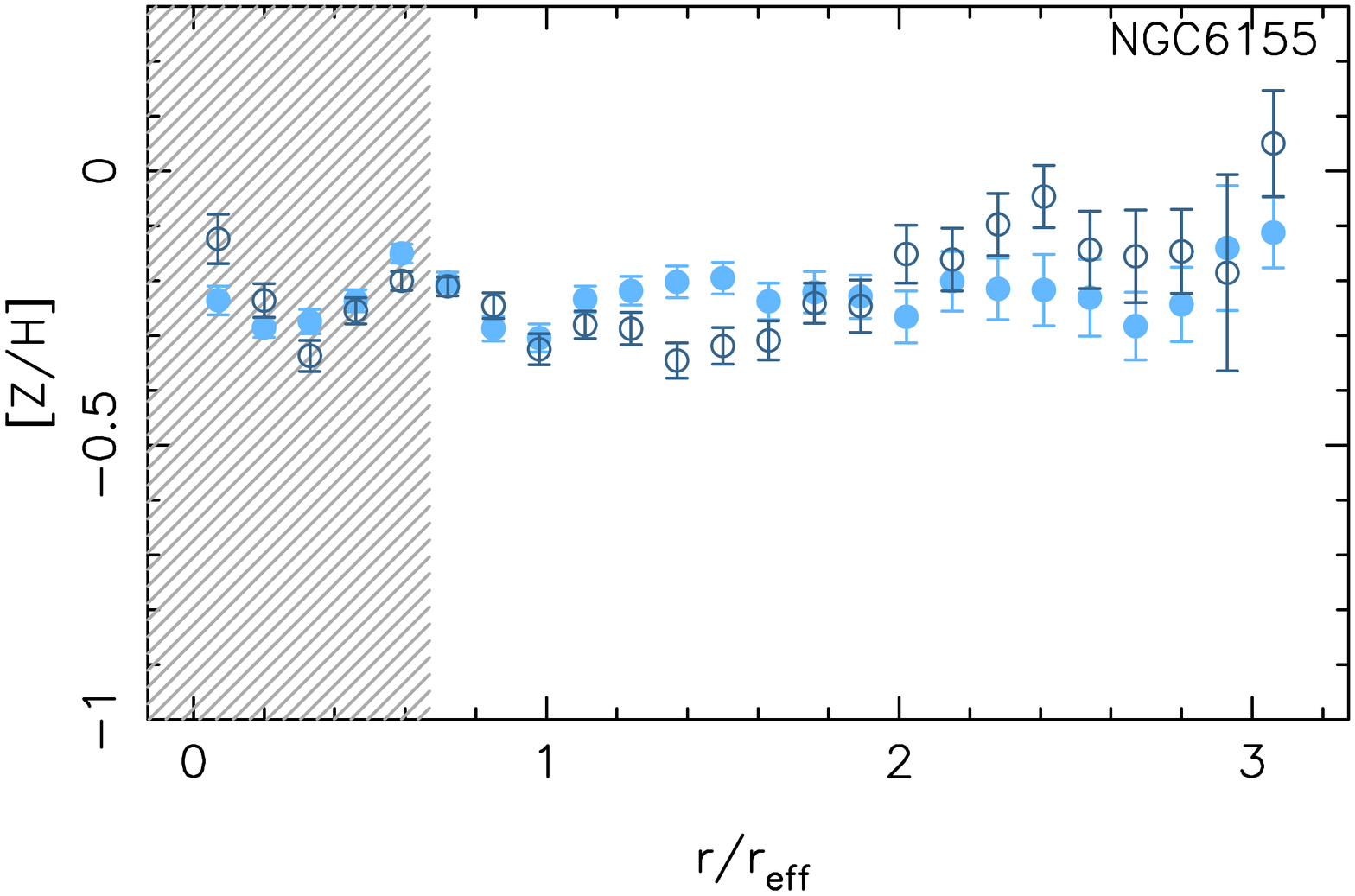}}
\resizebox{0.22\textwidth}{!}{\includegraphics[angle=0]{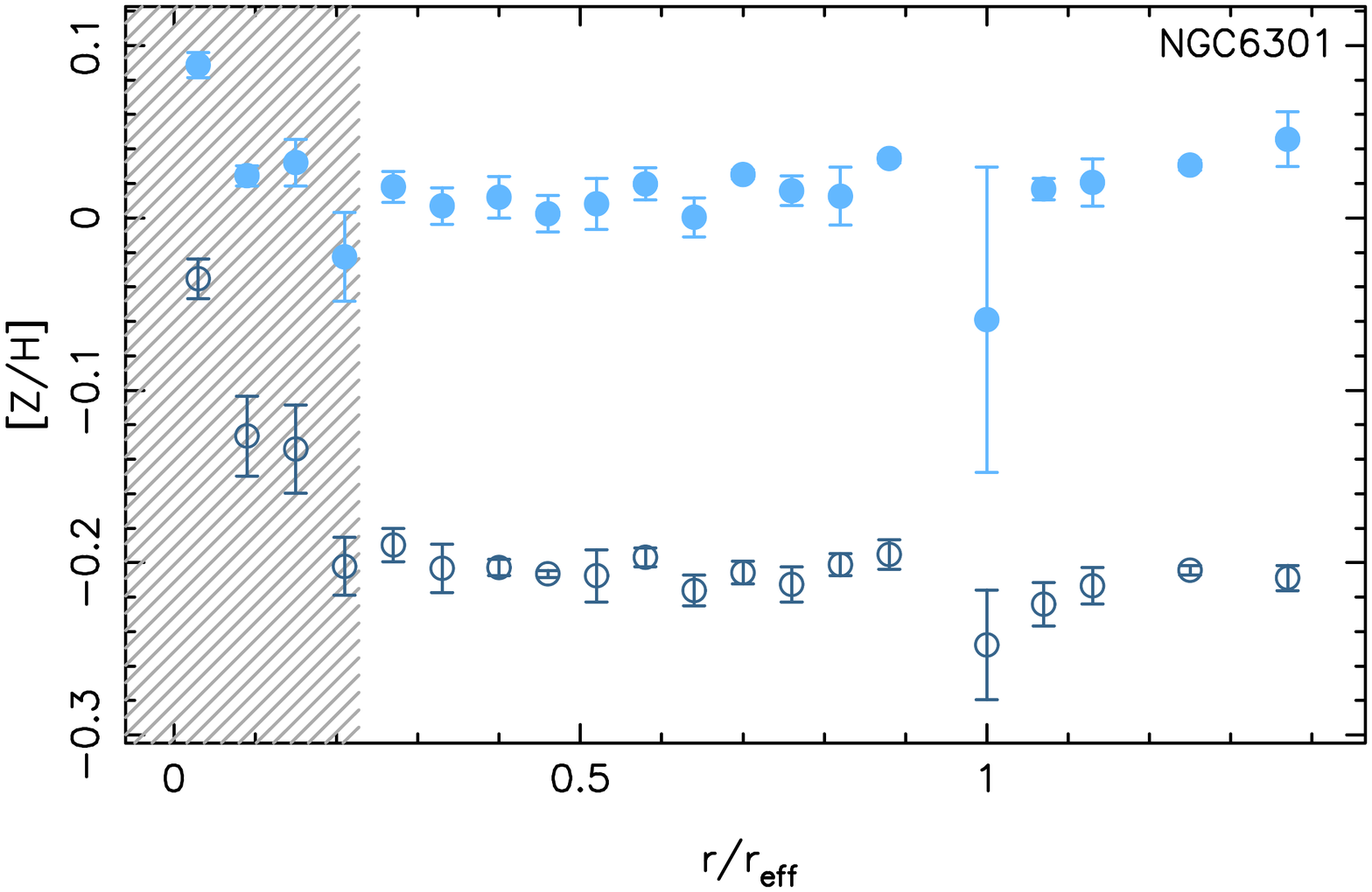}}
\resizebox{0.22\textwidth}{!}{\includegraphics[angle=0]{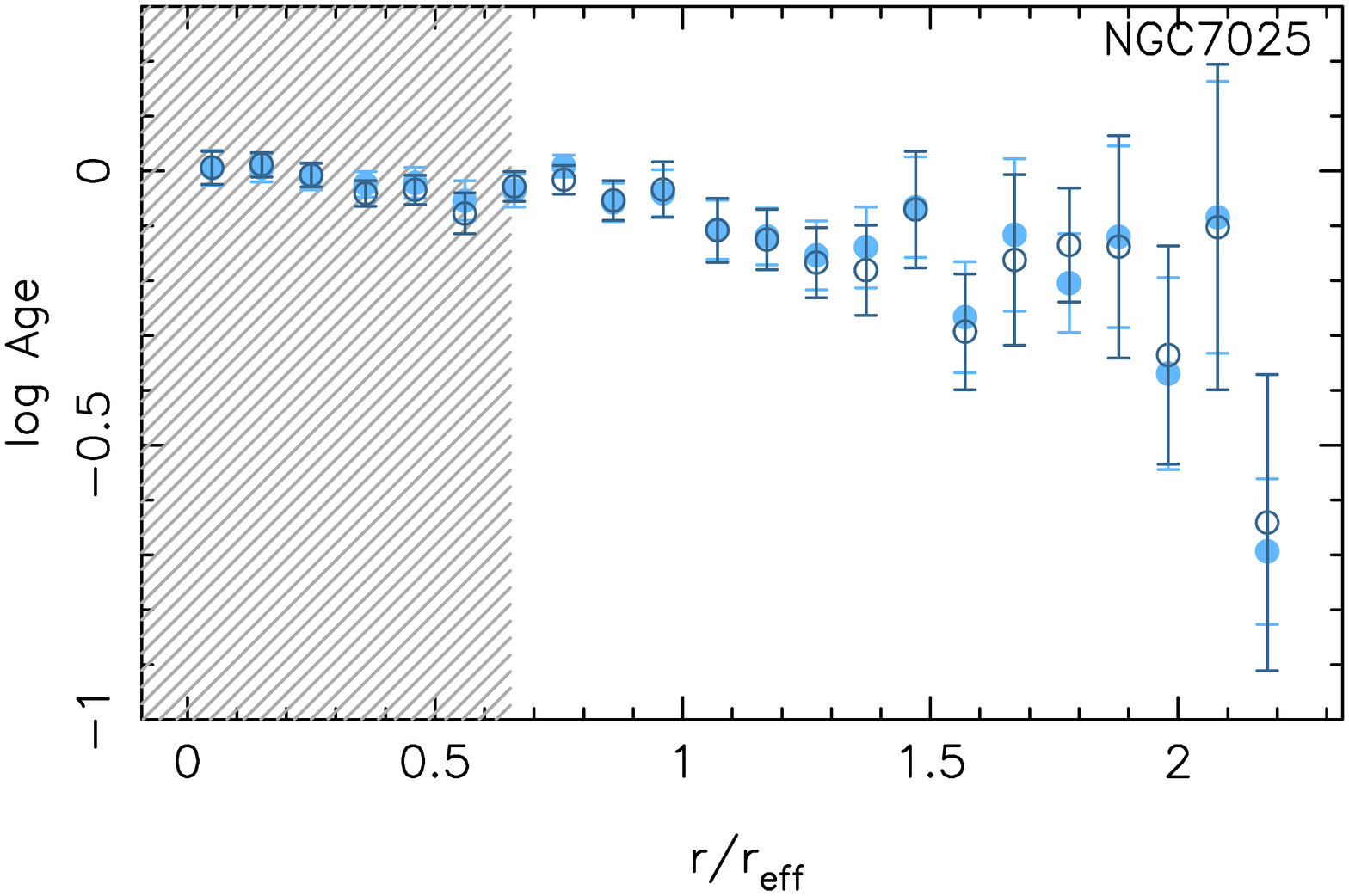}}
\resizebox{0.22\textwidth}{!}{\includegraphics[angle=0]{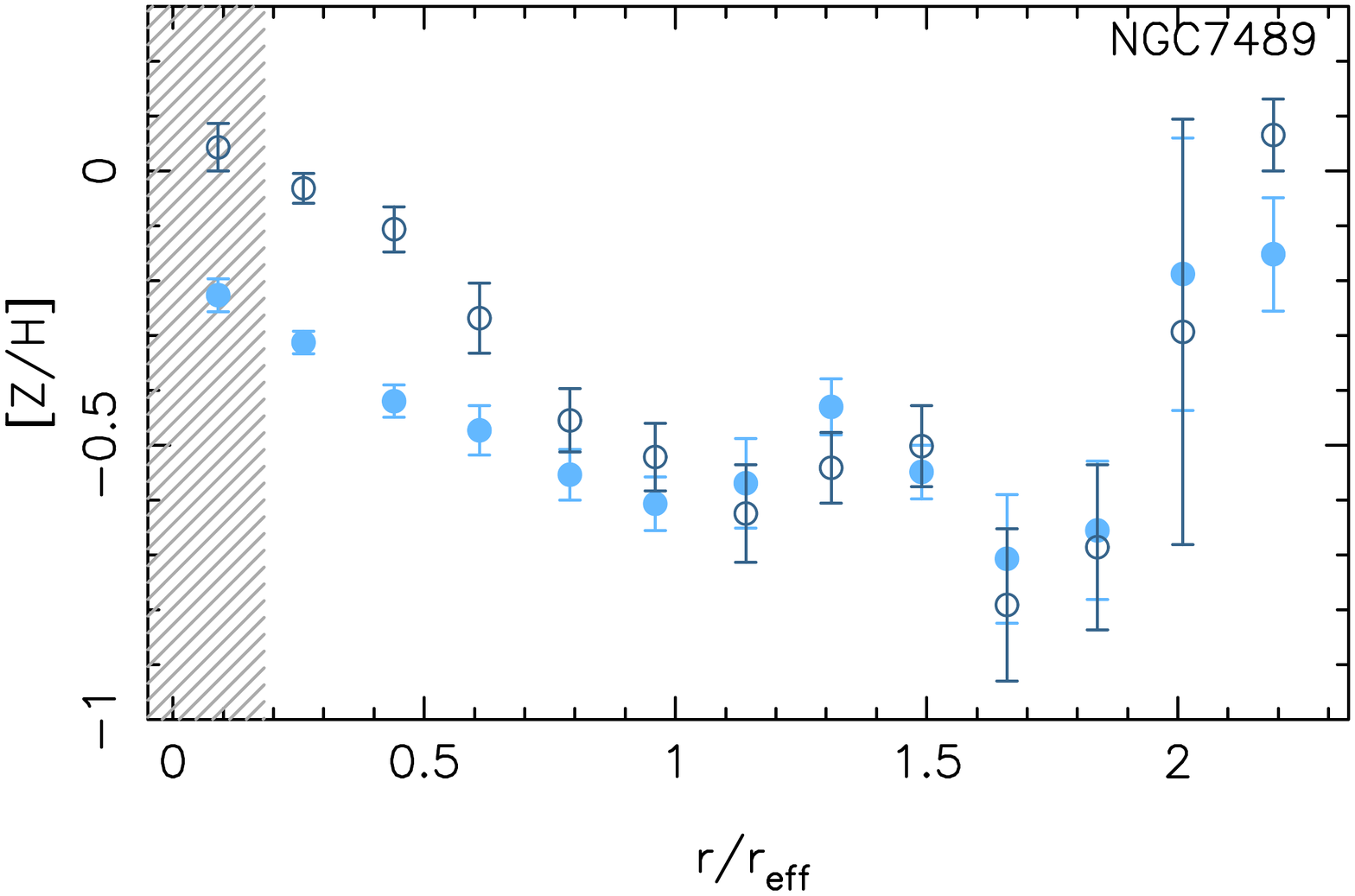}}
\resizebox{0.22\textwidth}{!}{\includegraphics[angle=0]{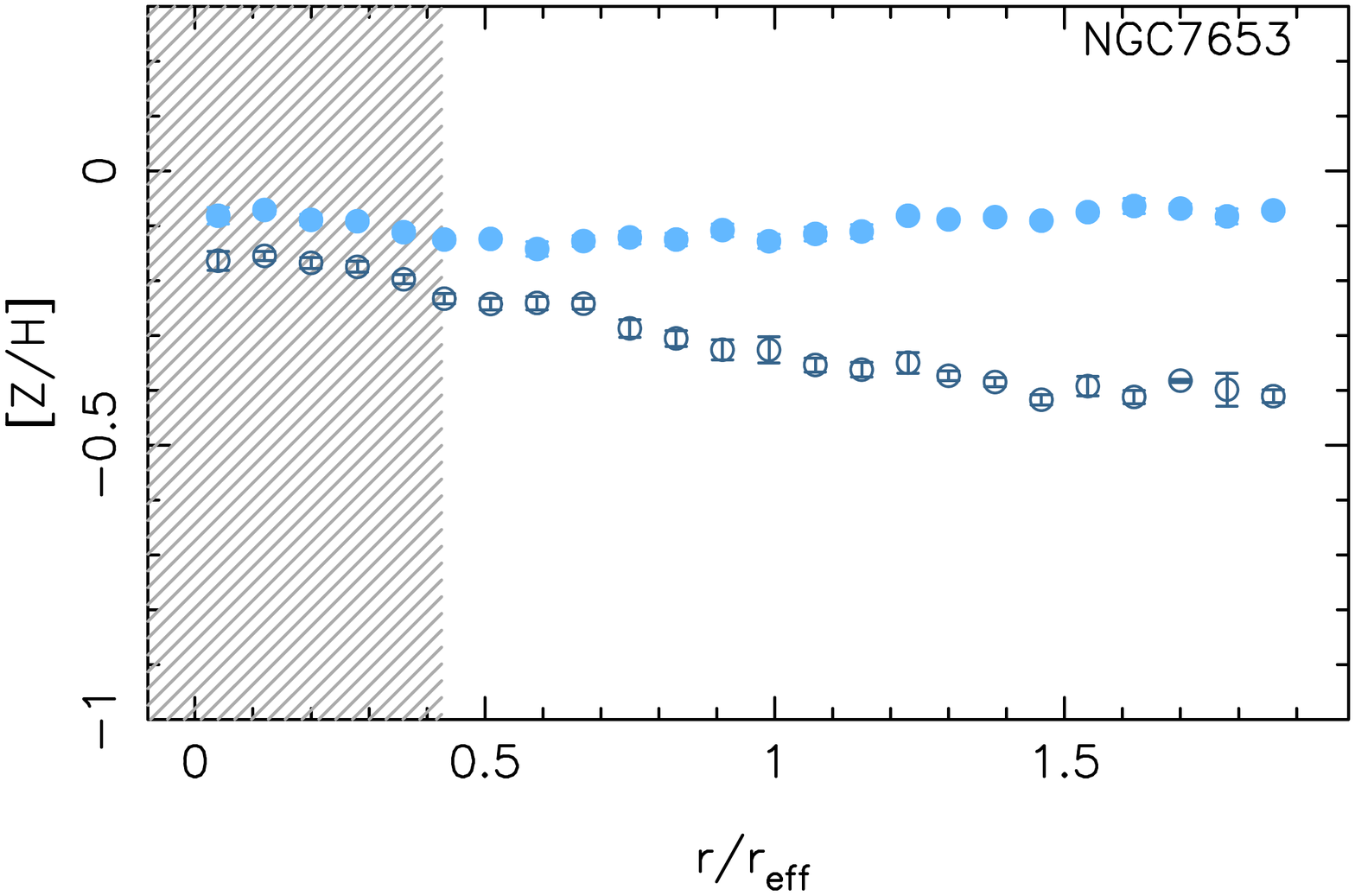}}
\resizebox{0.22\textwidth}{!}{\includegraphics[angle=0]{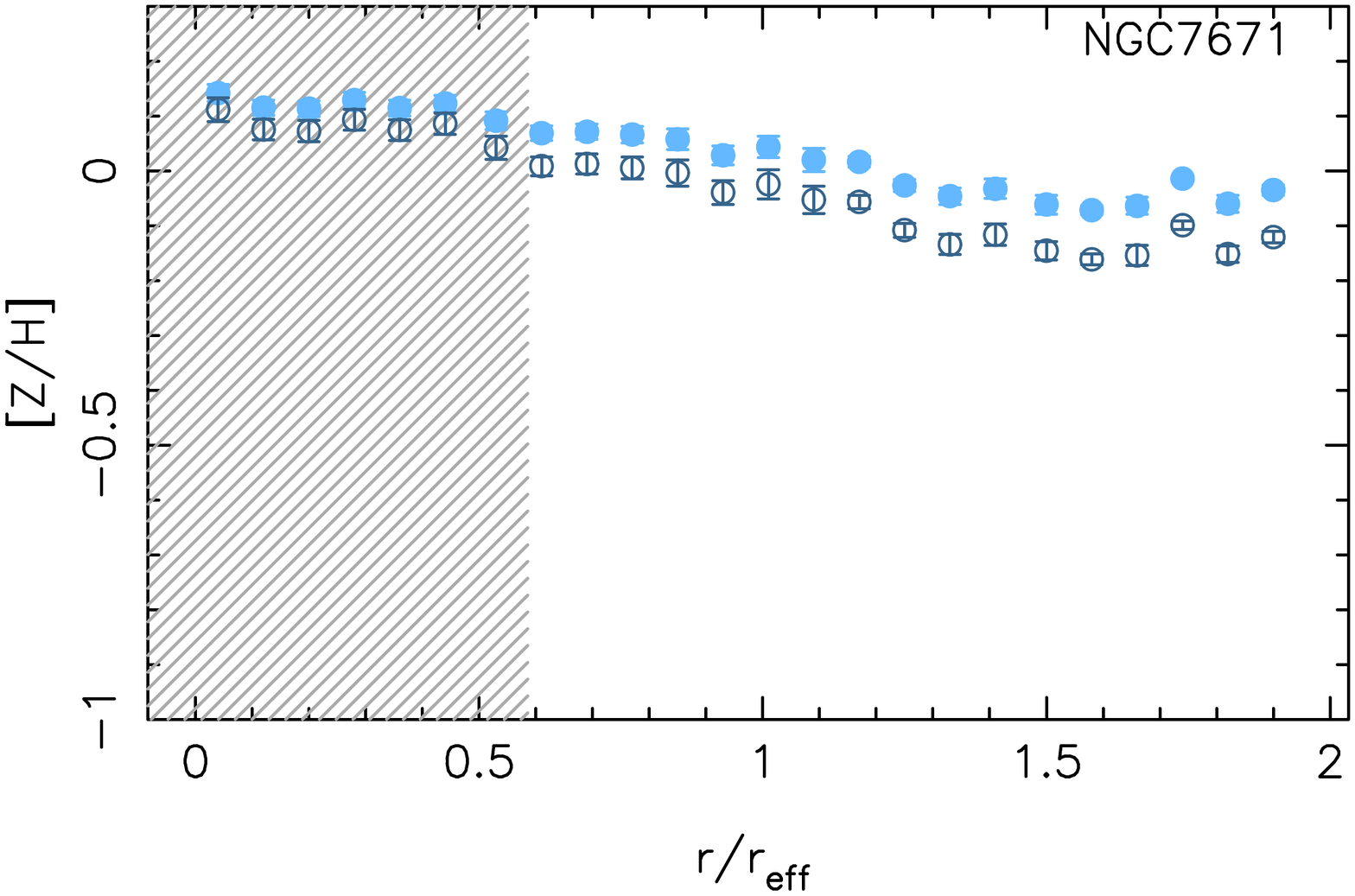}}
\resizebox{0.22\textwidth}{!}{\includegraphics[angle=0]{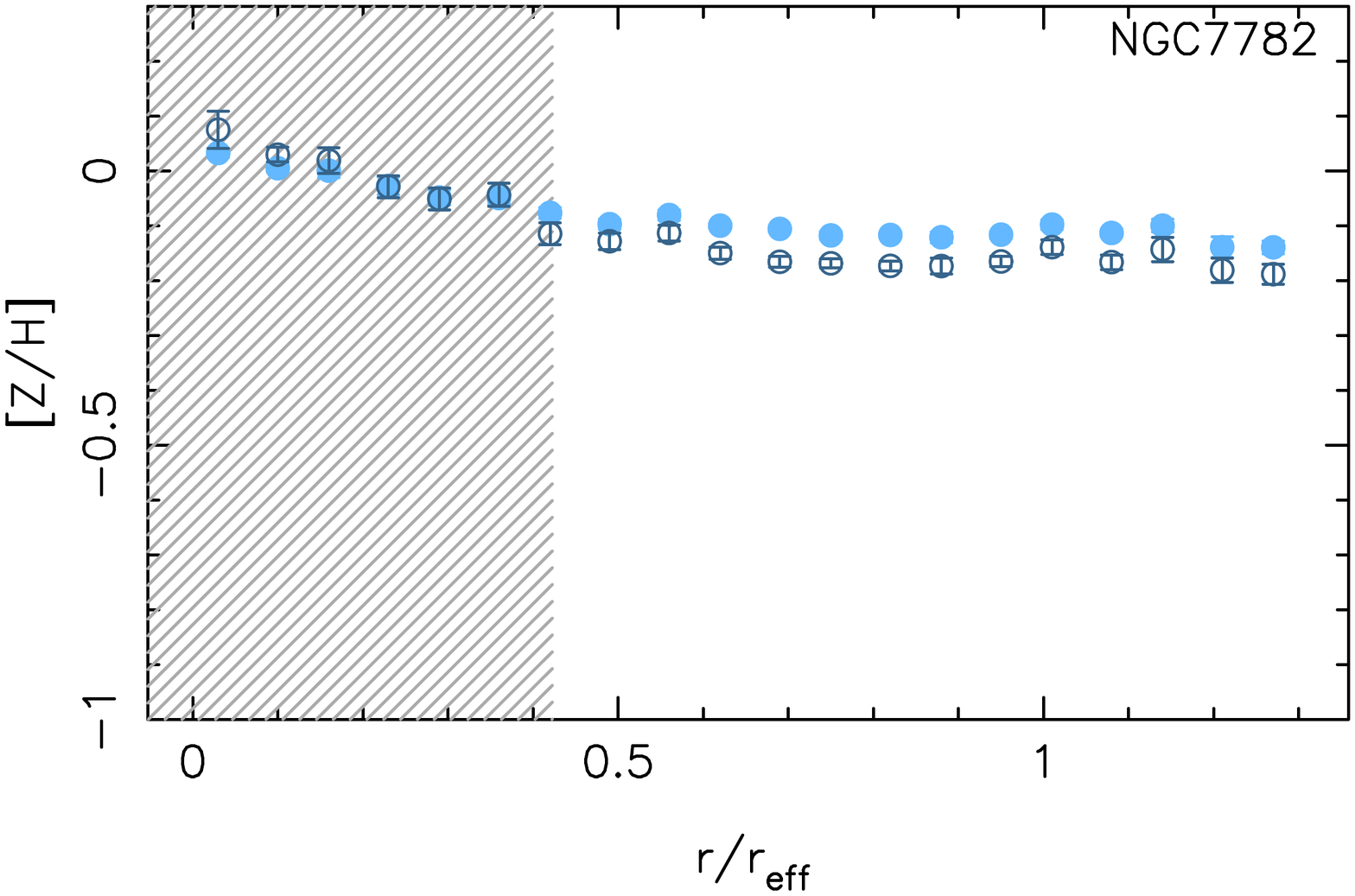}}
\resizebox{0.22\textwidth}{!}{\includegraphics[angle=0]{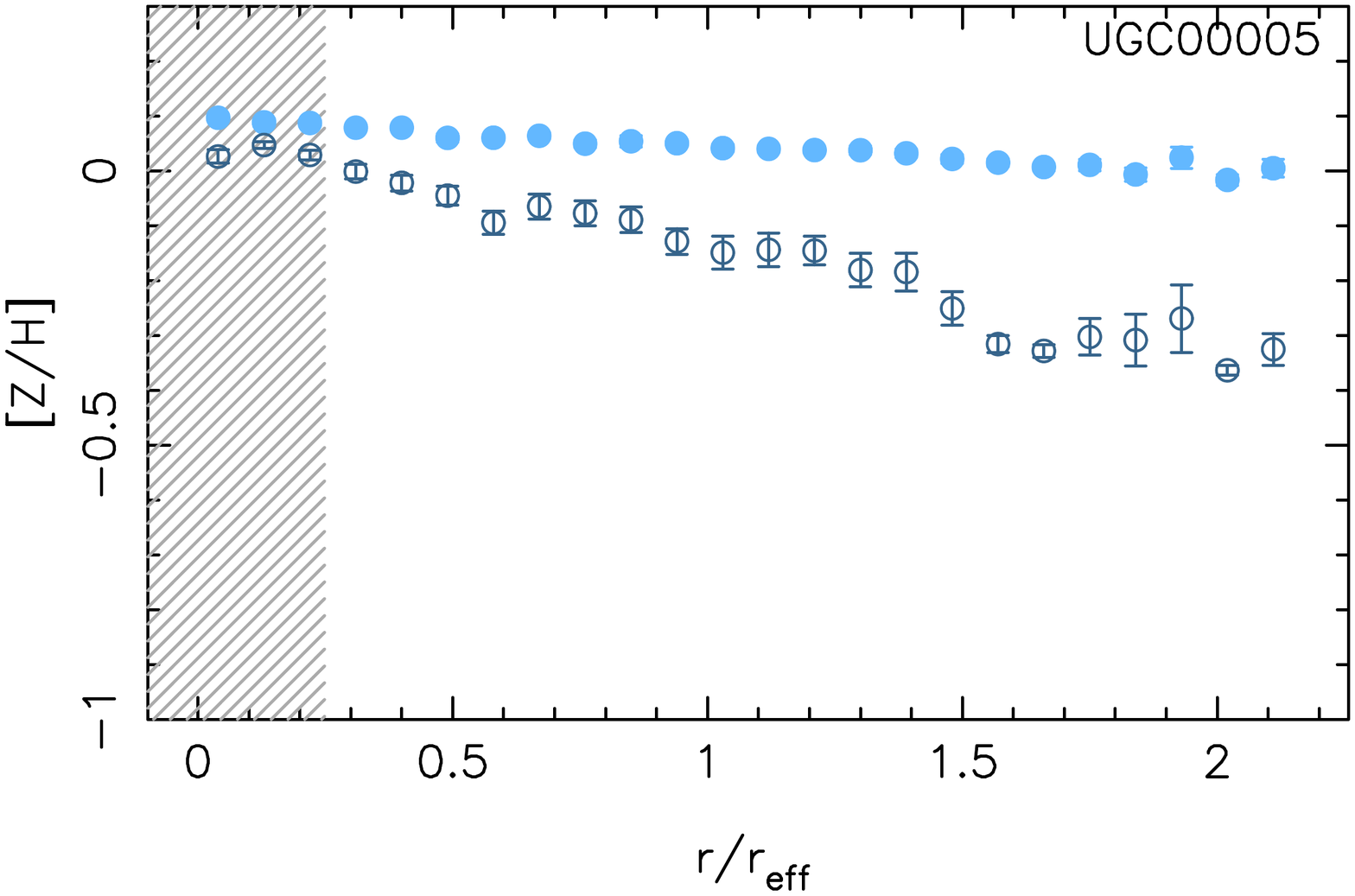}}
\resizebox{0.22\textwidth}{!}{\includegraphics[angle=0]{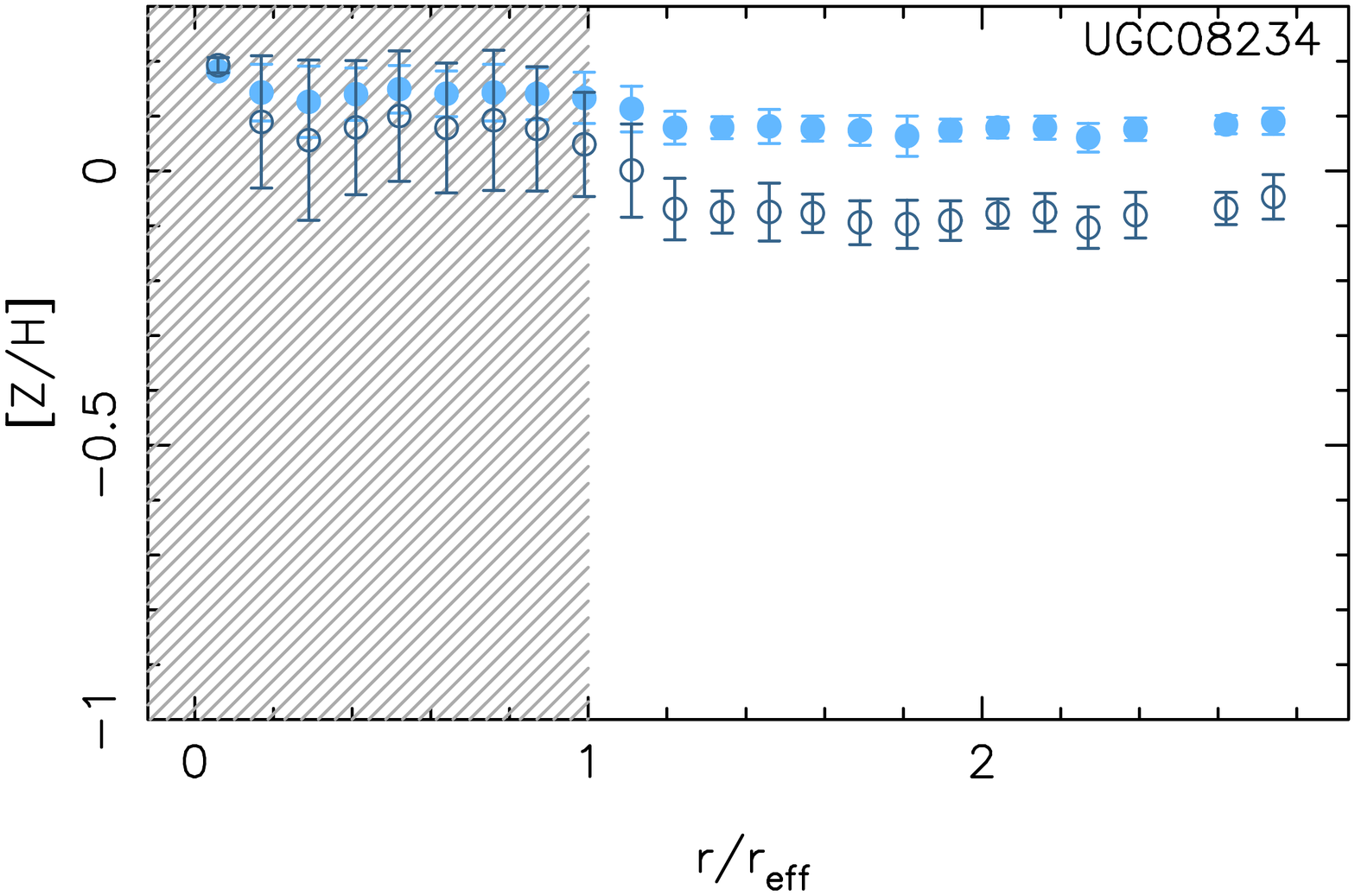}}
\resizebox{0.22\textwidth}{!}{\includegraphics[angle=0]{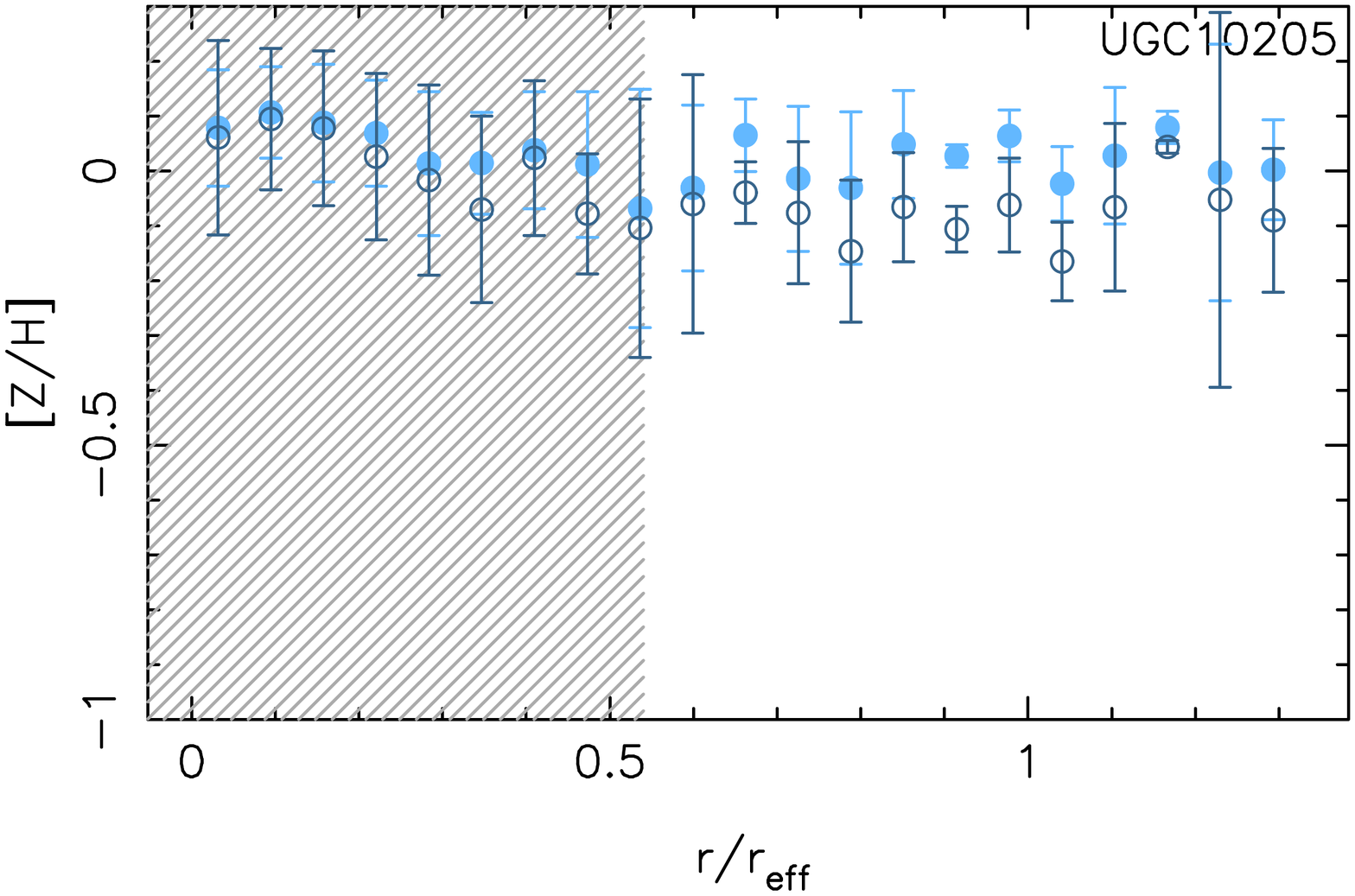}}
\resizebox{0.22\textwidth}{!}{\includegraphics[angle=0]{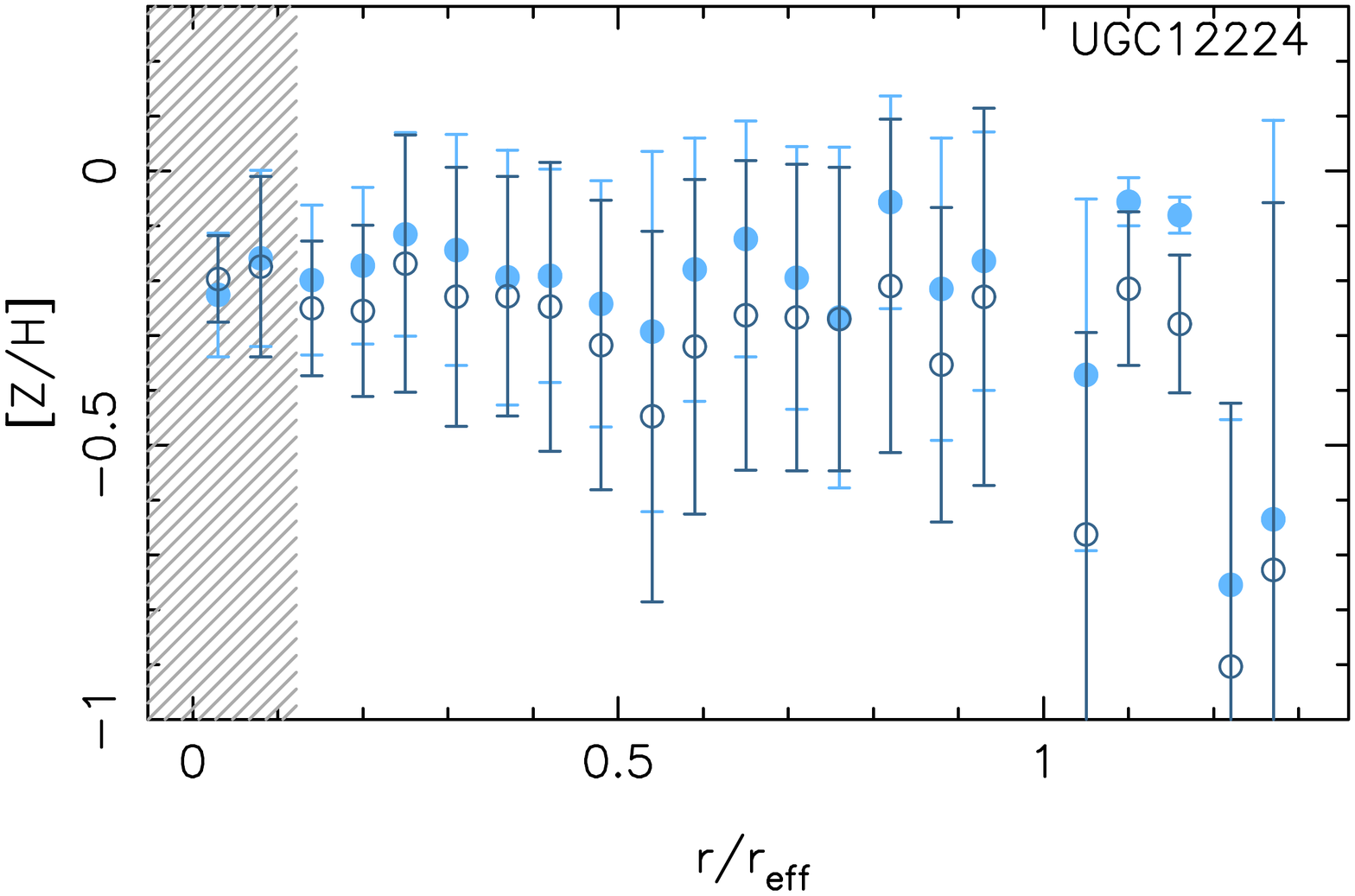}}
\resizebox{0.22\textwidth}{!}{\includegraphics[angle=0]{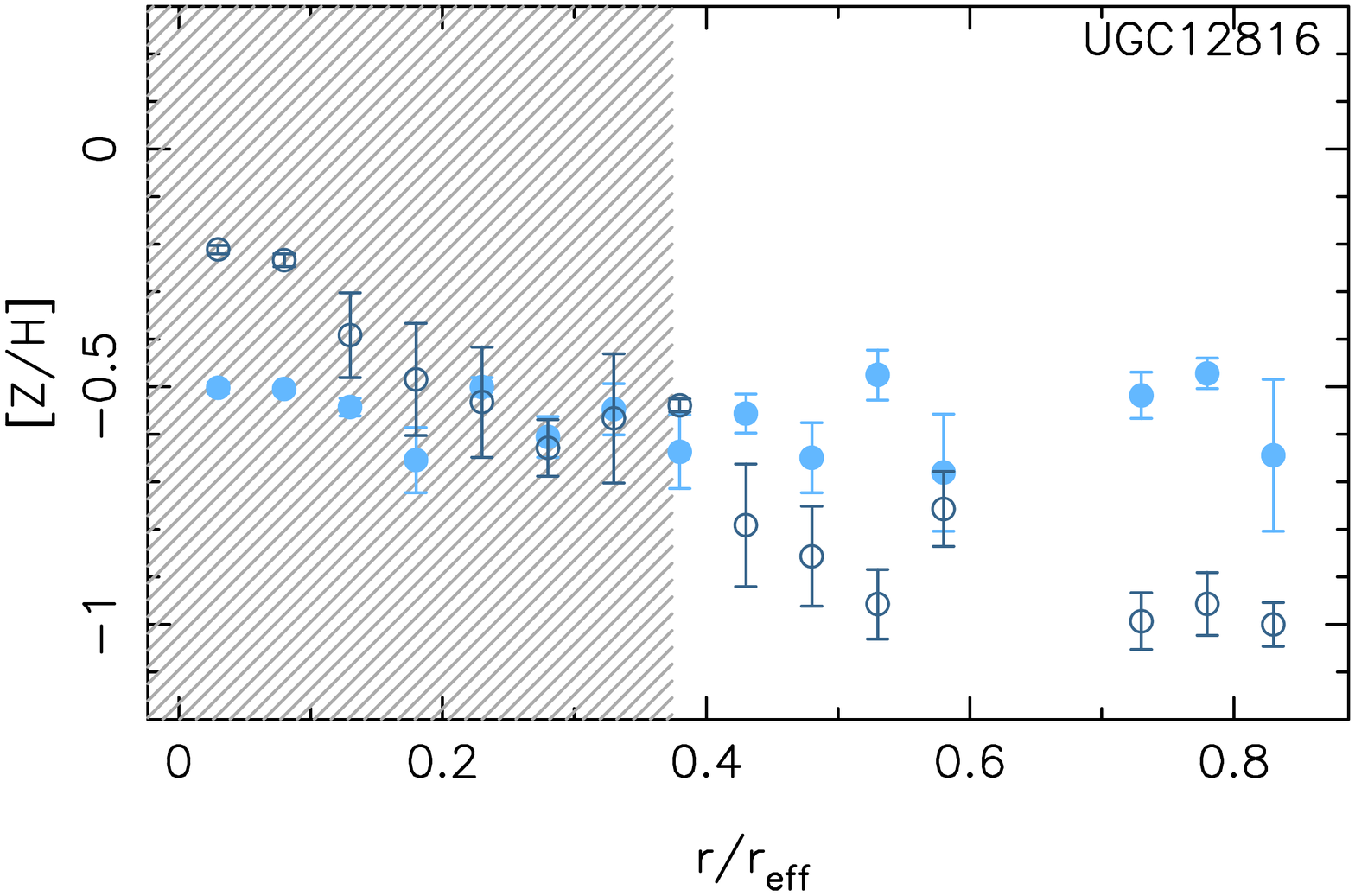}}
\caption{Luminosity- (light red) and Mass- (dark red) metallicity gradients
for our sample of barred galaxies. \label{fig:metgrads2}}
\end{figure*}

\begin{figure*}
\centering
\resizebox{0.22\textwidth}{!}{\includegraphics[angle=0]{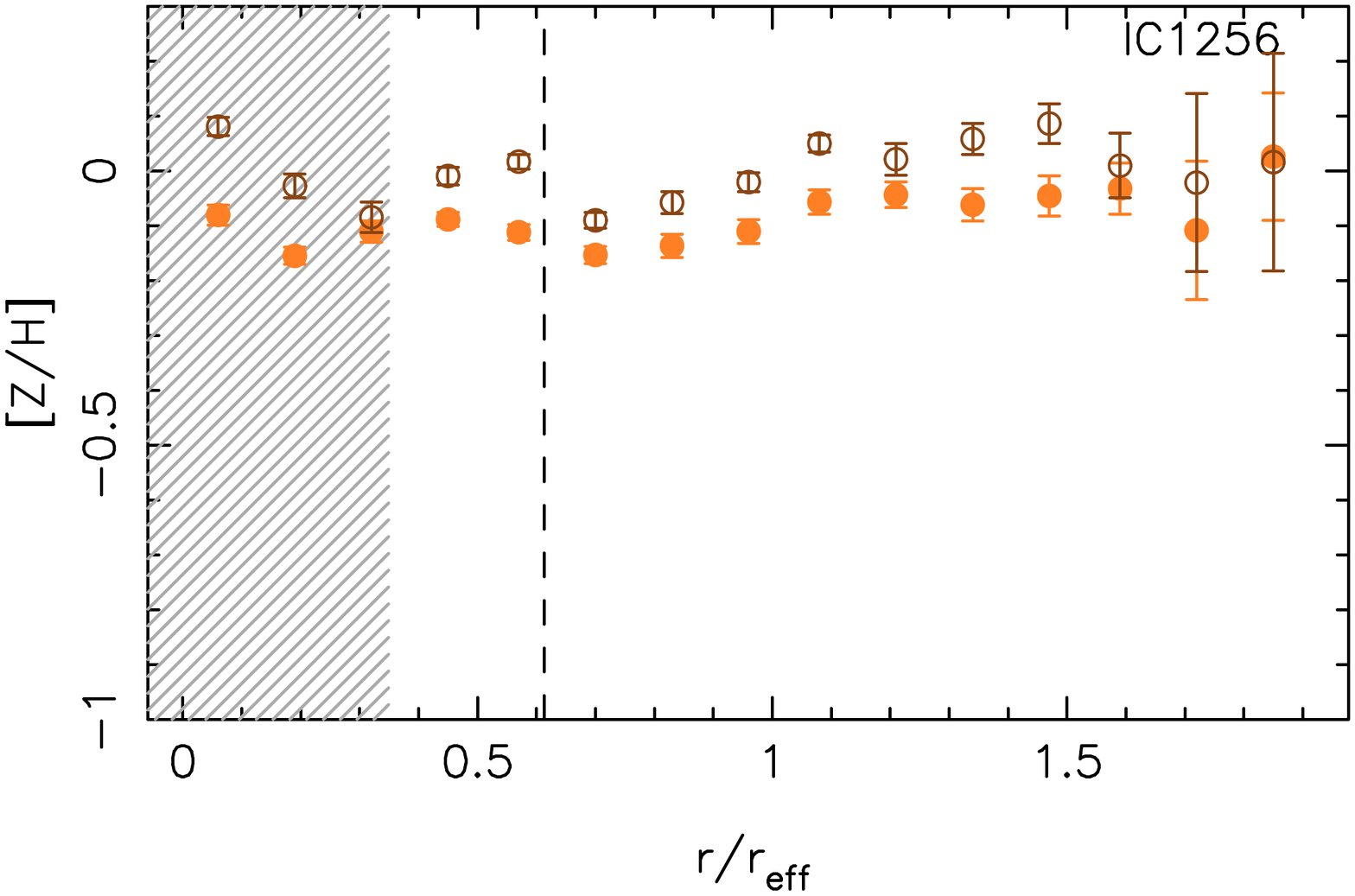}}
\resizebox{0.22\textwidth}{!}{\includegraphics[angle=0]{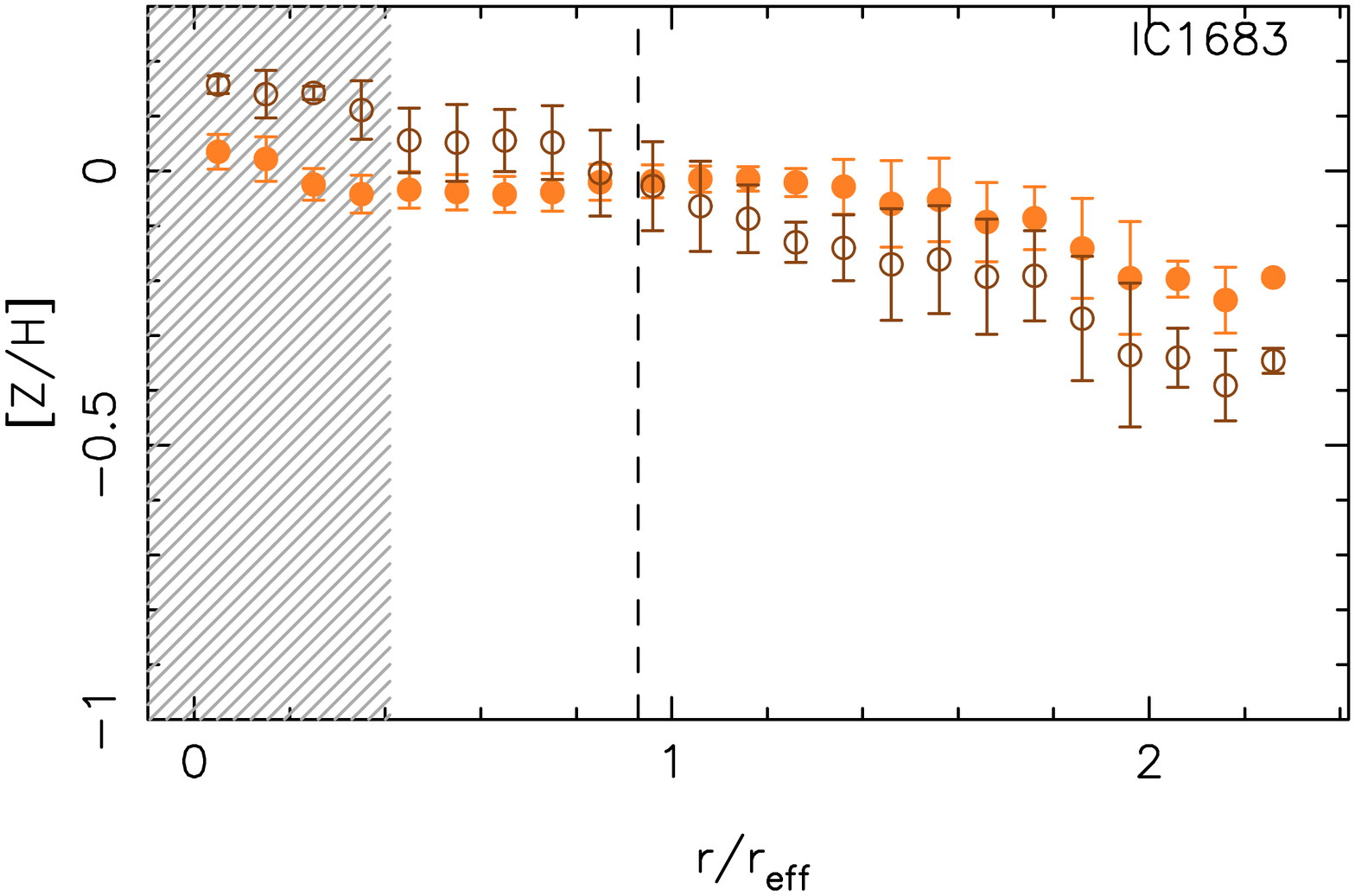}}
\resizebox{0.22\textwidth}{!}{\includegraphics[angle=0]{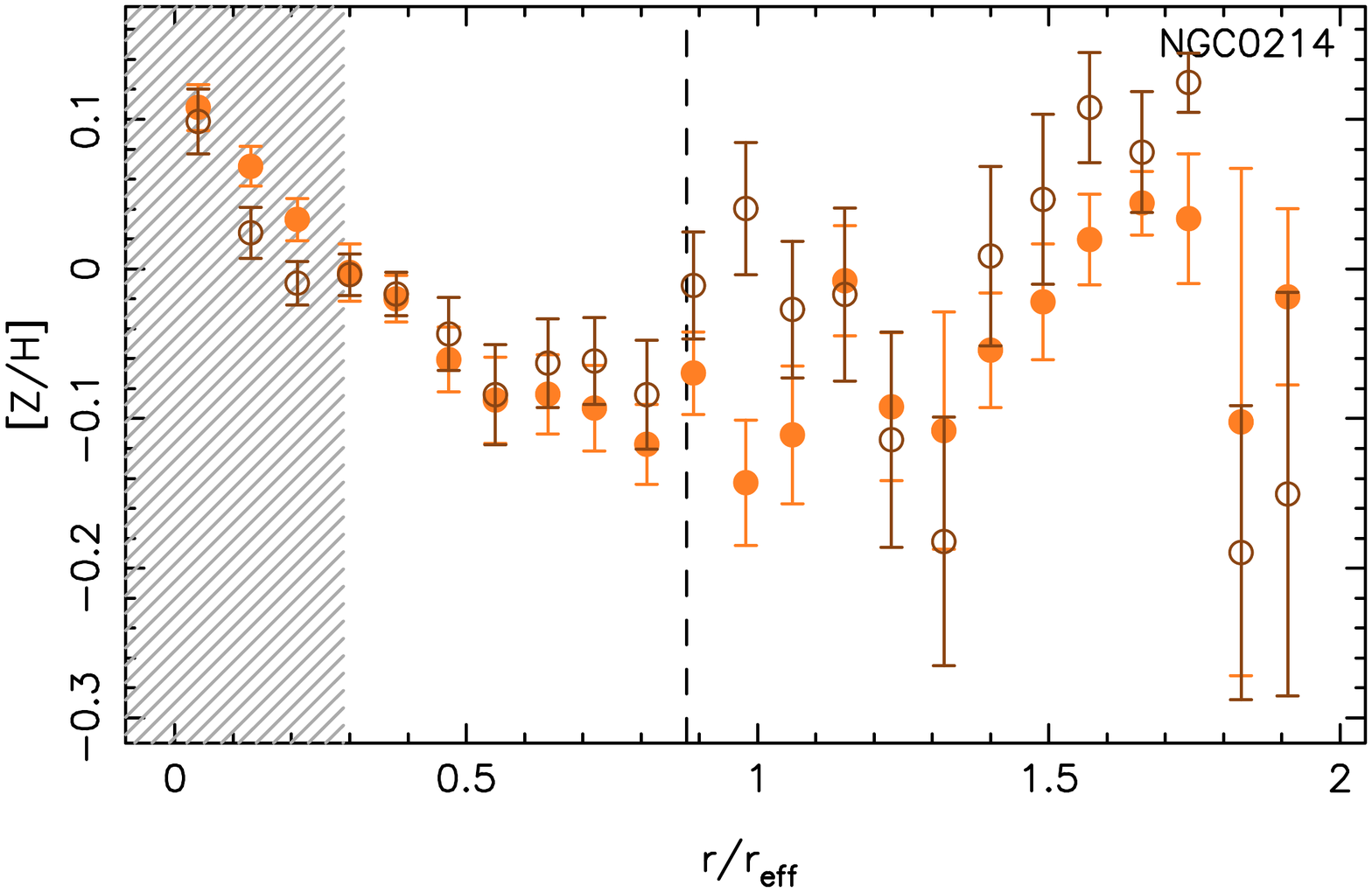}}
\resizebox{0.22\textwidth}{!}{\includegraphics[angle=0]{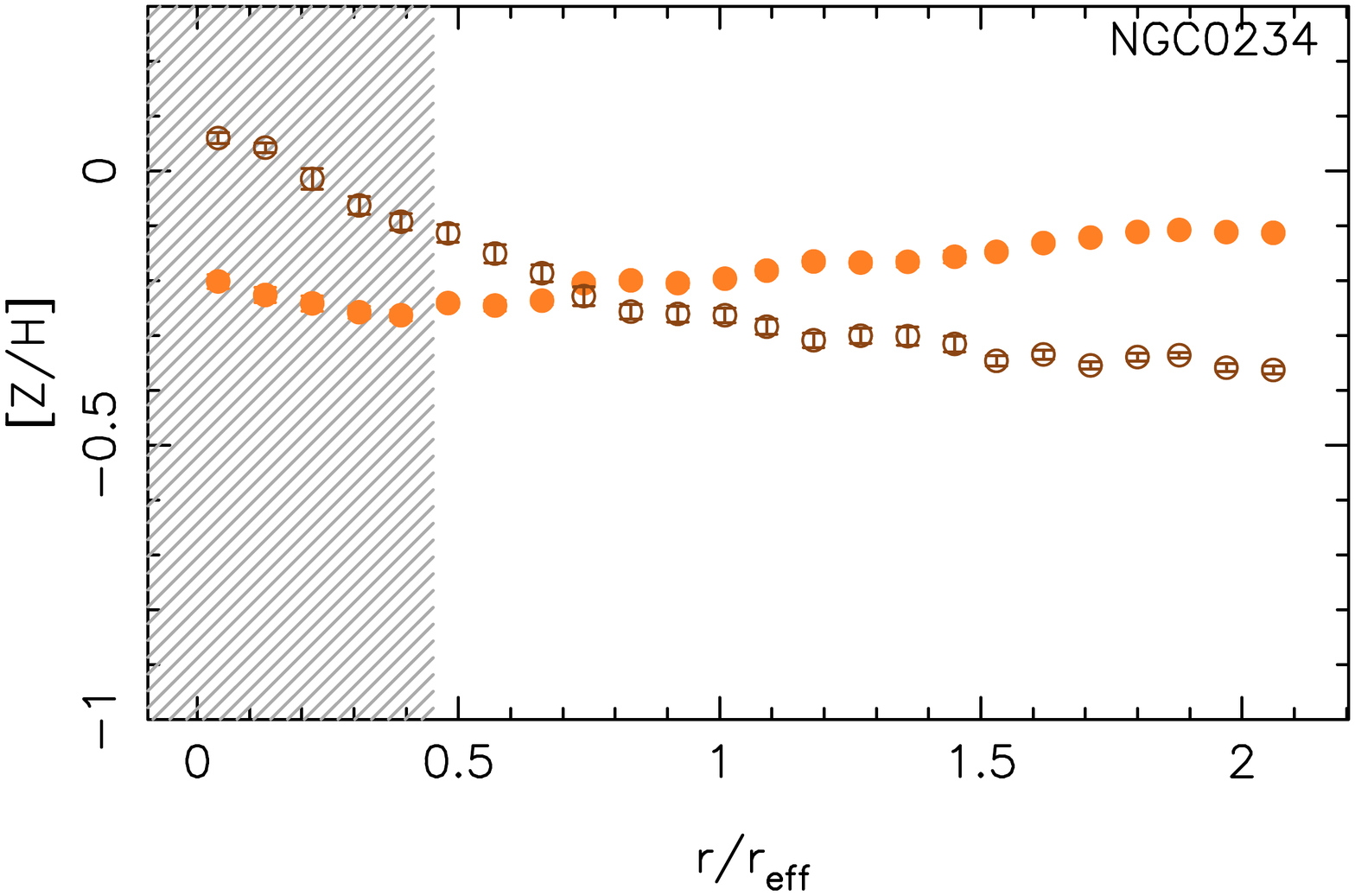}}
\resizebox{0.22\textwidth}{!}{\includegraphics[angle=0]{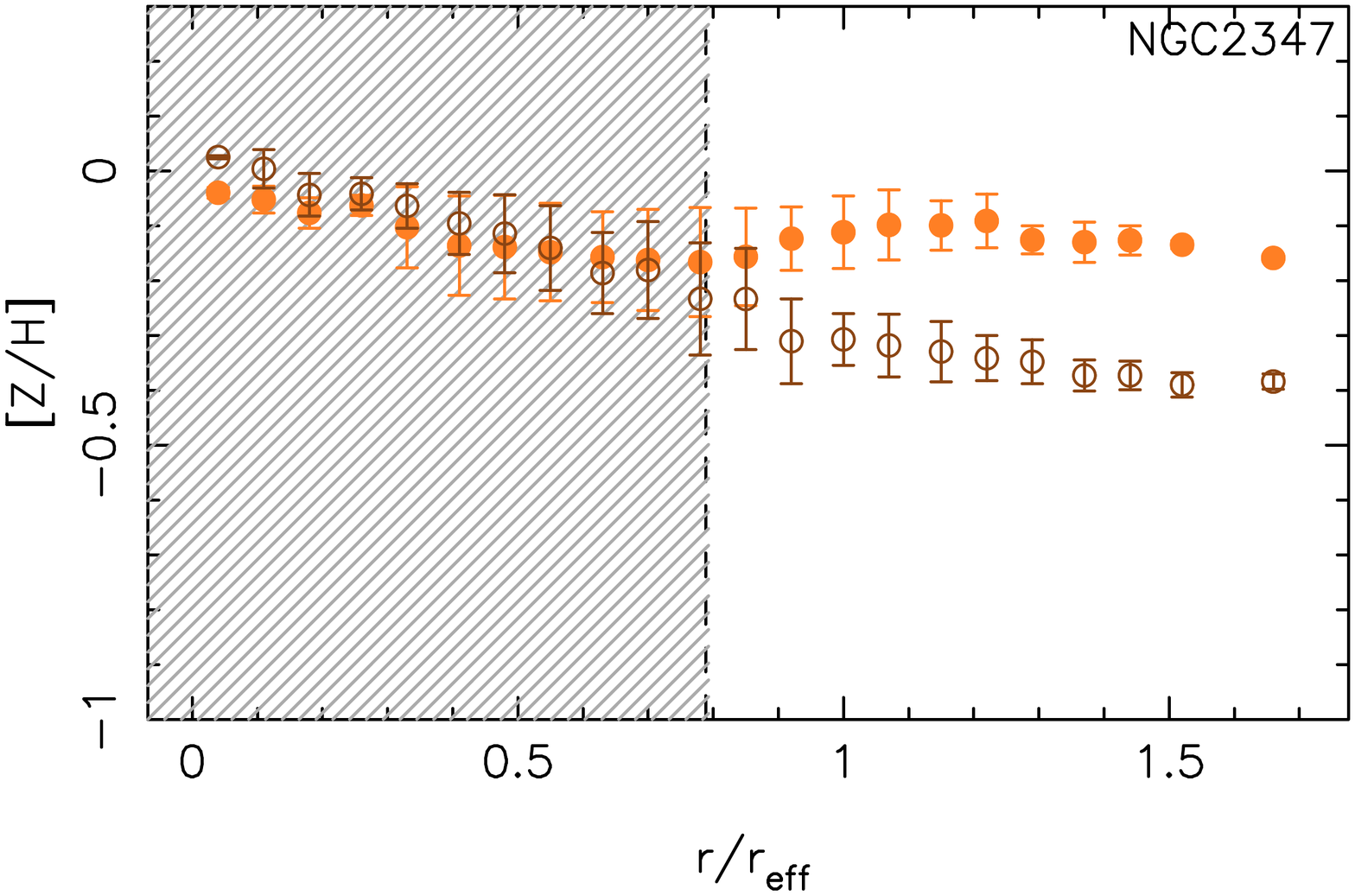}}
\resizebox{0.22\textwidth}{!}{\includegraphics[angle=0]{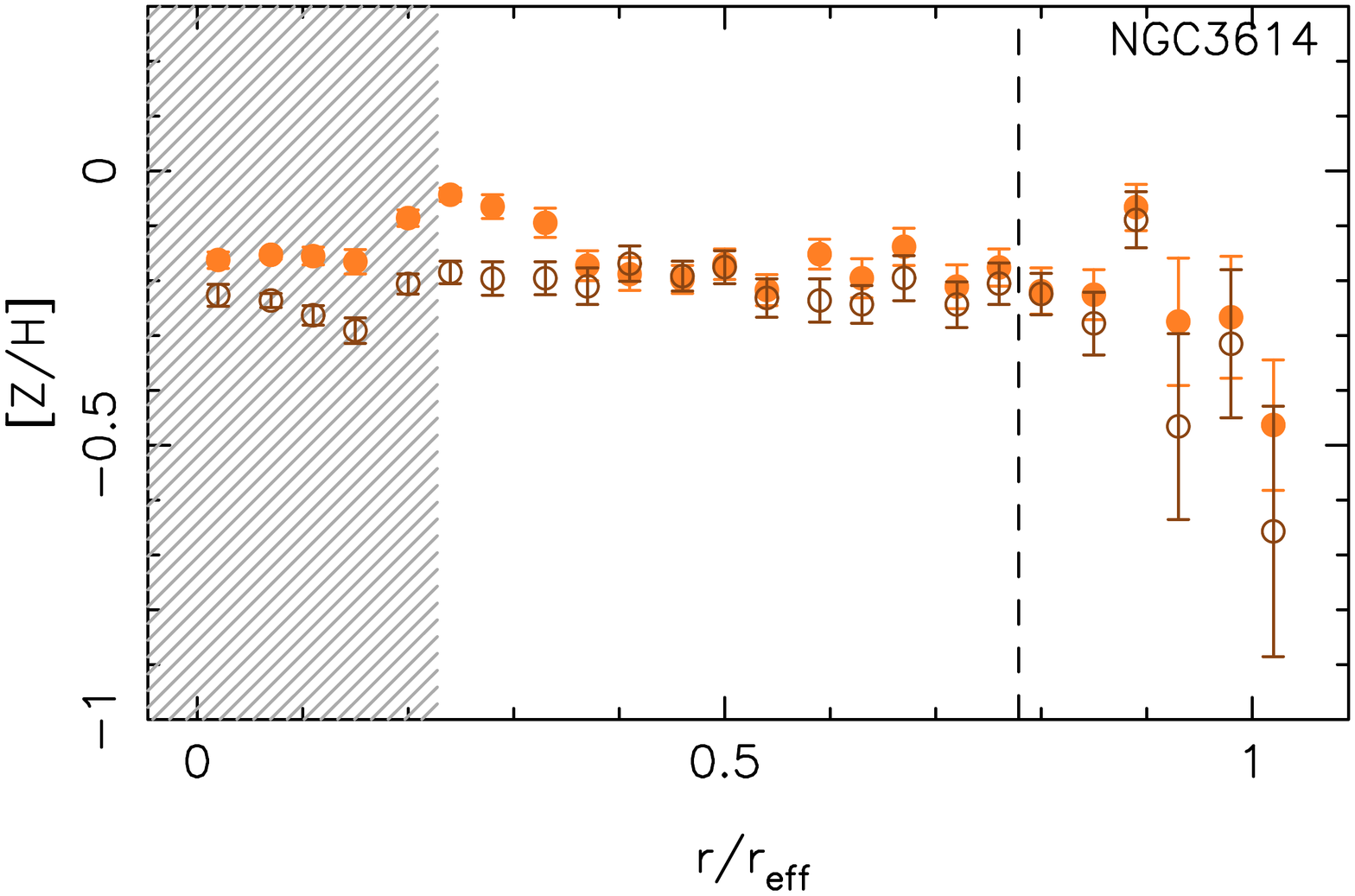}}
\resizebox{0.22\textwidth}{!}{\includegraphics[angle=0]{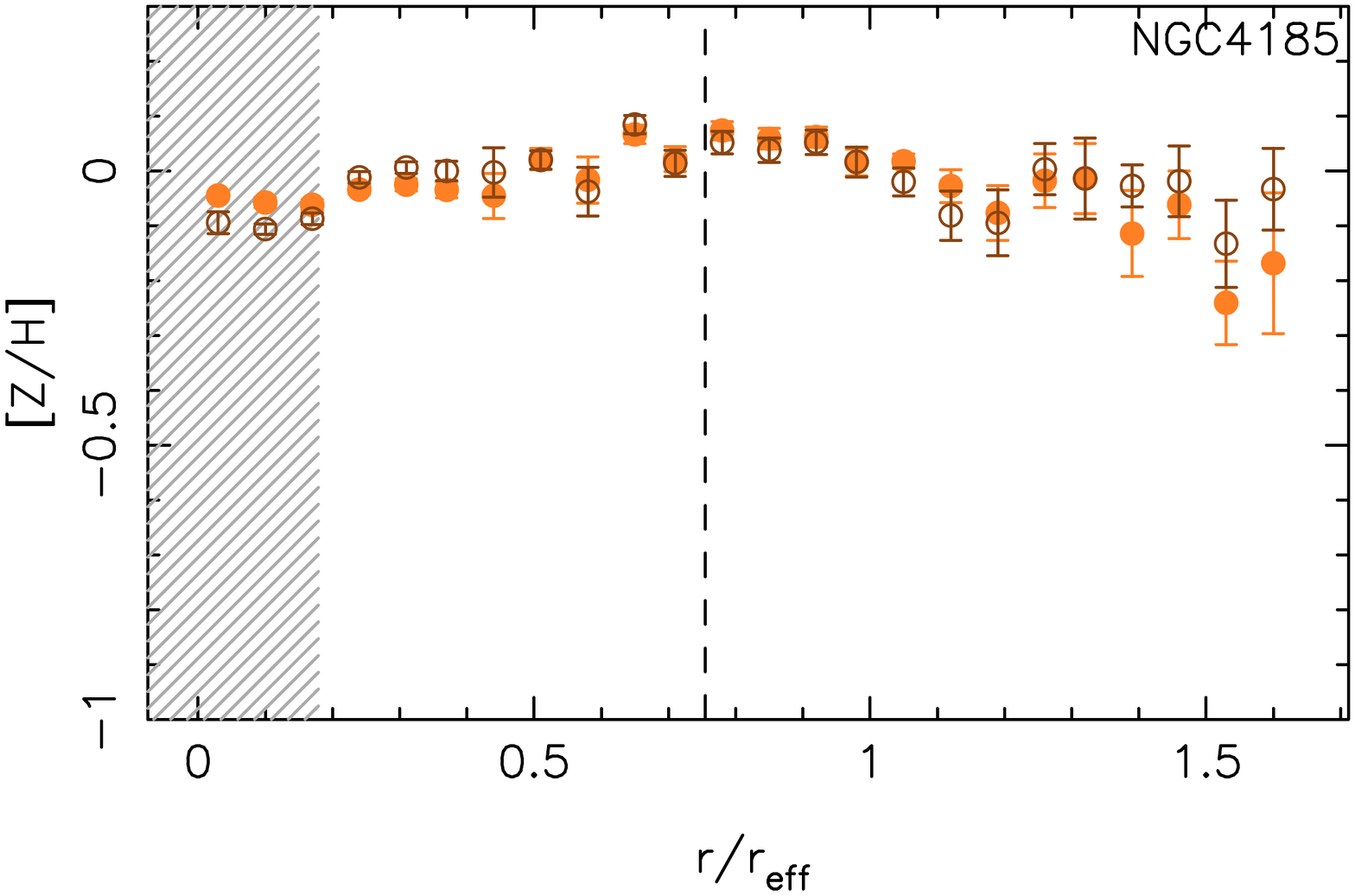}}
\resizebox{0.22\textwidth}{!}{\includegraphics[angle=0]{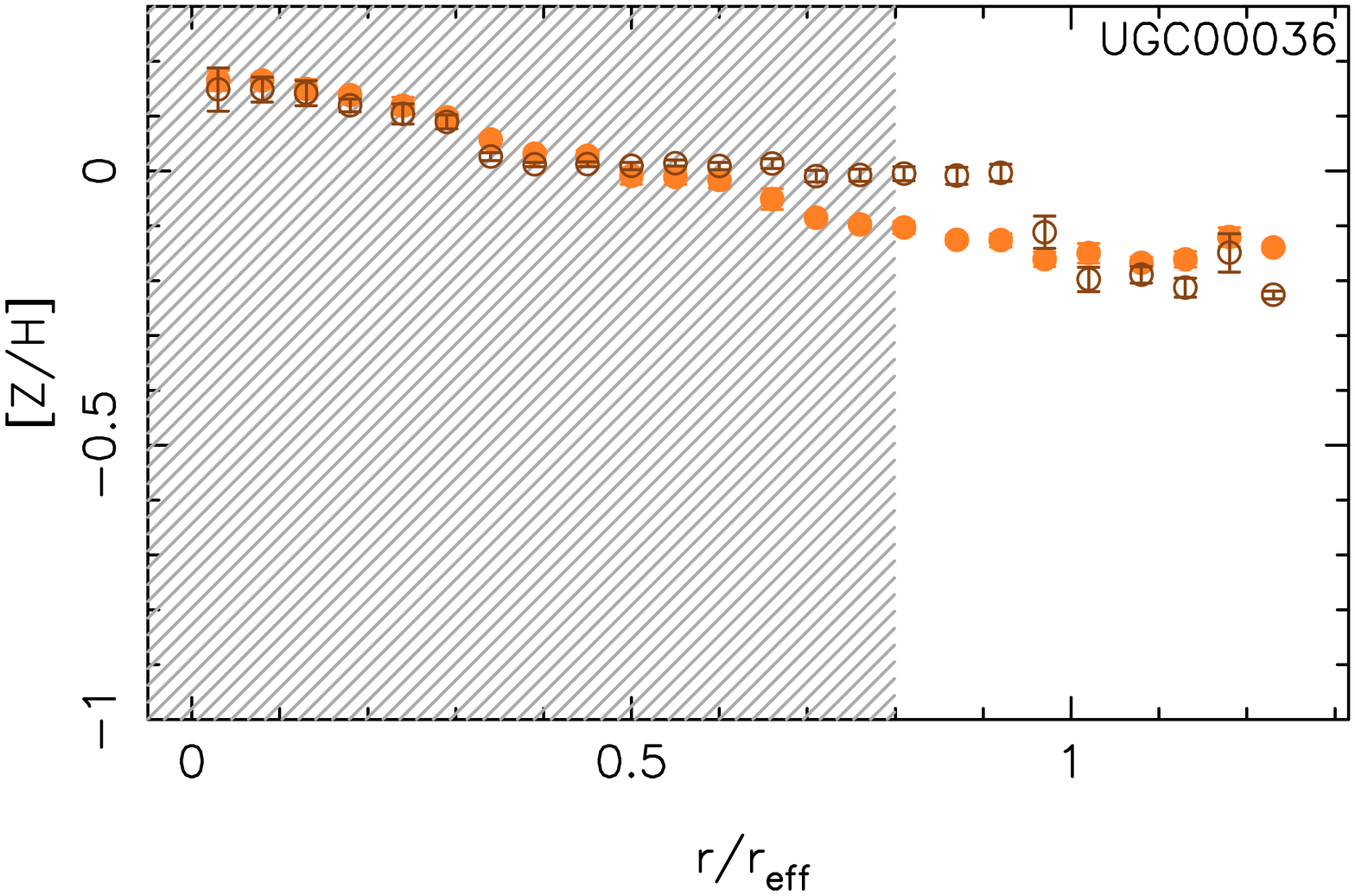}}
\resizebox{0.22\textwidth}{!}{\includegraphics[angle=0]{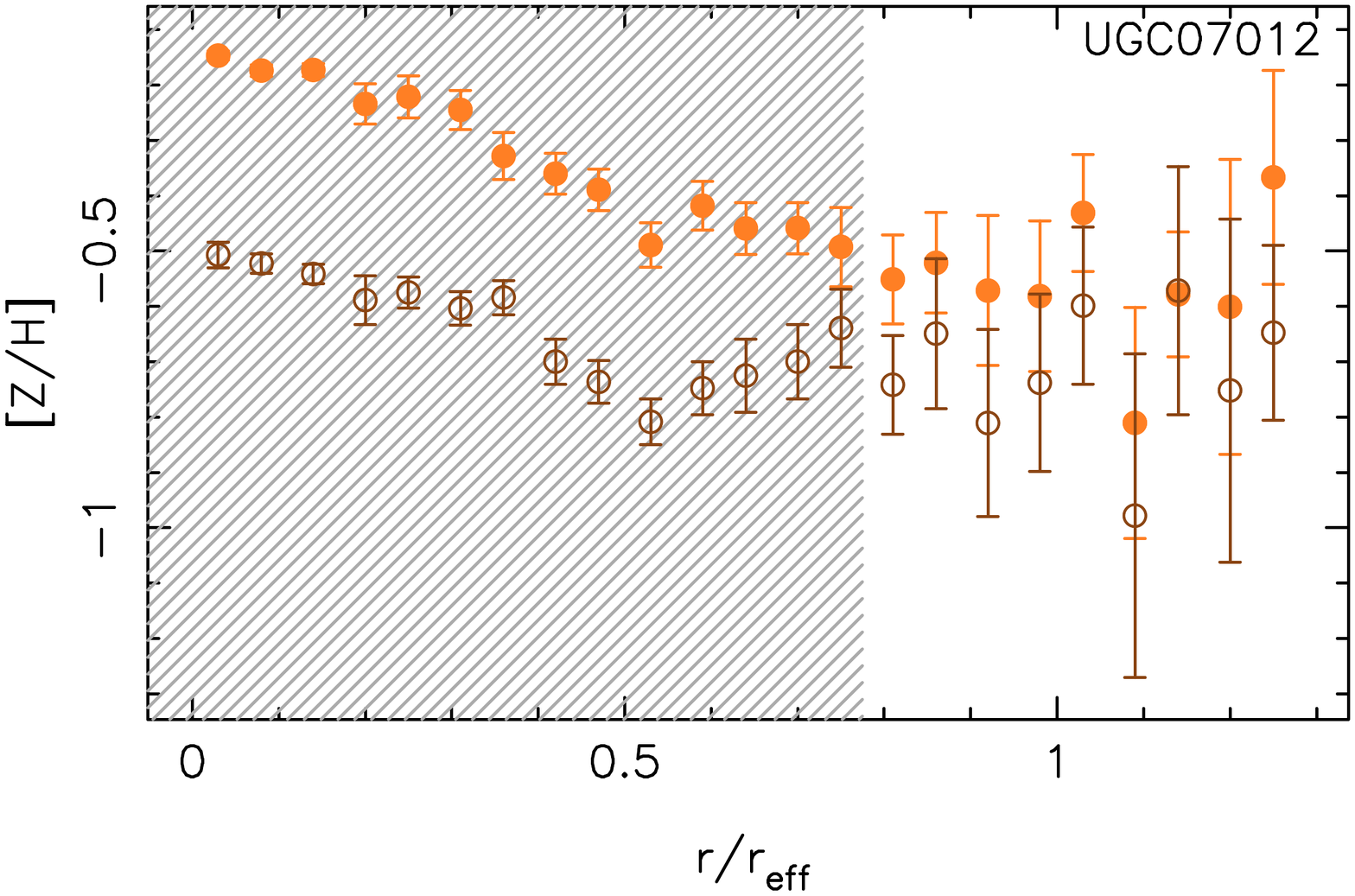}}
\caption{Light- (light yellow) and mass-weighted (dark yellow) metallicity gradient
for the galaxies morphologically classified as weakly barred \label{fig:metgrads3}}
\end{figure*}
\end{document}